\pdfoutput=1
\newif\iffull \fullfalse

\documentclass[acmsmall, screen, authorversion]{acmart}

\setcopyright{rightsretained}
\acmDOI{10.1145/3633280}
\acmYear{2024}
\copyrightyear{2024}
\acmSubmissionID{popl24main-p20-p}
\acmJournal{PACMPL}
\acmVolume{8}
\acmNumber{POPL}
\acmArticle{5}
\acmMonth{1}
\received{2023-07-11}
\received[accepted]{2023-11-07}

\usepackage{proof}
\usepackage{mathtools}
\mathtoolsset{showonlyrefs=true}

\usepackage{multirow}
\usepackage{wrapfig}

\usepackage{bussproofs}
\usepackage{stackengine}

\usepackage{tikz}
\usetikzlibrary{automata,positioning,arrows.meta}

\usepackage{thelanguage}
\newcommand*{\defeq}{\triangleq}
\newcommand{\rulename}[1]{(\textsc{#1})}
\newcommand{\infersc}[3][]{\infer[\!\!\text{\rulename{#1}}]{#2}{#3}}

\begin{document}

\title{Answer Refinement Modification: Refinement Type System for Algebraic Effects and Handlers}

\author{Fuga Kawamata}
\orcid{0009-0003-4147-9572}
\affiliation{%
  \institution{Waseda University}
  \city{Tokyo}
  \country{Japan}
}
\email{maple-river@fuji.waseda.jp}

\author{Hiroshi Unno}
\orcid{0000-0002-4225-8195}
\affiliation{%
  \institution{University of Tsukuba}
  \city{Tsukuba}
  \country{Japan}
}
\email{uhiro@cs.tsukuba.ac.jp}

\author{Taro Sekiyama}
\orcid{0000-0001-9286-230X}
\affiliation{%
  \institution{National Institute of Informatics}
  \city{Tokyo}
  \country{Japan}
}
\email{ryukilon@gmail.com}

\author{Tachio Terauchi}
\orcid{0000-0001-5305-4916}
\affiliation{%
  \institution{Waseda University}
  \city{Tokyo}
  \country{Japan}
}
\email{terauchi@waseda.jp}

\begin{abstract}
Algebraic effects and handlers are a mechanism to structure
programs with computational effects in a modular way. They
are recently gaining popularity and being adopted in practical languages,
such as OCaml.
Meanwhile, there has been substantial progress in program verification via {\em
refinement type systems}.  While a variety of
refinement type systems have been proposed, thus far there has not been a
satisfactory refinement type system for
algebraic effects and handlers.  In this paper, we fill the void by proposing a
novel refinement type system for languages with algebraic effects and handlers.
The expressivity and usefulness of algebraic effects and handlers come
from their ability to manipulate \emph{delimited continuations}, but delimited continuations also complicate programs'
control flow and make their verification harder.
To address the complexity,
we introduce a novel concept that we call {\em answer refinement modification} (ARM for
short), which allows the refinement type system to precisely track what effects occur and in what order when a program is executed, and reflect such information as modifications to the refinements in the types of delimited continuations.
We formalize our type system that supports ARM (as well as answer \emph{type} modification, or ATM) and prove its soundness. Additionally, as a proof of concept, we have
extended the refinement type system to a subset of OCaml 5 which comes with a built-in support for effect handlers,
implemented a type checking and inference algorithm for the extension,
and evaluated it on a number of benchmark programs that use algebraic effects and handlers.
The evaluation demonstrates that ARM is conceptually simple and practically useful.

Finally, a natural alternative to directly reasoning about a program with delimited continuations is to apply a {\em continuation passing style} (CPS) transformation that transforms the program to a pure program without delimited continuations.  We investigate this alternative in the paper, and show that the approach is indeed possible by proposing a novel CPS transformation for algebraic effects and handlers that enjoys bidirectional (refinement-)type-preservation.  We show that there are pros and cons with this approach, namely, while one can use an existing refinement type checking and inference algorithm that can only (directly) handle pure programs, there are issues such as
needing type annotations in source programs and
making the inferred types less informative to a user.
\end{abstract}



\maketitle

\section{Introduction} \label{sec:intro}

Algebraic effects \cite{Plotkin03} and handlers \cite{Plotkin09,Plotkin13} are a mechanism
to structure programs with computational effects
in a modular way.
Algebraic effects represent abstracted computational effects
and handlers specify their behaviors using delimited continuations.
The ability to use delimited continuations makes algebraic effects and handlers highly expressive,
allowing them to describe prominent computational effects
such as exceptions, nondeterminism, mutable states, backtracking, and
cooperative multithreading.
Additionally, algebraic effects and handlers are recently gaining quite a recognition in practice
and are adopted in popular programming languages,
such as OCaml \cite{Sivaramakrishnan21}.

Meanwhile, there has been substantial progress in program verification via {\em refinement type systems}~\cite{DBLP:conf/pldi/FreemanP91,DBLP:conf/popl/XiP99,Rondon08,Unno09,DBLP:conf/popl/Terauchi10,Bengston11,DBLP:conf/vmcai/ZhuJ13,DBLP:conf/icfp/VazouSJVJ14,DBLP:conf/popl/SwamyHKRDFBFSKZ16,DBLP:conf/pldi/VekrisCJ16,DBLP:journals/pacmpl/0001ST18,DBLP:conf/lics/Nanjo0KT18,Sekiyama23}. Such type systems allow the user to express a precise specification for a program as a type embedding logic formulas and their type checking (sometimes even type inference) \mbox{(semi-)}algorithms (semi-)automatically check whether the program conforms to the specification.  While a variety of refinement type systems have been proposed for various classes of programming languages and features, including functional languages~\cite{DBLP:conf/pldi/FreemanP91,Rondon08,DBLP:conf/icfp/VazouSJVJ14}, object-oriented languages~\cite{DBLP:conf/pldi/VekrisCJ16}, and delimited control operators~\cite{Sekiyama23}, there has not been a satisfactory refinement type system for programming languages with algebraic effects and handlers.

In this work, we propose a new refinement type system
for algebraic effects and handlers.
A challenge with the precise verification in the presence of algebraic effects and handlers is the presence of the \emph{delimited continuations}: they are the key ingredient of algebraic effects and handlers to implement a variety of computational effects, but they also complicate programs' control flow and make it difficult to statically discern what effects occur in what order.
To address this challenge,
we propose a novel concept that we call \emph{answer refinement modification} (ARM for short),
inspired by \emph{answer type modification} (ATM) employed in certain type systems for delimited control operators such as \texttt{shift} and \texttt{reset} \cite{Danvy90,Asai09}.
Similarly to ATM that can statically track how the use of delimited control operators influence the types of expressions, ARM can statically track how the use of algebraic effect operations (and the execution of the corresponding handlers) influence the refinements in the types of expressions, where the latter, as in prior refinement type systems, is used to precisely describe the \emph{values}, rather than just their ordinary (i.e., non-refinement) types, computed by the expressions.  Thus, our novel refinement type system supporting ARM can be used to precisely reason about programs with algebraic effects and handlers.

ATM and ARM are closely related: in fact, our refinement type system supports ATM, that is, our system allows the whole types and not just the refinements in them to be modified.
As far as we know, the only prior (ordinary or refinement) type system for algebraic effects and handlers that supports ATM or ARM is a recent system of \citet{Cong22}.
However, their system does not support refinement types (and so, obviously, no ARM), and moreover, even when compared as mechanisms for ordinary type systems, their ATM is less expressive than ours. We refer to Section~\ref{sec:related} for detailed comparison.

While our system supports the full ATM,
from the perspective of program verification, ARM alone, that is, allowing only modification in type refinements, is useful.
Indeed, as in other refinement-type-based approaches, our aim is verification of programs typed in \emph{ordinary} background type systems
(such as the type systems of OCaml 5 and Koka~\cite{Leijen14} that do not support ATM),
not to make more programs typable by extending the background type systems with ATM.
As a proof of concept, we have extended the refinement type system and implemented a corresponding type checking and inference algorithm for a subset of OCaml 5 which comes with a built-in support for effect handlers, and evaluated it on a number of benchmark programs that use algebraic effects and handlers.
The evaluation demonstrates that ARM is conceptually simple and practically useful.

Finally, a natural alternative to directly reasoning about a program with delimited continuations is to apply a continuation passing style (CPS) transformation that transforms the program to a pure program without delimited continuations. We investigate this alternative in the paper, and show that the approach is indeed possible by proposing a novel CPS transformation for algebraic effects and handlers that enjoys bidirectional (refinement-)type-preservation.
Bidirectional type-preservation means that an expression is well-typed in the source language if and only if
its CPS-transformed result is well-typed in the target language.
This implies that we can use existing refinement type systems without support for effect handlers to verify programs with effect handlers by applying our CPS transformation.
However, like other CPS transformations~\cite{Plotkin75,Danvy90,Appel92,Hillerstrom17,Cong18}, ours makes global changes to the program and can radically change its structure, making it difficult for the programmer to recast the type checking and inference results back to the original program. Also, the CPS transformation is type directed and requires the program to be annotated by types conforming to our new type system, albeit only needing type ``structures'' without concrete refinement predicates.
Moreover, in some cases, CPS-transformed expressions need extra parameters or higher-order predicate polymorphism to be typed as precisely as the source expressions, because the CPS transformation introduces higher-order continuation arguments.
Nonetheless, our CPS transformation is novel, and we foresee that it would provide new interesting insights, as CPS transformations often do~\cite{Danvy90}, and be a useful tool for future studies on refinement type systems and effect handlers.

Our main contributions are summarized as follows.
\begin{itemize}
    \item We show a sound refinement type system
        for algebraic effects and handlers,
        where ARM plays an important role.
    \item We have implemented the refinement type system for a subset of OCaml language
        with effect handlers, and evaluate it on a number of programs that use effect handlers.
    \item We define a bidirectionally-type-preserving CPS transformation
        which can be used to verify programs with effect handlers,
        and discuss pros and cons between direct type checking using our system and
        indirect type checking via the CPS transformation.
\end{itemize}

The rest of the paper is organized as follows.
In Section~\ref{sec:overview}, we briefly explain algebraic effects and handlers
and ATM, and then describe the motivation for ARM and our system.
Section~\ref{sec:language} presents our language.
We define its syntax, semantics and type system,
present some typing examples, and show type safety of the language.
Section~\ref{sec:impl} explains the implementation of the system.
In Section~\ref{sec:cps}, we provide the CPS transformation
and discuss pros and cons between the direct type checking via our type system
and the indirect type checking via CPS transformation.
Finally, we describe related works in Section~\ref{sec:related}
and conclude the paper in Section~\ref{sec:conclusion}.

\section{Overview} \label{sec:overview}

We briefly overview algebraic effects and handlers, ATM, and ARM.

\subsection{Algebraic Effects and Handlers} \label{sec:overview/algeff}

\newcommand{\rmmax}{\mathrm{max}}

Algebraic effects and handlers enable users to define their own effects in a
modular way.  The modularity stems from separating the use of effects from their
implementations: effects are performed via {\em operations} and implemented via
{\em effect handlers} (or handlers for short).  For example, consider the
following program where $h_{\op[d]} \defeq \{ x_r \mapsto x_r, \ 
\op[decide](x, k) \mapsto \rmmax~(k~\exptrue)~(k~\expfalse) \}$:
\begin{align}
    \expwith{
		h_{\op[d]}
	}{
        \explet{a}{\expif{\op[decide]~()}{10}{20}}{
            \explet{b}{\expif{\op[decide]~()}{1}{2}}{
                a - b
            }
        }
    }
\end{align}
It calls an operation $\op[decide]$, which takes
the unit value $()$ and returns a Boolean value, to choose one of two integer
values and then calculates the difference between the chosen values.
Because operation calls invoke effects in algebraic effects, the operations work as interfaces of the effects.

An implementation of an effect is given by an effect handler.
The program installs the handler $h_{\op[d]}$ for $\op[decide]$ using the
handling construct.  In general, a handling construct takes the form
$\expwith{h}{e}$, which means that a handler $h$ defines interpretations of
operations performed during the evaluation of the expression $e$; we call the
expression $e$ a \emph{handled expression}.
A handler consists of a single \emph{return clause} and zero or more
\emph{operation clauses}.
A return clause takes the form $x_r \mapsto e_r$, which determines the value of
the handling construct by evaluating expression $e_r$ with variable $x_r$
that denotes the value of the handled expression.
In the example, because the return clause is $x_r \mapsto x_r$, the handling
construct simply returns the value of the handled expression.
An operation clause takes the form $\op[op](x, k) \mapsto e$. It defines the
interpretation of the operation $\op[op]$ to be expression $e$ with variable $x$
that denotes the arguments to the operation. When the handled expression calls
the operation $\op[op]$, the remaining computation up to the handling construct
is suspended and instead the body $e$ of the operation clause evaluates.
Therefore, effect handlers behave like exception handlers by regarding operation
calls as raising exceptions.  However, effect handlers are equipped with the
additional ability to resume the suspended computation.
The suspended remaining computation, called a \emph{delimited
continuation}, is functionalized, and the body $e$ of the operation clause can
refer to it via the variable $k$.

Let us take a closer look at the behavior of the above example.
Because the handled expression starts with the call to $\op[decide]$,
the operation clause for $\op[decide]$ given by $h_{\op[d]}$ evaluates.
The delimited continuation $K$ of the first call to $\op[decide]$ is
\[
 \expwith{ h_{\op[d]} }{
        (\explet{a}{\expif{\hole}{10}{20}}{
            \; \explet{b}{\expif{\op[decide]~()}{1}{2}}{
                \; a - b
            }
        })
    }
\]
where $\hole$ denotes the hole of the continuation.
We write $K[e]$ for the expression obtained by filling the hole in $K$ with expression $e$.
Then, the functional representation of the delimited continuation $K$ takes the
form $\lambda y. K[y]$, and it is substituted for $k$ in the body of the
operation clause.
Namely, the handling construct evaluates to $\rmmax~(v~\exptrue)~(v~\expfalse)$
where $v = \lambda y. K[y]$.
Note that the variable $x$ of the operation clause for $\op[decide]$ is replaced
by the unit value $()$, but it is not referenced.
The first argument $v~\exptrue$ to $\rmmax$
reduces to $K[\exptrue]$, that is,
\[
    \expwith{ h_{\op[d]} }{
        (\explet{a}{\expif{\colorbox{gray!50}{$\exptrue$}}{10}{20}}{
            \explet{b}{\expif{\op[decide]~()}{1}{2}}{
              \; a - b
            }
        })
    }
\]
(the grayed part represents the value by which the hole in $K$ is replaced).
Therefore, the expression $v~\exptrue$ evaluates to
$
    \expwith{ h_{\op[d]} }{
        (
            \explet{b}{\expif{\op[decide]~()}{1}{2}}{
                10 - b
            }
        )
    }
$~.
Again, $\op[decide]$ is called and the continuation
$
    K' \defeq \expwith{ h_{\op[d]} }{
		(\explet{b}{\expif{\hole}{1}{2}}{
			10 - b
		})
	}
$
is captured.
Then, the operation clause for $\op[decide]$ evaluates
after substituting $\lambda y. K'[y]$ for $k$.
The expression $(\lambda y. K'[y])~\exptrue$
evaluates to $K'[\exptrue]$, that is,
$
 \expwith{ h_{\op[d]} }{
    (\explet{b}{\expif{\colorbox{gray!50}{$\exptrue$}}{1}{2}}{10 - b})
 }
$
and then to
$
 \expwith{ h_{\op[d]} }{ 9 }
$~.
Here, the handled expression is a value.  Therefore, the return clause in the
handler evaluates after substituting the value $9$ for variable $x_r$. Because
the return clause in $h_{\op[d]}$ just returns $x_r$, the evaluation of
$(\lambda y. K'[y])~\exptrue$ results in $9$. Similarly, $(\lambda
y. K'[y])~\expfalse$ evaluates to $8$ (which is the result of binding $b$ to
$2$). Therefore, $\rmmax~((\lambda y. K'[y])~\exptrue)~((\lambda
y. K'[y])~\expfalse)$ evaluates to $\rmmax~9~8$ and then to $9$. In
a similar way, $v~\expfalse$ calculates $\rmmax~(20-1)~(20-2)$, that is,
evaluates to $19$, because $a$ is bound to $20$ and $b$ is bound to each of $1$
and $2$ depending on the result of the second invocation of $\op[decide]$.
Finally, the entire program evaluates to $19$, which is the result of
$\rmmax~(v~\exptrue)~(v~\expfalse)$, that is, $\rmmax~9~19$.

\subsection{Answer Type Modification and Answer Refinement Modification} \label{sec:overview/atm}

An \emph{answer type} is the type of the closest enclosing delimiter,
or the return type of a delimited continuation.
In the setting of algebraic effects and handlers,
delimiters are handling constructs.
For example, consider the following expression:
\begin{align}
 &\explet{x}{
		\expwith{
			\{ x_r \mapsto x_r, \ 
         \op((), k) \mapsto k~0 < k~1 \}
		}{1 + \op~()}
	}{c} ~.
\end{align}
The delimited continuation of $\op~()$ is
$
    K'' \defeq
	\expwith{
		\{ x_r \mapsto x_r, \ 
			\op((), k) \mapsto k~0 < k~1 \}
	}{1 + \hole}
$~.
At first glance, the answer type of $\op~()$ seems to be the integer type $\tyint$
since the handled computation in the continuation returns the integer $1+n$ for an integer $n$ given to fill the hole, and the return clause returns given values as they are.
In other words, from the perspective of $\op~()$,
the handling construct seems to give an integer value to the outer context $\explet{x}{\hole}{c}$.
However, after the operation call, the entire expression evaluates to
$
	\explet{x}{
		v''~0 < v''~1
	}{c}
$ where $v'' \defeq \lambda y. K''[y]$.
Now the handling construct becomes the expression $v''~0 < v''~1$, which gives a Boolean value to the outer context.
That is, the answer type changes to the Boolean type $\tybool$.
\emph{Answer type modification} (ATM) is a mechanism to track this dynamic change on answer types.

ATM is not supported in existing type systems for effect handlers~\cite{Plotkin13,Brady13,Bauer13,Kammar13,Bauer15,Leijen17,Lindley17}, with the exception of the one recently proposed by \citet{Cong22} (see Section~\ref{sec:related} for comparison with their work).
Such type systems require the answer types before and after an operation call to be unified (and so the example above will be rejected as ill-typed).
Nonetheless, useful programming with effect handlers is still possible without ATM (which is why they are implemented in popular languages like OCaml without ATM).\footnote{One could also argue that the absence of ATM is natural for algebraic effects and handlers because they are designed after concepts from universal algebra~\cite{DBLP:conf/fossacs/PlotkinP01,DBLP:journals/corr/abs-1807-05923}, and there, (algebraic) operations are usually expected to preserve types.}
For instance,
the program in Section~\ref{sec:overview/algeff} is well-typed
in existing (non-refinement) type systems for algebraic effects and handlers without ATM,
since the return type of the continuation $k$ in the $\op[decide]$ clause
(i.e., the answer type before the execution) is $\tyint$
and the return type of the $\op[decide]$ clause
(i.e., the answer type after the execution) is also $\tyint$.

However, even if answer types are not modified,
{\em actual values returned by delimited continuations usually change}.
Let us see the program in Section~\ref{sec:overview/algeff} again.
Focus on the first call to $\op[decide]$.
When this is called, the operation clause receives the continuation $v = \lambda y. K[y]$,
which returns $9$ if applied to $\exptrue$ and returns $19$ if applied to $\expfalse$,
as described previously.
Therefore, $v$ can be assigned the refinement type
$(y: \tybool) \rarr \tyrfn{z}{\tyint}{z = (\ternaryif{y}{9}{19})}$,
and thus the precise answer type before the execution is
$\tyrfn{z}{\tyint}{z = (\ternaryif{y}{9}{19})}$
where $y$ is the Boolean value passed to the continuation.
On the other hand, the clause for $\op[decide]$
returns integer $19$. Thus, the precise answer type after the operation call is $\tyrfn{z}{\tyint}{z = 19}$.
Now the refinement in the answer type becomes different before and after the operation call.
The same phenomenon happens in the second call to $\op[decide]$.
When the second call evaluates, the handler receives the continuation $\lambda y. K'[y]$.
It returns $a - 1$ if applied to $\exptrue$ and returns $a - 2$ if applied to $\expfalse$
(where $a$ is either $10$ or $20$ depending on the result of the first call to $\op[decide]$).
Thus, the answer type before the execution is $\tyrfn{z}{\tyint}{z = (\ternaryif{y}{a - 1}{a - 2})}$.
In contrast, the return value of the clause for $\op[decide]$ is $\rmmax~(a - 1)~(a - 2) = a - 1$,
so the answer type after the execution is $\tyrfn{z}{\tyint}{z = a - 1}$.
Here again, the refinement in the answer type changed by the operation call.
We call this change \emph{answer refinement modification} (ARM). Armed with ARM (pun intened), the refinement type system that we propose in this paper is able to assign the precise refinement type $\tyrfn{z}{\tyint}{z = 19}$ to the program, and more generally, the type $\tyrfn{z}{\tyint}{z = v - x}$ when the constants $10$, $20$, $1$, and $2$ are replaced by variables $u$, $v$, $x$, and $y$ respectively with the assumption $u \leq v \wedge x \leq y$ (such an assumption on free variables can be given by refinement types in the top-level type environment).
The example demonstrates that ARM is useful for precisely reasoning about programs with algebraic effects and handlers in refinement type systems.  Indeed, without ARM, the most precise refinement type that a type system could assign to the example would be $\tyrfn{z}{\tyint}{z \in \{8,9,18,19\}}$.

As another illuminating example, we show that ARM provides a new approach to the classic \emph{strong update} problem~\cite{DBLP:conf/pldi/FosterTA02}.  It is well known that algebraic effects and handlers can implement mutable references by operations $\op[set]$ and $\op[get]$, that respectively destructively updates and reads a mutable reference, and a handler that implements the operations by state-passing (see, e.g., \cite{Pretnar15}).
On programs with such a standard implementation of mutable references by algebraic effects and handlers, our refinement type system is able to reason flow-sensitively and derive refinement types that cannot be obtained with ordinary flow-insensitive reasoning.
For instance, consider the following program 
where $h \defeq \{
    x_r \mapsto \lambda s. x_r, \ 
    \op[set](x, k) \mapsto \lambda s.k~()~x, \ 
    \op[get](x, k) \mapsto \lambda s. k~s~s \}$:
\begin{gather}
    (\expwith{h}{
        \ (\op[set]~3;\,
        \explet{n}{\op[get]~()}{
            \op[set]~5;\,
            \explet{m}{\op[get]~()}{
                n + m
            }
        })
    })~0
\end{gather}
Thanks to ARM, our type system can give the program the most precise type $\tyrfn{z}{\tyint}{z = 8}$,
which would not be possible in a type system without ARM as it would conflate the two calls to $\op[set]$ and fail to reason that the first $\op[get]~()$ returns $3$ whereas the second $\op[get]~()$ returns $5$.  Roughly, ARM accomplishes the flow-sensitive reasoning about the changes in the state by tracking changes in the refinements in the answer types, albeit in a \emph{backward} fashion as shown in Section~\ref{sec:language/exmaples}.

Using this ability of ARM, we can also verify that effectful operations are used
in a specific order. For example, consider operations $\op[open]$,
$\op[close]$, $\op[read]$, and $\op[write]$ for file manipulation being
implemented using effect handlers.
The use of these operations should conform to the regular scheme
$(\op[open] \ (\op[read] \mid \op[write])^\ast \ \op[close])^\ast$.
Our refinement type system can check if a program meets this requirement. For instance, consider the
following recursive function:
\begin{gather}
 \expfun{x}{~\expwhile{\star}{\op[open]~x;~\expwhile{\star}{\explet{y}{\op[read]()}{\op[write]~(y\texttt{\textasciicircum"X"})}};~\op[close]~()}}
\end{gather}
where $\expwhile{\star}{c}$ loops computation $c$ and terminates
nondeterministically, and the binary operation $(\texttt{\textasciicircum})$
concatenates given strings (operation $\op[read]$ is supposed to return a string).\footnote{For simplicity, we assume that the clause of $\op[open]$
creates an object for a specified file and stores it in a reference implemented
by an effect handler, and the clauses of the other operations refer to the stored object to
manipulate the file.}
The function repeats opening the specified file $x$ and closing it after reading
from and writing to the file zero or more times.
Thus, this function follows the discipline of the file manipulation
operations.
We will show in Section~\ref{sec:language/exmaples} how ARM enables us to check
it formally and detect the invalid use of the operations if any.

\section{Language} \label{sec:language}

This section presents our language with algebraic effects and handlers.
The semantics is formalized using evaluation contexts like in \citet{Leijen17},
and the type system is a novel refinement type system with ARM (and ATM).

\subsection{Syntax and Semantics} \label{sec:language/syntax-semantics}

The upper half of Figure~\ref{fig:syntax-eval} shows the syntax of our language.
It indicates that expressions are split into values, ranged over by $v$, and computations, ranged over by $c$,
as in the fine-grain call-by-value style of \citet{Levy03}.
Values, which are effect-free expressions in a canonical form,
consist of variables $x$, primitive values $p$,
and (recursive) functions $\exprec{f}{x}{c}$
where variable $f$ denotes the function itself for recursive calls in the body $c$.
If $f$ does not occur in $c$, we simply write $\lambda x. c$.
Computations, which are possibly effectful expressions,
consist of six kinds of constructs.
A value-return $\expret{v}$ lifts a value $v$ to a computation.
An operation call $\op~v$ performs the operation $\op$ with the
argument $v$.
A function application $v_1~v_2$, conditional branch $\expif{v}{c_1}{c_2}$,
and let-expression $\explet{x}{c_1}{c_2}$ are standard.
Note that functions, arguments, and conditional expressions are
restricted to values, but this does not reduce expressivity because, e.g.,
a conditional branch $\expif{c}{c_1}{c_2}$ can be expressed as
$\explet{x}{c}{\expif{x}{c_1}{c_2}}$ using a fresh variable $x$.
A handling construct $\expwith{h}{c}$ handles operations performed during the
evaluation of the handled computation $c$ using the clauses in the handler $h$.
A handler $\{ \expret{x_r} \mapsto c_r, \repi{\op_i(x_i, k_i) \mapsto c_i} \}$
has a return clause $\expret{x_r} \mapsto c_r$ where the variable $x_r$ denotes
the value of the handled computation $c$, and an operation clause
$\op_i(x_i, k_i) \mapsto c_i$ for each operation $\op_i$ where the variables
$x_i$ and $k_i$ denote the argument to $\op_i$ and the continuation from the
invocation of $\op_i$, respectively.
The notions of free variables and substitution are defined as usual.
We write $c[v/x]$ for the computation obtained by substituting the value $v$ for
the variable $x$ in the computation $c$.  We use similar notation to substitute
values for variables in types and substitute types for type variables.

\newif\ifshowsemantics \showsemanticstrue
\ifshowsemantics{
\begin{figure}
 \raggedright
 \textbf{Syntax}
 \[
 \begin{array}{r@{\ \ }c@{\ \ }lr@{\ \ }c@{\ \ }lr@{\ \ }c@{\ \ }l}
   p &::=& \exptrue \mid \expfalse \mid \cdots &
   v &::=& x \mid p \mid \exprec{f}{x}{c} &
   K &::=& [\ ] \mid \explet{x}{K}{c} \\
   c &::=& \multicolumn{7}{@{}l}{\expret{v} \mid \op~v \mid v_1~v_2
   \mid \expif{v}{c_1}{c_2} \mid \explet{x}{c_1}{c_2}
   \mid \expwith{h}{c}
   } \\
   h &::=& \multicolumn{7}{@{}l}{ \{ \expret{x_r} \mapsto c_r, \repi{\op_i(x_i, k_i) \mapsto c_i} \} } \\
  \end{array}
 \]
    \textbf{Evaluation rules} \quad \fbox{$c \eval c'$}
    \begin{gather}
        \infersc[E-Let]{\explet{x}{c_1}{c_2} \eval \explet{x}{c'_1}{c_2}}
        {c_1 \eval c'_1}
        \quad
        \infersc[E-LetRet]{\explet{x}{\expret{v}}{c_2} \eval c_2[v/x]}
        {}
        \\[1.5ex]
        \infersc[E-IfT]{\expif{\exptrue}{c_1}{c_2} \eval c_1}
        {}
        \quad
        \infersc[E-IfF]{\expif{\expfalse}{c_1}{c_2} \eval c_2}
        {}
        \\[1.5ex]
        \infersc[E-App]{(\exprec{f}{x}{c})~v \eval c[v/x][(\exprec{f}{x}{c})/f]}
        {}
        \quad
        \infersc[E-Prim]{p~v \eval \zeta(p, v)}
        {}
        \\[1.5ex]
        \text{below, let } h = \{ \expret{x_r} \mapsto c_r, \repi{\op_i(x_i, k_i) \mapsto c_i} \}
        \\
        \infersc[E-Hndl]{\expwith{h}{c} \eval \expwith{h}{c'}}
        {c \eval c'}
        \\[1.5ex]
        \infersc[E-HndlRet]{\expwith{h}{\expret{v}} \eval c_r[v/x_r]}
        {}
        \\[1.5ex]
        \infersc[E-HndlOp]{\expwith{h}{K[\op_i~v]} \eval c_i[v/x_i][(\expfun{y}{\expwith{h}{K[\expret{y}]}})/k_i]}
        {}
    \end{gather}
    \caption{Syntax and evaluation rules.}
    \label{fig:syntax-eval}
\end{figure}

The semantics of the language is defined by the evaluation relation $\eval$,
which is the smallest binary relation over computations satisfying the
evaluation rules in the lower half of Figure~\ref{fig:syntax-eval}.
The evaluation of a let-expression $\explet{x}{c_1}{c_2}$ begins by
evaluating the computation $c_1$.
When $c_1$ returns a value, the computation $c_2$ evaluates after substituting
the return value for $x$.
The evaluation rules for conditional branching and function application are
standard.
The result of applying a primitive value relies on the metafunction $\zeta$,
which maps pairs of a primitive value and an argument value to computations.
For a handling construct $\expwith{h}{c}$, the handled computation $c$
evaluates first.
When $c$ returns a value, the body of the return clause in the handler $h$
evaluates with the return value.
If the evaluation of $c$ encounters an operation call $\op_i~v$,
its delimited continuation, which is represented as a pure evaluation context $K$ defined in Figure~\ref{fig:syntax-eval}, is captured.
Then, the body $c_i$
of the operation clause $\op_i(x_i, k_i) \mapsto c_i$ for $\op_i$ in the handler
$h$ evaluates after substituting the argument $v$ and the function
$\lambda y. \expwith{h}{K[\expret{y}]}$ for variables $x_i$ and $k_i$, respectively.
Note that the function
substituted for $k_i$ wraps
the delimited continuation $K[\expret{y}]$ by the handling construct
with the handler $h$.
It means that the operation calls in $K[\expret{y}]$ are handled by the handler $h$.
Our semantics assumes that the handler $h$
provides operation clauses for all the operations performed by the handled
computation $c$.
}
\else{
The semantics of the language is defined by the evaluation relation $\eval$,
which is a binary relation over computations.
It is mostly identical to the semantics given by \citet{Pretnar15}; we refer to the supplementary material for its definition.
The only difference is that Pretnar's semantics allows handlers
that do not involve clauses for operations performed by a handled
computation---a call to such an operation is forwarded to outer
handlers---whereas our semantics assumes handlers to contain clauses for all such operations.
}\fi
Our type system ensures that this assumption holds on any well-typed computations.
However, our language can also implement the forwarding semantics by encoding: given a handler
that does not contain an operation clause for $\op$, we add to the handler
an operation clause $\op(x,k) \mapsto \explet{y}{\op~x}{k~y}$.\footnote{We employ the semantics without forwarding in the body of the paper to simplify the typing rule for handling constructs. The supplementary material shows an extended typing rule for handling constructs that natively supports forwarding.}

\subsection{Type System} \label{sec:language/type-system}

\begin{wrapfigure}[11]{R}{0.65\textwidth}
    \vspace*{-4ex}
    $\begin{array}{rc@{\ \ }c@{\ }l@{\qquad}rc@{\ \ }c@{\ \ }l}
        \text{term} &
        t &::=& x \mid \ldots
        &
        \text{formula} &
        \phi &::=& A(\rep{t}) \mid \ldots
        \\
        \text{predicate} &
        A &::=& X \mid \ldots
        &
        \text{base type} &
        B &::=& \tybool \mid \ldots
    \end{array}$
    $\begin{array}{rc@{\ \ }c@{\ }l}
        \text{value type} &
        T &::=& \tyrfn{x}{B}{\phi} \mid (x: T) \rarr C
        \\
        \text{computation type} &
        C &::=& \tycomp{\Sigma}{T}{S}
        \\
        \text{operation signature} &
        \Sigma &::=& \{ \repi{\op_i : \forall \rep{X_i: \rep{B}_i}. F_i} \}
        \\
        &
        F &::=& (x: T_1) \rarr ((y: T_2) \rarr C_1) \rarr C_2
        \\
        \text{control effect} &
        S &::=& \square \mid \tyctl{x}{C_1}{C_2}
        \\
        \text{typing context} &
        \Gamma &::=& \emptyset \mid \Gamma, x: T \mid \Gamma, X: \rep{B}
    \end{array}$
    \caption{Type syntax.}
    \label{fig:type-syntax}
\end{wrapfigure}

Figure~\ref{fig:type-syntax} shows the syntax of types.
As in prior refinement type systems \cite{Bengston11, Rondon08, Unno09}, our type system allows a type
specification for values of base types, ranged over by $B$, such as $\tybool$
and $\tyint$, to be refined using logic formulas, ranged over by $\phi$.
Our type system is parameterized over a logic. We assume that the logic is a
predicate logic where: terms, denoted by $t$, include variables $x$;
predicates, denoted by $A$, include predicate variables $X$; and each primitive
value $p$ can be represented as a term.
Throughout the paper, we use the over-tilde notation to denote a sequence of
entities.  For example, $\rep{t}$ represents a sequence $t_1, \cdots, t_n$ of
some terms $t_1, \ldots, t_n$, and then $A(\rep{t})$ represents a formula
$A(t_1, \cdots, t_n)$.
We also assume that base types include at least the Boolean type $\tybool$.

Types consist of value and computation types, which are assigned to values and
computations, respectively.
A value type, denoted by $T$, is either a refinement type $\tyrfn{x}{B}{\phi}$,
which is assigned to a value $v$ of base type $B$ such that the formula
$\phi[v/x]$ is true, or a dependent function type $(x: T) \rarr C$, which is
assigned to a function that, given an argument $v$ of the type $T$, performs
the computation specified by the type $C[v/x]$.
We abbreviate $\tyfun{x}{T}{C}$ as $\tyfunshort{T}{C}$ if $x$ does not occur in $C$, and $\tyrfn{z}{B}{\exptrue}$ as $B$.

A computation type is formed by three components: an operation signature,
which specifies operations that a computation may perform;
a value type, which specifies the value that the computation returns
if any; and a control effect, which specifies how the computation modifies the
answer type via operation call.

Control effects, denoted by $S$, are inspired by the formalism of \citet{Sekiyama23} who extended control effects in simple typing~\cite{Materzok11} to dependent typing.
A control effect is either pure or impure.
The pure control effect $\square$ means that a computation calls no operation.
An impure control effect is given to a computation that may perform operations,
specifying how the execution of the computation modifies its answer type.
Impure control effects take the form $\tyctl{x}{C_1}{C_2}$ where variable $x$ is
bound in computation type $C_1$.
We write $\tyctlMB{C_1}{C_2}$ when $x$ does not occur in $C_1$.
In what follows, we first illustrate impure control effects in the simple,
nondependent form $\tyctlMB{C_1}{C_2}$ and then extend to the fully dependent
form $\tyctl{x}{C_1}{C_2}$ that can specify the behavior of captured
continuations using the input (denoted by $x$) to the continuations.

A control effect $\tyctlMB{C_1}{C_2}$ represents the answer type of a program
changes from type $C_1$ to type $C_2$.
When it is assigned to a computation $c$, the initial answer type $C_1$
specifies how the continuation of the computation $c$ up to the closest handing
construct behaves, and the final answer type $C_2$ specifies what can be
guaranteed for the \emph{meta-context}, i.e., the context of the closest
handling construct.
To see the idea more concretely, revisit the first example in
Section~\ref{sec:overview/atm}:
\begin{align}
	&\explet{x}{
		\expwith{
			\{ x_r \mapsto x_r, \ 
				\op((), k) \mapsto k~0 < k~1 \}
		}{1 + \op~()}
	}{c}
  ~.
\end{align}
Let $h$ be the handler in the example.
Focusing on the operation call $\op~()$, we can find that it captures the
continuation $\expwith{h}{1 + \hole}$.
Because the continuation behaves as if it is a pure function returning integers,
the initial answer type of $\op~()$ turns out to be the computation type
$\tycompMB{\tyint}{\square}$ (we omit $\Sigma$ for a while; it will be explained shortly).
Furthermore, by the operation call, the handling construct
$\expwith{h}{1+\op~()}$ is replaced with the body $k~0 < k~1$ of $\op$'s clause
in $h$ and the functional representation $v$ of the continuation is substituted
for $k$.
It means that the meta-context $\explet{x}{\hole}{c}$ of the operation call
takes the computation $v~0 < v~1$, which is of type
$\tycompMB{\tybool}{\square}$ (note that $v~0 < v~1$ is pure because $v$ is a
pure function).
Therefore, the final answer type of $\op~()$ is $\tycompMB{\tybool}{\square}$.
As a result, the impure control effect of $\op~()$ is
$\tyctlMB{\tycompMB{\tyint}{\square}}{\tycompMB{\tybool}{\square}}$.

\citet{Sekiyama23} extended the simple form of impure control effects to a
dependent form $\tyctl{x}{C_1}{C_2}$, where the initial answer type $C_1$ can
depend on inputs, denoted by variable $x$, to continuations.
For instance, consider the continuation $\expwith{h}{1 + \hole}$ captured in the
above example.
When passed an integer $n$, it returns $1+n$.
Using the dependent form of impure control effects, we can describe such
behavior by the control effect
$\tyctl{x}{\tycompMB{\tyrfn{y}{\tyint}{y=x+1}}{\square}}{\tycompMB{\tybool}{\square}}$,
where $x$ represents the input to the continuation and the refinement type
$\tyrfn{y}{\tyint}{y=x+1}$ precisely specifies the return value of the
continuation for input $x$.
The type of $x$ is matched with the continuation's input type.
Since the continuation of $\op~()$ takes integers, the type assigned to $x$ is $\tyint$.
In general, given a computation type $\tycompMB{T}{\tyctl{x}{C_1}{C_2}}$, the
type $T$ is assigned to the variable $x$ because it corresponds to the input type of
the continuations of computations given that computation type.
The type information refined by dependent impure control effects is exploited in
typechecking operation clauses.
In the example, our type system typechecks the body of $\op$'s clause by
assigning the function type
$\tyfun{x}{\tyint}{\tycompMB{\tyrfn{y}{\tyint}{y=x+1}}{\square}}$
to the continuation variable $k$.
Then, since the body is $k~0 < k~1$, its type---i.e., the final answer
type---can be refined to $\tycompMB{\tyrfn{z}{\tybool}{z=\exptrue}}{\square}$.
Hence, the type system can assign  control effect
$\tyctl{x}{\tycompMB{\tyrfn{y}{\tyint}{y=x+1}}{\square}}{\tycompMB{\tyrfn{z}{\tybool}{z=\exptrue}}{\square}}$
to the operation call and ensure that the meta-context takes $\exptrue$ finally
(if the handling construct terminates).
We will demonstrate the expressivity and usefulness of dependent control effects
in more detail in Section~\ref{sec:language/exmaples}.

Operation signatures, denoted by $\Sigma$, are sets of pairs of an operation
name and a type scheme.
We write $\repi{\cdot}$ to denote a sequence of entities indexed by $i$.
The type scheme associated with an operation $\op$ is in the form
$\forall \rep{X: \rep{B}}. (x: T_1) \rarr ((y: T_2) \rarr C_1) \rarr C_2$,
where the types $T_1$ and $T_2$ are the input and output types, respectively, of the operation and the types $C_1$ and $C_2$ are
the initial and final answer types, respectively, of the operation call for $\op$.
Recall that the initial answer type $C_1$ corresponds to the return type
of delimited continuations captured by the call to $\op$, and that the
continuations take the return values of the operation call.
Therefore, the function type $(y: T_2) \rarr C_1$ represents the type of the
captured delimited continuations. Note that the variable $y$ denotes values
passed to the continuations.
Furthermore, the final answer type $C_2$ corresponds to the type of
the operation clause for $\op$ in the closest enclosing handler.
Therefore, the operation clause $\op(x, k) \mapsto c$ in the
handler is typed by checking that the body $c$ is of the type $C_2$ with the assumption that
argument variable $x$ is of the type $T_1$ and the continuation variable $k$ is
of the type $(y: T_2) \rarr C_1$.
A notable point of the type scheme is that it can be parameterized over
predicates. The predicate variables $\rep{X}$ abstract over the predicates, and the annotations $\rep{B}$ represent
the (base) types of the arguments to the predicates.
This allows calls to the same
operation in different contexts to have different control effects, which is
crucial for precisely verifying programs with algebraic effects and handlers as
we will show in Section~\ref{sec:language/exmaples}.
It is also noteworthy that operation signatures include not only operation names
but also type schemes as in \citet{Kammar13} and \citet{Kammar17}.
It allows an operation to have different types depending on the contexts
where it is used.
Another approach is to include only operation names and assumes that unique types
are assigned to them globally as in, e.g., \citet{Bauer13} and \citet{Leijen17}.
We decided to assign types to operations locally because it makes the type
system more flexible in that the types of operations can be refined depending on
contexts if needed.

Typing contexts $\Gamma$ are lists of variable bindings $x: T$
and predicate variable bindings $X: \rep{B}$.
We write $\Gamma, \phi$ for $\Gamma, x: \tyrfn{z}{B}{\phi}$
where $x$ and $z$ are fresh.
The notions of free variables, free predicate variables,
and predicate substitution are defined as usual.

\fullfalse
\iffull{
\begin{figure}
    \begin{gather}
        \infer{\jdwf{}{\emptyset}}
        {}
        \quad
        \infer{\jdwf{}{\Gamma, x: T}}
        {
            \jdwf{}{\Gamma} &
            x \notin \dom(\Gamma) &
            \jdwf{\Gamma}{T}
        }
        \quad
        \infer{\jdwf{}{\Gamma, X: \rep{B}}}
        {
            \jdwf{}{\Gamma} &
            X \notin \dom(\Gamma)
        }
        \qquad
        \infer{\jdwf{\Gamma}{\tyrfn{x}{B}{\phi}}}
        {\Gamma, x: B \vdash \phi}
        \quad
        \infer{\jdwf{\Gamma}{(x: T) \rarr C}}
        {
            \jdwf{\Gamma, x: T}{C}
        }
        \\
        \infer{\jdwf{\Gamma}{\tycomp{\Sigma}{T}{S}}}
        {
            \jdwf{\Gamma}{\Sigma} &
            \jdwf{\Gamma}{T} &
            \jdwf{\Gamma \mid T}{S}
        }
        \qquad
        \infer{\jdwf{\Gamma}{\{ \repi{\op_i : \forall \rep{X_i: \rep{B}_i}. F_i} \}}}
        {\repi{\jdwf{\Gamma, \rep{X_i: \rep{B}_i}}{F_i}}}
        \qquad
        \infer{\jdwf{\Gamma \mid T}{\square}}
        {\jdwf{}{\Gamma}}
        \quad
        \infer{\jdwf{\Gamma \mid T}{\tyctl{x}{C_1}{C_2}}}
        {
            \jdwf{\Gamma, x: T}{C_1} &
            \jdwf{\Gamma}{C_2}
        }
    \end{gather}
    \caption{Well-formedness of typing contexts and types.}
    \label{fig:wf}
\end{figure}

Well-formedness of typing contexts, value types, and computation types, whose
judgments are in the forms $\jdwf{}{\Gamma}$, $\jdwf{\Gamma}{T}$, and
$\jdwf{\Gamma}{C}$, respectively, are defined straightforwardly as shown in
Figure~\ref{fig:wf}.
We write $\dom(\Gamma)$ for the set of variables
and predicate variables bound in the typing context $\Gamma$.
The well-formedness of a computation type $\tycomp{\Sigma}{T}{S}$ rests on the
well-formedness of its components, that is, the operation signature $\Sigma$,
the value type $T$, and the control effect $S$.
An operation signature $\Sigma$ is well-formed under a typing context $\Gamma$,
written as $\jdwf{\Gamma}{\Sigma}$, if the type schemes associated with
operations in $\Sigma$ are well-formed under $\Gamma$.
Note that, for every type scheme $\forall \rep{X_i: \rep{B}_i}. F$, the qualified type $F$ is a function type.
The well-formedness of a control effect $S$ in terms of value type $T$, written as $\jdwf{\Gamma \mid T}{S}$ with a typing context $\Gamma$,
is derived by either of the last two rules in Figure~\ref{fig:wf}.
When the control effect $S$ is an impure effect $\tyctl{x}{C_1}{C_2}$, the type
$T$ is assigned to the variable $x$ because $x$ denotes values passed to
continuations of computations of the type $\tycomp{\Sigma}{T}{S}$ and the passed
values are the return values of the computations.
Additionally, we assume that the logic for refinements is equipped with
well-formedness judgments of formulas $\jdwf{\Gamma}{\phi}$ and of predicates
$\jdwf{\Gamma}{A : \rep{B}}$.
The properties assumed on these judgments are stated in the supplementary
material.
}\else{
Well-formedness of typing contexts, value types, and computation types, whose
judgments are in the forms $\jdwf{}{\Gamma}$, $\jdwf{\Gamma}{T}$, and
$\jdwf{\Gamma}{C}$, respectively, are defined straightforwardly by following \citet{Sekiyama23}.
We refer to the supplementary material for detail.
}\fi

\begin{figure}
    \raggedright
    \textbf{Typing rules for values} \quad \fbox{$\jdty{\Gamma}{v}{T}$}
    \begin{gather}
        \infersc[T-CVar]{\jdty{\Gamma}{x}{\tyrfn{z}{B}{z = x}}}
        {
            \jdwf{}{\Gamma} &
            \Gamma(x) = \tyrfn{z}{B}{\phi}
        }
        \quad
        \infersc[T-Var]{\jdty{\Gamma}{x}{\Gamma(x)}}
        {
            \jdwf{}{\Gamma} &
            \forall y, B, \phi. \Gamma(x) \neq \tyrfn{y}{B}{\phi}
        }
        \quad
        \infersc[T-Prim]{\jdty{\Gamma}{p}{\ty(p)}}
        {\jdwf{}{\Gamma}}
        \\[-2ex]
        \infersc[T-Fun]{\jdty{\Gamma}{\exprec{f}{x}{c}}{(x: T) \rarr C}}
        {\jdty{\Gamma, f: (x: T) \rarr C, x: T}{c}{C}}
        \quad
        \infersc[T-VSub]{\jdty{\Gamma}{v}{T_2}}
        {
            \jdty{\Gamma}{v}{T_1} &
            \jdsub{\Gamma}{T_1}{T_2} &
            \jdwf{\Gamma}{T_2}
        }
    \end{gather}
    \textbf{Typing rules for computations} \quad \fbox{$\jdty{\Gamma}{c}{C}$}
    \begin{gather}
        \infersc[T-Ret]{\jdty{\Gamma}{\expret{v}}{\tycomp{\emptyset}{T}{\square}}}
        {\jdty{\Gamma}{v}{T}}
        \quad
        \infersc[T-App]{\jdty{\Gamma}{v_1~v_2}{C[v_2/x]}}
        {
            \jdty{\Gamma}{v_1}{(x: T) \rarr C} &
            \jdty{\Gamma}{v_2}{T}
        }
        \\[1ex]
        \infersc[T-If]{\jdty{\Gamma}{\expif{v}{c_1}{c_2}}{C}}
        {\begin{gathered}
            \jdty{\Gamma}{v}{\tyrfn{x}{\tybool}{\phi}} \\[-.5ex]
            \jdty{\Gamma, v = \exptrue}{c_1}{C} \quad
            \jdty{\Gamma, v = \expfalse}{c_2}{C}
        \end{gathered}}
        \quad
        \infersc[T-CSub]{\jdty{\Gamma}{c}{C_2}}
        {
            \jdty{\Gamma}{c}{C_1} &
            \jdsub{\Gamma}{C_1}{C_2} &
            \jdwf{\Gamma}{C_2}
        }
        \\[1.5ex]
        \infersc[T-LetP]{\jdty{\Gamma}{\explet{x}{c_1}{c_2}}{\tycomp{\Sigma}{T_2}{\square}}}
        {\begin{gathered}
            \jdty{\Gamma}{c_1}{\tycomp{\Sigma}{T_1}{\square}} \\
            \jdty{\Gamma, x: T_1}{c_2}{\tycomp{\Sigma}{T_2}{\square}} \\
            x \notin \fv(T_2) \cup \fv(\Sigma)
        \end{gathered}}
        \quad
        \infersc[T-LetIp]{\jdty{\Gamma}{\explet{x}{c_1}{c_2}}{\tycomp{\Sigma}{T_2}{\tyctl{y}{C_{21}}{C_{12}}}}}
        {\begin{gathered}
            \jdty{\Gamma}{c_1}{\tycomp{\Sigma}{T_1}{\tyctl{x}{C}{C_{12}}}} \\
            \jdty{\Gamma, x: T_1}{c_2}{\tycomp{\Sigma}{T_2}{\tyctl{y}{C_{21}}{C}}} \\
            x \notin \fv(T_2) \cup \fv(\Sigma) \cup (\fv(C_{21}) \setminus \{y\})
        \end{gathered}}
        \\[-1.5ex]
        \infersc[T-Op]{\jdty{\Gamma}{\op~v}{
          \tycomp{\Sigma}{T_2[\rep{A/X}][v/x]}
          {(\tyctl{y}{C_1}{C_2})[\rep{A/X}][v/x]}
        }}
        {\begin{gathered}
            \Sigma \ni \op: \forall \rep{X: \rep{B}}. (x: T_1) \rarr ((y: T_2) \rarr C_1) \rarr C_2 \quad
            \jdwf{\Gamma}{\Sigma} \quad
            \rep{\jdty{\Gamma}{A}{\rep{B}}} \quad
            \jdty{\Gamma}{v}{T_1[\rep{A/X}]}
        \end{gathered}}
        \\[1.5ex]
        \infersc[T-Hndl]{\jdty{\Gamma}{\expwith{h}{c}}{C_2}}
        {\begin{gathered}
            h = \{ \expret{x_r} \mapsto c_r, \repi{\op_i(x_i, k_i) \mapsto c_i} \} \quad
            \jdty{\Gamma}{c}{\tycomp{\Sigma}{T}{\tyctl{x_r}{C_1}{C_2}}} \\[-.5ex]
            \jdty{\Gamma, x_r: T}{c_r}{C_1} \quad
            \bigrepi{\jdty{\Gamma, \rep{X_i: \rep{B}_i}, x_i: T_{1i}, k_i: (y_i: T_{2i}) \rarr C_{1i}}{c_i}{C_{2i}}} \\[-.5ex]
            \Sigma = \{ \repi{\op_i: \forall \rep{X_i: \rep{B}_i}. (x_i: T_{1i}) \rarr ((y_i: T_{2i}) \rarr C_{1i}) \rarr C_{2i}} \}
        \end{gathered}}
    \end{gather}
    \caption{Typing rules.}
    \label{fig:typing}
\end{figure}

Typing judgements for values and computations are in the forms $\jdty{\Gamma}{v}{T}$ and $\jdty{\Gamma}{c}{C}$,
respectively.
Figure~\ref{fig:typing} shows the typing rules.
By \rulename{T-CVar}, a variable $x$ of a refinement type is assigned a type
which states that the value of this type is exactly $x$.
For a variable of a non-refinement type (i.e., a function type in our language),
the rule \rulename{T-Var} assigns the type associated with the variable in the
typing context.
The rule \rulename{T-Prim} uses the mapping $\ty$ to type primitive values $p$.
We assume that $\ty$ assigns an appropriate value type to every primitive value.
We refer to the supplementary material for the formalization of the assumption.
The rule \rulename{T-Fun} for functions, \rulename{T-App} for function
applications, and \rulename{T-If} for conditional branches are standard in
refinement type systems (with support for value-dependent refinements).
The rules \rulename{T-VSub} and \rulename{T-CSub} allow values and computations,
respectively, to be typed at supertypes of their types.
We will define subtyping shortly.
By \rulename{T-Ret}, a value-return $\expret{v}$ has a computation type
where the operation signature is empty, the return value type is the type of $v$,
and the control effect is pure.

To type a let-expression $\explet{x}{c_1}{c_2}$, either the rule \rulename{T-LetP} or \rulename{T-LetIp} is used. Both of them
require that the types of the sub-expressions $c_1$ and $c_2$ have the same
operation signature $\Sigma$ and then assign $\Sigma$ to the type of the entire
let-expression.
The typing context for $c_2$ is extended by $x: T_1$ with the value
type $T_1$ of $c_1$, but $x$ cannot occur in $\Sigma$ and $T_2$ (as well as $C_{21}$ in \rulename{T-LetIp}) to prevent
the leakage of $x$ from its scope.
On the other hand, the two rules differ in how they treat control effects.
When both of the control effects of $c_1$ and $c_2$ are pure, the rule \rulename{T-LetP} is used. It states that the control effect of the entire let-expression is also pure.
When both are impure, the rule \rulename{T-LetIp} is used. It states that the control effect of the let-expression results in an impure control effect that is composed of the control effects of $c_1$ and $c_2$.
Note that, even when one of the control effects of $c_1$ and $c_2$ is pure and the other is
impure, we can view both of them as impure effects via subtyping
because it allows converting a pure control effect to an impure control
effect, as shown later.
We first explain how the composition works in the non-dependent form. Let
the control effect of $c_1$ be $\tyctlMB{C_{11}}{C_{12}}$ and that of $c_2$ be $\tyctlMB{C_{21}}{C_{22}}$,
and assume that a control effect $\tyctlMB{C_1}{C_2}$ is assigned to the
let-expression.
First, recall that the type $C_1$ expresses the return type of the continuation
of the let-expression up to the closest handling construct and that the closest
handling construct is replaced by a computation of the type $C_2$.
Based on this idea, the types $C_1$ and $C_2$ can be determined as follows.
First, because the delimited continuation of the let-expression is matched with
that of the computation $c_2$, the initial answer type $C_{21}$ of $c_2$
expresses the return type of the delimited continuation of the let-expression.
Therefore, the type $C_1$ should be matched with the type $C_{21}$.
Second, because the closest handling construct enclosing the let-expression is the same
as the one enclosing the sub-computation $c_1$, the type $C_2$ should be matched
with the final answer type $C_{12}$ of $c_1$.
Therefore, the control effect $\tyctlMB{C_1}{C_2}$ should be matched with
$\tyctlMB{C_{21}}{C_{12}}$, as stated in \rulename{T-LetIp}.
Furthermore, the rule \rulename{T-LetIp} requires that the
initial answer type $C_{11}$ of $c_1$ to be the same as the final answer type
$C_{22}$ of $c_2$.
This requirement is explained as follows.
First, the computation $c_1$ expects its delimited continuation to behave as
specified by the type $C_{11}$.
The delimited continuation of $c_1$ first evaluates the succeeding computation
$c_2$.
The final answer type $C_{22}$ of $c_2$ expresses that the closest
handling construct enclosing $c_2$ behaves as specified by the type $C_{22}$.
Because the closest handling construct enclosing $c_2$ corresponds to the
top-level handling construct in the delimited continuation of $c_1$, the
type $C_{11}$ should be matched with the type $C_{22}$.
We now extend to the fully dependent form.
From the discussion thus far, we can let
the control effects of $c_1$, $c_2$, and the let-expression be
$\tyctl{x_1}{C}{C_{12}}$, $\tyctl{x_2}{C_{21}}{C}$, and
$\tyctl{y}{C_{21}}{C_{12}}$ respectively, for some variables $x_1$, $x_2$, and $y$.
Then, the constraints on the names of these variables are determined as follows.
First, the input to the delimited continuation of $c_1$, which is denoted by the variable $x_1$,
should be matched with the evaluation result of $c_1$.
Then, since the let-expression binds the variable $x$ to the evaluation result of $c_1$,
the variable $x_1$ is matched with $x$.
Second, because the delimited continuation of $c_2$ is matched with
that of the let-expression, the inputs to them should be matched with each other.
They are denoted by the variables $x_2$ and $y$ respectively, and hence the variable $x_2$ is matched with $y$.

The rule \rulename{T-Hndl} for handling constructs $\expwith{h}{c}$
is one of the most important rules of our system.
It assumes that the handled computation $c$ is of a type
$\tycomp{\Sigma}{T}{\tyctl{x_r}{C_1}{C_2}}$, where the control effect is impure.
Even when $c$ is pure (i.e., performs no operation), it can have an impure control effect via subtyping.
Because the type of the handling construct represents how the expression is
viewed from the context, it should be matched with the final answer type $C_2$
of the handled computation $c$.
The premises in the second line define typing disciplines that the clauses in the installed handler $h$ have to satisfy.
First, let us consider the return clause $\expret{x_r} \mapsto c_r$.
Because the variable $x_r$ denotes the return value of the handled
computation $c$, the value type $T$ of $c$ is assigned to $x_r$.
Moreover, since the return clause is executed after evaluating $c$, the body
$c_r$ is the delimited continuation of $c$.
Therefore, the type of $c_r$ should be matched with the initial answer type $C_1$
of $c$.
Because the variable $x_r$ bound in the return clause can be viewed as the input
to the delimited continuation $c_r$, it should be matched with the
variable $x_r$ bound in the impure control effect $\tyctl{x_r}{C_1}{C_2}$.
Operation clauses are typed using the corresponding type schemes in the
operation signature $\Sigma$, as explained above.
Note that the rule also requires the installed handler $h$ to include a clause
for each of the operations in $\Sigma$, i.e., those that $c$ may perform.

The rule \rulename{T-Op} for operation calls is
another important rule.
Consider an operation call $\op~v$.
The rule assumes that an enclosing handler addresses the operation $\op$
by requiring that an operation signature $\Sigma$ assigned to the operation call
include the operation $\op$ with a type scheme
$\forall \rep{X: \rep{B}}. (x: T_1) \rarr ((y: T_2) \rarr C_1) \rarr C_2$,
and instantiates the predicate
variables $\rep{X}$ in the type scheme with well-formed predicates $\rep{A}$ to
reflect the contextual information of the operation call.
Then, it checks that the argument $v$ has the input type $T_1[\rep{A/X}]$ of the
operation.
Finally, the rule assigns the output type $T_2[\rep{A/X}][v/x]$ of the operation
as the value type of the operation call, and $C_1[\rep{A/X}][v/x]$ and
$C_2[\rep{A/X}][v/x]$ as the initial and final answer types of the operation call, respectively
(note that the types $T_2$, $C_1$, and $C_2$ are parameterized over predicates and arguments).

\begin{figure}
    \raggedright
    \textbf{Subtyping rules} \quad \fbox{$\jdsub{\Gamma}{T_1}{T_2}$} \ 
    \fbox{$\jdsub{\Gamma}{\Sigma_1}{\Sigma_2}$} \ \fbox{$\jdsub{\Gamma}{C_1}{C_2}$} \ 
    \fbox{$\jdsub{\Gamma \mid T}{S_1}{S_2}$}
    \begin{gather}
        \infersc[S-Rfn]{\jdsub{\Gamma}{\tyrfn{x}{B}{\phi_1}}{\tyrfn{x}{B}{\phi_2}}}
        {\Gamma, x: B \vDash \phi_1 \implies \phi_2}
        \quad
        \infersc[S-Fun]{\jdsub{\Gamma}{(x: T_1) \rarr C_1}{(x: T_2) \rarr C_2}}
        {
            \jdsub{\Gamma}{T_2}{T_1} &
            \jdsub{\Gamma, x: T_2}{C_1}{C_2}
        }
        \\[1ex]
        \infersc[S-Sig]{\jdsub{\Gamma}{\{ \repi{\op_i: \forall \rep{X_i: \rep{B}_i}. F_{1i}}, \repi{\op'_i: \forall \rep{X'_i: \rep{B'}_i}. F'_i} \}}
            {\{ \repi{\op_i: \forall \rep{X_i: \rep{B}_i}. F_{2i}} \}}}
        {\repi{\jdsub{\Gamma, \rep{X_i: \rep{B}_i}}{F_{1i}}{F_{2i}}}}
        \\[1ex]
        \infersc[S-Comp]{\jdsub{\Gamma}{\tycomp{\Sigma_1}{T_1}{S_1}}{\tycomp{\Sigma_2}{T_2}{S_2}}}
        {
            \jdsub{\Gamma}{\Sigma_2}{\Sigma_1} &
            \jdsub{\Gamma}{T_1}{T_2} &
            \jdsub{\Gamma \mid T_1}{S_1}{S_2}
        }
        \qquad
        \infersc[S-Pure]{\jdsub{\Gamma \mid T}{\square}{\square}}
        {}
        \\[1ex]
        \infersc[S-ATM]{\jdsub{\Gamma \mid T}{\tyctl{x}{C_{11}}{C_{12}}}{\tyctl{x}{C_{21}}{C_{22}}}}
        {
            \jdsub{\Gamma, x: T}{C_{21}}{C_{11}} &
            \jdsub{\Gamma}{C_{12}}{C_{22}}
        }
        \quad
        \infersc[S-Embed]{\jdsub{\Gamma \mid T}{\square}{\tyctl{x}{C_1}{C_2}}}
        {
            \jdsub{\Gamma, x: T}{C_1}{C_2} &
            x \notin \fv(C_2)
        }
    \end{gather}
 \mbox{} \\[-3ex]
    \caption{Subtyping rules.}
    \label{fig:subty}
\end{figure}

The type system defines four kinds of subtyping judgments:
$\jdsub{\Gamma}{T_1}{T_2}$ for value types,
$\jdsub{\Gamma}{C_1}{C_2}$ for computations types,
$\jdsub{\Gamma}{\Sigma_1}{\Sigma_2}$ for operation signatures, and
$\jdsub{\Gamma \mid T}{S_1}{S_2}$ for control effects.
Figure~\ref{fig:subty} shows the subtyping rules.
The subtyping rules for control effects are adopted from the work of \citet{Sekiyama23},
which extends subtyping for control effects given by \citet{Materzok11} to
dependent typing.
The rules \rulename{S-Rfn} and \rulename{S-Fun} for value types are standard.
The judgement $\valid{\Gamma}{\phi}$ in \rulename{S-Rfn} means
the semantic validity of the formula $\phi$ under the assumption $\Gamma$.
Subtyping between operation signatures is determined by \rulename{S-Sig}.
This rule is based on the observation that an operation signature $\Sigma$
represents the types of operation clauses in handlers, as seen in
\rulename{T-Hndl}.
Then, the rule \rulename{S-Sig} can be viewed as defining a subtyping relation between
the types of handlers (except for return clauses): a handler for operations in
$\Sigma_1$ can be used as one for operations in $\Sigma_2$ if every operation
$\op$ in $\Sigma_2$ is included in $\Sigma_1$ (i.e., the handler has an
operation clause for every $\op$ in $\Sigma_2$) and the type scheme of $\op$ in
$\Sigma_1$ is a subtype of the type scheme of $\op$ in $\Sigma_2$ (i.e., the
operation clause for $\op$ in the handler works as one for $\op$ in $\Sigma_2$).
Given a computation type $C_1 \defeq \tycomp{\Sigma_1}{T_1}{S_1}$ and its
supertype $C_2 \defeq \tycomp{\Sigma_2}{T_2}{S_2}$, a handler for operations
performed by the computations of the type $C_2$ (i.e., the operations in
$\Sigma_2$) is required to be able to handle operations performed by the
computations of the type $C_1$ (i.e., the operations in $\Sigma_1$) because the
subtyping allows deeming the computations of $C_1$ to be of $C_2$.
The safety of such handling is ensured by requiring $\Sigma_2 <: \Sigma_1$.
In the rule \rulename{S-Comp}, the first premise represents this requirement.
The second premise $\jdsub{\Gamma}{T_1}{T_2}$ in \rulename{S-Comp} allows
viewing the return values of the computations of the type $C_1$ as those of the
type $C_2$.
The third premise $\jdsub{\Gamma \mid T_1}{S_1}{S_2}$ expresses that the use of
effects by the computations of the type $C_1$ is subsumed by the use of effects
allowed by the type $C_2$.
It is derived by the last three rules: \rulename{S-Pure},
\rulename{S-ATM}, and \rulename{S-Embed}.
The rule \rulename{S-Pure} just states reflexivity of the pure control effect.
If both $S_1$ and $S_2$ are impure, the rule \rulename{S-ATM} is applied.
Because initial answer types represent the assumptions of computations
on their contexts, \rulename{S-ATM} allows strengthening the assumptions by
being contravariant in them.
By contrast, because final answer types represent the guarantees of how
enclosing handling constructs behave, \rulename{S-ATM} allows weakening the
guarantees by being covariant in them.
Note that the typing context for the initial answer types is extended
with the binding $x : T_1$ because they may reference the inputs to the
continuations via the variable $x$ and the inputs are of the type
$T_1$.
Finally, the rule \rulename{S-Embed} allows converting the pure control effect
to an impure control effect $\tyctl{x}{C_1}{C_2}$.
Because a computation $c$ with the pure control effect performs no operation,
what is guaranteed for the behavior of the handling construct enclosing $c$
coincides with what is assumed on $c$'s delimited continuation.
Because the guarantee and assumption are specified by the types $C_2$ and
$C_1$, respectively, if $C_1$ is matched with $C_2$---more generally, the
``assumption'' $C_1$ implies the ``guarantee'' $C_2$---the pure computation
$c$ can be viewed as the computation with the impure control effect
$\tyctl{x}{C_1}{C_2}$.
The first premise in \rulename{S-Embed} formalizes this idea.
Note that, because the variable $x$ is bound in the type $C_1$, the rule
\rulename{S-Embed} disallows $x$ to occur in the type $C_2$.

Finally, we state the type safety of our system. Its proof, via progress and subject reduction, is given in the supplementary material.
We define $\eval^*$ as the reflexive, transitive closure of the one-step evaluation relation $\eval$.
\begin{theorem}[type safety] \label{thm:safety}
    If\, $\jdty{\emptyset}{c}{\tycomp{\Sigma}{T}{S}}$ and $c \eval^* c'$, then one of the following holds:
    (1) $c' = \expret{v}$ for some $v$ such that $\jdty{\emptyset}{v}{T}$;
    (2) $c' = K[\op~v]$ for some $K$, $\op$, and $v$ such that $\op \in \dom(\Sigma)$; or
    (3) $c' \eval c''$ for some $c''$ such that $\jdty{\emptyset}{c''}{\tycomp{\Sigma}{T}{S}}$~.
\end{theorem}

\newcommand{\cbody}{c_{\mathit{body}}}

\subsection{Examples} \label{sec:language/exmaples}

In this section, we demonstrate how our type system verifies programs with
algebraic effects and handlers by showing typing derivations of a few
examples.
Here, we abbreviate a pure computation type $\tycomp{\{\}}{T}{\square}$ to $T$
and omit the empty typing context from typing and subtyping judgments.
For simplicity, we often write $c_1~c_2$ for an expression
$\explet{x_1}{c_1}{\explet{x_2}{c_2}{x_1~x_2}}$ where $x_1$ does not occur in
$c_2$.
Furthermore, we deal with a pure computation as if it is a value.
For example, we write $\expret{c}$ for a computation $\explet{x}{c}{\expret{x}}$
if $c$ is pure (e.g., as $\expret{a-b}$).

\subsubsection{Example 1: Nondeterministic Computation} \label{sec:language/examples/nondet}
We first revisit the example presented in Section~\ref{sec:overview/algeff}.
In our language, it can be expressed as follows:
\begin{align}
    \expwith{h}{
        (&\explet{a}{(\explet{y}{\op[decide]~()}{\expif{y}{\expret{10}}{\expret{20}}})}{ \\
            &\explet{b}{(\explet{y'}{\op[decide]~()}{\expif{y'}{\expret{1}}{\expret{2}}})}{
                \expret{a - b}
            }
        })
    }
\end{align}
where
$
 h \defeq \{ \expret{x_r} \mapsto \expret{x_r},
         \op[decide](x, k) \mapsto \explet{r_t}{k~\exptrue}{
            \explet{r_f}{k~\expfalse}{\mathrm{max}~r_t~r_f}
            } \}
$ ~.
As seen before, executing this program results in $19$.
Our system can assign the most precise type $\tyrfn{z}{\tyint}{z = 19}$ to this program.
We now show the typing process to achieve this.
In what follows, we write $\Gamma_{\rep{x}}$ for the typing context binding the
variables $\rep{x}$ with some appropriate types $\rep{B}$.
In particular, these variables have these base types:
$x_r : \tyint$, $a : \tyint$, $b : \tyint$, $y : \tybool$, and $y' : \tybool$.

First, consider the types assigned to the clauses in the handler $h$.
The return clause can be typed as
$\jdty{\Gamma_{x_r}}{\expret{x_r}}{\tyrfn{z}{\tyint}{z = x_r}}$.
The clause for $\op[decide]$ can be typed as follows:
\begin{align} \label{eqn:typ-decide-clause}
    \jdty{
       \Gamma
    }{
        \explet{r_t}{k~\exptrue}{
            \explet{r_f}{k~\expfalse}{\mathrm{max}~r_t~r_f}
        }
    }{\tyrfn{z}{\tyint}{\phi}}
\end{align}
where $\Gamma \defeq X: (\tyint, \tybool), x: \tyunit,
k: (y: \tybool) \rarr \tyrfn{z}{\tyint}{X(z, y)}$,
$\phi \defeq \forall r_t r_f.
X(r_t, \exptrue) \land X(r_f, \expfalse) \implies z = \mathtt{max}(r_t, r_f)$, and $\mathtt{max}$ is a term-level function that returns the larger of given two integers.
In this typing judgment, the predicate variable $X$ abstracts over relationships
between inputs $y$ and outputs $z$ of delimited continuations captured by calls to
$\op[decide]$, and the refinement formula $\phi$ summarizes what the operation
clause computes.
Therefore, the operation signature $\Sigma$ of the type of the handled
computation, $\cbody$ in what follows, can be given as
follows:
\begin{align}
    \Sigma \,\defeq\, \{ &\op[decide]: \forall X: (\tyint, \tybool).
        (x: \tyunit) \rarr ((y: \tybool) \rarr \tyrfn{z}{\tyint}{X(z, y)})
        \rarr \tyrfn{z}{\tyint}{\phi}
    \} ~. \!
\end{align}

Therefore, we can conclude that the program is typable as desired by the following derivation
\begin{prooftree}
 \AxiomC{\Shortstack{
 {$\jdty{\Gamma_{x_r}}{\expret{x_r}}{\tyrfn{z}{\tyint}{z = x_r}}$ \qquad
  (Judgment (\refeq{eqn:typ-decide-clause}))}
 {(I) \ \ %
  $\jdty{}{\cbody}{\tycomp{\Sigma}{\tyint}{\tyctl{x_r}{\tyrfn{z}{\tyint}{z = x_r}}{\tyrfn{z}{\tyint}{z = 19}}}}$}
 }}
 \RightLabel{\rulename{T-Hndl}}
 \UnaryInfC{$\jdty{}{\expwith{h}{\cbody}}{\tyrfn{z}{\tyint}{z = 19}}$}
\end{prooftree}
if the premise (I) for $\cbody$ holds.
We derive it by \rulename{T-LetIp}, obtaining a derivation of the form
\begin{prooftree}
 \AxiomC{
   \stackanchor{
     (II) \ \ %
     $\jdty{}{(\explet{y}{\op[decide]~()}{\kw{if} \ y \ \cdots})}{\tycomp{\Sigma}{\tyint}{\tyctl{a}{C_1}{\tyrfn{z}{\tyint}{z = 19}}}}$
   }{
     (III) \ \ %
     $\jdty{\Gamma_{a}}{\explet{b}{\cdots}{\expret{a - b}}}{\tycomp{\Sigma}{\tyint}{\tyctl{x_r}{\tyrfn{z}{\tyint}{z = x_r}}{C_1}}}$
   }
 }
 \RightLabel{\rulename{T-LetIp}}
 \UnaryInfC{
   (I) \ \ $\jdty{}{\cbody}{\tycomp{\Sigma}{\tyint}{\tyctl{x_r}{\tyrfn{z}{\tyint}{z = x_r}}{\tyrfn{z}{\tyint}{z = 19}}}}$
 }
\end{prooftree}
for some type $C_1$.

We start by examining judgement (III) because its derivation gives the
constraints to identify the type $C_1$.
By \rulename{T-LetIp} again, we can derive
\begin{prooftree}
 \AxiomC{\stackanchor{
  (III-1) \ \ %
  $\jdty{\Gamma_{a}}{(\explet{y'}{\op[decide]~()}{\kw{if} \ y' \ \cdots})}{\tycomp{\Sigma}{\tyint}{\tyctl{b}{C_2}{C_1}}}$
 }{
  (III-2) \ \ %
  $\jdty{\Gamma_{a,b}}{\expret{a - b}}{\tycomp{\Sigma}{\tyint}{\tyctl{x_r}{\tyrfn{z}{\tyint}{z = x_r}}{C_2}}}$
 }}
 \RightLabel{\rulename{T-LetIp}}
 \UnaryInfC{
  (III) \ \ %
  $\jdty{\Gamma_{a}}{\explet{b}{\cdots}{\expret{a - b}}}{\tycomp{\Sigma}{\tyint}{\tyctl{x_r}{\tyrfn{z}{\tyint}{z = x_r}}{C_1}}}$
 }
\end{prooftree}
with the premises (III-1) and (III-2) and some type $C_2$.
Judgment (III-2) is derivable by
\begin{prooftree}
 \AxiomC{\stackanchor{
  $\jdty
   {\Gamma_{a,b}}
   {\expret{a - b}}
   {\tyrfn{z}{\tyint}{z = a - b}}$
 }{
  (III-2-S) \ \ %
  $\jdsub
   {\Gamma_{a,b}}
   {\tyrfn{z}{\tyint}{z = a - b}}
   {\tycomp{\Sigma}{\tyint}{\tyctl{x_r}{\tyrfn{z}{\tyint}{z = x_r}}{C_2}}}$
 }}
 \RightLabel{\rulename{T-Sub}}
 \UnaryInfC{
  (III-2) \ \ %
  $\jdty
   {\Gamma_{a,b}}
   {\expret{a - b}}
   {\tycomp{\Sigma}{\tyint}{\tyctl{x_r}{\tyrfn{z}{\tyint}{z = x_r}}{C_2}}}$
 }
\end{prooftree}
with the derivation of the subtyping judgment (III-2-S):
\begin{prooftree}
 \AxiomC{\stackanchor{
  $\jdsub{\Gamma_{a,b}}{\Sigma}{\emptyset}$
  \qquad
  $\jdsub{\Gamma_{a,b}}{\tyrfn{z}{\tyint}{z = a - b}}{\tyint}$
 }{
  $\jdsub
   {\Gamma_{a,b} \mid \tyrfn{z}{\tyint}{z = a - b}}
   {\square}
   {\tyctl{x_r}{\tyrfn{z}{\tyint}{z = x_r}}{C_2}}$
 }}
 \RightLabel{\rulename{S-Comp}}
 \UnaryInfC{
  (III-2-S) \ \ %
  $\jdsub
   {\Gamma_{a,b}}
   {\tyrfn{z}{\tyint}{z = a - b}}
   {\tycomp{\Sigma}{\tyint}{\tyctl{x_r}{\tyrfn{z}{\tyint}{z = x_r}}{C_2}}}$
 }
\end{prooftree}
The first two subtyping premises are derivable trivially.
We can derive the third one by letting
$
 C_2 {\,\defeq\,} \tyrfn{z}{\tyint}{z = a - b}
$
because:
\begin{prooftree}
 \AxiomC{
  $\valid
   {\Gamma_{a,b}, x_r: \tyrfn{z}{\tyint}{z = a - b}, z: \tyint}
   {(z = x_r) \implies (z = a - b)}$
 }
 \RightLabel{\rulename{S-Rfn}}
 \UnaryInfC{
  $\jdsub
   {\Gamma_{a,b}, x_r: \tyrfn{z}{\tyint}{z = a - b}}
   {\tyrfn{z}{\tyint}{z = x_r}}
   {\tyrfn{z}{\tyint}{z = a - b}}$
 }
 \RightLabel{\rulename{S-Embed}}
 \UnaryInfC{
  $\jdsub
  {\Gamma_{a,b} \mid \tyrfn{z}{\tyint}{z = a - b}}
  {\square}
  {\tyctl{x_r}{\tyrfn{z}{\tyint}{z = x_r}}{\colorbox{gray!50}{$\tyrfn{z}{\tyint}{z = a - b}$}}}$
 }
\end{prooftree}
where the grayed part is denoted by $C_2$ in the original premise.
We note that our type inference algorithm automatically infers such a type by constraint solving (cf.~Section~\ref{sec:impl}).
Next, judgment (III-1) is derivable by
\begin{prooftree}
 \AxiomC{\stackanchor{
  (III-1-1) \ \ %
  $\jdty
   {\Gamma_{a}}
   {\op[decide]~()}
   {\tycomp{\Sigma}{\tybool}{\tyctl{y'}{C_3}{C_1}}}$
 }{
  (III-1-2) \ \ %
  $\jdty
   {\Gamma_{a,y'}}
   {\kw{if} \ y' \ \cdots}
   {\tycomp{\Sigma}{\tyint}{\tyctl{b}{C_2}{C_3}}}$
 }}
 \RightLabel{\rulename{T-LetIp}}
 \UnaryInfC{
 (III-1) \ \ %
 $\jdty{\Gamma_{a}}{(\explet{y'}{\op[decide]~()}{\kw{if} \ y' \ \cdots})}{\tycomp{\Sigma}{\tyint}{\tyctl{b}{C_2}{C_1}}}$
 }
\end{prooftree}
with the premises (III-1-1) and (III-1-2) and some type $C_3$.
By letting
$
 C_3 {\,\defeq\,} \tyrfn{z}{\tyint}{z = (\ternaryif{y'}{(a-1)}{(a-2)})}
$,
we can derive judgment (III-1-2):
\begin{prooftree}
 \AxiomC{\stackanchor{
  $\jdty{\Gamma_{a,y'}}
        {\kw{if} \ y' \ \cdots}
        {\tyrfn{z}{\tyint}{z = (\ternaryif{y'}{1}{2})}}$
 }{
  $\jdsub{\Gamma_{a,y'}}
         {\tyrfn{z}{\tyint}{z = (\ternaryif{y'}{1}{2})}}
         {\tycomp{\Sigma}{\tyint}{\tyctl{b}{C_2}{C_3}}}$
 }}
 \RightLabel{\rulename{T-Sub}}
 \UnaryInfC{
 (III-1-2) \ \ %
 $\jdty
  {\Gamma_{a,y'}}
  {\kw{if} \ y' \ \cdots}
  {\tycomp{\Sigma}{\tyint}{\tyctl{b}{C_2}{C_3}}}$
 }
\end{prooftree}
It is easy to see that the first typing premise holds.
We can derive the second subtyping premise similarly to subtyping
judgment (III-2-S), namely, by \rulename{S-Comp} with the following derivation
for the subtyping on control effects:
\begin{prooftree}
 \AxiomC{
  $\valid
   {\Gamma_{a,y'}, b: \tyrfn{z}{\tyint}{z = (\ternaryif{y'}{1}{2})}}
   {(z = a - b) \implies (z = (\ternaryif{y'}{(a-1)}{(a-2)}))}$
 }
 \RightLabel{\rulename{S-Rfn}}
 \UnaryInfC{
  $\jdsub
   {\Gamma_{a,y'}, b: \tyrfn{z}{\tyint}{z = (\ternaryif{y'}{1}{2})}}
   {C_2}
   {C_3}$
 }
 \RightLabel{\rulename{S-Embed}}
 \UnaryInfC{$\jdsub{\Gamma_{a,y'} \mid \tyrfn{z}{\tyint}{z = (\ternaryif{y'}{1}{2})}}{\square}{\tyctl{b}{C_2}{C_3}}$}
\end{prooftree}
Judgment (III-1-1) is derived by \rulename{T-Op}, but for that,
we need to instantiate the predicate variable $X$ in the type scheme of $\op[decide]$ in $\Sigma$ with
a predicate $A$ such that the constraint $C_3 = \tyrfn{z}{\tyint}{A(z, y')}$ imposed by \rulename{T-Op} is met.
Let
$
 A \,\defeq\, \lambda (z, y). z = (\ternaryif{y}{(a-1)}{(a-2)})
$, which satisfies the constraint trivially.
Then, by letting
$
 C_1 {\,\defeq\,} \tyrfn{z}{\tyint}{\phi}[A/X]~,
$
we have the following derivation:
\begin{prooftree}
 \AxiomC{
 $\jdty
  {\Gamma_{a}}
  {()}
  {\tyunit}$
 }
 \RightLabel{\rulename{T-Op}}
 \UnaryInfC{
 (III-1-1) \ \ %
 $\jdty{\Gamma_{a}}{\op[decide]~()}{\tycomp{\Sigma}{\tybool}{\tyctl{y'}{\tyrfn{z}{\tyint}{A(z, y')}}{\tyrfn{z}{\tyint}{\phi}[A/X]}}}$
 }
\end{prooftree}

Finally, we examine judgment (II).
It is derivable by
\begin{prooftree}
 \AxiomC{\stackanchor{
  (II-1) \ \ %
  $\jdty
   {}
   {\op[decide]~()}
   {\tycomp{\Sigma}{\tybool}{\tyctl{y}{C_4}{\tyrfn{z}{\tyint}{z = 19}}}}$
 }{
  (II-2) \ \ %
  $\jdty
   {\Gamma_{y}}
   {\kw{if} \ y \ \cdots}
   {\tycomp{\Sigma}{\tyint}{\tyctl{a}{C_1}{C_4}}}$
 }}
 \RightLabel{\rulename{T-LetIp}}
 \UnaryInfC{
 (II) \ \ %
 $\jdty{}{(\explet{y}{\op[decide]~()}{\kw{if} \ y \ \cdots})}{\tycomp{\Sigma}{\tyint}{\tyctl{a}{C_1}{\tyrfn{z}{\tyint}{z = 19}}}}$
 }
\end{prooftree}
with the premises (II-1) and (II-2) and some type $C_4$.
Judgement (II-2) is derivable similarly to (III-1-2) by letting
$
 C_4 {\,\defeq\,} \tyrfn{z}{\tyint}{z = (\ternaryif{y}{9}{19})}~.
$
For judgment (II-1), we instantiate the predicate variable $X$ in the first call to $\op[decide]$
with the predicate
$
 A' \,\defeq\, \lambda (z, y). z = (\ternaryif{y}{9}{19}) ~.
$
Then, we can derive the judgment by the following derivation:
\begin{prooftree}
 \AxiomC{\stackanchor{
  $\jdty
   {}
   {\op[decide]~()}
   {\tycomp{\Sigma}{\tybool}{\tyctl{y}{\tyrfn{z}{\tyint}{A'(z, y)}}{\tyrfn{z}{\tyint}{\phi[A'/X]}}}}$
 }{
  $\jdsub
   {}
   {\tycomp{\Sigma}{\tybool}{\tyctl{y}{C_4}{\tyrfn{z}{\tyint}{\phi[A'/X]}}}}
   {\tycomp{\Sigma}{\tybool}{\tyctl{y}{C_4}{\tyrfn{z}{\tyint}{z = 19}}}}$
 }}
 \RightLabel{\rulename{T-Sub}}
 \UnaryInfC{
 (II-1) \ \ %
 $\jdty{}{\op[decide]~()}{\tycomp{\Sigma}{\tybool}{\tyctl{y}{C_4}{\tyrfn{z}{\tyint}{z = 19}}}}$
 }
\end{prooftree}
(note that $C_4 = \tyrfn{z}{\tyint}{A'(z, y)}$)
where the first premise is derived by \rulename{T-Op} and the second one
holds because the formula $\phi[A'/X]$ is semantically equivalent to the formula $z = 19$.

We note that the predicate variable in the type scheme of $\op[decide]$ is important to typing this example.
The delimited continuations captured by the two calls to $\op[decide]$ behave differently. Namely, they respectively behave according to the predicates $A(u,v)$ and $A'(u,v)$ where $u$ is the integer output given the Boolean input $v$.
By using predicate variables, our type system gives a single type scheme to an operation that abstracts over such different behaviors.\footnote{An alternative approach is to use intersection types (i.e., allow a set of types to be given to an operation).  But, we find our approach more uniform and modular as it is able to give a single compact type scheme and enables operation signatures to be unaware of in which contexts operations are called.}

\subsubsection{Example 2: State} \label{sec:language/examples/state}
We next revisit the second example from
Section~\ref{sec:overview/algeff}.
Recall the example, which is the following program:
\begin{gather}
    (\expwith{h}{
        \ (\op[set]~3;\,
        \explet{n}{\op[get]~()}{
            \op[set]~5;\,
            \explet{m}{\op[get]~()}{
                n + m
            }
        })
    })~0
\end{gather}
where $h \defeq \{
  x_r \mapsto \lambda s. x_r, \ 
  \op[set](x, k) \mapsto \lambda s.k~()~x, \ 
  \op[get](x, k) \mapsto \lambda s. k~s~s \}$~.
For this example, we use
the following syntactic sugars: $c_1;\,c_2 \,\defeq\, \explet{x}{c_1}{c_2}$
(where $x$ does not occur in $c_2$) and $\lambda x. v \,\defeq\, \lambda x. \expret{v}$.
Then,
the program is in our language.
This program uses two operations: $\op[set]$, which updates the state value, and
$\op[get]$, which returns the current state value. The handling construct returns a function that
maps any integer value to the value $8$;
arguments to the function are initial state values, but they are not used
because the function begins by initializing the state.
Applying the function to the initial state value $0$, the whole program returns $8$.

This program is expected to be of the type
$\tyrfn{z}{\tyint}{z = 8}$.
The rest of this section explains how the type system assigns this type
to the program.
First, the operation signature $\Sigma$ for the handler $h$ can be defined as follows:
\[\begin{array}{@{}l@{\ \ }l@{\ \ }l}
 \Sigma &\defeq&
   \{
        \op[set]: \forall X: (\tyint, \tyint).\,
            (x: \tyint) \rarr (\tyunit \rarr ((s: \tyint) \rarr \tyrfn{z}{\tyint}{X(z, s)})) \\
    && \qquad\qquad\qquad\qquad\qquad\qquad\qquad      \rarr ((s: \tyint) \rarr \tyrfn{z}{\tyint}{X(z, x)}), \\
   && \ \, \op[get]: \forall X: (\tyint, \tyint, \tyint).\,
            \tyunit \rarr ((y: \tyint) \rarr ((s: \tyint) \rarr \tyrfn{z}{\tyint}{X(z, s, y)})) \\
    && \qquad\qquad\qquad\qquad\qquad\qquad\qquad      \rarr ((s: \tyint) \rarr \tyrfn{z}{\tyint}{X(z, s, s)})
    \ \}
  \end{array}
\]
Then, each sub-computation in the handled computation can be typed as follows:
\[\begin{array}{r@{\ }l}
    \jdty{&}{\op[set]~3}{\tycomp{\Sigma}{\tyint}{
        \tyctl{\_}{
            (s: \tyint) \rarr \tyrfn{z}{\tyint}{z = s + 5}
        }{
            (s: \tyint) \rarr \tyrfn{z}{\tyint}{z = 3 + 5}
        }
    }} \\
    \jdty{&}{\op[get]~()}{\tycomp{\Sigma}{\tyint}{
        \tyctl{n}{
            (s: \tyint) \rarr \tyrfn{z}{\tyint}{z = n + 5}
        }{
            (s: \tyint) \rarr \tyrfn{z}{\tyint}{z = s + 5}
        }
    }} \\
    \jdty{n: \tyint&}{\op[set]~5}{\tycomp{\Sigma}{\tyint}{
        \tyctl{\_}{
            (s: \tyint) \rarr \tyrfn{z}{\tyint}{z = n + s}
        }{
            (s: \tyint) \rarr \tyrfn{z}{\tyint}{z = n + 5}
        }
    }} \\
    \jdty{n: \tyint&}{\op[get]~()}{\tycomp{\Sigma}{\tyint}{
        \tyctl{m}{
            (s: \tyint) \rarr \tyrfn{z}{\tyint}{z = n + m}
        }{
            (s: \tyint) \rarr \tyrfn{z}{\tyint}{z = n + s}
        }
    }} \\
    \jdty{n: \tyint, &m: \tyint}{\expret{n + m}}{\\ \tycomp{\Sigma}{\tyrfn{x_r&}{\tyint}{x_r = n + m}}{
        \tyctl{x_r}{
            (s: \tyint) \rarr \tyrfn{z}{\tyint}{z = x_r}
        }{
            (s: \tyint) \rarr \tyrfn{z}{\tyint}{z = n + m}
        }
    }}
  \end{array}
\]
The first four judgments are derived by \rulename{T-Op} with appropriate
instantiation of the type schemes of $\op[set]$ and $\op[get]$.
The last judgement is derived using \rulename{S-Embed} as in the first example.
The type of the handled computation is derived from these computation types, taking the following form:
\[
    \tycomp{\Sigma}{\tyint}{
        \tyctl{x_r}{
            (s: \tyint) \rarr \tyrfn{z}{\tyint}{z = x_r}
        }{
            (s: \tyint) \rarr \tyrfn{z}{\tyint}{z = 8}
        }
    } ~.
\]
Therefore, by \rulename{T-Hndl}, the type of the handling construct is
$(s: \tyint) \rarr \tyrfn{z}{\tyint}{z = 8}$, and by \rulename{T-App}, the type of the whole program is $\tyrfn{z}{\tyint}{z = 8}$
as promised.

\subsubsection{Example 3: File Manipulation} \label{sec:language/examples/file}
Finally, we consider the last example in Section~\ref{sec:overview/algeff} that
manipulates a specified file.
Because the example uses nondeterministic while-loop constructs
$\expwhile{\star}{c}$, we informally extend our language with them.\footnote{An
alternative is to encode the while-loop constructs in our language by supposing
that the termination of a while-loop construct is determined by some function
parameter $f : \tyfunshort{\tyunit}{\tybool}$.}
The semantics of the while-loop constructs is given by the reduction rules
$\expwhile{\star}{c} \eval c;\, \expwhile{\star}{c}$ and $\expwhile{\star}{c}
\eval \expret{()}$, and the typing rule is given as follows:
\begin{prooftree}
 \AxiomC{$\jdty{\Gamma}{c}{\tycomp{\Sigma}{\tyunit}{\tyctlMB{C}{C}}}$}
 \RightLabel{\rulename{T-Loop}}
 \UnaryInfC{$\jdty{\Gamma}{\expwhile{\star}{c}}{\tycomp{\Sigma}{\tyunit}{\tyctlMB{C}{C}}}$}
\end{prooftree}
Note that it is easy to adapt the type safety to this extension.

Recall that the example for file manipulation is the following function:
\begin{gather}
 v \,\defeq\, \expfun{x}{~\expwhile{\star}{\op[open]~x;\ \expwhile{\star}{\explet{y}{\op[read]()}{\op[write]~(y\texttt{\textasciicircum"X"})}};\ \op[close]~()}} ~.
\end{gather}
The regular scheme stipulating the valid use of the file operations is
$(\op[open] \ (\op[read] \mid \op[write])^\ast \ \op[close])^\ast$,
which is equivalent to the automaton to the right.
\begin{wrapfigure}[4]{R}{0.38\textwidth}
\vspace*{-5ex}
\begin{tikzpicture}[>=Stealth,shorten >=1pt,node distance=2cm,on grid,auto]
  \node[state,initial,initial text=,accepting] (q0) {$q0$};
  \node[state,right=of q0]       (q1) {$q1$};
   \path[->]
   (q0) edge[bend left,above] node {$\op[open]$}  (q1)
   (q1) edge[bend left,below] node {$\op[close]$} (q0)
        edge[loop right]      node[align=left] {$\op[read]$\\$\op[write]$} (q1);
\end{tikzpicture}
\end{wrapfigure}

Our idea to verify the correctness of the file manipulation is to encode the
automaton states as program states, simulate the state transitions in the
automaton by state-passing, and check that the file operations are used only in
appropriate states.
Let $Q0 \,\defeq\, 0$ and $Q1 \,\defeq\, 1$; they represent the automaton states
$q0$ and $q1$, respectively.
We suppose that an effect handler implements the file operations $\op[open]$ ,
$\op[close]$, $\op[read]$, and $\op[write]$ in a state-passing style for states
$Q0$ and $Q1$.
Then, the type scheme of each file operation can be given as an instance of the
following template:
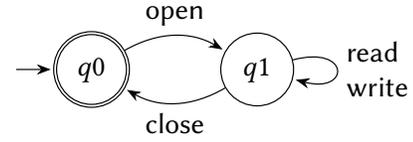
\[
 F(T_\mathrm{in}, T_\mathrm{out}, Q_\mathrm{pre}, Q_\mathrm{post}) \defeq T_\mathrm{in} \rarr (T_\mathrm{out} \rarr (\tyrfn{x}{\tyint}{x = Q_\mathrm{post}} \rarr C)) \rarr (\tyrfn{x}{\tyint}{x = Q_\mathrm{pre}} \rarr C)
\]
where the parameters $T_\mathrm{in}$ and $T_\mathrm{out}$ are the input and
output types, respectively, of the operation, and $Q_\mathrm{pre}$ and
$Q_\mathrm{post}$ are the states before and after, respectively, performing the
operation.
We do not specify the final answer type $C$ concretely here because it is not
important.
Using this template, an operation signature $\Sigma$ of the file operations is
given as
\[\begin{array}{l@{\ \ }l@{\ }l@{\ }l@{\ \ }l@{\ }l@{\ }l@{\ \ }l}
 \{ &
   \op[open]  &:& F(\tystr,\tyunit,Q0,Q1),  &
   \op[close] &:& F(\tyunit,\tyunit,Q1,Q0), \\ &
   \op[read]  &:& F(\tyunit,\tystr,Q1,Q1), &
   \op[write] &:& F(\tystr,\tyunit,Q1,Q1)
  & \} ~.
  \end{array}
\]
Note that the state transitions represented in $\Sigma$ are matched with those in
the automaton.
Let
\[
 S(Q_\mathrm{pre},Q_\mathrm{post}) \,\defeq\, \tyctlMB{(\tyrfn{x}{\tyint}{x = Q_\mathrm{post}} \rarr C)}{(\tyrfn{x}{\tyint}{x = Q_\mathrm{pre}} \rarr C)} ~.
\]
Given an effect handler $h$ conforming to $\Sigma$ and
a computation $c$ with control effect $S(n_\mathrm{pre},n_\mathrm{post})$ for some $n_\mathrm{pre}$ and $n_\mathrm{post}$,
if a handling construct $\expwith{h}{c}$ is well typed, the body of $h$'s return clause is typed at
$\tyrfn{x}{\tyint}{x = n_\mathrm{post}} \rarr C$---i.e., the computation $c$
terminates at the state $n_\mathrm{post}$---and the handling construct
$\expwith{h}{c}$ is typed at $\tyrfn{x}{\tyint}{x = n_\mathrm{pre}} \rarr
C$---i.e., it requires $n_\mathrm{pre}$ as the initial state to start the computation $c$.
Therefore, if $n_\mathrm{pre} = n_\mathrm{post} = Q0$, then it is guaranteed
that the file operations are used in a valid manner.
Furthermore, even if $c$ is non-terminating, our type system can ensure that it
does not use the file operations in an invalid manner.
For example, suppose that $c$ is a computation $\op[close]~();\ \Omega$ where $\Omega$ is a
diverging computation. If it is well typed, its final answer type is
$\tyrfn{x}{\tyint}{x = Q1} \rarr C$, which indicates that $\expwith{h}{c}$
requires $Q1$ as the initial state.  It is clearly inconsistent with the
above automation.
As another instance, suppose that $c$ involves a computation $\cdots ;\ \op[close]~();\ \op[write]~"X"; \cdots$.
This is illegal because it tries to call $\op[write]$ after $\op[close]$ without
$\op[open]$.
Our type system rejects it because the initial answer type
$\tyrfn{x}{\tyint}{x = Q0} \rarr C$ of $\op[close]~()$ is not matched with the
final answer type $\tyrfn{x}{\tyint}{x = Q1} \rarr C$ of $\op[write]~"X"$ while
they must be matched for the computation to be well typed.

We end this section by showing that the example function $v$ can be typed at
$\tystr \rarr \tycomp{\Sigma}{\tyunit}{S(Q0,Q0)}$,
which means that $v$'s body uses the file operations appropriately.
Note that, for any typing context $\Gamma'$ and file operation $\op$,
if
$\op : F(T_1,T_2,n_\mathrm{pre},n_\mathrm{post}) \in \Sigma$ and
$\jdty{\Gamma'}{v}{T_1}$,
then
$\jdty{\Gamma'}{\op~v}{\tycomp{\Sigma}{T_2}{S(n_\mathrm{pre},n_\mathrm{post})}}$
by \rulename{T-Op}.
Let $\Gamma \defeq x: \tystr$.
For the inner while-loop construct, we have the following typing derivation:
\begin{prooftree}
     \AxiomC{
       \stackanchor{
         $\jdty{\Gamma}{ \op[read]() }{ \tycomp{\Sigma}{\tystr}{S(Q1,Q1)} }$
       }{
         $\jdty{\Gamma,y:\tystr}{ \op[write]~(y\texttt{\textasciicircum"X"}) }{ \tycomp{\Sigma}{\tyunit}{S(Q1,Q1)} }$
       }
     }
     \RightLabel{\rulename{T-LetIp}}
   \UnaryInfC{$\jdty{\Gamma}{ \explet{y}{\op[read]()}{\op[write]~(y\texttt{\textasciicircum"X"})} }{ \tycomp{\Sigma}{\tyunit}{S(Q1,Q1)} }$}
   \RightLabel{\rulename{T-Loop}}
 \UnaryInfC{$\jdty{\Gamma}{ \expwhile{\star}{\explet{y}{\op[read]()}{\op[write]~(y\texttt{\textasciicircum"X"})}} }{ \tycomp{\Sigma}{\tyunit}{S(Q1,Q1)} }$}
\end{prooftree}
Thus, the sub-computations of the outer while-loop construct can be typed as follows:
\[\begin{array}{r@{\ }l@{\ }l}
 \jdty{\Gamma&}{\op[open]~x&}{\tycomp{\Sigma}{\tyunit}{S(Q0,Q1)}}
 \\
 \jdty{\Gamma&}{\expwhile{\star}{\explet{y}{\op[read]()}{\op[write]~(y\texttt{\textasciicircum"X"})}}&}{\tycomp{\Sigma}{\tyunit}{S(Q1,Q1)}}
 \\
 \jdty{\Gamma&}{\op[close]~()&}{\tycomp{\Sigma}{\tyunit}{S(Q1,Q0)}} ~.
  \end{array}
\]
where the control effects express how the state changes according to the
operation calls.
By \rulename{T-LetIp}, \rulename{T-Loop}, and \rulename{T-Fun}, they then imply that $v$ is typed at
$\tystr \rarr \tycomp{\Sigma}{\tyunit}{S(Q0,Q0)}$ as desired.

\subsection{Discussion} \label{sec:language/discussions}
In this section, we discuss the current limitations and future extensions of our system.

\subsubsection{Abstraction of Effects}
Our type system has no mechanism for abstraction of effects.
Therefore, if we cannot know possible effects of the handled computation in advance
(e.g., as in $\lambda f. \expwith{h}{(f~())}$, where the effects of the handled computation $f~()$ are determined by function parameter $f$),
we have to fix its effects (both the operation signature and the control effect).
A possible way to address this issue is to incorporate some mechanism to abstract effects.
For operation signatures, effect polymorphism as in the existing effect systems
for algebraic effects and handlers~\cite{Leijen17,Lindley17}, is a promising solution.
However, adapting it to our system is not trivial.
Effect polymorphism enables specifying a part of an operation signature as a
parameter, and handling constructs implicitly forward operations in the
parameter.
The problem is that \emph{our type system modifies the type schemes of forwarded
operations} (see the supplementary material for detail). Therefore, even though
the type schemes are involved in an operation signature parameter, we need to
track how they are modified.
We leave addressing this challenge for future work.
For control effects, we conjecture that bounded polymorphism can be used
to abstract control effects while respecting the necessary sub-effecting constraints.

\subsubsection{Combination with Other Computational Effects}
Algebraic effects and handlers are sometimes used with other computational effects.
For example, when implementing a scheduler with algebraic effects and handlers,
an imperative queue is often used to keep suspended continuations, like in an example from \citet{MulticoreOCamlrepo}.
Even though some computational effects can be simulated
by algebraic effects and handlers themselves,
it is often convenient to address them as primitive operations for efficiency.
Our system does not support such primitive computational effects.
It is left for future work to combine these features in one system.

\subsubsection{Shallow Handlers}
The handlers we adopt in this work are called \emph{deep handlers}~\cite{Kammar13},
which is the most widely used variant.
Another variant of algebraic effect handlers is \emph{shallow handlers}~\cite{Hillerstrom18},
which
we do not address in the present work.
Just as deep handlers are related to shift0/reset0,
shallow handlers are related to control0/prompt0~\cite{Pirog19}.
Therefore, the type system for control0/prompt0 with ATM~\cite{Ishio22} may be adapted
to develop a refinement type system for shallow handlers,
as we have developed our refinement type system for deep handlers
based on the type systems for shift0/reset0 with ATM~\cite{Materzok11, Sekiyama23}.

\subsubsection{Recursive Computation Types}
Some useful programs with algebraic effect handlers are ill typed in our system due to the lack of support for recursive computation types.
For example, consider the following program:
\begin{align}
    \exprec{f}{n}{
        \expwith{
            h
        }
        {\expif{n = 0}{\op[Err]~\text{\texttt{"error"}}}{f~(n - 1)}}
    }
\end{align}
where $h \defeq \{ \op[Err](msg, k) \mapsto
\op[Err]~(\text{\texttt{sprintf "called at \%d. \%s"}}~n~msg) \}$~.
This recursive function handles the error in each function call,
producing its own stack trace.
It cannot be typed without recursive types
because the type of the handled computations appears recursively as its answer type.
To see this, assume that the type of the handled computation (i.e., the conditional branch)
is assigned a type $\tycomp{\Sigma}{T}{\tyctlMB{C_1}{C_2}}$
(here we consider only simple types for simplicity).
Then, the type of the handling construct (i.e., the body of the function) is $C_2$,
which implies that the overall function has type $\tyint \rarr C_2$.
And so, the recursive call to the function $f~(n - 1)$ also has type $C_2$.
Then, the type $C_2$ should be a subtype of $\tycomp{\Sigma}{T}{\tyctlMB{C_1}{C_2}}$
since $f~(n - 1)$ is the else-branch of the conditional branch.
However, we cannot derive $\jdsub{\Gamma}{C_2}{\tycomp{\Sigma}{T}{\tyctlMB{C_1}{C_2}}}$
(for some $\Gamma$) in our system
because while the type on the left-hand side is $C_2$ itself,
$C_2$ appears as the answer type in the control effect of the type on the right-hand side.
On the other hand, using recursive types, we can give this function the following type
(again, we consider only simple types for simplicity):
$\tyint \rarr \mu \alpha. \tycomp{\Sigma_\alpha}{T}{\tyctlMB{T}{\alpha}}$
where $\Sigma_\alpha \defeq \{ \op[Err] : \tystr \rarr (T \rarr T) \rarr \alpha \}$ and $T$ is an arbitrary value type.
Type $\mu \alpha. C$ denotes a recursive computation type
where the type variable $\alpha$ refers to the whole type itself.
The control effect of this type is recursively nested,
which reflects the fact that the handling construct is recursively nested
due to the recursive call to the function.

\subsubsection{Type Polymorphic Effect Operations}
Consider the following program that evaluates to $[[21]]$:
\begin{align}
    &\expwith{\{ x_r \mapsto x_r,
    \op[wrap]((), k) \mapsto [k~()]\}}
    {(\op[wrap]~(); \op[wrap]~(); 21)}
\end{align}
This does not type-check in our current system because a type polymorphic operation signature like $\Sigma \defeq \{ \op[wrap]: \forall \alpha. \tyunit \rarr (\tyunit \rarr \tycompMB{\alpha}{S}) \rarr \tycompMB{\alpha~\tylist}{S'} \}$ is required.  It is, however, easy to extend our type system to support type polymorphic operation signatures to handle such examples. Specifically, in the typing of operation clauses $c_i$ in the \rulename{T-Hndl} rule, one would generalize type variables, and in the \rulename{T-Op} rule, one would instantiate type polymorphism.

\subsubsection{Modularity (or Abstraction) versus Preciseness (or Concreteness)}
In our system, operation signatures are of the form $\op_i : T_i \rarr (T_i' \rarr C_i) \rarr C_i'$ where the types $C_i$ and $C_i'$ represent behavior of the effect handler. In other words, the signature reveals specific implementation details regarding effect handlers. This design, from our perspective of precise specification and verification, is valuable. Indeed, our type system can formally specify and verify the assume-guarantee-like contracts between the handler and operation-call sides.
However, from the perspective of modularity and abstraction, this design choice is not the optimal one. In fact, one of the purposes of effect handlers is to abstract away the specifics so that one could later choose a different implementation.

To ensure that the handler implementation details do not leak in the operation signatures, one can introduce computation type polymorphism: The types $C_i$ and $C_i'$ in the operation signature above will be replaced by computation type variables, thus hiding the details.
However, completely hiding the information of handler implementations in this way implies that we are not providing and verifying a detailed specification requirement for the handler implementations.

Practically speaking, rather than the two extremes, we believe that it is engineering-wise desirable to allow for a gradient between modularity (abstraction) and preciseness (concreteness) and to describe and verify types at the appropriate level of detail depending on the use case. Introducing all the polymorphisms discussed in this section might achieve this goal, but we plan to investigate whether it is indeed the case by specifying and verifying various real-world programs. In our view, the issue of how to describe types at an appropriate level of abstraction, as discussed above, is an important open problem not just for algebraic effects but for general control operators and, more broadly, for effectful computation.

\section{Implementation} \label{sec:impl}

\subsection{Description of Our Implementation} \label{sec:impl/impl}

In this section, we describe our prototype implementation of
a refinement type checking and inference system, \textsc{RCaml}\footnote{available at \url{https://github.com/hiroshi-unno/coar}}.
It takes a program written in a subset of the OCaml 5 language
(including algebraic data types, pattern matching, recursive functions, exceptions, mutable references\footnote{Strong updates~\cite{DBLP:conf/pldi/FosterTA02} are not supported. \label{footnote:strong-update}},
let-polymorphism, and effect handlers)
and a refinement-type specification for the function of interest.
It first (1) obtains an ML-typed AST of the program
using OCaml's compiler library,
(2) infers refinement-free operation signatures and control effects,
(3) generates refinement constraints for the program and its specification as Constrained Horn Clauses (CHCs) (see e.g., the work of \citet{Bjorner2015a}),
and finally (4) solves these constraints
to verify if the program satisfies the specification.
The steps (3) and (4), where the refinement type checking is reduced to CHC solving,
follow existing standard approaches such as those proposed by \citet{Rondon08} and \citet{Unno09}.
The inference of (refinement-free) operation signatures is similar to
that of record types using row variables,
and is mutually recursive with the inference of control effects.
It is based on the inference
of control effects for shift0/reset0~\cite{Materzok11}.
As we split the steps of CHC generation and
solving,
we can use different solvers as the backend CHC solver depending on benchmarks.
In this experiment, we used two kinds of CHC solvers:
\textsc{Spacer}~\cite{Komuravelli13} that is based on Property Directed Reachability (PDR)~\cite{Bradley11,Een11},
and \textsc{PCSat}~\cite{Unno2021} that is based on template-based CEGIS~\cite{Solar-Lezama06,Unno2021} with Z3~\cite{Moura2008} as an SMT solver.

Because inputs to the implementation are OCaml programs
that are type-checked by OCaml's type checker which does not allow ATM,
the underlying OCaml types corresponding to the answer types cannot be modified.
However, as remarked before in Section~\ref{sec:intro}, our aim is to verify \emph{existing} programs with algebraic effects and handlers,
and, as remarked before, our ARM, that allows only modification in the refinements, is useful for that purpose.

Our implementation supports several kinds of polymorphism.
In addition to the standard let-polymorphism on types,
it supports refinement predicate polymorphism.
The implementation extends the formal system by allowing {\em bounded} predicate polymorphism in which abstracted predicates can be bounded by constraints on them, and further allows predicate-polymorphic types to be assigned to let-bound terms.
However, because the implementation can infer predicate-polymorphic types only at let-bindings,
we used a different approach, which we will discuss in Section~\ref{sec:impl/eval},
to simulate predicate polymorphism in operation signatures.

Another notable point is that our implementation deals with operations and exceptions uniformly.
That is, exception raising is treated as an operation invocation
and it can be handled by a certain kind of effect handlers which have clauses for exceptions
(the exception clauses are included in the effect handlers of OCaml by default).

\subsection{Evaluation} \label{sec:impl/eval}

We performed a preliminary experiments to evaluate our method
on some benchmark programs that use algebraic effect handlers.
The benchmarks are based on example programs
from \citet{Bauer15} and the repository of the Eff language~\cite{Effrepo}.
We gathered the effect handlers in those examples
and created benchmark programs each of which uses one of the effect handlers.
We also added a refinement type specification of the function of interest to each benchmark.
(Other auxiliary functions are not given such extra information,
and so their types are \emph{inferred automatically} even for recursive functions.)
Most benchmarks could be solved automatically without the annotations,
but some need them as hints.
We discuss the details at the end of this section.
It is also notable that,
although the examples presented in Section~\ref{sec:language/exmaples} focus on
the specifications specialized in concrete, constant values such as
$\tyrfn{z}{\tyint}{z = 19}$ for Example~1, the benchmarks include programs that demonstrate that our type system and
implementation can address more general specifications.%
For instance, the specification for the benchmark \texttt{choose-max-SAT.ml},
which is a general version of Example 1
where the constants 10, 20, 1, and 2 are replaced by
parameters $u$, $v$, $x$, and $y$, respectively, of a function \texttt{main} to be verified,
is as follows:
\[
    \jdty{}{\mathtt{main}}{
        (u: \tyint) \rarr (v: \tyrfnshort{z \t: \tyint}{z \ge u})
        \rarr (x: \tyint) \rarr (y: \tyrfnshort{z \t: \tyint}{z \ge x})
        \rarr \tyrfnshort{z \t: \tyint}{z = v - x}
    }
\]
We refer to the supplementary material
for the source code and the specifications of our benchmarks.
All the experiments were conducted on
Intel Xeon Platinum 8360Y, 256~GB RAM.

\begin{table}
    \caption{Evaluation results}
    \label{tab:eval}
    \footnotesize
    \begin{tabular}{lcrcr}
        \toprule
        \multirow[c]{2}{*}{file name} & \multicolumn{2}{c}{\textsc{Spacer}} & \multicolumn{2}{c}{\textsc{PCSat}} \\
        & result correct? & time (sec.) & result correct? & time (sec.) \\
        \midrule
        \texttt{amb-1-SAT.ml} & Yes & 0.55 & Yes & 15.30 \\
        \texttt{amb-1-UNSAT.ml} & Yes & 0.72 & Yes & 63.62 \\
        \texttt{amb-2-SAT.ml} & Yes & 2.31 & Yes & 31.48 \\
        \texttt{amb-2-UNSAT.ml} & Yes & 2.26 & - & timeout$^\dagger$ \\
        \texttt{amb-3-SAT.ml} & Yes & 3.20 & Yes & 182.41 \\
        \texttt{amb-3-simpl-SAT.ml} & Yes & 1.71 & Yes & 16.79 \\
        \texttt{bfs-SAT.ml} & No$^{*1}$ & 1.67 & - & timeout$^{*1}$ \\
        \texttt{bfs-UNSAT.ml} & Yes & 2.00 & - & timeout$^\dagger$ \\
        \texttt{bfs-simpl-SAT.ml} & No$^{*1}$ & 2.22 & - & timeout$^{*1}$ \\
        \texttt{choose-all-SAT.ml} & Yes & 16.23 & - & timeout$^\dagger$ \\
        \texttt{choose-all-UNSAT.ml} & Yes & 12.56 & - & timeout$^\dagger$ \\
        \texttt{choose-max-SAT.ml} & Yes & 23.08 & - & timeout$^\dagger$ \\
        \texttt{choose-max-UNSAT.ml} & Yes & 15.97 & - & timeout$^\dagger$ \\
        \texttt{choose-sum-SAT.ml} & Yes & 1.54 & - & timeout$^\dagger$ \\
        \texttt{choose-sum-UNSAT.ml} & Yes & 7.99 & Yes & 15.00 \\
        \texttt{deferred-1-SAT.ml} & Yes & 0.46 & Yes & 4.49 \\
        \texttt{deferred-1-UNSAT.ml} & Yes & 0.27 & Yes & 4.09 \\
        \texttt{deferred-2-SAT.ml} & Yes & 0.43 & Yes & 4.38 \\
        \texttt{distribution-SAT.ml} & Abort$^\div$ & - & - & timeout$^{*2}$ \\
        \texttt{distribution-UNSAT.ml} & Abort$^\div$ & - & - & timeout$^{*2}$ \\
        \texttt{expectation-SAT.ml} & Yes & 0.51 & Yes & 7.25 \\
        \texttt{expectation-UNSAT.ml} & Yes & 1.45 & Yes & 7.33 \\
        \texttt{io-read-1-SAT.ml} & Yes & 0.43 & Yes & 13.90 \\
        \texttt{io-read-1-UNSAT.ml} & Yes & 0.41 & Yes & 12.21 \\
        \texttt{io-read-2-SAT.ml} & Yes & 0.56 & Yes & 21.10 \\
        \texttt{io-read-3-SAT.ml} & Yes & 0.54 & Yes & 14.88 \\
        \texttt{io-write-1-SAT.ml} & Yes & 0.32 & Yes & 8.48 \\
        \texttt{io-write-1-UNSAT.ml} & Yes & 0.32 & Yes & 8.76 \\
        \texttt{io-write-2-SAT.ml} & Yes & 0.46 & Yes & 11.33 \\
        \texttt{io-write-2-UNSAT.ml} & Yes & 0.68 & Yes & 11.65 \\
        \texttt{modulus-SAT.ml} & Yes & 14.23 & Yes & 11.89 \\
        \texttt{modulus-UNSAT.ml} & Yes & 26.56 & Yes & 11.91 \\
        \texttt{queue-1-SAT.ml} & Yes & 0.78 & Yes & 19.22 \\
        \texttt{queue-1-UNSAT.ml} & Yes & 0.52 & Yes & 16.93 \\
        \texttt{queue-2-SAT.ml} & Yes & 0.89 & Yes & 22.63 \\
        \texttt{round-robin-SAT.ml} & Yes & 0.96 & - & timeout$^\dagger$ \\
        \texttt{round-robin-UNSAT.ml} & Yes & 0.73 & - & timeout$^\dagger$ \\
        \texttt{safe-div-1-SAT.ml} & Abort$^\div$ & - & Yes & 2.71 \\
        \texttt{safe-div-1-UNSAT.ml} & Abort$^\div$ & - & Yes & 2.73 \\
        \texttt{safe-div-2-SAT.ml} & Abort$^\div$ & - & Yes & 2.55 \\
        \texttt{safe-div-2-UNSAT.ml} & Abort$^\div$ & - & Yes & 3.58 \\
        \texttt{select-SAT.ml} & - & timeout$^{\ddagger}$ & Yes & 13.28 \\
        \texttt{select-UNSAT.ml} & - & timeout$^{\ddagger}$ & Yes & 13.26 \\
        \texttt{shift-SAT.ml} & Yes & 0.28 & Yes & 2.92 \\
        \texttt{shift-UNSAT.ml} & Yes & 1.25 & Yes & 3.93 \\
        \texttt{state-SAT.ml} & - & timeout$^{\ddagger}$ & Yes & 33.69 \\
        \texttt{state-UNSAT.ml} & Yes & 0.63 & Yes & 13.56 \\
        \texttt{state-easy-SAT.ml} & Yes & 0.90 & Yes & 35.54 \\
        \texttt{transaction-SAT.ml} & - & timeout$^{\ddagger}$ & Yes & 15.36 \\
        \texttt{transaction-UNSAT.ml} & - & timeout$^{\ddagger}$ & Yes & 15.77 \\
        \texttt{yield-SAT.ml} & Yes & 1.51 & Yes & 17.57 \\
        \texttt{yield-UNSAT.ml} & Yes & 1.52 & - & timeout$^\dagger$ \\
        \bottomrule
    \end{tabular}
\end{table}

Table~\ref{tab:eval} shows the results of the evaluation.
The files that are suffixed with \texttt{-SAT} are expected to result in ``SAT'',
that is, the programs are expected to be typed
with the refinement types given as their specifications.
The other files (suffixed with \texttt{-UNSAT}) are expected to result in ``UNSAT'',
that is, the programs are expected not to be typed
with the given refinement types.
For each program, we conducted verification in two configurations
((1) \textsc{Spacer}, and (2) \textsc{PCSat}).
The field ``time'' indicates the time spent in the whole process of the verification.
We set the timeout to 600 seconds.
Our implementation successfully answered correct result for most programs.
For instance, we show the benchmark \texttt{io-write-2-SAT.ml} as an example
(where \texttt{@annot\_MB} is an effect annotation written in the underlying OCaml type,
explained in the last paragraph of this section):
\begin{verbatim}
let[@annot_MB "(unit -> ({Write: s} |> unit / s3 => s3)) -> unit * int list"]
  accumulate (body: unit -> unit) = match_with body () {
    retc = (fun v -> (v, []));  exnc = raise;
    effc = fun (type a) (e: a eff) -> match e with
      | Write x -> Some (fun (k: (a, _) continuation) ->
        let (v, xs) = continue k () in (v, x :: xs) ) }
let write_all l = accumulate (fun () ->
  let rec go li = match li with
    | [] -> () | s :: ss -> let _ = perform (Write s) in go ss
  in go l )
\end{verbatim}
It iterates over a list \texttt{l} to pass its elements to the operation \texttt{Write},
and the handler for \texttt{Write} accumulates the passed elements into another list.
It is checked against the following specification:
\[
    \jdty{}{\mathtt{write\_all}}{
        \tyrfn{z}{\kwty{int~list}}{z \ne []} \rarr
        \tyrfn{z}{\tyunit \times \kwty{int~list}}{\forall u, v.\, z = (u, v) \Rarr v \ne []}
    }
\]
That is, if the iterated list is not empty, the accumulated list is not, either.
Our implementation successfully answered that \texttt{write\_all} satisfies the specification,
with the following inferred type:
\begin{align}
    (l: \tyrfn{z}{\kwty{int~list}}{z \ne []}) \rarr
    \tyrfn{z}{\tyunit \times \tyrfn{z'}{\tyint}{l \ne []}~\kwty{list}}{\phi}
\end{align}
where $\phi \defeq \exists t: \kwty{int~list}. (t = [] \lor z.2 \ne []) \land t \ne [] \land l \ne []$ and $z.2$ means the second element of the pair $z$.
ARM is indispensable for this example because
the initial answer type of the body of the function \texttt{go} should be
$\tyrfn{z}{\tyunit \times \kwty{int~list}}{z.2 = []}$
(since it should be matched with the type of the return clause of the handler),
while its final answer type should be
$\tyrfn{z}{\tyunit \times \kwty{int~list}}{z.2 \ne []}$~.
We also present another interesting example (\texttt{queue-2-SAT.ml}) in detail
in the supplementary material.

The benchmarks that were not verified correctly in both configurations are
\texttt{bfs(-simpl)-SAT.ml} (marked with $*1$)
and \texttt{distribution-(UN)SAT.ml} (marked with $*2$).
They need some specific features which the implementation does not support.
The formers need
an invariant which states that there exists an element of a list
that satisfies a certain property.
The latter needs recursive predicates
in the type of an integer list, which states a property about the sum of the elements of the list.
These issues are orthogonal to the main contributions of this paper;
they are about the expressiveness of the background theory used for refinement predicates, to which our novel refinement type system is agnostic.
Also, \texttt{bfs(-simpl)-SAT.ml} uses mutable references
which our implementation does not handle in a flow-sensitive manner
(as mentioned in the footnote~\ref{footnote:strong-update}).
One solution to this issue is to encode references with an effect handler as in Section~\ref{sec:overview/algeff},
but our implementation does not do such encoding automatically. More advanced support for native effects including references is left for future work, as discussed in Section~\ref{sec:language/discussions}.

We discuss pros and cons between the two configurations.
First, \textsc{Spacer} does not support division operator,
and so it cannot verify some programs that use division (marked with $\div$,
aborting with the message ``\texttt{Z3 Error: Uninterpreted 'div' in <null>}'').
Also, some programs can be solved in one configuration but not in the other.
Among those solved by \textsc{Spacer} but not by \textsc{PCSat} (marked with $\dagger$),
\texttt{round-robin-(UN)SAT.ml} timed out during the simplification of its constraints.
For the remaining programs, their constraints tend to contain predicate variables
that take a large number of arguments,
which makes it hard for \textsc{PCSat} to find solutions.
Conversely, the programs solved by \textsc{PCSat} but not by \textsc{Spacer} (marked with $\ddagger$)
involve constraints where some predicate variables occur many times,
which leads to complicated solutions that are difficult for \textsc{Spacer} to solve.

It is worth noting that
our benchmarks do not rely on refinement type annotation in most places,
even for recursive functions and recursive ADTs.
However, a few kinds of annotations are still needed.
First, as mentioned in Section~\ref{sec:language/discussions},
our type system does not support effect polymorphism.
Therefore, we added effect annotations to function-type arguments
which may perform operations when executed, as the one given to the benchmark \texttt{io-write-2-SAT.ml} using \texttt{@annot\_MB}.
These annotations are written in the underlying OCaml types,
that is, we did not specify concrete refinements in the annotations.
Second,
we provided refinement type annotations for two small parts of \texttt{state-SAT.ml},
because otherwise it could not be verified within the timeout period in both configurations.
Third, because our implementation
infers predicate-polymorphic types only at let-bindings,
we added \emph{ghost parameters} to some operations and functions
to
infer precise refinement types of them which are not let-bound
but need some abstraction of refinements.
Ghost parameters are parameters which are used to express dependencies in dependent type checking,
but have no impact on the dynamic execution of the program so they can be removed at runtime.
In automated verification, completely inferring predicate variables requires
higher-order predicate constraints, which are not expressible with CHC.
Therefore, we provided ghost parameters
to make it possible to reduce the verification to CHC solving.
For example, the following is a part of \texttt{state-SAT.ml}:
\begin{verbatim}
let rec counter c =
    let i = perform (Lookup c) in
    if i = 0 then c else (perform (Update (c, i - 1)); counter (c + 1))
in counter 0
\end{verbatim}
which is handled by a handler that simulates a mutable reference
similar to that of Example 2 in Section~\ref{sec:language/examples/state}.
Here, we pass the variable $c$ to the operation \texttt{Lookup} and \texttt{Update}
as the ghost parameter.
In the formal system presented in Section~\ref{sec:language/type-system}
where predicate polymorphism is available in operation signatures,
we can give \texttt{Update} the type
\begin{align}
    \forall X\t: (\tyint, \tyint).\,
        (x\t: \tyint) \rarr (\tyunit &\rarr ((s\t: \tyint) \rarr \tyrfnshort{z\t:\tyint}{X(z, s)}))
    \rarr ((s\t: \tyint) \rarr \tyrfnshort{z\t:\tyint}{X(z, x)})
\end{align}
in the same way as Example 2 in Section~\ref{sec:language/examples/state},
and instantiate the predicate variable $X$ with $\lambda (z, s). z = c + 1 + s$
to correctly verify \texttt{state-SAT.ml}.
On the other hand, in the implementation, since predicate polymorphism is not available in operation signatures,
the handler needs to know the concrete predicate which replaces $X$.
However, the predicate contains $c$, which the handler cannot know
without receiving some additional information.
Therefore, we need to add the ghost parameter $c$ to \texttt{Update}
(and the same for \texttt{Lookup}).
This time we added them manually,
but one possible approach for automating insertion of ghost parameters is
to adopt the technique proposed by \citet{Unno13}.
We conjecture that a similar technique can be used
for our purpose.

\section{CPS Transformation} \label{sec:cps}

\subsection{Definitions and Properties} \label{sec:cps/def}

This section presents the crux of our CPS transformation that translate the
language defined in Section~\ref{sec:language} to a $\lambda$-calculus without
effect handlers.
Readers interested in the complete definitions of the target language and the
CPS transformation are referred to the supplementary material.

\begin{figure}
 \raggedright
 \textbf{Evaluation, Typing and subtyping rules} \quad \fbox{$c \eval c'$} \ \fbox{$\jdty{\Gamma}{c}{\tau}$} \ \fbox{$\jdsub{\Gamma}{\tau_1}{\tau_2}$}
    \begin{gather}
        {(c: \tau) \eval c}
        {}
        \quad\ \ %
        {(\Lambda \alpha. c)~\tau \eval c[\tau/\alpha]}
        {}
        \quad\ \ %
        {(\Lambda \rep{X: \rep{B}}. c)~\rep{A} \eval c[\rep{A/X}]}
        {}
        \quad\ \ %
        {\{ \repi{\op_i = v_i} \}\#\op_i \eval v_i}
        {}
        \\
        \infer{\jdty{\Gamma}{c~\tau}{\tau'[\tau/\alpha]}}
        {
            \jdty{\Gamma}{c}{\forall \alpha. \tau'} &
            \jdwf{\Gamma}{\tau}
        }
        \qquad
        \infer{\jdsub{\Gamma}{\{ \repi{\op_i: \tau_{1i}}, \repi{\op'_i: \tau'_i} \}}{\{ \repi{\op_i:  \tau_{2i}} \}}}
        {\repi{\jdsub{\Gamma}{\tau_{1i}}{\tau_{2i}}}}
        \\[1ex]
        \infer{\jdsub{\Gamma}{\forall \alpha. \tau_1}{\forall \beta. \tau_2}}
        {
            \jdsub{\Gamma, \beta}{\tau_1[\tau/\alpha]}{\tau_2} &
            \jdwf{\Gamma, \beta}{\tau}&
            \beta \notin \fv(\forall \alpha. \tau_1)
        }
    \end{gather}
    \caption{The operational semantics and the type system of the target language (excerpt).}
    \label{fig:cps-target-excerpt}
\end{figure}

The target language of the CPS transformation is a polymorphic
$\lambda$-calculus with records and recursion.
Its program and type syntax are defined as follows:
\[\begin{array}{rcl}
 v &::=& x \mid p \mid \exprec{f:\tau_1}{x:\tau_2}{c} \mid \Lambda \rep{X: \rep{B}}. c \mid \{ \repi{\op_i = v_i} \} \mid \Lambda \alpha. c \\
 c &::=& v \mid c~v \mid \expif{v}{c_1}{c_2} \mid c~\rep{A} \mid v\#\op \mid c~\tau \mid (c : \tau) \\
 \tau &::=& \tyrfn{x}{B}{\phi} \mid (x: \tau_1) \rarr \tau_2 \mid \forall \rep{X: \rep{B}}. \tau
        \mid \{ \repi{\op_i : \tau_i} \} \mid \alpha \mid \forall \alpha. \tau
  \end{array}
\]
In the target language, values are not strictly separated from computations as
those in the source language; for example, functions in function applications
can be computations.
The metavariables $\alpha$ and $\beta$ range over type variables.  Expressions
$\Lambda \alpha. c$ and $c~\tau$ are a type abstraction and application,
respectively.
Type polymorphism is introduced to express the pure control effect in the target
language using \emph{answer type polymorphism}~\cite{Thielecke03}.
Expressions $\{ \repi{\op_i = v_i} \}$ and $v\#\op$ are a record literal and
projection, respectively.
We use operation names as record labels for the target language to encode
handlers using records.
Our CPS transformation produces programs with type annotations for proving
bidirectional type-preservation. Recursive functions with type annotations and
type ascriptions $(c : \tau)$ are used to annotate programs.
We abbreviate $\exprec{f:T_1}{x:T_2}{c}$ to $\lambda x: T_2. c$ if $f$ does not
occur in $c$.
Types are defined in a standard manner.
Typing contexts $\Gamma$ are extended to include type variables.
The operational semantics is almost standard.
Figure~\ref{fig:cps-target-excerpt} shows four evaluation rules.
Type ascriptions simply drop the ascribed type $\tau$.
Type applications substitute a given type $\tau$ for the bound type variable $\alpha$.
Predicate applications are similar.
Record projections with $\op_i$ extract the associated field $v_i$.
The type
system is also standard,
presented in Figure~\ref{fig:cps-target-excerpt}.  We write $\jdwf{\Gamma}{\tau}$ to
state that all the free variables (including type and predicate ones) in
the type $\tau$ are bound in the typing context $\Gamma$.
The subtyping for record types allows supertypes to forget some fields in
subtypes, and the types of each corresponding field in two record types to be in
the subtyping relation (we deem record types, as well as records, to be
equivalent up to permutation of fields).
The subtyping rule for type polymorphism is a weaker variant of the containment
rule for polymorphic types~\cite{Mitchell88}.
It is introduced to emulate \rulename{S-Embed} in the target language.

\begin{figure}
 \[\begin{array}{rcl}
        \cps{\tyrfn{x}{B}{\phi}} \ \ \defeq \ \ \tyrfn{x}{B}{\phi} \quad && \quad
        \cps{(x: T) \rarr C} \ \ \defeq \ \ (x: \cps{T}) \rarr \cps{C} \\
        \cps{\tycomp{\Sigma}{T}{\tyctl{x}{C_1}{C_2}}} &\defeq&
            \forall \_. \cps{\Sigma} \rarr ((x: \cps{T}) \rarr \cps{C_1}) \rarr \cps{C_2} \\
        \cps{\tycomp{\Sigma}{T}{\square}} &\defeq&
            \forall \alpha. \cps{\Sigma} \rarr (\cps{T} \rarr \alpha) \rarr \alpha \\
        \cps{\{ \repi{\op_i : \forall \rep{X_i: \rep{B}_i}. F_i} \}} &\defeq&
            \{ \repi{\op_i : \forall \rep{X_i: \rep{B}_i}. \cps{F_i}^\mathcal{F}} \} \\
        \cps{(x: T_1) \rarr ((y: T_2) \rarr C_1) \rarr C_2}^\mathcal{F} &\defeq&
            (x: \cps{T_1}) \rarr \cps{((y: T_2) \rarr C_1)} \rarr \cps{C_2}
   \end{array}
 \]
    \begin{align}
        \cps{(\op^{\rep{\mathit{A}}}~v)^{\tycomp{\Sigma}{T}{\tyctl{y}{C_1}{C_2}}}} &\defeq
            \stLambda \alpha. \stlambda h:\cps{\Sigma}. \stlambda k:(y: \cps{T} \rarr \cps{C_1}).
            h\#\op~\rep{A}~\cps{v}~(\lambda y': \cps{T}. k~y') \\
        \cps{(\expwith{h}{c})^{C}} &\defeq \cps{c} \stapp \cps{C} \stapp \cps{h^{\mathit{ops}}} \stapp \cps{h^{\mathit{ret}}} \\[-.5ex]
            \text{where} \hspace{-25pt} & \hspace{25pt} \left\{ \begin{aligned}
            h &= \{ \expret{x_r^{T_r}} \mapsto c_r, \repi{\op_i^{\rep{X_i: \rep{B_i}}}(x_i^{T_{x_i}}, k_i^{T_{k_i}}) \mapsto c_i} \} \\[-.5ex]
            \cps{h^{\mathit{ops}}} &\defeq
                \{ \repi{\op_i = \Lambda \rep{X_i: \rep{B_i}}. \lambda x_i:\cps{T_{x_i}}. \lambda k_i:\cps{T_{k_i}}. \cps{c_i}} \} \\[-.5ex]
            \cps{h^{\mathit{ret}}} &\defeq
                \lambda x_r:\cps{T_r}. \cps{c_r}
        \end{aligned} \right.
    \end{align}
    \caption{CPS transformation of types and expressions (excerpt).}
    \label{fig:cps-trans-excerpt}
\end{figure}

We show the key part of the CPS transformation in
Figure~\ref{fig:cps-trans-excerpt}.
The upper half presents the transformation of types.
The transformation of value types is straightforward.  Operation signatures are
transformed into record types, which means that operation clauses in a handler
are transformed into a record.
The transformations of computation types indicate that computations are
transformed into functions that receive two value parameters: handlers and
continuations.
If the control effect is pure, the answer types of computations become
polymorphic in CPS.
This treatment of control effects is different from that of \citet{Materzok11},
who define CPS transformation for control effects in the
simply typed setting.
Their CPS transformation transforms, in our notation,
a computation type $\tycompMB{T}{\square}$ into the type $\cps{T}$, and
a type $\tycompMB{T}{\tyctlMB{C_1}{C_2}}$ into the type $\tyfunshort{(\tyfunshort{\cps{T}}{\cps{C_1}})}{\cps{C_2}}$ (note that they address neither operation signatures nor dependent typing).
Because the latter takes continuations whereas the former does not, CPS
transformation needs to know where pure computations are converted into impure
ones (via subtyping).
To address this issue, Materzok and Biernacki's CPS transformation focuses on
typing derivations in the source language rather than expressions.
However, because our aim is at reducing the typing of programs with
algebraic effects and handlers to that of programs without them,
we cannot assume typing derivations in the source language to be available.
By treating two kinds of control effects uniformly using answer type
polymorphism, our CPS transformation can focus only on expressions (with type
annotations).

The lower half of Figure~\ref{fig:cps-trans-excerpt} shows the key cases of the
transformation of expressions.
We separate abstractions and applications in the target language
into \emph{static} and \emph{dynamic} ones, as in the work of \citet{Hillerstrom17},
for proving the preservation of the operational semantics
(Theorem~\ref{thm:cps-sim}).
Redexes represented by static applications are known as \emph{administrative redexes}, inserted and reduced at compile (CPS-transformed) time.
By contrast, redexes represented by dynamic applications are reduced at run time because they originate in the source program.
Constructors for static expressions are denoted by the overline notation,
like $\stlambda$, $\stLambda$, and $\stapp$.
We use the ``at'' symbol explicitly as an infix operator of static applications
for clarification.
Non-overlined abstractions and applications are dynamic ones,
which are treated as ordinary expressions.
Also, for backward type-preservation (Theorem~\ref{thm:cps-backward-excerpt}), we
extend the source language with type annotations.
For example, in an operation call $(\op^{\rep{\mathit{A}}}~v)^{\tycomp{\Sigma}{T}{\tyctl{y}{C_1}{C_2}}}$,
$\rep{\mathit{A}}$ are predicates used to instantiate the type scheme of the operation $\op$,
and
$\tycomp{\Sigma}{T}{\tyctl{y}{C_1}{C_2}}$ is the type of the operation call $\op~v$.
Without type annotations, CPS-transformed expressions may have a type that
cannot be transformed back to a type in the source language.
An operation call $(\op^{\rep{\mathit{A}}}~v)^{\tycomp{\Sigma}{T}{\tyctl{y}{C_1}{C_2}}}$
is transformed into a function that seeks the corresponding
operation clause in a given handler and then applies it to a given sequence of
predicates, argument, and continuation.
Note that the continuation is in the $\eta$-expanded form
because, for the preservation of the operational semantics,
we need a dynamic lambda abstraction that corresponds to the continuation $\expfun{y}{\expwith{h}{K[\expret{y}]}}$
introduced in the rule \rulename{E-HndlOp} of the source language.
An expression $\expwith{h}{c}$ is transformed into a function that applies the
CPS-transformed handled computation to the record of the CPS-transformed
operation clauses and the CPS-transformed return clause (because the return
clause works as the continuation of $c$).
The transformation preserves operational semantics bidirectionally in the following way:
\newcommand{\stappTop}{\stapp \tau \stapp \{\} \stapp (\lambda x: \tau. x)}
\begin{theorem}[simulation] \label{thm:cps-sim}
    Let $\equiv_\beta$ be the smallest congruence relation over expressions in the target language that satisfies $(\stlambda x: \tau. c) \stapp v \equiv_\beta c[v/x]$ and $(\stLambda \alpha. c) \stapp \tau \equiv_\beta c[\tau/\alpha]$.
    If\, $c \eval^* \expret{v}$, then
    $\cps{c} \stappTop \eval^+ v'$ for some $v'$ such that $\cps{v} \equiv_\beta v'$.
    Also, if\, $\cps{c} \stappTop \eval^+ v'$, then
    $c \eval^* \expret{v}$ and $\cps{v} \equiv_\beta v'$ for some $v$.
\end{theorem}
\noindent
(Note that $\tau$ can be any type since types are irrelevant to the operational semantics.)
The first half states that
if a computation $c$ in the source language evaluates to a value-return of $v$,
the transformed computation $\cps{c}$ applied to a type, an empty handler $\{\}$,
and a trivial continuation $\lambda x: \tau. x$ evaluates to the transformed value $\cps{v}$.
Similarly, the second half states the reverse direction.

Now, we state forward and backward type-preservation of the CPS transformation.
\begin{theorem}[Forward type-preservation] \label{thm:cps-forward-excerpt}
    The following holds:
        (1) If\, $\jdty{\Gamma}{v}{T}$ then $\jdty{\cps{\Gamma}}{\cps{v}}{\cps{T}}$.
        (2) If\, $\jdty{\Gamma}{c}{C}$ then $\jdty{\cps{\Gamma}}{\cps{c}}{\cps{C}}$.
\end{theorem}

\begin{theorem}[Backward type-preservation] \label{thm:cps-backward-excerpt}
    The following holds:
     (1) If\, $\jdty{\emptyset}{\cps{v}}{\tau}$, then
           there exists some $T$ such that
           $\jdty{\emptyset}{v}{T}$ and
           $\jdsub{\emptyset}{\cps{T}}{\tau}$.
     (2) If\, $\jdty{\emptyset}{\cps{c}}{\tau}$, then
           there exists some $C$ such that
           $\jdty{\emptyset}{c}{C}$ and
           $\jdsub{\emptyset}{\cps{C}}{\tau}$.
\end{theorem}
\noindent
Theorem~\ref{thm:cps-backward-excerpt} is implied immediately by backward type
preservation of the CPS transformation for \emph{open} expressions.
See the supplementary material for the statement for open expressions.
Theorem~\ref{thm:cps-backward-excerpt} indicates that it is possible to reduce
typechecking in our source language to that in a language without effect
handlers.
That is, if ones want to verify whether an expression $c$ has type $C$,
they can obtain the same result as the direct verification
by first applying CPS transformation to $c$ and $C$, and
then checking whether $\cps{c}$ has type $\cps{C}$
with a refinement type verification tool that does not support algebraic effect handlers.

Type annotations in the source language are necessary to restrict the image of the transformation.
Without them, a CPS-transformed program may be of a type $\tau$ that cannot be transformed to a type in the source language inversely (i.e., there exists no type $C$ in the source language satisfying $\cps{C} = \tau$).
For example, consider $\stLambda \alpha. \stlambda h. \stlambda k. k~0$,
the CPS form (without annotations) of expression $\expret{0}$.
Without annotations, we can pick arbitrary types as the type of $h$.
Therefore, it can have type
$\forall \alpha. \tybool \rarr (\tyint \rarr \alpha) \rarr \alpha$.
However, there is no type $C$ in the source language such that $\cps{C} = \forall \alpha. \tybool \rarr (\tyint \rarr \alpha) \rarr \alpha$.
Even worse, the source language has no type that is a \emph{subtype} of the type of the CPS form
since $\tybool$ and record types are incomparable with each other.
Another example is $\lambda x. \stLambda \alpha. \stlambda h. \stlambda k. k~x$,
the CPS form (again, without annotations) of expression $\lambda x. \expret{x}$.
Its type can be
$(\tyint \rarr \tyint) \rarr \forall \alpha. \{\} \rarr ((\tyint \rarr \tyint) \rarr \alpha) \rarr \alpha$,
that is, $x$ can be of type $\tyint \rarr \tyint$.
However, there is no value type $T$ in the source language
such that $\cps{T}$ is a subtype of $\tyint \rarr \tyint$.
Note that since a function type in the source language is in the form $(x: T_x) \rarr C$,
the right hand side of the arrow in the CPS-transformed function type must be in the form
$\forall \alpha. \{ \cdots \} \rarr \cdots$, which does not match with $\tyint$.
Therefore, without type annotations, Theorem~\ref{thm:cps-backward-excerpt} does not hold.

\newcommand{\er}{\mathit{er}}

While our formalization requires concrete refinement type annotations in the source language,
actually we can relax this restriction
by using predicate variables as placeholders instead of concrete refinements in type annotations.
This is because type annotations are only
for prohibiting occurrences of types with unintended \emph{structures}, not for restricting refinements.
Those predicate variables are instantiated after CPS transformation
with concrete predicates inferred by generating and solving CHC constraints
that contain these predicate variables from the CPS-transformed expression.
Formally, by allowing occurrences of predicate variables
in type annotations of both the source and target language,
and introducing predicate variable substitution $\sigma$,
we can state that $\cps{\sigma(c)} = \sigma(\cps{c})$.
This means that,
for an expression $c$ that is annotated with types containing predicate variables,
both of the followings result in the same expression:
(1) first instantiating the predicate variables in $c$ with concrete refinements,
and then CPS-transforming it (i.e., CPS-transforming the concretely-annotated expression),
and
(2) first CPS-transforming $c$,
and then instantiating the predicate variables in the CPS-transformed expression
with the concrete refinements.
In other words, concrete refinements are irrelevant to the CPS transformation.
This irrelevance is ensured by the fact that
refinements can depend only on first-order values because it means that handler variables $h$ and continuation variables $k$,
which occur only in CPS-transformed expressions, cannot be used in instantiated refinements.
The reason why we have defined the CPS transformation with concrete refinements
is just to state Theorem~\ref{thm:cps-forward-excerpt} and Theorem~\ref{thm:cps-backward-excerpt}.

\subsection{Comparison between the Direct Verification and the Indirect Verification} \label{sec:cps/comparison}

In this section, we compare the direct verification using our refinement system
presented in Section~\ref{sec:language}
with the indirect verification via the CPS transformation presented above.
One of the differences is that the direct verification requires
special support of verification tools for algebraic effect handlers,
while the indirect one can be done by existing tools without such support.
On the other hand, the indirect verification has some disadvantages.
First, in most cases, CPS-transformed programs tend to be complicated and be in
the forms quite different from the source programs.  This complexity incurred in
the indirect typechecking may lead to confusing error messages when the
typechecking fails.
Transforming the inferred complex types back to the types of the source language would be helpful,
but it is unclear whether we can do this
because the inferred types of the CPS-transformed expressions do not necessarily correspond to
the CPS-transformed types of the source expressions,
as stated in Section~\ref{sec:cps/def}.
By contrast, because the direct typechecking deals with the
structures of the source programs as they are, error messages can be made more
user-friendly.
Second, our CPS transformation needs a non-negligible amount of type
annotations---type annotations are necessary in let-expressions, conditional
branches, and recursive functions as well as operation calls and handling
constructs.  In practice, it is desired to infer as many types as possible.
However, it seems quite challenging to define a CPS transformation that
enjoys backward type-preservation and needs no, or few, type annotations.
One of the possible approaches for addressing type annotations in more automated way is
to use the underlying simple type system of our refinement type system
for algebraic effect handlers.
As mentioned in Section~\ref{sec:cps/def},
concrete refinements are not necessary for type annotations.
Therefore, we can generate type annotations for an expression
using its simple type inferred by the underlying type system.

We also compare these two approaches based on an experiment.
We used some direct style (DS) programs
(i.e., programs using algebraic effect handlers),
and for each program, we applied our CPS transformation manually
and ran the verification on both DS one and CPS one.
Additionally, we also compared them with optimized CPS programs
where administrative redexes were reduced.
We used the same implementation as the one in Section~\ref{sec:impl} with the configuration of \textsc{Spacer}.
We added annotations of source programs to only top-level, closed first-order expressions,
but the correctness of the verification can be justified by the preservation of dynamic semantics.

\begin{wraptable}[11]{R}{0.6\textwidth}
    \setlength{\arraycolsep}{2pt}
    \vspace*{-1.8ex}
    \caption{Evaluation results of CPS transformation}
    \label{tab:cps}
    \begin{tabular}{lcrcrcr}
        \toprule
        \multirow[c]{2}{*}{program} & \multicolumn{2}{c}{DS} & \multicolumn{2}{c}{CPS} & \multicolumn{2}{c}{CPS (opt)} \\
        & \checkmark? & time & \checkmark? & time & \checkmark? & time \\
        \midrule
        \texttt{amb-2} &Yes & 1.30 & Yes & 1.32 & Yes & 0.91 \\
        \texttt{choose-easy} &Yes & 0.26 & Yes & 0.27 & Yes & 0.22 \\
        \texttt{choose-sum} &Yes & 2.18 & Yes & 1.79 & Yes & 12.87 \\
        \texttt{io-read-2} &Yes & 0.66 & No & 1.29 & No & 0.62 \\
        \texttt{simple} &Yes & 0.11 & Yes & 0.16 & Yes & 0.14 \\
        \bottomrule
    \end{tabular}
\end{wraptable}

Table~\ref{tab:cps} shows the results of the experiment.
The columns ``\checkmark?'' show whether the verification result is correct.
The columns ``time'' are in seconds.
Some programs have no big difference in verification time
among the three variants,
but there are two notable things.
First, optimized CPS version of \texttt{choose-sum} took more time than the other versions.
This seems because the size of the program became larger by the optimization.
The CPS \texttt{choose-sum} program contains some branching expressions
and each branch uses variables representing its continuation and the outer handler.
By reducing administrative redexes in the program, these variables are instantiated
with a concrete continuation and handler,
that is, the continuation and handler are copied to each branch,
which results in larger size of the program and its constraints generated during the verification.
Second, CPS version of \texttt{io-read-2} could not be verified correctly.
One possible reason is lack of support for higher-order predicate polymorphism.
Since CPS programs explicitly pass around continuations,
their types tend to be higher-order.
Then, in some cases, higher-order predicate polymorphism becomes necessary by CPS transformation.

\section{Related Work} \label{sec:related}

\subsection{Algebraic Effects and Handlers} \label{sec:related/algeff}
Algebraic effect handlers introduced by \citet{Plotkin13} turned out to be
greatly expressive, which have inspired researchers and programming language
designers and leads to a variety of
implementations~\cite{Bauer15,Leijen17,Lindley17,Sivaramakrishnan21,Brady13,Kammar13}.
For advanced verification of algebraic effects and handlers, \citet{Ahman17}
proposed a dependent type system for algebraic effects and
handlers. \citet{Brady13} introduced algebraic effect handlers to Idris, a
dependently typed programming language.
In contrast to our system, these systems do not allow initial answer types to depend
on values passed to continuations.  \citet{Ahman15} investigated an algebraic
treatment of computational effects with refinement types, but their language
is not equipped with effect handlers.  To our knowledge, there is no research focusing
on refinement type systems with support for algebraic effect handlers and their
implementations for automated verification.

\citet{Cong22} provided a type system with ATM for algebraic effect handlers
in a simply typed setting.
Compared with ours, their system is limited in a few points.
First, it allows programs to use only one operation.
Second, the operation can be invoked two or more times only when it is handled
by an effect handler where the result types of the return and operation clause
are the same.
This limitation is particularly critical for our aim, program verification,
because it means that there is no way to track the state of continuations that
changes with the execution of programs.
For instance, the examples presented in Section~\ref{sec:language/exmaples}
cannot be verified under such a restriction because they include multiple calls
to an operation and each call changes the state of continuations.
Our type system has none of these limitations---it supports multiple operations
and an unlimited number of calls to operations even under a handler with clauses
of different types.
The key idea of our system to allow such a handler is to introduce the
abstraction of operation clauses over predicates.
By this abstraction, our type system can represent how the same operation clause
behaves differently under different continuations.

Our CPS transformation is based on \citet{Hillerstrom17}. They defined a CPS transformation from a language
with effect handlers but without dependent/refinement types,
and proved that it enjoys forward type-preservation, but they, and others, such
as \citet{Cong22}, who studied CPS transformation for effect handlers, did not
consider the backward direction.
Their transformation also assumes that programs are fully annotated with types.

\subsection{Type Systems for Other Delimited Control Operators} \label{sec:related/control}

ATM was proposed by \citet{Danvy90} to type more expressions
with the delimited control operators shift/reset.
\citet{Cong18} proposed a dependent type system for shift/reset,
where initial answer types cannot depend on
values passed to continuations.
A type system with ATM for another set of delimited control operators shift0/reset0,
is developed by \citet{Materzok11}.
They proposed a new subtyping relation that allow lifting pure expressions to impure ones.
Based on their work,
\citet{Sekiyama23} proposed a refinement type system for shift0/reset0.
Their type system utilizes ATM for reasoning about traces (sequences of events) precisely.
In their system, initial answer types \emph{can} depend on
values passed to continuations.
Our control effects are inspired by their work, but
they use the dependency of control effects mainly for reasoning about traces while we use it for refining properties of values.
Their target operators shift0/reset0 are closely related to
our target operators, algebraic effect handlers~\cite{Forster17,Pirog19}.
However, naively applying their approach to algebraic effect handlers does not
enable precise verification.  A critical difference between shift0/reset0 and
algebraic effect handlers is that, while shift0/reset0 allows deciding the usage
of captured delimited continuations per each call site of the
continuation-capture operator shift0, algebraic effect handlers require all the
calls to the same operation under a handler $h$ to be interpreted by the same
operation clause in $h$. This hinders precise verification of the use of
continuations per each operation call. Our type system solves this problem by
abstracting the type schemes of operations over predicates.

\section{Conclusion} \label{sec:conclusion}

We developed a sound refinement type system
for algebraic effects and handlers,
which adopts the concept of ATM (especially, ARM)
to capture how the use of effects and the handling of them influence the results of computations.
This enables precise analysis of programs
with algebraic effects and handlers.
We also implemented the type checking and inference algorithm for a subset of OCaml 5 and
demonstrated the usefulness of ARM.
Additionally, we defined a bidirectionally-type-preserving CPS transformation
from our language with effect handlers to the language without effect handlers.
It enables the reuse of existing refinement type checkers to verify programs with effect handlers, but makes programs to be verified complicated and requires them to be fully annotated.
One possible direction for future work is to incorporate
temporal verification as in \citet{Sekiyama23} into algebraic effects and handlers.
Also, it is interesting to apply ARM to other variants of effect handlers, such as lexically scoped effect handlers~\cite{ZhangM19,BiernackiPPS20}.

\section*{Data-Availability Statement}
Our artifact is available in the GitHub repository, at \url{https://github.com/hiroshi-unno/coar}.
The experimental results shown in Table~\ref{tab:eval} and Table~\ref{tab:cps}
can be reproduced by following the instructions in \texttt{popl24ae/README.md} of the repository.

\begin{acks}
We are grateful to anonymous reviewers for their helpful and useful comments on the paper,
especially regarding its presentation.
We also thank Yiyang Guo and Kanaru Isoda for their contribution to our implementation.
This work was supported by \grantsponsor{JSPS}{JSPS}{} KAKENHI Grant Numbers
\grantnum{JSPS}{JP19K20247}, 
\grantnum{JSPS}{JP22K17875}, 
\grantnum{JSPS}{JP20H00582}, 
\grantnum{JSPS}{JP20H04162}, 
\grantnum{JSPS}{JP22H03564}, 
\grantnum{JSPS}{JP20H05703}, 
\grantnum{JSPS}{JP20K20625}, 
and
\grantnum{JSPS}{JP22H03570}  
as well as
\grantsponsor{JST}{JST}{} CREST Grant Number \grantnum{JST}{JPMJCR21M3}.
\end{acks}

\bibliographystyle{ACM-Reference-Format}
\bibliography{main}


\begin{thebibliography}{58}


\ifx \showCODEN    \undefined \def \showCODEN     #1{\unskip}     \fi
\ifx \showDOI      \undefined \def \showDOI       #1{#1}\fi
\ifx \showISBNx    \undefined \def \showISBNx     #1{\unskip}     \fi
\ifx \showISBNxiii \undefined \def \showISBNxiii  #1{\unskip}     \fi
\ifx \showISSN     \undefined \def \showISSN      #1{\unskip}     \fi
\ifx \showLCCN     \undefined \def \showLCCN      #1{\unskip}     \fi
\ifx \shownote     \undefined \def \shownote      #1{#1}          \fi
\ifx \showarticletitle \undefined \def \showarticletitle #1{#1}   \fi
\ifx \showURL      \undefined \def \showURL       {\relax}        \fi
\providecommand\bibfield[2]{#2}
\providecommand\bibinfo[2]{#2}
\providecommand\natexlab[1]{#1}
\providecommand\showeprint[2][]{arXiv:#2}

\bibitem[Ahman(2017)]%
        {Ahman17}
\bibfield{author}{\bibinfo{person}{Danel Ahman}.}
  \bibinfo{year}{2017}\natexlab{}.
\newblock \showarticletitle{Handling Fibred Algebraic Effects}.
\newblock \bibinfo{journal}{\emph{Proc. ACM Program. Lang.}}
  \bibinfo{volume}{2}, \bibinfo{number}{POPL}, Article \bibinfo{articleno}{7}
  (\bibinfo{date}{dec} \bibinfo{year}{2017}), \bibinfo{numpages}{29}~pages.
\newblock
\urldef\tempurl%
\url{https://doi.org/10.1145/3158095}
\showDOI{\tempurl}


\bibitem[Ahman and Plotkin(2015)]%
        {Ahman15}
\bibfield{author}{\bibinfo{person}{Danel Ahman} {and} \bibinfo{person}{Gordon
  Plotkin}.} \bibinfo{year}{2015}\natexlab{}.
\newblock \showarticletitle{Refinement types for algebraic effects}. In
  \bibinfo{booktitle}{\emph{Abstracts of the 21st Meeting `Types for Proofs and
  Programs' (TYPES)}}. \bibinfo{publisher}{Institute of Cybernetics, Tallinn
  University of Technology}, \bibinfo{pages}{10--11}.
\newblock


\bibitem[Appel(1992)]%
        {Appel92}
\bibfield{author}{\bibinfo{person}{Andrew~W. Appel}.}
  \bibinfo{year}{1992}\natexlab{}.
\newblock \bibinfo{booktitle}{\emph{Compiling with Continuations}}.
\newblock \bibinfo{publisher}{Cambridge University Press}.
\newblock
\urldef\tempurl%
\url{https://doi.org/10.1017/CBO9780511609619}
\showDOI{\tempurl}


\bibitem[Asai(2009)]%
        {Asai09}
\bibfield{author}{\bibinfo{person}{Kenichi Asai}.}
  \bibinfo{year}{2009}\natexlab{}.
\newblock \showarticletitle{On typing delimited continuations: three new
  solutions to the printf problem}.
\newblock \bibinfo{journal}{\emph{Higher-Order and Symbolic Computation}}
  \bibinfo{volume}{22}, \bibinfo{number}{3} (\bibinfo{date}{01 Sep}
  \bibinfo{year}{2009}), \bibinfo{pages}{275--291}.
\newblock
\showISSN{1573-0557}
\urldef\tempurl%
\url{https://doi.org/10.1007/s10990-009-9049-5}
\showDOI{\tempurl}


\bibitem[Bauer(2018)]%
        {DBLP:journals/corr/abs-1807-05923}
\bibfield{author}{\bibinfo{person}{Andrej Bauer}.}
  \bibinfo{year}{2018}\natexlab{}.
\newblock \showarticletitle{What is algebraic about algebraic effects and
  handlers?}
\newblock \bibinfo{journal}{\emph{CoRR}}  \bibinfo{volume}{abs/1807.05923}
  (\bibinfo{year}{2018}).
\newblock
\showeprint[arXiv]{1807.05923}
\urldef\tempurl%
\url{http://arxiv.org/abs/1807.05923}
\showURL{%
\tempurl}


\bibitem[Bauer and Pretnar(2013)]%
        {Bauer13}
\bibfield{author}{\bibinfo{person}{Andrej Bauer} {and} \bibinfo{person}{Matija
  Pretnar}.} \bibinfo{year}{2013}\natexlab{}.
\newblock \showarticletitle{An Effect System for Algebraic Effects and
  Handlers}. In \bibinfo{booktitle}{\emph{Algebra and Coalgebra in Computer
  Science}}, \bibfield{editor}{\bibinfo{person}{Reiko Heckel} {and}
  \bibinfo{person}{Stefan Milius}} (Eds.). \bibinfo{publisher}{Springer Berlin
  Heidelberg}, \bibinfo{address}{Berlin, Heidelberg}, \bibinfo{pages}{1--16}.
\newblock
\showISBNx{978-3-642-40206-7}


\bibitem[Bauer and Pretnar(2015)]%
        {Bauer15}
\bibfield{author}{\bibinfo{person}{Andrej Bauer} {and} \bibinfo{person}{Matija
  Pretnar}.} \bibinfo{year}{2015}\natexlab{}.
\newblock \showarticletitle{Programming with algebraic effects and handlers}.
\newblock \bibinfo{journal}{\emph{Journal of Logical and Algebraic Methods in
  Programming}} \bibinfo{volume}{84}, \bibinfo{number}{1}
  (\bibinfo{year}{2015}), \bibinfo{pages}{108--123}.
\newblock
\showISSN{2352-2208}
\urldef\tempurl%
\url{https://doi.org/10.1016/j.jlamp.2014.02.001}
\showDOI{\tempurl}
\newblock
\shownote{Special Issue: The 23rd Nordic Workshop on Programming Theory (NWPT
  2011) Special Issue: Domains X, International workshop on Domain Theory and
  applications, Swansea, 5-7 September, 2011}.


\bibitem[Bengtson et~al\mbox{.}(2011)]%
        {Bengston11}
\bibfield{author}{\bibinfo{person}{Jesper Bengtson},
  \bibinfo{person}{Karthikeyan Bhargavan}, \bibinfo{person}{C\'{e}dric
  Fournet}, \bibinfo{person}{Andrew~D. Gordon}, {and} \bibinfo{person}{Sergio
  Maffeis}.} \bibinfo{year}{2011}\natexlab{}.
\newblock \showarticletitle{Refinement Types for Secure Implementations}.
\newblock \bibinfo{journal}{\emph{ACM Trans. Program. Lang. Syst.}}
  \bibinfo{volume}{33}, \bibinfo{number}{2}, Article \bibinfo{articleno}{8}
  (\bibinfo{date}{feb} \bibinfo{year}{2011}), \bibinfo{numpages}{45}~pages.
\newblock
\showISSN{0164-0925}
\urldef\tempurl%
\url{https://doi.org/10.1145/1890028.1890031}
\showDOI{\tempurl}


\bibitem[Biernacki et~al\mbox{.}(2020)]%
        {BiernackiPPS20}
\bibfield{author}{\bibinfo{person}{Dariusz Biernacki}, \bibinfo{person}{Maciej
  Pir{\'{o}}g}, \bibinfo{person}{Piotr Polesiuk}, {and} \bibinfo{person}{Filip
  Sieczkowski}.} \bibinfo{year}{2020}\natexlab{}.
\newblock \showarticletitle{Binders by day, labels by night: effect instances
  via lexically scoped handlers}.
\newblock \bibinfo{journal}{\emph{Proc. {ACM} Program. Lang.}}
  \bibinfo{volume}{4}, \bibinfo{number}{{POPL}} (\bibinfo{year}{2020}),
  \bibinfo{pages}{48:1--48:29}.
\newblock
\urldef\tempurl%
\url{https://doi.org/10.1145/3371116}
\showDOI{\tempurl}


\bibitem[Bj{\o}rner et~al\mbox{.}(2015)]%
        {Bjorner2015a}
\bibfield{author}{\bibinfo{person}{Nikolaj Bj{\o}rner}, \bibinfo{person}{Arie
  Gurfinkel}, \bibinfo{person}{Kenneth~L. McMillan}, {and}
  \bibinfo{person}{Andrey Rybalchenko}.} \bibinfo{year}{2015}\natexlab{}.
\newblock \showarticletitle{{Horn} Clause Solvers for Program Verification}. In
  \bibinfo{booktitle}{\emph{Fields of Logic and Computation {II}: Essays
  Dedicated to Yuri Gurevich on the Occasion of His 75th Birthday}},
  Vol.~\bibinfo{volume}{9300}. \bibinfo{pages}{24--51}.
\newblock


\bibitem[Bradley(2011)]%
        {Bradley11}
\bibfield{author}{\bibinfo{person}{Aaron~R. Bradley}.}
  \bibinfo{year}{2011}\natexlab{}.
\newblock \showarticletitle{SAT-Based Model Checking without Unrolling}. In
  \bibinfo{booktitle}{\emph{Proceedings of the 12th International Conference on
  Verification, Model Checking, and Abstract Interpretation}} (Austin, TX, USA)
  \emph{(\bibinfo{series}{VMCAI'11})}. \bibinfo{publisher}{Springer-Verlag},
  \bibinfo{address}{Berlin, Heidelberg}, \bibinfo{pages}{70–87}.
\newblock
\showISBNx{9783642182747}


\bibitem[Brady(2013)]%
        {Brady13}
\bibfield{author}{\bibinfo{person}{Edwin Brady}.}
  \bibinfo{year}{2013}\natexlab{}.
\newblock \showarticletitle{Programming and Reasoning with Algebraic Effects
  and Dependent Types}. In \bibinfo{booktitle}{\emph{Proceedings of the 18th
  ACM SIGPLAN International Conference on Functional Programming}} (Boston,
  Massachusetts, USA) \emph{(\bibinfo{series}{ICFP '13})}.
  \bibinfo{publisher}{Association for Computing Machinery},
  \bibinfo{address}{New York, NY, USA}, \bibinfo{pages}{133–144}.
\newblock
\showISBNx{9781450323260}
\urldef\tempurl%
\url{https://doi.org/10.1145/2500365.2500581}
\showDOI{\tempurl}


\bibitem[Cong and Asai(2018)]%
        {Cong18}
\bibfield{author}{\bibinfo{person}{Youyou Cong} {and} \bibinfo{person}{Kenichi
  Asai}.} \bibinfo{year}{2018}\natexlab{}.
\newblock \showarticletitle{Handling Delimited Continuations with Dependent
  Types}.
\newblock \bibinfo{journal}{\emph{Proc. ACM Program. Lang.}}
  \bibinfo{volume}{2}, \bibinfo{number}{ICFP}, Article \bibinfo{articleno}{69}
  (\bibinfo{date}{jul} \bibinfo{year}{2018}), \bibinfo{numpages}{31}~pages.
\newblock
\urldef\tempurl%
\url{https://doi.org/10.1145/3236764}
\showDOI{\tempurl}


\bibitem[Cong and Asai(2022)]%
        {Cong22}
\bibfield{author}{\bibinfo{person}{Youyou Cong} {and} \bibinfo{person}{Kenichi
  Asai}.} \bibinfo{year}{2022}\natexlab{}.
\newblock \showarticletitle{Understanding Algebraic Effect Handlers via
  Delimited Control Operators}. In \bibinfo{booktitle}{\emph{Trends in
  Functional Programming - 23rd International Symposium, {TFP} 2022, Virtual
  Event, March 17-18, 2022, Revised Selected Papers}}
  \emph{(\bibinfo{series}{Lecture Notes in Computer Science},
  Vol.~\bibinfo{volume}{13401})}, \bibfield{editor}{\bibinfo{person}{Wouter
  Swierstra} {and} \bibinfo{person}{Nicolas Wu}} (Eds.).
  \bibinfo{publisher}{Springer}, \bibinfo{pages}{59--79}.
\newblock
\urldef\tempurl%
\url{https://doi.org/10.1007/978-3-031-21314-4\_4}
\showDOI{\tempurl}


\bibitem[Danvy and Filinski(1990)]%
        {Danvy90}
\bibfield{author}{\bibinfo{person}{Olivier Danvy} {and}
  \bibinfo{person}{Andrzej Filinski}.} \bibinfo{year}{1990}\natexlab{}.
\newblock \showarticletitle{Abstracting Control}. In
  \bibinfo{booktitle}{\emph{Proceedings of the 1990 ACM Conference on LISP and
  Functional Programming}} (Nice, France) \emph{(\bibinfo{series}{LFP '90})}.
  \bibinfo{publisher}{Association for Computing Machinery},
  \bibinfo{address}{New York, NY, USA}, \bibinfo{pages}{151–160}.
\newblock
\showISBNx{089791368X}
\urldef\tempurl%
\url{https://doi.org/10.1145/91556.91622}
\showDOI{\tempurl}


\bibitem[de~Moura and Bj{\o}rner(2008)]%
        {Moura2008}
\bibfield{author}{\bibinfo{person}{Leonardo de Moura} {and}
  \bibinfo{person}{Nikolaj Bj{\o}rner}.} \bibinfo{year}{2008}\natexlab{}.
\newblock \showarticletitle{Z3: An Efficient {SMT} Solver}. In
  \bibinfo{booktitle}{\emph{TACAS'08}} (Budapest, Hungary, March29 -- April 6),
  Vol.~\bibinfo{volume}{4963}. \bibinfo{publisher}{Springer Berlin Heidelberg},
  \bibinfo{pages}{337--340}.
\newblock


\bibitem[Een et~al\mbox{.}(2011)]%
        {Een11}
\bibfield{author}{\bibinfo{person}{Niklas Een}, \bibinfo{person}{Alan
  Mishchenko}, {and} \bibinfo{person}{Robert Brayton}.}
  \bibinfo{year}{2011}\natexlab{}.
\newblock \showarticletitle{Efficient Implementation of Property Directed
  Reachability}. In \bibinfo{booktitle}{\emph{Proceedings of the International
  Conference on Formal Methods in Computer-Aided Design}} (Austin, Texas)
  \emph{(\bibinfo{series}{FMCAD '11})}. \bibinfo{publisher}{FMCAD Inc},
  \bibinfo{address}{Austin, Texas}, \bibinfo{pages}{125–134}.
\newblock
\showISBNx{9780983567813}


\bibitem[Forster et~al\mbox{.}(2017)]%
        {Forster17}
\bibfield{author}{\bibinfo{person}{Yannick Forster}, \bibinfo{person}{Ohad
  Kammar}, \bibinfo{person}{Sam Lindley}, {and} \bibinfo{person}{Matija
  Pretnar}.} \bibinfo{year}{2017}\natexlab{}.
\newblock \showarticletitle{On the Expressive Power of User-Defined Effects:
  Effect Handlers, Monadic Reflection, Delimited Control}.
\newblock \bibinfo{journal}{\emph{Proc. ACM Program. Lang.}}
  \bibinfo{volume}{1}, \bibinfo{number}{ICFP}, Article \bibinfo{articleno}{13}
  (\bibinfo{date}{aug} \bibinfo{year}{2017}), \bibinfo{numpages}{29}~pages.
\newblock
\urldef\tempurl%
\url{https://doi.org/10.1145/3110257}
\showDOI{\tempurl}


\bibitem[Foster et~al\mbox{.}(2002)]%
        {DBLP:conf/pldi/FosterTA02}
\bibfield{author}{\bibinfo{person}{Jeffrey~S. Foster}, \bibinfo{person}{Tachio
  Terauchi}, {and} \bibinfo{person}{Alexander Aiken}.}
  \bibinfo{year}{2002}\natexlab{}.
\newblock \showarticletitle{Flow-Sensitive Type Qualifiers}. In
  \bibinfo{booktitle}{\emph{Proceedings of the 2002 {ACM} {SIGPLAN} Conference
  on Programming Language Design and Implementation (PLDI), Berlin, Germany,
  June 17-19, 2002}}, \bibfield{editor}{\bibinfo{person}{Jens Knoop} {and}
  \bibinfo{person}{Laurie~J. Hendren}} (Eds.). \bibinfo{publisher}{{ACM}},
  \bibinfo{pages}{1--12}.
\newblock
\urldef\tempurl%
\url{https://doi.org/10.1145/512529.512531}
\showDOI{\tempurl}


\bibitem[Freeman and Pfenning(1991)]%
        {DBLP:conf/pldi/FreemanP91}
\bibfield{author}{\bibinfo{person}{Timothy~S. Freeman} {and}
  \bibinfo{person}{Frank Pfenning}.} \bibinfo{year}{1991}\natexlab{}.
\newblock \showarticletitle{Refinement Types for {ML}}. In
  \bibinfo{booktitle}{\emph{Proceedings of the {ACM} SIGPLAN'91 Conference on
  Programming Language Design and Implementation (PLDI), Toronto, Ontario,
  Canada, June 26-28, 1991}}, \bibfield{editor}{\bibinfo{person}{David~S.
  Wise}} (Ed.). \bibinfo{publisher}{{ACM}}, \bibinfo{pages}{268--277}.
\newblock
\urldef\tempurl%
\url{https://doi.org/10.1145/113445.113468}
\showDOI{\tempurl}


\bibitem[Hillerstr{\"o}m and Lindley(2018)]%
        {Hillerstrom18}
\bibfield{author}{\bibinfo{person}{Daniel Hillerstr{\"o}m} {and}
  \bibinfo{person}{Sam Lindley}.} \bibinfo{year}{2018}\natexlab{}.
\newblock \showarticletitle{Shallow Effect Handlers}. In
  \bibinfo{booktitle}{\emph{Programming Languages and Systems}},
  \bibfield{editor}{\bibinfo{person}{Sukyoung Ryu}} (Ed.).
  \bibinfo{publisher}{Springer International Publishing},
  \bibinfo{address}{Cham}, \bibinfo{pages}{415--435}.
\newblock
\showISBNx{978-3-030-02768-1}


\bibitem[Hillerstr{\"o}m et~al\mbox{.}(2017)]%
        {Hillerstrom17}
\bibfield{author}{\bibinfo{person}{Daniel Hillerstr{\"o}m},
  \bibinfo{person}{Sam Lindley}, \bibinfo{person}{Robert Atkey}, {and}
  \bibinfo{person}{K.~C. Sivaramakrishnan}.} \bibinfo{year}{2017}\natexlab{}.
\newblock \showarticletitle{{Continuation Passing Style for Effect Handlers}}.
  In \bibinfo{booktitle}{\emph{2nd International Conference on Formal
  Structures for Computation and Deduction (FSCD 2017)}}
  \emph{(\bibinfo{series}{Leibniz International Proceedings in Informatics
  (LIPIcs)}, Vol.~\bibinfo{volume}{84})},
  \bibfield{editor}{\bibinfo{person}{Dale Miller}} (Ed.).
  \bibinfo{publisher}{Schloss Dagstuhl--Leibniz-Zentrum fuer Informatik},
  \bibinfo{address}{Dagstuhl, Germany}, \bibinfo{pages}{18:1--18:19}.
\newblock
\showISBNx{978-3-95977-047-7}
\showISSN{1868-8969}
\urldef\tempurl%
\url{https://doi.org/10.4230/LIPIcs.FSCD.2017.18}
\showDOI{\tempurl}


\bibitem[Ishio and Asai(2022)]%
        {Ishio22}
\bibfield{author}{\bibinfo{person}{Chiaki Ishio} {and} \bibinfo{person}{Kenichi
  Asai}.} \bibinfo{year}{2022}\natexlab{}.
\newblock \showarticletitle{Type System for Four Delimited Control Operators}.
  In \bibinfo{booktitle}{\emph{Proceedings of the 21st ACM SIGPLAN
  International Conference on Generative Programming: Concepts and
  Experiences}} (Auckland, New Zealand) \emph{(\bibinfo{series}{GPCE 2022})}.
  \bibinfo{publisher}{Association for Computing Machinery},
  \bibinfo{address}{New York, NY, USA}, \bibinfo{pages}{45–58}.
\newblock
\showISBNx{9781450399203}
\urldef\tempurl%
\url{https://doi.org/10.1145/3564719.3568691}
\showDOI{\tempurl}


\bibitem[Kammar et~al\mbox{.}(2013)]%
        {Kammar13}
\bibfield{author}{\bibinfo{person}{Ohad Kammar}, \bibinfo{person}{Sam Lindley},
  {and} \bibinfo{person}{Nicolas Oury}.} \bibinfo{year}{2013}\natexlab{}.
\newblock \showarticletitle{Handlers in Action}. In
  \bibinfo{booktitle}{\emph{Proceedings of the 18th ACM SIGPLAN International
  Conference on Functional Programming}} (Boston, Massachusetts, USA)
  \emph{(\bibinfo{series}{ICFP '13})}. \bibinfo{publisher}{Association for
  Computing Machinery}, \bibinfo{address}{New York, NY, USA},
  \bibinfo{pages}{145–158}.
\newblock
\showISBNx{9781450323260}
\urldef\tempurl%
\url{https://doi.org/10.1145/2500365.2500590}
\showDOI{\tempurl}


\bibitem[Kammar and Pretnar(2017)]%
        {Kammar17}
\bibfield{author}{\bibinfo{person}{Ohad Kammar} {and} \bibinfo{person}{Matija
  Pretnar}.} \bibinfo{year}{2017}\natexlab{}.
\newblock \showarticletitle{No value restriction is needed for algebraic
  effects and handlers}.
\newblock \bibinfo{journal}{\emph{Journal of Functional Programming}}
  \bibinfo{volume}{27} (\bibinfo{year}{2017}), \bibinfo{pages}{e7}.
\newblock
\urldef\tempurl%
\url{https://doi.org/10.1017/S0956796816000320}
\showDOI{\tempurl}


\bibitem[Komuravelli et~al\mbox{.}(2013)]%
        {Komuravelli13}
\bibfield{author}{\bibinfo{person}{Anvesh Komuravelli}, \bibinfo{person}{Arie
  Gurfinkel}, \bibinfo{person}{Sagar Chaki}, {and} \bibinfo{person}{Edmund~M.
  Clarke}.} \bibinfo{year}{2013}\natexlab{}.
\newblock \showarticletitle{Automatic Abstraction in SMT-Based Unbounded
  Software Model Checking}. In \bibinfo{booktitle}{\emph{Proceedings of the
  25th International Conference on Computer Aided Verification - Volume 8044}}
  (Saint Petersburg, Russia) \emph{(\bibinfo{series}{CAV 2013})}.
  \bibinfo{publisher}{Springer-Verlag}, \bibinfo{address}{Berlin, Heidelberg},
  \bibinfo{pages}{846–862}.
\newblock
\showISBNx{9783642397981}


\bibitem[Leijen(2014)]%
        {Leijen14}
\bibfield{author}{\bibinfo{person}{Daan Leijen}.}
  \bibinfo{year}{2014}\natexlab{}.
\newblock \showarticletitle{Koka: Programming with Row Polymorphic Effect
  Types}. In \bibinfo{booktitle}{\emph{Proceedings 5th Workshop on
  Mathematically Structured Functional Programming, MSFP@ETAPS 2014, Grenoble,
  France, 12 April 2014}} \emph{(\bibinfo{series}{{EPTCS}},
  Vol.~\bibinfo{volume}{153})}, \bibfield{editor}{\bibinfo{person}{Paul~Blain
  Levy} {and} \bibinfo{person}{Neel Krishnaswami}} (Eds.).
  \bibinfo{pages}{100--126}.
\newblock
\urldef\tempurl%
\url{https://doi.org/10.4204/EPTCS.153.8}
\showDOI{\tempurl}


\bibitem[Leijen(2017)]%
        {Leijen17}
\bibfield{author}{\bibinfo{person}{Daan Leijen}.}
  \bibinfo{year}{2017}\natexlab{}.
\newblock \showarticletitle{Type Directed Compilation of Row-Typed Algebraic
  Effects}. In \bibinfo{booktitle}{\emph{Proceedings of the 44th ACM SIGPLAN
  Symposium on Principles of Programming Languages}} (Paris, France)
  \emph{(\bibinfo{series}{POPL '17})}. \bibinfo{publisher}{Association for
  Computing Machinery}, \bibinfo{address}{New York, NY, USA},
  \bibinfo{pages}{486–499}.
\newblock
\showISBNx{9781450346603}
\urldef\tempurl%
\url{https://doi.org/10.1145/3009837.3009872}
\showDOI{\tempurl}


\bibitem[Levy et~al\mbox{.}(2003)]%
        {Levy03}
\bibfield{author}{\bibinfo{person}{PaulBlain Levy}, \bibinfo{person}{John
  Power}, {and} \bibinfo{person}{Hayo Thielecke}.}
  \bibinfo{year}{2003}\natexlab{}.
\newblock \showarticletitle{Modelling environments in call-by-value programming
  languages}.
\newblock \bibinfo{journal}{\emph{Information and Computation}}
  \bibinfo{volume}{185}, \bibinfo{number}{2} (\bibinfo{year}{2003}),
  \bibinfo{pages}{182--210}.
\newblock
\showISSN{0890-5401}
\urldef\tempurl%
\url{https://doi.org/10.1016/S0890-5401(03)00088-9}
\showDOI{\tempurl}


\bibitem[Lindley et~al\mbox{.}(2017)]%
        {Lindley17}
\bibfield{author}{\bibinfo{person}{Sam Lindley}, \bibinfo{person}{Conor
  McBride}, {and} \bibinfo{person}{Craig McLaughlin}.}
  \bibinfo{year}{2017}\natexlab{}.
\newblock \showarticletitle{Do Be Do Be Do}. In
  \bibinfo{booktitle}{\emph{Proceedings of the 44th ACM SIGPLAN Symposium on
  Principles of Programming Languages}} (Paris, France)
  \emph{(\bibinfo{series}{POPL '17})}. \bibinfo{publisher}{Association for
  Computing Machinery}, \bibinfo{address}{New York, NY, USA},
  \bibinfo{pages}{500–514}.
\newblock
\showISBNx{9781450346603}
\urldef\tempurl%
\url{https://doi.org/10.1145/3009837.3009897}
\showDOI{\tempurl}


\bibitem[Materzok and Biernacki(2011)]%
        {Materzok11}
\bibfield{author}{\bibinfo{person}{Marek Materzok} {and}
  \bibinfo{person}{Dariusz Biernacki}.} \bibinfo{year}{2011}\natexlab{}.
\newblock \showarticletitle{Subtyping Delimited Continuations}. In
  \bibinfo{booktitle}{\emph{Proceedings of the 16th ACM SIGPLAN International
  Conference on Functional Programming}} (Tokyo, Japan)
  \emph{(\bibinfo{series}{ICFP '11})}. \bibinfo{publisher}{Association for
  Computing Machinery}, \bibinfo{address}{New York, NY, USA},
  \bibinfo{pages}{81–93}.
\newblock
\showISBNx{9781450308656}
\urldef\tempurl%
\url{https://doi.org/10.1145/2034773.2034786}
\showDOI{\tempurl}


\bibitem[Mitchell(1988)]%
        {Mitchell88}
\bibfield{author}{\bibinfo{person}{John~C. Mitchell}.}
  \bibinfo{year}{1988}\natexlab{}.
\newblock \showarticletitle{Polymorphic type inference and containment}.
\newblock \bibinfo{journal}{\emph{Information and Computation}}
  \bibinfo{volume}{76}, \bibinfo{number}{2} (\bibinfo{year}{1988}),
  \bibinfo{pages}{211--249}.
\newblock
\showISSN{0890-5401}
\urldef\tempurl%
\url{https://doi.org/10.1016/0890-5401(88)90009-0}
\showDOI{\tempurl}


\bibitem[{Multicore OCaml}(2022)]%
        {MulticoreOCamlrepo}
\bibfield{author}{\bibinfo{person}{{Multicore OCaml}}.}
  \bibinfo{year}{2022}\natexlab{}.
\newblock \bibinfo{booktitle}{\emph{OCaml effects examples}}.
\newblock
\urldef\tempurl%
\url{https://github.com/ocaml-multicore/effects-examples}
\showURL{%
Retrieved November 5, 2022 from \tempurl}


\bibitem[Nanjo et~al\mbox{.}(2018)]%
        {DBLP:conf/lics/Nanjo0KT18}
\bibfield{author}{\bibinfo{person}{Yoji Nanjo}, \bibinfo{person}{Hiroshi Unno},
  \bibinfo{person}{Eric Koskinen}, {and} \bibinfo{person}{Tachio Terauchi}.}
  \bibinfo{year}{2018}\natexlab{}.
\newblock \showarticletitle{A Fixpoint Logic and Dependent Effects for Temporal
  Property Verification}. In \bibinfo{booktitle}{\emph{Proceedings of the 33rd
  Annual {ACM/IEEE} Symposium on Logic in Computer Science, {LICS} 2018,
  Oxford, UK, July 09-12, 2018}}, \bibfield{editor}{\bibinfo{person}{Anuj
  Dawar} {and} \bibinfo{person}{Erich Gr{\"{a}}del}} (Eds.).
  \bibinfo{publisher}{{ACM}}, \bibinfo{pages}{759--768}.
\newblock
\urldef\tempurl%
\url{https://doi.org/10.1145/3209108.3209204}
\showDOI{\tempurl}


\bibitem[Pir{\'o}g et~al\mbox{.}(2019)]%
        {Pirog19}
\bibfield{author}{\bibinfo{person}{Maciej Pir{\'o}g}, \bibinfo{person}{Piotr
  Polesiuk}, {and} \bibinfo{person}{Filip Sieczkowski}.}
  \bibinfo{year}{2019}\natexlab{}.
\newblock \showarticletitle{{Typed Equivalence of Effect Handlers and Delimited
  Control}}. In \bibinfo{booktitle}{\emph{4th International Conference on
  Formal Structures for Computation and Deduction (FSCD 2019)}}
  \emph{(\bibinfo{series}{Leibniz International Proceedings in Informatics
  (LIPIcs)}, Vol.~\bibinfo{volume}{131})},
  \bibfield{editor}{\bibinfo{person}{Herman Geuvers}} (Ed.).
  \bibinfo{publisher}{Schloss Dagstuhl--Leibniz-Zentrum fuer Informatik},
  \bibinfo{address}{Dagstuhl, Germany}, \bibinfo{pages}{30:1--30:16}.
\newblock
\showISBNx{978-3-95977-107-8}
\showISSN{1868-8969}
\urldef\tempurl%
\url{https://doi.org/10.4230/LIPIcs.FSCD.2019.30}
\showDOI{\tempurl}


\bibitem[Plotkin(1975)]%
        {Plotkin75}
\bibfield{author}{\bibinfo{person}{Gordon Plotkin}.}
  \bibinfo{year}{1975}\natexlab{}.
\newblock \showarticletitle{Call-by-name, call-by-value and the
  $\lambda$-calculus}.
\newblock \bibinfo{journal}{\emph{Theoretical Computer Science}}
  \bibinfo{volume}{1}, \bibinfo{number}{2} (\bibinfo{year}{1975}),
  \bibinfo{pages}{125--159}.
\newblock
\showISSN{0304-3975}
\urldef\tempurl%
\url{https://doi.org/10.1016/0304-3975(75)90017-1}
\showDOI{\tempurl}


\bibitem[Plotkin and Power(2003)]%
        {Plotkin03}
\bibfield{author}{\bibinfo{person}{Gordon Plotkin} {and} \bibinfo{person}{John
  Power}.} \bibinfo{year}{2003}\natexlab{}.
\newblock \showarticletitle{Algebraic Operations and Generic Effects}.
\newblock \bibinfo{journal}{\emph{Applied Categorical Structures}}
  \bibinfo{volume}{11}, \bibinfo{number}{1} (\bibinfo{date}{01 Feb}
  \bibinfo{year}{2003}), \bibinfo{pages}{69--94}.
\newblock
\showISSN{1572-9095}
\urldef\tempurl%
\url{https://doi.org/10.1023/A:1023064908962}
\showDOI{\tempurl}


\bibitem[Plotkin and Pretnar(2009)]%
        {Plotkin09}
\bibfield{author}{\bibinfo{person}{Gordon Plotkin} {and}
  \bibinfo{person}{Matija Pretnar}.} \bibinfo{year}{2009}\natexlab{}.
\newblock \showarticletitle{Handlers of Algebraic Effects}. In
  \bibinfo{booktitle}{\emph{Programming Languages and Systems}},
  \bibfield{editor}{\bibinfo{person}{Giuseppe Castagna}} (Ed.).
  \bibinfo{publisher}{Springer Berlin Heidelberg}, \bibinfo{address}{Berlin,
  Heidelberg}, \bibinfo{pages}{80--94}.
\newblock
\showISBNx{978-3-642-00590-9}


\bibitem[Plotkin and Power(2001)]%
        {DBLP:conf/fossacs/PlotkinP01}
\bibfield{author}{\bibinfo{person}{Gordon~D. Plotkin} {and}
  \bibinfo{person}{John Power}.} \bibinfo{year}{2001}\natexlab{}.
\newblock \showarticletitle{Adequacy for Algebraic Effects}. In
  \bibinfo{booktitle}{\emph{Foundations of Software Science and Computation
  Structures, 4th International Conference, {FOSSACS} 2001 Held as Part of the
  Joint European Conferences on Theory and Practice of Software, {ETAPS} 2001
  Genova, Italy, April 2-6, 2001, Proceedings}} \emph{(\bibinfo{series}{Lecture
  Notes in Computer Science}, Vol.~\bibinfo{volume}{2030})},
  \bibfield{editor}{\bibinfo{person}{Furio Honsell} {and}
  \bibinfo{person}{Marino Miculan}} (Eds.). \bibinfo{publisher}{Springer},
  \bibinfo{pages}{1--24}.
\newblock
\urldef\tempurl%
\url{https://doi.org/10.1007/3-540-45315-6\_1}
\showDOI{\tempurl}


\bibitem[Plotkin and Pretnar(2013)]%
        {Plotkin13}
\bibfield{author}{\bibinfo{person}{Gordon~D Plotkin} {and}
  \bibinfo{person}{Matija Pretnar}.} \bibinfo{year}{2013}\natexlab{}.
\newblock \showarticletitle{{Handling Algebraic Effects}}.
\newblock \bibinfo{journal}{\emph{{Logical Methods in Computer Science}}}
  \bibinfo{volume}{{Volume 9, Issue 4}} (\bibinfo{date}{Dec.}
  \bibinfo{year}{2013}).
\newblock
\urldef\tempurl%
\url{https://doi.org/10.2168/LMCS-9(4:23)2013}
\showDOI{\tempurl}


\bibitem[Pretnar(2015)]%
        {Pretnar15}
\bibfield{author}{\bibinfo{person}{Matija Pretnar}.}
  \bibinfo{year}{2015}\natexlab{}.
\newblock \showarticletitle{An Introduction to Algebraic Effects and Handlers.
  Invited tutorial paper}.
\newblock \bibinfo{journal}{\emph{Electronic Notes in Theoretical Computer
  Science}}  \bibinfo{volume}{319} (\bibinfo{year}{2015}),
  \bibinfo{pages}{19--35}.
\newblock
\showISSN{1571-0661}
\urldef\tempurl%
\url{https://doi.org/10.1016/j.entcs.2015.12.003}
\showDOI{\tempurl}
\newblock
\shownote{The 31st Conference on the Mathematical Foundations of Programming
  Semantics (MFPS XXXI)}.


\bibitem[Pretnar(2022)]%
        {Effrepo}
\bibfield{author}{\bibinfo{person}{Matija Pretnar}.}
  \bibinfo{year}{2022}\natexlab{}.
\newblock \bibinfo{booktitle}{\emph{Eff}}.
\newblock
\urldef\tempurl%
\url{https://github.com/matijapretnar/eff}
\showURL{%
Retrieved November 5, 2022 from \tempurl}


\bibitem[Rondon et~al\mbox{.}(2008)]%
        {Rondon08}
\bibfield{author}{\bibinfo{person}{Patrick~M. Rondon}, \bibinfo{person}{Ming
  Kawaguci}, {and} \bibinfo{person}{Ranjit Jhala}.}
  \bibinfo{year}{2008}\natexlab{}.
\newblock \showarticletitle{Liquid Types}. In
  \bibinfo{booktitle}{\emph{Proceedings of the 29th ACM SIGPLAN Conference on
  Programming Language Design and Implementation}} (Tucson, AZ, USA)
  \emph{(\bibinfo{series}{PLDI '08})}. \bibinfo{publisher}{Association for
  Computing Machinery}, \bibinfo{address}{New York, NY, USA},
  \bibinfo{pages}{159–169}.
\newblock
\showISBNx{9781595938602}
\urldef\tempurl%
\url{https://doi.org/10.1145/1375581.1375602}
\showDOI{\tempurl}


\bibitem[Sekiyama and Unno(2023)]%
        {Sekiyama23}
\bibfield{author}{\bibinfo{person}{Taro Sekiyama} {and}
  \bibinfo{person}{Hiroshi Unno}.} \bibinfo{year}{2023}\natexlab{}.
\newblock \showarticletitle{Temporal Verification with Answer-Effect
  Modification: Dependent Temporal Type-and-Effect System with Delimited
  Continuations}.
\newblock \bibinfo{journal}{\emph{Proc. ACM Program. Lang.}}
  \bibinfo{volume}{7}, \bibinfo{number}{POPL}, Article \bibinfo{articleno}{71}
  (\bibinfo{date}{jan} \bibinfo{year}{2023}), \bibinfo{numpages}{32}~pages.
\newblock
\urldef\tempurl%
\url{https://doi.org/10.1145/3571264}
\showDOI{\tempurl}


\bibitem[Sivaramakrishnan et~al\mbox{.}(2021)]%
        {Sivaramakrishnan21}
\bibfield{author}{\bibinfo{person}{KC Sivaramakrishnan},
  \bibinfo{person}{Stephen Dolan}, \bibinfo{person}{Leo White},
  \bibinfo{person}{Tom Kelly}, \bibinfo{person}{Sadiq Jaffer}, {and}
  \bibinfo{person}{Anil Madhavapeddy}.} \bibinfo{year}{2021}\natexlab{}.
\newblock \showarticletitle{Retrofitting Effect Handlers onto OCaml}. In
  \bibinfo{booktitle}{\emph{Proceedings of the 42nd ACM SIGPLAN International
  Conference on Programming Language Design and Implementation}} (Virtual,
  Canada) \emph{(\bibinfo{series}{PLDI 2021})}. \bibinfo{publisher}{Association
  for Computing Machinery}, \bibinfo{address}{New York, NY, USA},
  \bibinfo{pages}{206–221}.
\newblock
\showISBNx{9781450383912}
\urldef\tempurl%
\url{https://doi.org/10.1145/3453483.3454039}
\showDOI{\tempurl}


\bibitem[Solar-Lezama et~al\mbox{.}(2006)]%
        {Solar-Lezama06}
\bibfield{author}{\bibinfo{person}{Armando Solar-Lezama},
  \bibinfo{person}{Liviu Tancau}, \bibinfo{person}{Rastislav Bodik},
  \bibinfo{person}{Sanjit Seshia}, {and} \bibinfo{person}{Vijay Saraswat}.}
  \bibinfo{year}{2006}\natexlab{}.
\newblock \showarticletitle{Combinatorial Sketching for Finite Programs}. In
  \bibinfo{booktitle}{\emph{Proceedings of the 12th International Conference on
  Architectural Support for Programming Languages and Operating Systems}} (San
  Jose, California, USA) \emph{(\bibinfo{series}{ASPLOS XII})}.
  \bibinfo{publisher}{Association for Computing Machinery},
  \bibinfo{address}{New York, NY, USA}, \bibinfo{pages}{404–415}.
\newblock
\showISBNx{1595934510}
\urldef\tempurl%
\url{https://doi.org/10.1145/1168857.1168907}
\showDOI{\tempurl}


\bibitem[Swamy et~al\mbox{.}(2016)]%
        {DBLP:conf/popl/SwamyHKRDFBFSKZ16}
\bibfield{author}{\bibinfo{person}{Nikhil Swamy}, \bibinfo{person}{Catalin
  Hritcu}, \bibinfo{person}{Chantal Keller}, \bibinfo{person}{Aseem Rastogi},
  \bibinfo{person}{Antoine Delignat{-}Lavaud}, \bibinfo{person}{Simon Forest},
  \bibinfo{person}{Karthikeyan Bhargavan}, \bibinfo{person}{C{\'{e}}dric
  Fournet}, \bibinfo{person}{Pierre{-}Yves Strub}, \bibinfo{person}{Markulf
  Kohlweiss}, \bibinfo{person}{Jean~Karim Zinzindohoue}, {and}
  \bibinfo{person}{Santiago~Zanella B{\'{e}}guelin}.}
  \bibinfo{year}{2016}\natexlab{}.
\newblock \showarticletitle{Dependent types and multi-monadic effects in {F}}.
  In \bibinfo{booktitle}{\emph{Proceedings of the 43rd Annual {ACM}
  {SIGPLAN-SIGACT} Symposium on Principles of Programming Languages, {POPL}
  2016, St. Petersburg, FL, USA, January 20 - 22, 2016}},
  \bibfield{editor}{\bibinfo{person}{Rastislav Bod{\'{\i}}k} {and}
  \bibinfo{person}{Rupak Majumdar}} (Eds.). \bibinfo{publisher}{{ACM}},
  \bibinfo{pages}{256--270}.
\newblock
\urldef\tempurl%
\url{https://doi.org/10.1145/2837614.2837655}
\showDOI{\tempurl}


\bibitem[Terauchi(2010)]%
        {DBLP:conf/popl/Terauchi10}
\bibfield{author}{\bibinfo{person}{Tachio Terauchi}.}
  \bibinfo{year}{2010}\natexlab{}.
\newblock \showarticletitle{Dependent types from counterexamples}. In
  \bibinfo{booktitle}{\emph{Proceedings of the 37th {ACM} {SIGPLAN-SIGACT}
  Symposium on Principles of Programming Languages, {POPL} 2010, Madrid, Spain,
  January 17-23, 2010}}, \bibfield{editor}{\bibinfo{person}{Manuel~V.
  Hermenegildo} {and} \bibinfo{person}{Jens Palsberg}} (Eds.).
  \bibinfo{publisher}{{ACM}}, \bibinfo{pages}{119--130}.
\newblock
\urldef\tempurl%
\url{https://doi.org/10.1145/1706299.1706315}
\showDOI{\tempurl}


\bibitem[Thielecke(2003)]%
        {Thielecke03}
\bibfield{author}{\bibinfo{person}{Hayo Thielecke}.}
  \bibinfo{year}{2003}\natexlab{}.
\newblock \showarticletitle{From control effects to typed continuation
  passing}. In \bibinfo{booktitle}{\emph{Conference Record of {POPL} 2003: The
  30th {SIGPLAN-SIGACT} Symposium on Principles of Programming Languages, New
  Orleans, Louisisana, USA, January 15-17, 2003}},
  \bibfield{editor}{\bibinfo{person}{Alex Aiken} {and} \bibinfo{person}{Greg
  Morrisett}} (Eds.). \bibinfo{publisher}{{ACM}}, \bibinfo{pages}{139--149}.
\newblock
\urldef\tempurl%
\url{https://doi.org/10.1145/604131.604144}
\showDOI{\tempurl}


\bibitem[Unno and Kobayashi(2009)]%
        {Unno09}
\bibfield{author}{\bibinfo{person}{Hiroshi Unno} {and} \bibinfo{person}{Naoki
  Kobayashi}.} \bibinfo{year}{2009}\natexlab{}.
\newblock \showarticletitle{Dependent Type Inference with Interpolants}. In
  \bibinfo{booktitle}{\emph{Proceedings of the 11th ACM SIGPLAN Conference on
  Principles and Practice of Declarative Programming}} (Coimbra, Portugal)
  \emph{(\bibinfo{series}{PPDP '09})}. \bibinfo{publisher}{Association for
  Computing Machinery}, \bibinfo{address}{New York, NY, USA},
  \bibinfo{pages}{277–288}.
\newblock
\showISBNx{9781605585680}
\urldef\tempurl%
\url{https://doi.org/10.1145/1599410.1599445}
\showDOI{\tempurl}


\bibitem[Unno et~al\mbox{.}(2018)]%
        {DBLP:journals/pacmpl/0001ST18}
\bibfield{author}{\bibinfo{person}{Hiroshi Unno}, \bibinfo{person}{Yuki
  Satake}, {and} \bibinfo{person}{Tachio Terauchi}.}
  \bibinfo{year}{2018}\natexlab{}.
\newblock \showarticletitle{Relatively complete refinement type system for
  verification of higher-order non-deterministic programs}.
\newblock \bibinfo{journal}{\emph{Proc. {ACM} Program. Lang.}}
  \bibinfo{volume}{2}, \bibinfo{number}{{POPL}} (\bibinfo{year}{2018}),
  \bibinfo{pages}{12:1--12:29}.
\newblock
\urldef\tempurl%
\url{https://doi.org/10.1145/3158100}
\showDOI{\tempurl}


\bibitem[Unno et~al\mbox{.}(2013)]%
        {Unno13}
\bibfield{author}{\bibinfo{person}{Hiroshi Unno}, \bibinfo{person}{Tachio
  Terauchi}, {and} \bibinfo{person}{Naoki Kobayashi}.}
  \bibinfo{year}{2013}\natexlab{}.
\newblock \showarticletitle{Automating Relatively Complete Verification of
  Higher-Order Functional Programs}.
\newblock \bibinfo{journal}{\emph{SIGPLAN Not.}} \bibinfo{volume}{48},
  \bibinfo{number}{1} (\bibinfo{date}{jan} \bibinfo{year}{2013}),
  \bibinfo{pages}{75–86}.
\newblock
\showISSN{0362-1340}
\urldef\tempurl%
\url{https://doi.org/10.1145/2480359.2429081}
\showDOI{\tempurl}


\bibitem[Unno et~al\mbox{.}(2021)]%
        {Unno2021}
\bibfield{author}{\bibinfo{person}{Hiroshi Unno}, \bibinfo{person}{Tachio
  Terauchi}, {and} \bibinfo{person}{Eric Koskinen}.}
  \bibinfo{year}{2021}\natexlab{}.
\newblock \showarticletitle{Constraint-Based Relational Verification}. In
  \bibinfo{booktitle}{\emph{CAV'21}}. \bibinfo{publisher}{Springer Berlin
  Heidelberg}, \bibinfo{pages}{742--766}.
\newblock


\bibitem[Vazou et~al\mbox{.}(2014)]%
        {DBLP:conf/icfp/VazouSJVJ14}
\bibfield{author}{\bibinfo{person}{Niki Vazou}, \bibinfo{person}{Eric~L.
  Seidel}, \bibinfo{person}{Ranjit Jhala}, \bibinfo{person}{Dimitrios
  Vytiniotis}, {and} \bibinfo{person}{Simon L.~Peyton Jones}.}
  \bibinfo{year}{2014}\natexlab{}.
\newblock \showarticletitle{Refinement types for Haskell}. In
  \bibinfo{booktitle}{\emph{Proceedings of the 19th {ACM} {SIGPLAN}
  international conference on Functional programming, Gothenburg, Sweden,
  September 1-3, 2014}}, \bibfield{editor}{\bibinfo{person}{Johan Jeuring}
  {and} \bibinfo{person}{Manuel M.~T. Chakravarty}} (Eds.).
  \bibinfo{publisher}{{ACM}}, \bibinfo{pages}{269--282}.
\newblock
\urldef\tempurl%
\url{https://doi.org/10.1145/2628136.2628161}
\showDOI{\tempurl}


\bibitem[Vekris et~al\mbox{.}(2016)]%
        {DBLP:conf/pldi/VekrisCJ16}
\bibfield{author}{\bibinfo{person}{Panagiotis Vekris},
  \bibinfo{person}{Benjamin Cosman}, {and} \bibinfo{person}{Ranjit Jhala}.}
  \bibinfo{year}{2016}\natexlab{}.
\newblock \showarticletitle{Refinement types for TypeScript}. In
  \bibinfo{booktitle}{\emph{Proceedings of the 37th {ACM} {SIGPLAN} Conference
  on Programming Language Design and Implementation, {PLDI} 2016, Santa
  Barbara, CA, USA, June 13-17, 2016}},
  \bibfield{editor}{\bibinfo{person}{Chandra Krintz} {and}
  \bibinfo{person}{Emery~D. Berger}} (Eds.). \bibinfo{publisher}{{ACM}},
  \bibinfo{pages}{310--325}.
\newblock
\urldef\tempurl%
\url{https://doi.org/10.1145/2908080.2908110}
\showDOI{\tempurl}


\bibitem[Xi and Pfenning(1999)]%
        {DBLP:conf/popl/XiP99}
\bibfield{author}{\bibinfo{person}{Hongwei Xi} {and} \bibinfo{person}{Frank
  Pfenning}.} \bibinfo{year}{1999}\natexlab{}.
\newblock \showarticletitle{Dependent Types in Practical Programming}. In
  \bibinfo{booktitle}{\emph{{POPL} '99, Proceedings of the 26th {ACM}
  {SIGPLAN-SIGACT} Symposium on Principles of Programming Languages, San
  Antonio, TX, USA, January 20-22, 1999}},
  \bibfield{editor}{\bibinfo{person}{Andrew~W. Appel} {and}
  \bibinfo{person}{Alex Aiken}} (Eds.). \bibinfo{publisher}{{ACM}},
  \bibinfo{pages}{214--227}.
\newblock
\urldef\tempurl%
\url{https://doi.org/10.1145/292540.292560}
\showDOI{\tempurl}


\bibitem[Zhang and Myers(2019)]%
        {ZhangM19}
\bibfield{author}{\bibinfo{person}{Yizhou Zhang} {and}
  \bibinfo{person}{Andrew~C. Myers}.} \bibinfo{year}{2019}\natexlab{}.
\newblock \showarticletitle{Abstraction-safe effect handlers via tunneling}.
\newblock \bibinfo{journal}{\emph{Proc. {ACM} Program. Lang.}}
  \bibinfo{volume}{3}, \bibinfo{number}{{POPL}} (\bibinfo{year}{2019}),
  \bibinfo{pages}{5:1--5:29}.
\newblock
\urldef\tempurl%
\url{https://doi.org/10.1145/3290318}
\showDOI{\tempurl}


\bibitem[Zhu and Jagannathan(2013)]%
        {DBLP:conf/vmcai/ZhuJ13}
\bibfield{author}{\bibinfo{person}{He Zhu} {and} \bibinfo{person}{Suresh
  Jagannathan}.} \bibinfo{year}{2013}\natexlab{}.
\newblock \showarticletitle{Compositional and Lightweight Dependent Type
  Inference for {ML}}. In \bibinfo{booktitle}{\emph{Verification, Model
  Checking, and Abstract Interpretation, 14th International Conference, {VMCAI}
  2013, Rome, Italy, January 20-22, 2013. Proceedings}}
  \emph{(\bibinfo{series}{Lecture Notes in Computer Science},
  Vol.~\bibinfo{volume}{7737})}, \bibfield{editor}{\bibinfo{person}{Roberto
  Giacobazzi}, \bibinfo{person}{Josh Berdine}, {and} \bibinfo{person}{Isabella
  Mastroeni}} (Eds.). \bibinfo{publisher}{Springer}, \bibinfo{pages}{295--314}.
\newblock
\urldef\tempurl%
\url{https://doi.org/10.1007/978-3-642-35873-9\_19}
\showDOI{\tempurl}


\end{thebibliography}


\begin{thebibliography}{1}
\providecommand{\natexlab}[1]{#1}
\providecommand{\url}[1]{\texttt{#1}}
\expandafter\ifx\csname urlstyle\endcsname\relax
  \providecommand{\doi}[1]{doi: #1}\else
  \providecommand{\doi}{doi: \begingroup \urlstyle{rm}\Url}\fi

\bibitem[Danvy and Nielsen(2003)]{Danvy03}
Olivier Danvy and Lasse~R. Nielsen.
\newblock A first-order one-pass cps transformation.
\newblock \emph{Theoretical Computer Science}, 308\penalty0 (1):\penalty0
  239--257, 2003.
\newblock ISSN 0304-3975.
\newblock \doi{https://doi.org/10.1016/S0304-3975(02)00733-8}.
\newblock URL
  \url{https://www.sciencedirect.com/science/article/pii/S0304397502007338}.

\end{thebibliography}

\end{document}


\tableofcontents

\listoftheorems[swapnumber]


\section{Typing Rule for Operation Forwarding} \label{sec:language/forwarding}
The typing rule for handling constructs presented in Section 3.2 of the main paper 
assumes that a handler covers all the operations performed
by the handled expression.
In this section, we present another typing rule for handling constructs
to allow \emph{operation forwarding}, that is, allow unhandled operations to be forwarded to outer handlers automatically.
The idea of the typing rule is simple: we derive it from an implementation of operation forwarding.
As mentioned in Section~3.1 of the main paper, 
operation forwarding can be implemented in a calculus without forwarding by adding to a handler
an operation clause $\op(x, k) \mapsto \explet{y}{\op~x}{k~y}$ for each forwarded operation $\op$.
Therefore, we can derive the new typing rule
from the typing of the added clauses. The following is the thus derived new typing rule for handling constructs which natively supports operation forwarding:
\[
\infer{\jdty{\Gamma}{\expwith{h}{c}}{C_2}}
{\begin{gathered}
    h = \{ \expret{x_r} \mapsto c_r, \repi{\op_i(x_i, k_i) \mapsto c_i} \} \quad
    \jdty{\Gamma}{c}{\tycomp{\Sigma}{T}{\tyctl{x_r}{C_1}{C_2}}} \\[-.5ex]
    \jdty{\Gamma, x_r: T}{c_r}{C_1} \quad
    \bigrepi{\jdty{\Gamma, \rep{X_i: \rep{B}_i}, x_i: T_{1i}, k_i: (y_i: T_{2i}) \rarr C_{1i}}{c_i}{C_{2i}}} \\[-.5ex]
    \bigrepi{ \Sigma \ni \op_i: \forall \rep{X_i: \rep{B}_i}. (x_i: T_{1i}) \rarr ((y_i: T_{2i}) \rarr C_{1i}) \rarr C_{2i} }
    \qquad \mathit{Ops}_{\mathrm{fwd}} = \dom(\Sigma) \setminus \dom(h) \\[-.5ex]
    \bigrepi[\op \in \mathit{Ops}_{\mathrm{fwd}}]{\begin{gathered}
        \begin{multlined}
            \Sigma \ni \op : \forall \rep{X^{\op}: \rep{B^{\op}}}. (x^{\op}: T_1^{\op}) \rarr \\
                ((y^{\op}: T_2^{\op}) \rarr
                    \tycomp{\Sigma'}{T_0^{\op}}{\tyctl{z^{\op}}{C_0^{\op}}{C_1^{\op}}}) \rarr
                \tycomp{\Sigma'}{T_0^{\op}}{\tyctl{z^{\op}}{C_0^{\op}}{C_2^{\op}}}
        \end{multlined} \\
        \Sigma' \ni \op : \forall \rep{X^{\op}: \rep{B^{\op}}}. (x^{\op}: T_1^{\op}) \rarr ((y^{\op}: T_2^{\op}) \rarr C_1^{\op}) \rarr C_2^{\op} \quad
        y^{\op} \notin C_0^{\op} \setminus \{ z^{\op} \}
    \end{gathered}
    }
\end{gathered}}
\]
where $\dom(\Sigma)$ denotes the set of the operations associated by $\Sigma$
and $\dom(h)$ denotes the set of the operations handled by $h$,
that is, the set $\{ \repi{\op_i} \}$.
The first two lines are the same as \rulename{T-Hndl}.
The third line is also similar to the last premise of \rulename{T-Hndl},
but here $\Sigma$ is allowed to contain operations other than those handled by $h$.
$\mathit{Ops}_{\mathrm{fwd}}$ is exactly the set of the unhandled (i.e., forwarded) operations.
The last part is the requirement for the forwarded operations,
which can be obtained from the typing derivations of $\op(x, k) \mapsto \explet{y}{\op~x}{k~y}$ as follows.
When we simulate the operation forwarding with the explicit clause,
the operation call $\op~x$ in the clause is handled
by an immediate outer handler (we denote it by $h'$ in what follows).
Therefore, its operation signature is different from $\Sigma$;
in fact, it corresponds to $\Sigma'$ in the rule.
Also, the answer types of the original operation calls of $\op$
(i.e., the answer types of the operation calls of $\op$ in the handled computation $c$)
should have $\Sigma'$ as their operation signatures,
because the final answer type corresponds to the type of the handling construct,
which is handled by the immediate outer handler $h'$.
Therefore, the types of the forwarded operations in $\Sigma$ contains $\Sigma'$
in their answer types.
In addition, the types
$T_0^{\op}$, $T_1^{\op}$, $T_2^{\op}$, $C_0^{\op}$, $C_1^{\op}$, and $C_2^{\op}$
appear multiple times in $\Sigma$ and $\Sigma'$, restricting the type schemes of the operations in $\mathit{Ops}_{\mathrm{fwd}}$.
This restriction can be understood as follows.
First, assume that the original operation call of $\op$ in $c$
has the operation signature $\Sigma$ such that
\[
    \Sigma \ni \op : T_1^{\op} \rarr
        (T_2^{\op} \rarr
            \tycomp{\Sigma'}{T_0^{\op}}{\tyctlMB{C_0^{\op}}{C_1^{\op}}}) \rarr
        \tycomp{\Sigma'}{T_{0A}^{\op}}{\tyctlMB{C_{0A}^{\op}}{C_2^{\op}}}
\]
for some $T_1^{\op}$, $T_2^{\op}$, $T_0^{\op}$, $C_0^{\op}$, $C_1^{\op}$,
$T_{0A}^{\op}$, $C_{0A}^{\op}$, $C_2^{\op}$, and $\Sigma'$, under a context $\Gamma$.
Here we consider only simple types for simplicity,
but a similar argument can be made for dependent and refinement types
by appropriately naming the variables like in the rule above.
Note that its answer types have $\Sigma'$ as described earlier,
and that we do not impose the restriction yet.
From the assumption, the clause $\explet{y}{\op~x}{k~y}$ should be typed
under the context $\Gamma, x: T_1^{\op},
k: T_2^{\op} \rarr \tycomp{\Sigma'}{T_0^{\op}}{\tyctlMB{C_0^{\op}}{C_1^{\op}}}$~.
Then, the input type of $\op$ in the clause should be the type of $x$, namely, $T_1^{\op}$,
and the output type of $\op$ should be the type of the variable $y$,
which turns out to be $T_2^{\op}$ from the type of $k$.
Therefore, the operation signature $\Sigma'$ for $\op~x$ should contain
$\op : T_1^{\op} \rarr (T_2^{\op} \rarr C_{1A}^{\op}) \rarr C_{2A}^{\op}$
for some $C_{1A}^{\op}$ and $C_{2A}^{\op}$~.
Then, according to the typing rules for operation calls and let-expressions,
it is required that $C_{1A}^{\op} = C_1^{\op}$,
and the type of $\explet{y}{\op~x}{k~y}$ is
$\tycomp{\Sigma'}{T_{0}^{\op}}{\tyctlMB{C_{0}^{\op}}{C_{2A}^{\op}}}$~.
Finally, since the type of the clause corresponds to
the final answer type of the operation $\op$ in $\Sigma$
(which is $\tycomp{\Sigma'}{T_{0A}^{\op}}{\tyctlMB{C_{0A}^{\op}}{C_2^{\op}}}$ from the assumption),
it should satisfy
$T_{0}^{\op} = T_{0A}^{\op}$, $C_{0}^{\op} = C_{0A}^{\op}$, and $C_{2A}^{\op} = C_{2}^{\op}$~.

\section{Detailed explanation of the benchmark} \label{sec:benchmark-details}

In this section, we present the result of the verification of
the benchmark \texttt{queue-2-SAT.ml} as an example.
The following is the main part of the program of \texttt{queue-2-SAT.ml}:
\begin{verbatim}
let[@annot_MB "int list ->
    (unit -> ({Get_next: s1, Add_to_queue: s2} |> int option / s => s)) ->
    int option"]
  queue initial (body :unit -> int option) =
    match_with body () {
      retc = (fun x -> (fun _ -> x));
      exnc = raise;
      effc = fun (type a) (e: a eff) -> match e with
        | Get_next _ctx -> Some (fun (k: (a, _) continuation) ->
            (fun queue -> match queue with
              | [] -> continue k None []
              | hd::tl -> continue k (Some hd) tl) )
        | Add_to_queue v -> Some (fun (k: (a, _) continuation) ->
            (fun queue -> continue k () (queue @ [v])) )
    } initial

let main init =
  queue init (fun () ->
    perform (Add_to_queue 42);
    let _ = perform (Get_next 1(*dummy*)) in
    perform (Get_next 2(*dummy*)) )
\end{verbatim}
This program uses two operations \texttt{Get\_next} and \texttt{Add\_to\_queue},
which are used to dequeue and enqueue elements respectively.
The function \texttt{queue} manages the queue.
It receives an initial queue \texttt{initial} and the function \texttt{body},
handling the operations performed in \texttt{body} in the state-passing manner
to simulate the behavior of the queue.
The first three lines of the program are the underlying simple type annotation,
which tells the function \texttt{queue}
that the argument \texttt{body} may perform the operations \texttt{Get\_next} and \texttt{Add\_to\_queue}
and that its control effect is impure.
This annotaion is necessary because our implementation does not support effect polymorphism
as mentioned in Section~4 of the main paper. 
The main function \texttt{main} first enqueue one element,
and then try to dequeue twice
(\texttt{Get\_next} returns \texttt{None} when the queue is empty).
Note that we added a ghost parameter \texttt{\_ctx} to \texttt{Get\_next},
which is used to distinguish its two occurrences.
We give \texttt{1} to the first occurrence of \texttt{Get\_next},
and \texttt{2} to the second.
This ghost parameter is crucial for the precise verification of this program,
described later in this section.

\newcommand{\ilist}{\kwty{ilist}}
\newcommand{\iopt}{\kwty{iopt}}
\newcommand{\expNil}{\mathtt{[]}}
\newcommand{\expNone}{\mathtt{None}}
\newcommand{\isCons}{\mathrm{isCons}}
\newcommand{\isSome}{\mathrm{isSome}}
\newcommand{\tail}{\mathrm{tail}}

We defined the following refinement type
as the specification for the main function \texttt{main}
(here after, we abbreviate the type $\kwty{int\ list}$ and $\kwty{int\ option}$
as $\ilist$ and $\iopt$ respectively):
\[
  \tyrfn{z}{\ilist}{z \ne \expNil} \rarr \tyrfn{z}{\iopt}{z \ne \expNone}
\]
That is, if the queue is initially not empty,
the last dequeue should return some value.

By running the verification of the program with the specification,
our implementation returns ``SAT'' as shown in Table~1 in the main paper, 
that is, the function \texttt{main} certainly has the type given as the specification.
Let us investigate more detail by seeing the inferred type of the function \texttt{queue}:
{\allowdisplaybreaks
\begin{align}
  &(init:\tyrfn{z}{\ilist}{z \ne \expNil}) \\
  &\rarr (\tyunit \rarr \tycomp{\Sigma}{\iopt}{\tyctl{x}{
    (\ilist \rarr \tyrfn{z}{\iopt}{\phi_1})
  }{
    (\tyrfn{z}{\ilist}{\phi_2} \rarr \tyrfn{z}{\iopt}{z \ne \expNone})
  }}) \!\!\! \\
  &\rarr \tyrfn{z}{\iopt}{z \ne \expNone}
\end{align}
\[
  \text{where} \begin{array}[t]{r@{}l}
    \Sigma \defeq \{
      &\mathtt{Add\_to\_queue}: \tyint \rarr (\tyunit \rarr \\
        &\quad ((q:\ilist) \rarr \tyrfn{z}{\iopt}{\phi_{41}}))
          \rarr (\tyrfn{z}{\ilist}{\phi_2} \rarr \tyrfn{z}{\iopt}{z \ne \expNone}), \\
      &\mathtt{Get\_next}: (ctx: \tyint) \rarr ((y: \iopt) \rarr \\
        &\quad ((q:\ilist) \rarr \tyrfn{z}{\iopt}{\phi_{31} \land \phi_{32}}))
          \rarr ((q:\ilist) \rarr \tyrfn{z}{\iopt}{\phi_{41} \land \phi_{42}})
    \}\end{array}
\]
\[\begin{array}[t]{r@{}l@{\quad}r@{}l}
  \phi_1 &\defeq \isSome(x) \Rarr z \ne \expNone &
  \phi_2 &\defeq init \ne \expNil \Rarr z \ne \expNil \\
  \phi_{31} &\defeq \isCons(q) \land \isSome(y) \Rarr z \ne \expNone &
  \phi_{32} &\defeq \isSome(y) \land ctx \ge 2 \Rarr z \ne \expNone \\
  \phi_{41} &\defeq \isCons(q) \land \isCons(\tail(q)) \Rarr z \ne \expNone &
  \phi_{42} &\defeq \isCons(q) \land ctx \ge 2 \Rarr z \ne \expNone
\end{array}\]
}
where $\isSome(x)$ holds if $x = \mathtt{Some}~v$ for some $v$,
$\isCons(x)$ holds if $x = v\mathtt{::}w$ for some $v$ and $w$,
and $\tail(x)$ returns the tail of the list $x$.
In the operation sigunature, we can find that
\texttt{Add\_to\_queue} changes the answer type from
$(q:\ilist) \rarr \tyrfn{z}{\iopt}{\phi_{41}}$
to
$\tyrfn{z}{\ilist}{\phi_2} \rarr \tyrfn{z}{\iopt}{z \ne \expNone}$.
Therefore, \verb|perform (Add_to_queue 42)| can be given the control effect
\[
  \tyctl{\_}{
    ((q:\ilist) \rarr \tyrfn{z}{\iopt}{\phi_{41}})
  }{
    (\tyrfn{z}{\ilist}{\phi_2} \rarr \tyrfn{z}{\iopt}{z \ne \expNone})
  }~.
\]
Similarly, in the operation sigunature,
\texttt{Get\_next} changes the answer type from
$(q:\ilist) \rarr \tyrfn{z}{\iopt}{\phi_{31} \land \phi_{32}}$
to
$(q:\ilist) \rarr \tyrfn{z}{\iopt}{\phi_{41} \land \phi_{42}}$~.
Here, since the refinements of these answer types contain a condition on $ctx$,
their truth depend on whether $ctx = 1\ (< 2)$ or $ctx = 2\ (\ge 2)$.
This enables assigning different control effects (i.e., different ARM)
to each occurrence of \texttt{Get\_next} depending on the context.
Namely, \verb|perform (Get_next 1)| can be given the control effect
\[
  \tyctl{y}{
    (q:\ilist) \rarr \tyrfn{z}{\iopt}{\phi_{31}}
  }{
    (q:\ilist) \rarr \tyrfn{z}{\iopt}{\phi_{41}}
  }
\]
since $ctx = 1$,
while \verb|perform (Get_next 2)| can be given the control effect
\[
  \tyctl{y}{
    \ilist \rarr \tyrfn{z}{\iopt}{\isSome(y) \Rarr z \ne \expNone}
  }{
    (q:\ilist) \rarr \tyrfn{z}{\iopt}{\isCons(q) \Rarr z \ne \expNone}
  }
\]
since $ctx = 2$.
Now, the control effect of the argument \texttt{body} can be obtained from
the composition of these three control effects,
which results in
\[
  \tyctl{x}{
    (\ilist \rarr \tyrfn{z}{\iopt}{\phi_1})
  }{
    (\tyrfn{z}{\ilist}{\phi_2} \rarr \tyrfn{z}{\iopt}{z \ne \expNone})
  }~.
\]
Then, the handling construct is assigned the final answer type of \texttt{body},
i.e., $\tyrfn{z}{\ilist}{\phi_2} \rarr \tyrfn{z}{\iopt}{z \ne \expNone}$,
and finally applying the non-empty initial queue to the handling construct returns
a value of type $\tyrfn{z}{\iopt}{z \ne \expNone}$ as expected.

\section{Definitions (other than those shown in the main paper) and Assumptions}

\subsection{Well-formedness of typing contexts, value types, and computation types}

\fbox{$\jdwf{}{\Gamma}$} \quad
\fbox{$\jdwf{\Gamma}{T}$} \quad \fbox{$\jdwf{\Gamma}{C}$} \quad
\fbox{$\jdwf{\Gamma}{\Sigma}$} \quad \fbox{$\jdwf{\Gamma \mid T}{S}$}
\begin{gather}
    \infersc[WE-Empty]{\jdwf{}{\emptyset}}
    {}
    \quad
    \infersc[WE-Var]{\jdwf{}{\Gamma, x: T}}
    {
        \jdwf{}{\Gamma} &
        x \notin \dom(\Gamma) &
        \jdwf{\Gamma}{T}
    }
    \quad
    \infersc[WE-PVar]{\jdwf{}{\Gamma, X: \rep{B}}}
    {
        \jdwf{}{\Gamma} &
        X \notin \dom(\Gamma)
    }
    \\
    \infersc[WT-Rfn]{\jdwf{\Gamma}{\tyrfn{x}{B}{\phi}}}
    {\Gamma, x: B \vdash \phi}
    \quad
    \infersc[WT-Fun]{\jdwf{\Gamma}{(x: T) \rarr C}}
    {
        \jdwf{\Gamma, x: T}{C}
    }
    \\
    \infersc[WT-Comp]{\jdwf{\Gamma}{\tycomp{\Sigma}{T}{S}}}
    {
        \jdwf{\Gamma}{\Sigma} &
        \jdwf{\Gamma}{T} &
        \jdwf{\Gamma \mid T}{S}
    }
    \qquad
    \infersc[WT-Sig]{\jdwf{\Gamma}{\{ \repi{\op_i : \forall \rep{X_i: \rep{B}_i}. F_i} \}}}
    {\repi{\jdwf{\Gamma, \rep{X_i: \rep{B}_i}}{F_i}}}
    \\
    \infersc[WT-Pure]{\jdwf{\Gamma \mid T}{\square}}
    {\jdwf{}{\Gamma}}
    \quad
    \infersc[WT-ATM]{\jdwf{\Gamma \mid T}{\tyctl{x}{C_1}{C_2}}}
    {
        \jdwf{\Gamma, x: T}{C_1} &
        \jdwf{\Gamma}{C_2}
    }
\end{gather}

\subsection{Assumptions on well-formedness judgments of formulas, well-formedness judgments of predicates, and semantic validity judgements of formulas}

\begin{assumption} \label{asm:formla} \quad
    \begin{itemize}
        \item If $\jdwf{\Gamma}{\phi}$, then $\jdwf{}{\Gamma}$.
        \item If $\jdwf{}{\Gamma}$, $z \notin \dom(\Gamma)$ and $\dom(\Gamma, z: B) \supseteq \fv(\phi)$, then $\jdwf{\Gamma, z: B}{\phi}$.
        \item If $\jdwf{}{\Gamma, x: T, \Gamma'}$ and $\jdty{\Gamma, \Gamma'}{A}{\rep{B}}$, then $\jdty{\Gamma, x: T, \Gamma'}{A}{\rep{B}}$.
        \item If $\jdwf{}{\Gamma, x: T, \Gamma'}$ and $\jdwf{\Gamma, \Gamma'}{\phi}$, then $\jdwf{\Gamma, x: T, \Gamma'}{\phi}$.
        \item If $\valid{\Gamma, \Gamma'}{\phi}$, then $\valid{\Gamma, x: T, \Gamma'}{\phi}$.
        \item If $\jdty{\Gamma}{v}{T}$ and $\jdty{\Gamma, x: T, \Gamma'}{A}{\rep{B}}$, then $\jdty{\Gamma, \Gamma'[v/x]}{A[v/x]}{\rep{B}}$.
        \item If $\jdty{\Gamma}{v}{T}$ and $\jdwf{\Gamma, x: T, \Gamma'}{\phi}$, then $\jdwf{\Gamma, \Gamma'[v/x]}{\phi[v/x]}$.
        \item If $\jdty{\Gamma}{v}{T}$ and $\valid{\Gamma, x: T, \Gamma'}{\phi}$, then $\valid{\Gamma, \Gamma'[v/x]}{\phi[v/x]}$.
        \item If $\jdty{\Gamma}{A}{\rep{B}}$ and $\jdty{\Gamma, X: \rep{B}, \Gamma'}{A'}{\rep{B'}}$, then $\jdty{\Gamma, \Gamma'[A/X]}{A'[A/X]}{\rep{B'}}$.
        \item If $\jdty{\Gamma}{A}{\rep{B}}$ and $\jdwf{\Gamma, X: \rep{B}, \Gamma'}{\phi}$, then $\jdwf{\Gamma, \Gamma'[A/X]}{\phi[A/X]}$.
        \item If $\jdty{\Gamma}{A}{\rep{B}}$ and $\valid{\Gamma, X: \rep{B}, \Gamma'}{\phi}$, then $\valid{\Gamma, \Gamma'[A/X]}{\phi[A/X]}$.
        \item If $\jdsub{\Gamma}{T_1}{T_2}$, $\jdwf{}{\Gamma, x:T_1, \Gamma'}$ and $\jdty{\Gamma, x:T_2, \Gamma'}{A}{\rep{B}}$, then $\jdty{\Gamma, x:T_1, \Gamma'}{A}{\rep{B}}$.
        \item If $\jdsub{\Gamma}{T_1}{T_2}$, $\jdwf{}{\Gamma, x:T_1, \Gamma'}$ and $\jdwf{\Gamma, x:T_2, \Gamma'}{\phi}$, then $\jdwf{\Gamma, x:T_1, \Gamma'}{\phi}$.
        \item If $\jdsub{\Gamma}{T_1}{T_2}$ and $\valid{\Gamma, x:T_2, \Gamma'}{\phi}$, then $\valid{\Gamma, x:T_1, \Gamma'}{\phi}$.
        \item If $x \notin \fv(\Gamma', \phi)$ and $\jdwf{\Gamma, x: T_0, \Gamma'}{\phi}$, then $\jdwf{\Gamma, \Gamma'}{\phi}$.
        \item If $\jdwf{\Gamma, x: (y: T_1) \rarr C_1, \Gamma'}{\phi}$, then $x \notin \fv(\Gamma', \phi)$.
        \item If $\vDash \phi$ and $\valid{\Gamma, \phi, \Gamma'}{\phi'}$, then $\valid{\Gamma, \Gamma'}{\phi'}$.
        \item If $\jdwf{\Gamma}{\phi}$, then $\Gamma \vDash \phi \Rarr \phi$.
        \item If $\Gamma \vDash \phi_1 \Rarr \phi_2$ and $\Gamma \vDash \phi_2 \Rarr \phi_3$, then $\Gamma \vDash \phi_1 \Rarr \phi_3$.
        \item If $\jdwf{\Gamma, x: \tyrfn{z}{B}{z = y}, \Gamma'}{\phi}$, then $\Gamma, x: \tyrfn{z}{B}{z = y}, \Gamma' \vDash \phi \implies \phi[y/x]$~.
        \item If $\jdwf{\Gamma, x: \tyrfn{z}{B}{z = y}, \Gamma'}{\phi}$, then $\Gamma, x: \tyrfn{z}{B}{z = y}, \Gamma' \vDash \phi[y/x] \implies \phi$~.
    \end{itemize}
\end{assumption}

\subsection{Assumptions on primitives}

\begin{assumption} \label{asm:prim} \quad
    \begin{itemize}
        \item $\jdwf{}{\ty(p)}$ for all $p$.
        \item If $\ty(p) = (x: T) \rarr C$, then
            $\zeta(p, v)$ is defined and $\jdty{}{\zeta(p, v)}{C[v/x]}$
            for all $v$ such that $\jdty{}{v}{T}$.
        \item If $\ty(p) = \tyrfn{z}{\tybool}{\phi}$, then $p = \exptrue$ or $p = \expfalse$.
    \end{itemize}
\end{assumption}

\section{Proof of Type Safety}

\subsection{Progress}

\begin{lemma}[Weakening] \label{lem:weaken} \quad
    \begin{enumerate}
        \item Assume that $\jdwf{}{\Gamma, x: T_0, \Gamma'}$.
            \begin{itemize}
                \item If $\jdwf{\Gamma, \Gamma'}{T}$, then $\jdwf{\Gamma, x: T_0, \Gamma'}{T}$.
                \item If $\jdwf{\Gamma, \Gamma'}{C}$, then $\jdwf{\Gamma, x: T_0, \Gamma'}{C}$.
                \item If $\jdwf{\Gamma, \Gamma'}{\Sigma}$, then $\jdwf{\Gamma, x: T_0, \Gamma'}{\Sigma}$.
                \item If $\jdwf{\Gamma, \Gamma' \mid T}{S}$, then $\jdwf{\Gamma, x: T_0, \Gamma' \mid T}{S}$.
            \end{itemize}
        \item Assume that $\jdwf{}{\Gamma, x: T_0, \Gamma'}$.
            \begin{itemize}
                \item If $\jdty{\Gamma, \Gamma'}{v}{T}$, then $\jdty{\Gamma, x: T_0, \Gamma'}{v}{T}$.
                \item If $\jdty{\Gamma, \Gamma'}{c}{C}$, then $\jdty{\Gamma, x: T_0, \Gamma'}{c}{C}$.
            \end{itemize}
        \item
            \begin{itemize}
                \item If $\jdsub{\Gamma, \Gamma'}{T_1}{T_2}$, then $\jdsub{\Gamma, x: T_0, \Gamma'}{T_1}{T_2}$.
                \item If $\jdsub{\Gamma, \Gamma'}{C_1}{C_2}$, then $\jdsub{\Gamma, x: T_0, \Gamma'}{C_1}{C_2}$.
                \item If $\jdsub{\Gamma, \Gamma'}{\Sigma_1}{\Sigma_2}$, then $\jdsub{\Gamma, x: T_0, \Gamma'}{\Sigma_1}{\Sigma_2}$.
                \item If $\jdsub{\Gamma, \Gamma' \mid T}{S_1}{S_2}$, then $\jdsub{\Gamma, x: T_0, \Gamma' \mid T}{S_1}{S_2}$.
            \end{itemize}
    \end{enumerate}
\end{lemma}
\begin{proof}
    By simultaneous induction on the derivations.
    The cases for \rulename{WT-Rfn}, \rulename{T-Op} and \rulename{S-Rfn} use Assumption \ref{asm:formla}.
\end{proof}

\begin{lemma}[Narrowing] \label{lem:narrow} \quad
    \begin{enumerate}
        \item Assume that $\jdsub{\Gamma}{T_1}{T_2}$ and $\jdwf{}{\Gamma, x:T_1, \Gamma'}$.
            \begin{itemize}
                \item If $\jdwf{\Gamma, x:T_2, \Gamma'}{T}$, then $\jdwf{\Gamma, x: T_1, \Gamma'}{T}$.
                \item If $\jdwf{\Gamma, x:T_2, \Gamma'}{C}$, then $\jdwf{\Gamma, x: T_1, \Gamma'}{C}$.
                \item If $\jdwf{\Gamma, x:T_2, \Gamma'}{\Sigma}$, then $\jdwf{\Gamma, x: T_1, \Gamma'}{\Sigma}$.
                \item If $\jdwf{\Gamma, x:T_2, \Gamma' \mid T}{S}$, then $\jdwf{\Gamma, x: T_1, \Gamma' \mid T}{S}$.
            \end{itemize}
        \item Assume that $\jdsub{\Gamma}{T_1}{T_2}$ and $\jdwf{}{\Gamma, x:T_1, \Gamma'}$.
            \begin{itemize}
                \item If $\jdty{\Gamma, x:T_2, \Gamma'}{v}{T}$, then $\jdty{\Gamma, x: T_1, \Gamma'}{v}{T}$.
                \item If $\jdty{\Gamma, x:T_2, \Gamma'}{c}{C}$, then $\jdty{\Gamma, x: T_1, \Gamma'}{c}{C}$.
            \end{itemize}
        \item Assume that $\jdsub{\Gamma}{T_1}{T_2}$.
            \begin{itemize}
                \item If $\jdsub{\Gamma, x:T_2, \Gamma'}{T_1'}{T_2'}$, then $\jdsub{\Gamma, x: T_1, \Gamma'}{T_1'}{T_2'}$.
                \item If $\jdsub{\Gamma, x:T_2, \Gamma'}{C_1}{C_2}$, then $\jdsub{\Gamma, x: T_1, \Gamma'}{C_1}{C_2}$.
                \item If $\jdsub{\Gamma, x:T_2, \Gamma'}{\Sigma_1}{\Sigma_2}$, then $\jdsub{\Gamma, x: T_1, \Gamma'}{\Sigma_1}{\Sigma_2}$.
                \item If $\jdsub{\Gamma, x:T_2, \Gamma' \mid T}{S_1}{S_2}$, then $\jdsub{\Gamma, x: T_1, \Gamma' \mid T}{S_1}{S_2}$.
            \end{itemize}
        \item If $\jdsub{\Gamma}{T_1}{T_2}$ and $\jdsub{\Gamma \mid T_2}{S_1}{S_2}$, then $\jdsub{\Gamma \mid T_1}{S_1}{S_2}$.
    \end{enumerate}
\end{lemma}
\begin{proof}
    By simultaneous induction on the derivations.
    The cases for \rulename{WT-Rfn}, \rulename{T-Op} and \rulename{S-Rfn} use Assumption \ref{asm:formla}.
\end{proof}

\begin{lemma}[Substitution] \label{lem:subst} \quad
    \begin{enumerate}
        \item Assume that $\jdty{\Gamma}{v}{T_0}$.
            \begin{itemize}
                \item If $\jdwf{}{\Gamma, x:T_0, \Gamma'}$, then $\jdwf{}{\Gamma, \Gamma'[v/x]}$.
                \item If $\jdwf{\Gamma, x:T_0, \Gamma'}{T}$, then $\jdwf{\Gamma, \Gamma'[v/x]}{T[v/x]}$.
                \item If $\jdwf{\Gamma, x:T_0, \Gamma'}{C}$, then $\jdwf{\Gamma, \Gamma'[v/x]}{C[v/x]}$.
                \item If $\jdwf{\Gamma, x:T_0, \Gamma'}{\Sigma}$, then $\jdwf{\Gamma, \Gamma'[v/x]}{\Sigma[v/x]}$.
                \item If $\jdwf{\Gamma, x:T_0, \Gamma' \mid T}{S}$, then $\jdwf{\Gamma, \Gamma'[v/x] \mid T[v/x]}{S[v/x]}$.
            \end{itemize}
        \item Assume that $\jdty{\Gamma}{v}{T_0}$.
            \begin{itemize}
                \item If $\jdty{\Gamma, x:T_0, \Gamma'}{v}{T}$, then $\jdty{\Gamma, \Gamma'[v/x]}{v[v/x]}{T[v/x]}$.
                \item If $\jdty{\Gamma, x:T_0, \Gamma'}{c}{C}$, then $\jdty{\Gamma, \Gamma'[v/x]}{c[v/x]}{C[v/x]}$.
            \end{itemize}
        \item Assume that $\jdty{\Gamma}{v}{T_0}$.
            \begin{itemize}
                \item If $\jdsub{\Gamma, x:T_0, \Gamma'}{T_1}{T_2}$, then $\jdsub{\Gamma, \Gamma'[v/x]}{T_1[v/x]}{T_2[v/x]}$.
                \item If $\jdsub{\Gamma, x:T_0, \Gamma'}{C_1}{C_2}$, then $\jdsub{\Gamma, \Gamma'[v/x]}{C_1[v/x]}{C_2[v/x]}$.
                \item If $\jdsub{\Gamma, x:T_0, \Gamma'}{\Sigma_1}{\Sigma_2}$, then $\jdsub{\Gamma, \Gamma'[v/x]}{\Sigma_1[v/x]}{\Sigma_2[v/x]}$.
                \item If $\jdsub{\Gamma, x:T_0, \Gamma' \mid T}{S_1}{S_2}$, then $\jdsub{\Gamma, \Gamma'[v/x] \mid T}{S_1[v/x]}{S_2[v/x]}$.
            \end{itemize}
    \end{enumerate}
\end{lemma}
\begin{proof}
    By simultaneous induction on the derivations.
    The cases for \rulename{WT-Rfn}, \rulename{T-Op} and \rulename{S-Rfn} use Assumption \ref{asm:formla}.
\end{proof}

\begin{lemma}[Predicate Substitution] \label{lem:subst-pred} \quad
    \begin{enumerate}
        \item Assume that $\jdty{\Gamma}{A}{\rep{B}}$.
            \begin{itemize}
                \item If $\jdwf{}{\Gamma, X: \rep{B}, \Gamma'}$, then $\jdwf{}{\Gamma, \Gamma'[A/X]}$.
                \item If $\jdwf{\Gamma, X: \rep{B}, \Gamma'}{T}$, then $\jdwf{\Gamma, \Gamma'[A/X]}{T[A/X]}$.
                \item If $\jdwf{\Gamma, X: \rep{B}, \Gamma'}{C}$, then $\jdwf{\Gamma, \Gamma'[A/X]}{C[A/X]}$.
                \item If $\jdwf{\Gamma, X: \rep{B}, \Gamma'}{\Sigma}$, then $\jdwf{\Gamma, \Gamma'[A/X]}{\Sigma[A/X]}$.
                \item If $\jdwf{\Gamma, X: \rep{B}, \Gamma' \mid T}{S}$, then $\jdwf{\Gamma, \Gamma'[A/X] \mid T[A/X]}{S[A/X]}$.
            \end{itemize}
        \item Assume that $\jdty{\Gamma}{A}{\rep{B}}$.
            \begin{itemize}
                \item If $\jdty{\Gamma, X: \rep{B}, \Gamma'}{v}{T}$, then $\jdty{\Gamma, \Gamma'[A/X]}{v[A/X]}{T[A/X]}$.
                \item If $\jdty{\Gamma, X: \rep{B}, \Gamma'}{c}{C}$, then $\jdty{\Gamma, \Gamma'[A/X]}{c[A/X]}{C[A/X]}$.
            \end{itemize}
        \item Assume that $\jdty{\Gamma}{A}{\rep{B}}$.
            \begin{itemize}
                \item If $\jdsub{\Gamma, X: \rep{B}, \Gamma'}{T_1}{T_2}$, then $\jdsub{\Gamma, \Gamma'[A/X]}{T_1[A/X]}{T_2[A/X]}$.
                \item If $\jdsub{\Gamma, X: \rep{B}, \Gamma'}{C_1}{C_2}$, then $\jdsub{\Gamma, \Gamma'[A/X]}{C_1[A/X]}{C_2[A/X]}$.
                \item If $\jdsub{\Gamma, X: \rep{B}, \Gamma'}{\Sigma_1}{\Sigma_2}$, then $\jdsub{\Gamma, \Gamma'[A/X]}{\Sigma_1[A/X]}{\Sigma_2[A/X]}$.
                \item If $\jdsub{\Gamma, X: \rep{B}, \Gamma' \mid T}{S_1}{S_2}$, then $\jdsub{\Gamma, \Gamma'[A/X] \mid T}{S_1[A/X]}{S_2[A/X]}$.
            \end{itemize}
    \end{enumerate}
\end{lemma}
\begin{proof}
    By simultaneous induction on the derivations.
    The cases for \rulename{WT-Rfn}, \rulename{T-Op} and \rulename{S-Rfn} use Assumption \ref{asm:formla}.
\end{proof}

\begin{lemma}[Remove unused type bindings] \label{lem:rm-unused} \quad
    \begin{itemize}
        \item If $x \notin \fv(\Gamma')$ and $\jdwf{}{\Gamma, x:T_0, \Gamma'}$, then $\jdwf{}{\Gamma, \Gamma'}$.
        \item If $x \notin \fv(\Gamma', T)$ and $\jdwf{\Gamma, x:T_0, \Gamma'}{T}$, then $\jdwf{\Gamma, \Gamma'}{T}$.
        \item If $x \notin \fv(\Gamma', C)$ and $\jdwf{\Gamma, x:T_0, \Gamma'}{C}$, then $\jdwf{\Gamma, \Gamma'}{C}$.
        \item If $x \notin \fv(\Gamma', \Sigma)$ and $\jdwf{\Gamma, x:T_0, \Gamma'}{\Sigma}$, then $\jdwf{\Gamma, \Gamma'}{\Sigma}$.
        \item If $x \notin \fv(\Gamma', T, S)$ and $\jdwf{\Gamma, x:T_0, \Gamma' \mid T}{S}$, then $\jdwf{\Gamma, \Gamma' \mid T}{S}$.
    \end{itemize}
\end{lemma}
\begin{proof}
    By simultaneous induction on the derivations.
    The case for \rulename{WT-Rfn} uses Assumption \ref{asm:formla}.
\end{proof}

\begin{lemma}[Variables of non-refinement types do not occur in types] \label{lem:notin-nonrfn} \quad
    \begin{itemize}
        \item If $\jdwf{}{\Gamma, x:(y:T_1) \rarr C_1, \Gamma'}$, then $x \notin \fv(\Gamma')$.
        \item If $\jdwf{\Gamma, x:(y:T_1) \rarr C_1, \Gamma'}{T}$, then $x \notin \fv(\Gamma', T)$.
        \item If $\jdwf{\Gamma, x:(y:T_1) \rarr C_1, \Gamma'}{C}$, then $x \notin \fv(\Gamma', C)$.
        \item If $\jdwf{\Gamma, x:(y:T_1) \rarr C_1, \Gamma'}{\Sigma}$, then $x \notin \fv(\Gamma', \Sigma)$.
        \item If $\jdwf{\Gamma, x:(y:T_1) \rarr C_1, \Gamma' \mid T}{S}$, then $x \notin \fv(\Gamma', T, S)$.
    \end{itemize}
\end{lemma}
\begin{proof}
    By simultaneous induction on the derivations.
    The case for \rulename{WT-Rfn} uses Assumption \ref{asm:formla}.
\end{proof}

\begin{lemma}[Remove non-refinement type bindings] \label{lem:rm-nonrfn} \quad
    \begin{itemize}
        \item If $\jdwf{}{\Gamma, x:(y:T_1) \rarr C_1, \Gamma'}$, then $\jdwf{}{\Gamma, \Gamma'}$.
        \item If $\jdwf{\Gamma, x:(y:T_1) \rarr C_1, \Gamma'}{T}$, then $\jdwf{\Gamma, \Gamma'}{T}$.
        \item If $\jdwf{\Gamma, x:(y:T_1) \rarr C_1, \Gamma'}{C}$, then $\jdwf{\Gamma, \Gamma'}{C}$.
        \item If $\jdwf{\Gamma, x:(y:T_1) \rarr C_1, \Gamma'}{\Sigma}$, then $\jdwf{\Gamma, \Gamma'}{\Sigma}$.
        \item If $\jdwf{\Gamma, x:(y:T_1) \rarr C_1, \Gamma' \mid T}{S}$, then $\jdwf{\Gamma, \Gamma' \mid T}{S}$.
    \end{itemize}
\end{lemma}
\begin{proof}
    Immediate by Lemma \ref{lem:notin-nonrfn} and \ref{lem:rm-unused}.
\end{proof}

\begin{lemma}[Well-formedness of typing contexts from other judgements] \label{lem:wfg} \quad
    \begin{enumerate}
        \item If $\jdwf{\Gamma}{T}$, then $\jdwf{}{\Gamma}$.
        \item If $\jdwf{\Gamma}{C}$, then $\jdwf{}{\Gamma}$.
        \item If $\jdwf{\Gamma}{\Sigma}$, then $\jdwf{}{\Gamma}$.
        \item If $\jdwf{\Gamma \mid T}{S}$, then $\jdwf{}{\Gamma}$.
    \end{enumerate}
\end{lemma}
\begin{proof}
    By simultaneous induction on the derivations.
\end{proof}

\begin{lemma}[Well-formedness of types from other judgements] \label{lem:wft} \quad
    \begin{enumerate}
        \item If $\jdty{\Gamma}{v}{T}$, then $\jdwf{\Gamma}{T}$.
        \item If $\jdty{\Gamma}{c}{C}$, then $\jdwf{\Gamma}{C}$.
    \end{enumerate}
\end{lemma}
\begin{proof}
    \def\currentprefix{wft}
    By simultaneous induction on the derivations.
    \begin{enumit}
        \item 
        \begin{description}
            \item[Case \rulename{T-CVar}:] We have
                \def\currentprefix{wft:cvar}
                \begin{enumrm}
                    \item\llabel{eq-v} $v = x$,
                    \item\llabel{eq-T} $T = \tyrfn{z}{B}{z = x}$,
                    \item\llabel{wf-G} $\jdwf{}{\Gamma}$, and
                    \item\llabel{eq-Gx} $\Gamma(x) = \tyrfn{z}{B}{\phi}$
                \end{enumrm}
                for some $z, x$, and $B$.
                W.l.o.g., we can assume that $z \notin \dom(\Gamma)$.
                Also, since \lref{eq-Gx} implies $x \in \dom(\Gamma)$,
                it holds that $\dom(\Gamma, x: B) \supseteq \fv(z = x)$.
                Then, by the Assumption \ref{asm:formla},
                we have $\jdwf{\Gamma, x: B}{z = x}$.
                By \rulename{WT-Rfn}, we have the conclusion.
            \item[Case \rulename{T-Var}:] We have
                \def\currentprefix{wft:var}
                \begin{enumrm}
                    \item\llabel{eq-v} $v = x$,
                    \item\llabel{eq-T} $T = \Gamma(x)$,
                    \item\llabel{wf-G} $\jdwf{}{\Gamma}$, and
                    \item\llabel{eq-Gx} $\Gamma(x) \neq \tyrfn{z}{B}{\phi}$ for all $z, B$, and $\phi$
                \end{enumrm}
                for some $x$.
                \lref{eq-T} implies that $\Gamma$ is of the form $\Gamma_1, x: T, \Gamma_2$
                for some $\Gamma_1$ and $\Gamma_2$.
                Therefore, by inverting \lref{wf-G} repeatedly, we have $\jdwf{\Gamma_1}{T}$.
                By Lemma \ref{lem:weaken} with \lref{wf-G}, we have the conclusion.
            \item[Case \rulename{T-Prim}:] We have
                \def\currentprefix{wft:prim}
                \begin{enumrm}
                    \item\llabel{eq-v} $v = p$,
                    \item\llabel{eq-T} $T = \ty(p)$, and
                    \item\llabel{wf-G} $\jdwf{}{\Gamma}$
                \end{enumrm}
                for some $p$.
                By Assumption \ref{asm:prim}, we have $\jdwf{}{\ty(p)}$.
                By Lemma \ref{lem:weaken} with \lref{wf-G}, we have the conclusion.
            \item[Case \rulename{T-Fun}:] We have
                \def\currentprefix{wft:fun}
                \begin{enumrm}
                    \item\llabel{eq-v} $v = \exprec{f}{x}{c}$,
                    \item\llabel{eq-T} $T = (x: T_0) \rarr C$, and
                    \item\llabel{ty-c} $\jdty{\Gamma, x: T_0}{c}{C}$
                \end{enumrm}
                for some $f, x, c, T_0$, and $C$.
                By the IH of \lref{ty-c}, we have $\jdwf{\Gamma, f:(x: T_0) \rarr C, x: T_0}{C}$.
                By Lemma \ref{lem:rm-nonrfn}, we have $\jdwf{\Gamma, x: T_0}{C}$.
                By \rulename{WT-Fun}, we have the conclusion.
            \item[Case \rulename{T-VSub}:] Immediate by inversion.
        \end{description}
        \item 
        \begin{description}
            \item[Case \rulename{T-Ret}:] We have
                \def\currentprefix{wft:ret}
                \begin{enumrm}
                    \item\llabel{eq-c} $c = \expret{v}$,
                    \item\llabel{eq-C} $C = \tycomp{\emptyset}{T}{\square}$, and
                    \item\llabel{ty-v} $\jdty{\Gamma}{v}{T}$
                \end{enumrm}
                for some $v$ and $T$.
                By the IH of \lref{ty-v}, we have $\jdwf{\Gamma}{T}$.
                By Lemma \ref{lem:wfg}, we have $\jdwf{}{\Gamma}$.
                Then, we have the conclusion by the following derivation:
                \[
                    \infer{\jdwf{\Gamma}{\tycomp{\emptyset}{T}{\square}}}{
                        \infer{\jdwf{\Gamma}{\emptyset}}{}
                        &
                        \jdwf{\Gamma}{T}
                        &
                        \infer{\jdwf{\Gamma \mid T}{\square}}
                        {\jdwf{}{\Gamma}}
                    }
                \]
            \item[Case \rulename{T-App}:] We have
                \def\currentprefix{wft:app}
                \begin{enumrm}
                    \item\llabel{eq-c} $c = v_1~v_2$,
                    \item\llabel{eq-C} $C = C_0[v_2/x]$,
                    \item\llabel{ty-v1} $\jdty{\Gamma}{v_1}{(x:T_0) \rarr C_0}$, and
                    \item\llabel{ty-v2} $\jdty{\Gamma}{v_2}{T_0}$
                \end{enumrm}
                for some $x, v_1, v_2, T_0$ and $C_0$.
                By the IH of \lref{ty-v1}, we have $\jdwf{\Gamma}{(x:T_0) \rarr C_0}$.
                By inversion, we have $\jdwf{\Gamma, x:T_0}{C_0}$.
                By Lemma \ref{lem:subst}, we have the conclusion.
            \item[Case \rulename{T-If}:] We have
                \def\currentprefix{wft:if}
                \begin{enumrm}
                    \item\llabel{eq-c} $c = \expif{v}{c_1}{c_2}$,
                    \item\llabel{ty-v} $\jdty{\Gamma}{v}{\tyrfn{x}{\tybool}{\phi}}$,
                    \item\llabel{ty-c1} $\jdty{\Gamma, v = \exptrue}{c_1}{C}$, and
                    \item\llabel{ty-c2} $\jdty{\Gamma, v = \expfalse}{c_2}{C}$
                \end{enumrm}
                for some $x, v, c_1, c_2$, and $\phi$.
                By the IH of \lref{ty-c1}, we have $\jdwf{\Gamma, v = \exptrue}{C}$.
                By Lemma \ref{lem:rm-unused}, we have the conclusion.
            \item[Case \rulename{T-CSub}:] Immediate by inversion.
            \item[Case \rulename{T-LetP}:] We have
                \def\currentprefix{wft:letp}
                \begin{enumrm}
                    \item\llabel{eq-c} $c = \explet{x}{c_1}{c_2}$,
                    \item\llabel{eq-C} $C = \tycomp{\Sigma}{T_2}{\square}$,
                    \item\llabel{ty-c1} $\jdty{\Gamma}{c_1}{\tycomp{\Sigma}{T_1}{\square}}$,
                    \item\llabel{ty-c2} $\jdty{\Gamma, x: T_1}{c_2}{\tycomp{\Sigma}{T_2}{\square}}$, and
                    \item\llabel{in-x} $x \notin \fv(T_2) \cup \fv(\Sigma)$
                \end{enumrm}
                for some $x, c_1, c_2, \Sigma, T_1$, and $T_2$.
                By the IHs of \lref{ty-c1} and \lref{ty-c2} respectively, we have
                \begin{itemize}
                    \item $\jdwf{\Gamma}{\tycomp{\Sigma}{T_1}{\square}}$ and
                    \item $\jdwf{\Gamma, x: T_1}{\tycomp{\Sigma}{T_2}{\square}}$.
                \end{itemize}
                By inversion, we have
                \begin{enumrm}[resume]
                    \item\llabel{wf-sig} $\jdwf{\Gamma}{\Sigma}$, and
                    \item\llabel{wf-T2} $\jdwf{\Gamma, x:T_1}{T_2}$.
                \end{enumrm}
                By Lemma \ref{lem:rm-unused} with \lref{in-x} \lref{wf-T2}, we have
                \begin{enumrm}[resume]
                    \item\llabel{wf-T2-2} $\jdwf{\Gamma}{T_2}$~.
                \end{enumrm}
                By Lemma \ref{lem:wfg} with \lref{wf-sig}, we have $\jdwf{}{\Gamma}$.
                From this fact and \lref{wf-sig} and \lref{wf-T2-2},
                we have the conclusion by the following derivation:
                \[
                    \infer{\jdwf{\Gamma}{\tycomp{\Sigma}{T_2}{\square}}}{
                        \jdwf{\Gamma}{\Sigma}
                        &
                        \jdwf{\Gamma}{T_2}
                        &
                        \infer{\jdwf{\Gamma \mid T_2}{\square}}
                        {\jdwf{}{\Gamma}}
                    }
                \]
            \item[Case \rulename{T-LetIp}:] We have
                \def\currentprefix{wft:leti}
                \begin{enumrm}
                    \item\llabel{eq-c} $c = \explet{x}{c_1}{c_2}$,
                    \item\llabel{eq-C} $C = \tycomp{\Sigma}{T_2}{\tyctl{z}{C_{21}}{C_{12}}}$,
                    \item\llabel{ty-c1} $\jdty{\Gamma}{c_1}{\tycomp{\Sigma}{T_1}{\tyctl{x}{C_0}{C_{12}}}}$,
                    \item\llabel{ty-c2} $\jdty{\Gamma, x: T_1}{c_2}{\tycomp{\Sigma}{T_2}{\tyctl{z}{C_{21}}{C_0}}}$, and
                    \item\llabel{in-x} $x \notin \fv(T_2) \cup \fv(\Sigma) \cup (\fv(C_{21}) \setminus \{z\})$
                \end{enumrm}
                for some $x, c_1, c_2, \Sigma, T_1, T_2, C_0, C_{12}$ and $C_{21}$.
                By the IHs of \lref{ty-c1} and \lref{ty-c2} respectively, we have
                \begin{itemize}
                    \item $\jdwf{\Gamma}{\tycomp{\Sigma}{T_1}{\tyctl{x}{C_0}{C_{12}}}}$ and
                    \item $\jdwf{\Gamma, x: T_1}{\tycomp{\Sigma}{T_2}{\tyctl{z}{C_{21}}{C_0}}}$.
                \end{itemize}
                By inversion, we have
                \begin{enumrm}[resume]
                    \item\llabel{wf-sig} $\jdwf{\Gamma}{\Sigma}$,
                    \item\llabel{wf-S1} $\jdwf{\Gamma \mid T_1}{\tyctl{x}{C_0}{C_{12}}}$,
                    \item\llabel{wf-T2} $\jdwf{\Gamma, x:T_1}{T_2}$, and
                    \item\llabel{wf-S2} $\jdwf{\Gamma, x:T_1 \mid T_2}{\tyctl{z}{C_{21}}{C_0}}$.
                \end{enumrm}
                By Lemma \ref{lem:rm-unused} with \lref{in-x} \lref{wf-T2}, we have
                \begin{enumrm}[resume]
                    \item\llabel{wf-T2-2} $\jdwf{\Gamma}{T_2}$~.
                \end{enumrm}
                By inversion with \lref{wf-S1} and \lref{wf-S2} respectively, we have
                \begin{enumrm}[resume]
                    \item\llabel{wf-C0} $\jdwf{\Gamma, x:T_1}{C_0}$,
                    \item\llabel{wf-C1} $\jdwf{\Gamma}{C_{12}}$, and
                    \item\llabel{wf-C2} $\jdwf{\Gamma, x:T_1, z:T_2}{C_{21}}$~.
                \end{enumrm}
                W.l.o.g., we can assume $x \neq z$.
                Then, \lref{in-x} implies $x \notin \fv(C_{21})$.
                Therefore, by Lemma \ref{lem:rm-unused} with \lref{wf-C2} and \lref{in-x},
                we have $\jdwf{\Gamma, z:T_2}{C_{21}}$.
                From this and \lref{wf-sig}, \lref{wf-T2-2}, and \lref{wf-C1},
                we have the conclusion by the following derivation:
                \[
                    \infer{\jdwf{\Gamma}{\tycomp{\Sigma}{T_2}{\tyctl{z}{C_{21}}{C_{12}}}}}{
                        \jdwf{\Gamma}{\Sigma}
                        &
                        \jdwf{\Gamma}{T_2}
                        &
                        \infer{\jdwf{\Gamma \mid T_2}{\tyctl{z}{C_{21}}{C_{12}}}}{
                            \jdwf{\Gamma, z:T_2}{C_{21}}
                            &
                            \jdwf{\Gamma}{C_{12}}
                        }
                    }
                \]
            \item[Case \rulename{T-Op}:] We have
                \def\currentprefix{wft:op}
                \begin{enumrm}
                    \item\llabel{eq-c} $c = \op~v$,
                    \item\llabel{eq-C} $C = \tycomp{\Sigma}{T_2[\rep{A/X}][v/x]}{\tyctl{y}{C_1[\rep{A/X}][v/x]}{C_2[\rep{A/X}][v/x]}}$,
                    \item\llabel{in-sig} $\Sigma \ni \op: \forall \rep{X: \rep{B}}. (x: T_1) \rarr ((y: T_2) \rarr C_1) \rarr C_2$,
                    \item\llabel{wf-sig} $\jdwf{\Gamma}{\Sigma}$,
                    \item\llabel{wf-A} $\rep{\jdty{\Gamma}{A}{\rep{B}}}$, and
                    \item\llabel{ty-v} $\jdty{\Gamma}{v}{T_1[\rep{A/X}]}$
                \end{enumrm}
                for some $x, y, v, \rep{X}, \rep{A}, \rep{\rep{B}}, \Sigma, T_1, T_2, C_1$ and $C_2$.
                By inversion of \lref{wf-sig} with \lref{in-sig}, we have
                \[
                    \jdwf{\Gamma, \rep{X: \rep{B}}}{(x: T_1) \rarr ((y: T_2) \rarr C_1) \rarr C_2}~.
                \]
                By more inversion and Lemma \ref{lem:rm-nonrfn}, we have
                \begin{itemize}
                    \item $\jdwf{\Gamma, \rep{X: \rep{B}}, x: T_1}{T_2}$,
                    \item $\jdwf{\Gamma, \rep{X: \rep{B}}, x: T_1, y: T_2}{C_1}$, and
                    \item $\jdwf{\Gamma, \rep{X: \rep{B}}, x: T_1}{C_2}$~.
                \end{itemize}
                By Lemma \ref{lem:subst-pred} with \lref{wf-A} and Lemma \ref{lem:subst} with \lref{ty-v},
                we have
                \begin{itemize}
                    \item $\jdwf{\Gamma}{T_2[\rep{A/X}][v/x]}$,
                    \item $\jdwf{\Gamma, y: T_2[\rep{A/X}][v/x]}{C_1[\rep{A/X}][v/x]}$, and
                    \item $\jdwf{\Gamma}{C_2[\rep{A/X}][v/x]}$~.
                \end{itemize}
                From these and \lref{wf-sig},
                we have the conclusion by the following derivation:
                \[
                    \infer{\jdwf{\Gamma}{\tycomp{\Sigma}{T_2[\rep{A/X}][v/x]}{\tyctl{y}{C_1[\rep{A/X}][v/x]}{C_2[\rep{A/X}][v/x]}}}}{
                        \jdwf{\Gamma}{\Sigma}
                        &
                        \jdwf{\Gamma}{T_2[\rep{A/X}][v/x]}
                        &
                        \infer{\jdwf{\Gamma \mid T_2[\rep{A/X}][v/x]}{\tyctl{y}{C_1[\rep{A/X}][v/x]}{C_2[\rep{A/X}][v/x]}}}{
                            \jdwf{\Gamma, y:T_2[\rep{A/X}][v/x]}{C_1[\rep{A/X}][v/x]}
                            &
                            \jdwf{\Gamma}{C_2[\rep{A/X}][v/x]}
                        }
                    }
                \]
            \item[Case \rulename{T-Hndl}:] We have
                \def\currentprefix{wft:hndl}
                \begin{enumrm}
                    \item\llabel{eq-c} $c = \expwith{h}{c_0}$,
                    \item\llabel{ty-c0} $\jdty{\Gamma}{c_0}{\tycomp{\Sigma}{T}{\tyctl{x_r}{C_1}{C}}}$
                \end{enumrm}
                for some $x_r, h, c_0, \Sigma, T$ and $C_1$.
                We have the conclusion by applying inversion twice to \lref{ty-c0}.
        \end{description}
    \end{enumit}
\end{proof}

\begin{lemma}[Canonical forms] \label{lem:cano} \quad
    \begin{enumerate}
        \item If $\jdty{}{v}{(x: T) \rarr C}$, then (i) $v = \exprec{f}{x}{c}$ for some $f, c$,
            or (ii) $v = p$ for some $p$ and $\zeta(p, v)$ is defined for all $v$ such that $\jdty{}{v}{T}$.
        \item If $\jdty{}{v}{\tyrfn{x}{\tybool}{\phi}}$, then $v = \exptrue$ or $v = \expfalse$.
    \end{enumerate}
\end{lemma}
\begin{proof}
    By induction on the derivations.
    \begin{enumit}
        \item 
        \begin{description}
            \item[Case \rulename{T-Fun}:] Obvious.
            \item[Case \rulename{T-Prim}:] Immediate from Assumption \ref{asm:prim}.
            \item[Case \rulename{T-VSub}:] By the IH and inversion of the subtyping judgment.
                The case for (ii) uses Lemma \ref{lem:wft}.
            \item[Otherwise:] Contradictory.
        \end{description}
        \item 
        \begin{description}
            \item[Case \rulename{T-Prim}:] Immediate from Assumption \ref{asm:prim}.
            \item[Case \rulename{T-VSub}:] By the IH and inversion of the subtyping judgment.
            \item[Otherwise:] Contradictory.
        \end{description}

    \end{enumit}
\end{proof}

\begin{theorem}[Progress] \label{thm:progress}
    If $\jdty{\emptyset}{c}{\tycomp{\Sigma}{T}{S}}$, then either
    \begin{itemize}
        \item $c = \expret{v}$ for some $v$ such that $\jdty{\emptyset}{v}{T}$,
        \item $c = K[\op~v]$ for some $K, \op$ and $v$ such that $\op \in \dom(\Sigma)$, or
        \item $c \eval c'$ for some $c'$.
    \end{itemize}
\end{theorem}
\begin{proof}
    \def\currentprefix{prog}
    By induction on the derivation.
    \begin{description}
        \item[Case \rulename{T-Ret} and \rulename{T-Op}:] Obvious.
        \item[Case \rulename{T-CSub}:] By the IH.
            Note that $\jdsub{}{\Sigma'}{\Sigma}$ implies $\dom(\Sigma') \supseteq \dom(\Sigma)$.
        \item[Case \rulename{T-App}:] We have
            \def\currentprefix{prog:app}
            \begin{enumrm}
                \item\llabel{eq-c} $c = v_1~v_2$,
                \item\llabel{ty-v1} $\jdty{}{v_1}{(x: T_1) \rarr \tycomp{\Sigma}{T}{S}}$, and
                \item\llabel{ty-v2} $\jdty{}{v_2}{T_1}$
            \end{enumrm}
            for some $v_1, v_2, x$, and $T_1$.
            By Lemma \ref{lem:cano} with \lref{ty-v1}, either one of the following two cases holds.
            \begin{itemize}
                \item $v_1 = \exprec{f}{x}{c_1}$ for some $f, c_1$: \\
                    By \rulename{E-App}, we have $(\exprec{f}{x}{c_1})~v_2 \eval c_1[v_2/x][(\exprec{f}{x}{c_1})/f]$~.
                \item $v_1 = p$ for some $p$ and $\zeta(p, v)$ is defined for all $v$ such that $\jdty{}{v}{T_1}$: \\
                    As \lref{ty-v2} holds, $\zeta(p, v_2)$ is defined.
                    Therefore, by \rulename{E-Prim} we have $p~v_2 \eval \zeta(p, v_2)$~.
            \end{itemize}
        \item[Case \rulename{T-If}:] We have
            \def\currentprefix{prog:if}
            \begin{enumrm}
                \item\llabel{eq-c} $c = \expif{v}{c_1}{c_2}$,
                \item\llabel{ty-v} $\jdty{}{v}{\tyrfn{x}{\tybool}{\phi}}$,
                \item\llabel{ty-c1} $\jdty{v = \exptrue}{c_1}{\tycomp{\Sigma}{T}{S}}$, and
                \item\llabel{ty-c2} $\jdty{v = \expfalse}{c_2}{\tycomp{\Sigma}{T}{S}}$
            \end{enumrm}
            for some $v, c_1, c_2, x$, and $\phi$.
            By Lemma \ref{lem:cano} with \lref{ty-v}, either one of the following two cases holds.
            \begin{itemize}
                \item $v = \exptrue$: By \rulename{E-IfT}, we have $\expif{\exptrue}{c_1}{c_2} \eval c_1$.
                \item $v = \expfalse$: By \rulename{E-IfF}, we have $\expif{\expfalse}{c_1}{c_2} \eval c_2$.
            \end{itemize}
        \item[Case \rulename{T-LetP}:] We have
            \def\currentprefix{prog:let}
            \begin{enumrm}
                \item\llabel{eq-c} $c = \explet{x}{c_1}{c_2}$, and
                \item\llabel{ty-c1} $\jdty{}{c_1}{\tycomp{\Sigma}{T_1}{\square}}$
            \end{enumrm}
            for some $x, c_1, c_2$, and $T_1$.
            By the IH of \lref{ty-c1}, either one of the following three cases holds.
            \begin{itemize}
                \item $c_1 = \expret{v_1}$ for some $v_1$: \\
                    By \rulename{E-LetRet}, we have $\explet{x}{\expret{v_1}}{c_2} \eval c_2[v_1/x]$.
                \item $c_1 = K_1[\op~v_1]$ for some $K_1, \op$ and $v_1$ s.t. $\op \in \dom(\Sigma)$: \\
                    We have the conclusion with $c = K[\op~v_1]$ where $K = \explet{x}{K_1}{c_1}$.
                \item $c_1 \eval c_1'$ for some $c_1'$: \\
                    By \rulename{E-Let}, we have $\explet{x}{c_1}{c_2} \eval \explet{x}{c_1'}{c_2}$.
            \end{itemize}
        \item[Case \rulename{T-LetIp}:] Similar to the case for \rulename{T-LetP}.
        \item[Case \rulename{T-Hndl}:] We have
            \def\currentprefix{prog:hndl}
            \begin{enumrm}
                \item\llabel{eq-c} $c = \expwith{h}{c_0}$,
                \item\llabel{eq-h} $h = \{ \expret{x_r} \mapsto c_r, \repi{\op_i(x_i, k_i) \mapsto c_i} \}$,
                \item\llabel{eq-sig} $\Sigma_0 = \{ \repi{\op_i: \forall \rep{X_i: \rep{B}_i}. (x_i: T_{1i}) \rarr ((y_i: T_{2i}) \rarr C_{1i}) \rarr C_{2i}} \}$, and
                \item\llabel{ty-c0} $\jdty{}{c_0}{\tycomp{\Sigma_0}{T_0}{\tyctl{x_r}{C_1}{(\tycomp{\Sigma}{T}{S})}}}$
            \end{enumrm}
            for some $c_0, x_r, c_r, \repi{\op_i}, \repi{x_i}, \repi{k_i}, \repi{c_i}, \repi{\rep{X_i}}, \repi{\rep{\rep{B}_i}}, \repi{T_{1i}}, \repi{T_{2i}}, \repi{C_{1i}}, \repi{C_{2i}}, \Sigma_0, T_0$, and $C_1$.
            By the IH of \lref{ty-c0}, either one of the following three cases holds.
            \begin{itemize}
                \item $c_0 = \expret{v_0}$ for some $v_0$: \\
                    By \rulename{E-HndlRet}, we have $\expwith{h}{\expret{v_0}} \eval c_r[v_0/x_r]$.
                \item $c_0 = K_0[\op~v_0]$ for some $K_0, \op$, and $v_0$ s.t. $\op \in \dom(\Sigma_0)$: \\
                    Since $\op \in \dom(\Sigma_0) = \{ \repi{\op_i} \}$,
                    there exists some $j$ such that $1 \le j \le |\dom(\Sigma)|$ and $\op = \op_j$.
                    Then, by \rulename{E-HndlOp} we have
                    \[
                        \expwith{h}{K_0[\op_j~v_0]} \eval c_j[v_0/x_j][\expfun{y}{\expwith{h}{K_0[\expret{y}]}}/k_j]~.
                    \]
                \item $c_0 \eval c_0'$ for some $c_0'$: \\
                    By \rulename{E-Hndl}, we have $\expwith{h}{c_0} \eval \expwith{h}{c_0'}$.
            \end{itemize}
    \end{description}
\end{proof}

\subsection{Subject Reduction}

\begin{lemma}[Remove tautology] \label{lem:rm-tauto} \quad
    \begin{enumerate}
        \item If $\jdwf{}{\Gamma, \phi, \Gamma'}$, then $\jdwf{}{\Gamma, \Gamma'}$.
        \item 
            \begin{itemize}
                \item If $\jdwf{\Gamma, \phi, \Gamma'}{T}$, then $\jdwf{\Gamma, \Gamma'}{T}$.
                \item If $\jdwf{\Gamma, \phi, \Gamma'}{C}$, then $\jdwf{\Gamma, \Gamma'}{C}$.
                \item If $\jdwf{\Gamma, \phi, \Gamma'}{\Sigma}$, then $\jdwf{\Gamma, \Gamma'}{\Sigma}$.
                \item If $\jdwf{\Gamma, \phi, \Gamma' \mid T}{S}$, then $\jdwf{\Gamma, \Gamma' \mid T}{S}$.
            \end{itemize}
        \item Assume that $\vDash \phi$.
            \begin{itemize}
                \item If $\jdty{\Gamma, \phi, \Gamma'}{v}{T}$, then $\jdty{\Gamma, \Gamma'}{v}{T}$.
                \item If $\jdty{\Gamma, \phi, \Gamma'}{c}{C}$, then $\jdty{\Gamma, \Gamma'}{c}{C}$.
            \end{itemize}
        \item Assume that $\vDash \phi$.
            \begin{itemize}
                \item If $\jdsub{\Gamma, \phi, \Gamma'}{T_1'}{T_2'}$, then $\jdsub{\Gamma, \Gamma'}{T_1'}{T_2'}$.
                \item If $\jdsub{\Gamma, \phi, \Gamma'}{C_1}{C_2}$, then $\jdsub{\Gamma, \Gamma'}{C_1}{C_2}$.
                \item If $\jdsub{\Gamma, \phi, \Gamma'}{\Sigma_1}{\Sigma_2}$, then $\jdsub{\Gamma, \Gamma'}{\Sigma_1}{\Sigma_2}$.
                \item If $\jdsub{\Gamma, \phi, \Gamma' \mid T}{S_1}{S_2}$, then $\jdsub{\Gamma, \Gamma' \mid T}{S_1}{S_2}$.
            \end{itemize}
    \end{enumerate}
\end{lemma}
\begin{proof} \quad
    \begin{enumit}
        \item Immediate by Lemma \ref{lem:rm-unused}.
        \item Immediate by Lemma \ref{lem:rm-unused}.
        \item By simultaneous induction on the derivations.
        \item By simultaneous induction on the derivations.
            The case for \rulename{S-Rfn} uses Assumption \ref{asm:formla}.
    \end{enumit}
\end{proof}

\begin{lemma}[Reflexivity of subtyping] \label{lem:refl} \quad
    \begin{enumerate}
        \item If $\jdwf{\Gamma}{T}$, then $\jdsub{\Gamma}{T}{T}$.
        \item If $\jdwf{\Gamma}{C}$, then $\jdsub{\Gamma}{C}{C}$.
        \item If $\jdwf{\Gamma}{\Sigma}$, then $\jdsub{\Gamma}{\Sigma}{\Sigma}$.
        \item If $\jdwf{\Gamma \mid T}{S}$, then $\jdsub{\Gamma \mid T}{S}{S}$.
    \end{enumerate}
\end{lemma}
\begin{proof}
    By simultaneous induction on the derivations.
    The case for \rulename{WT-Rfn} uses Assumption \ref{asm:formla}.
\end{proof}

\begin{lemma}[Transitivity of subtyping] \label{lem:trans} \quad
    \begin{enumerate}
        \item If $\jdsub{\Gamma}{T_1}{T_2}$ and $\jdsub{\Gamma}{T_2}{T_3}$, then $\jdsub{\Gamma}{T_1}{T_3}$.
        \item If $\jdsub{\Gamma}{C_1}{C_2}$ and $\jdsub{\Gamma}{C_2}{C_3}$, then $\jdsub{\Gamma}{C_1}{C_3}$.
        \item If $\jdsub{\Gamma}{\Sigma_1}{\Sigma_2}$ and $\jdsub{\Gamma}{\Sigma_2}{\Sigma_3}$, then $\jdsub{\Gamma}{\Sigma_1}{\Sigma_3}$.
        \item If $\jdsub{\Gamma \mid T}{S_1}{S_2}$ and $\jdsub{\Gamma \mid T}{S_2}{S_3}$, then $\jdsub{\Gamma \mid T}{S_1}{S_3}$.
    \end{enumerate}
\end{lemma}
\begin{proof}
    By simultaneous induction on the structure of $T_2, C_2, \Sigma_2$ and $S_2$.
    \begin{enumit}
        \item Case analysis on $\jdsub{\Gamma}{T_1}{T_2}$.
            \begin{description}
                \item[Case \rulename{S-Rfn}:] By inversion, Assumption \ref{asm:formla} and \rulename{S-Rfn}.
                \item[Case \rulename{S-Fun}:] By inversion, the IHs, Lemma \ref{lem:narrow}, and \rulename{S-Fun}.
            \end{description}
        \item By inversion of the both derivations, we have
            \def\currentprefix{trans:opsig}
            \begin{enumrm}
                \item\llabel{eq-sig1} $\Sigma_1 = \{ \repi{\op_i: \forall \rep{X_i: \rep{B}_i}. F_{1i}},
                    \repi{\op'_i: \forall \rep{X'_i: \rep{B'}_i}. F'_{1i}},
                    \repi{\op''_i: \forall \rep{X''_i: \rep{B''}_i}. F''_{1i}} \}$,
                \item\llabel{eq-sig2} $\Sigma_2 = \{ \repi{\op_i: \forall \rep{X_i: \rep{B}_i}. F_{2i}},
                    \repi{\op'_i: \forall \rep{X'_i: \rep{B'}_i}. F'_{2i}} \}$,
                \item\llabel{eq-sig3} $\Sigma_3 = \{ \repi{\op_i: \forall \rep{X_i: \rep{B}_i}. F_{2i}} \}$,
                \item\llabel{sub-f1} $\repi{\jdsub{\Gamma, \rep{X_i: \rep{B}_i}}{F_{1i}}{F_{2i}}}$,
                \item\llabel{sub-f2} $\repi{\jdsub{\Gamma, \rep{X_i: \rep{B}_i}}{F_{2i}}{F_{3i}}}$, and
                \item\llabel{sub-f1'} $\repi{\jdsub{\Gamma, \rep{X'_i: \rep{B'}_i}}{F'_{1i}}{F'_{2i}}}$~.
            \end{enumrm}
            By the IH with \lref{sub-f1} and \lref{sub-f2}, we have
            $\repi{\jdsub{\Gamma, \rep{X_i: \rep{B}_i}}{F_{1i}}{F_{3i}}}$~.
            By \rulename{S-Sig}, we have the conclusion.
        \item By inversion, the IHs, Lemma \ref{lem:narrow}, and \rulename{S-Comp}.
        \item Case analysis on $\jdsub{\Gamma}{S_1}{S_2}$.
        \begin{description}
            \item[Case \rulename{S-Pure}:]
                Since we have $S_1 = \square = S_2$,
                we have the conclusion immediately from $\jdsub{\Gamma}{S_2}{S_3}$.
            \item[Case \rulename{S-ATM}:] We have
                \def\currentprefix{trans:atm}
                \begin{enumrm}
                    \item\llabel{eq-S1} $S_1 = \tyctl{x}{C_{11}}{C_{12}}$,
                    \item\llabel{eq-S2} $S_2 = \tyctl{x}{C_{21}}{C_{22}}$,
                    \item\llabel{sub-C21} $\jdsub{\Gamma, x:T}{C_{21}}{C_{11}}$, and
                    \item\llabel{sub-C12} $\jdsub{\Gamma}{C_{12}}{C_{22}}$
                \end{enumrm}
                for some $x, C_{11}, C_{12}, C_{21}$, and $C_{22}$.
                Since \lref{eq-S2}, the only rule applicable to $\jdsub{\Gamma}{S_2}{S_3}$ is \rulename{S-ATM}.
                Therefore, by inversion we have
                \begin{enumrm}[resume]
                    \item\llabel{eq-S3} $S_3 = \tyctl{x}{C_{31}}{C_{32}}$,
                    \item\llabel{sub-C31} $\jdsub{\Gamma, x:T}{C_{31}}{C_{21}}$, and
                    \item\llabel{sub-C22} $\jdsub{\Gamma}{C_{22}}{C_{32}}$
                \end{enumrm}
                for some $C_{31}$ and $C_{32}$.
                By the IHs, we have
                \begin{itemize}
                    \item $\jdsub{\Gamma, x:T}{C_{31}}{C_{11}}$ and
                    \item $\jdsub{\Gamma}{C_{12}}{C_{32}}$~.
                \end{itemize}
                We have the conclusion by \rulename{S-ATM}.
            \item[Case \rulename{S-Embed}:] We have
            \def\currentprefix{trans:emb}
            \begin{enumrm}
                \item\llabel{eq-S1} $S_1 = \square$,
                \item\llabel{eq-S2} $S_2 = \tyctl{x}{C_{21}}{C_{22}}$,
                \item\llabel{sub-C21} $\jdsub{\Gamma, x:T}{C_{21}}{C_{22}}$, and
                \item\llabel{in-x} $x \notin \fv(C_{22})$
            \end{enumrm}
            for some $x, C_{21}$, and $C_{22}$.
            Since \lref{eq-S2}, the only rule applicable to $\jdsub{\Gamma}{S_2}{S_3}$ is \rulename{S-ATM}.
            Therefore, by inversion we have
            \begin{enumrm}[resume]
                \item\llabel{eq-S3} $S_3 = \tyctl{x}{C_{31}}{C_{32}}$,
                \item\llabel{sub-C31} $\jdsub{\Gamma, x:T}{C_{31}}{C_{21}}$, and
                \item\llabel{sub-C22} $\jdsub{\Gamma}{C_{22}}{C_{32}}$
            \end{enumrm}
            for some $C_{31}$ and $C_{32}$.
            W.l.o.g., we can assume that $x \notin \fv(C_{32})$.
            Then, by Lemma \ref{lem:weaken} with \lref{sub-C22}, we have
            \begin{enumrm}[resume]
                \item\llabel{sub-C22-2} $\jdsub{\Gamma, x:T}{C_{22}}{C_{32}}$~.
            \end{enumrm}
            By the IHs with \lref{sub-C21}, \lref{sub-C21} and \lref{sub-C22-2},
            we have $\jdsub{\Gamma, x:T}{C_{31}}{C_{32}}$.
            Then we have the conclusion by \rulename{S-Embed}.
        \end{description}
    \end{enumit}
    
\end{proof}

\begin{lemma}[Subtyping with equal variables] \label{lem:sub-eq}
    \quad
    \begin{itemize}
        \item If $\jdwf{\Gamma, x: \tyrfn{z}{B}{z = y}, \Gamma'}{T}$, then
            $\jdsub{\Gamma, x: \tyrfn{z}{B}{z = y}, \Gamma'}{T}{T[y/x]}$~.
        \item If $\jdwf{\Gamma, x: \tyrfn{z}{B}{z = y}, \Gamma'}{T}$, then
            $\jdsub{\Gamma, x: \tyrfn{z}{B}{z = y}, \Gamma'}{T[y/x]}{T}$~.
        \item If $\jdwf{\Gamma, x: \tyrfn{z}{B}{z = y}, \Gamma'}{C}$, then
            $\jdsub{\Gamma, x: \tyrfn{z}{B}{z = y}, \Gamma'}{C}{C[y/x]}$~.
        \item If $\jdwf{\Gamma, x: \tyrfn{z}{B}{z = y}, \Gamma'}{C}$, then
            $\jdsub{\Gamma, x: \tyrfn{z}{B}{z = y}, \Gamma'}{C[y/x]}{C}$~.
        \item If $\jdwf{\Gamma, x: \tyrfn{z}{B}{z = y}, \Gamma'}{\Sigma}$, then
            $\jdsub{\Gamma, x: \tyrfn{z}{B}{z = y}, \Gamma'}{\Sigma}{\Sigma[y/x]}$~.
        \item If $\jdwf{\Gamma, x: \tyrfn{z}{B}{z = y}, \Gamma'}{\Sigma}$, then
            $\jdsub{\Gamma, x: \tyrfn{z}{B}{z = y}, \Gamma'}{\Sigma[y/x]}{\Sigma}$~.
        \item If $\jdwf{\Gamma, x: \tyrfn{z}{B}{z = y}, \Gamma' \mid T}{S}$, then
            $\jdsub{\Gamma, x: \tyrfn{z}{B}{z = y}, \Gamma' \mid T}{S}{S[y/x]}$~.
        \item If $\jdwf{\Gamma, x: \tyrfn{z}{B}{z = y}, \Gamma' \mid T}{S}$, then
            $\jdsub{\Gamma, x: \tyrfn{z}{B}{z = y}, \Gamma' \mid T}{S[y/x]}{S}$~.
    \end{itemize}
\end{lemma}
\begin{proof}
    By simultaneous induction on the derivations.
    The case for \rulename{WT-Rfn} uses Assumption~\ref{asm:formla}.
    The cases for \rulename{WT-Fun} and \rulename{WT-Comp} uses Lemma~\ref{lem:narrow}.
\end{proof}

\begin{lemma}[Inversion] \label{lem:inv} \quad
    \begin{enumerate}
        \item If $\jdty{\Gamma}{p}{T}$, then
            \begin{itemize}
                \item $\jdsub{\Gamma}{\ty(p)}{T}$, and
                \item $\jdty{\Gamma}{p}{\ty(p)}$.
            \end{itemize}
        \item If $\jdty{\Gamma}{\exprec{f}{x}{c}}{(x: T) \rarr C}$, then
            there exist some $T_0$ and $C_0$ such that
            \begin{itemize}
                \item $\jdty{\Gamma}{\exprec{f}{x}{c}}{(x: T_0) \rarr C_0}$,
                \item $\jdsub{\Gamma}{(x: T_0) \rarr C_0}{(x: T) \rarr C}$, and
                \item $\jdty{\Gamma, f: (x: T_0) \rarr C_0, x: T_0}{c}{C_0}$.
            \end{itemize}
        \item If $\jdty{\Gamma}{\expret{v}}{\tycomp{\Sigma}{T}{S}}$, then
            there exist some $T'$ such that
            \begin{itemize}
                \item $\jdsub{\Gamma}{T'}{T}$,
                \item $\jdty{\Gamma}{v}{T'}$, and
                \item $\jdsub{\Gamma \mid T'}{\square}{S}$.
            \end{itemize}
        \item If $\jdty{\Gamma}{\op~v}{\tycomp{\Sigma}{T}{S}}$, then
            there exist some $\rep{X}, \rep{\rep{B}}, \rep{A}, x, y, T_1, T_2, C_1, C_2, C_{01}$, and $C_{02}$ such that
            \begin{itemize}
                \item $S = \tyctl{y}{C_{01}}{C_{02}}$,
                \item $\Sigma \ni \op: \forall \rep{X: \rep{B}}. (x: T_1) \rarr ((y: T_2) \rarr C_1) \rarr C_2$,
                \item $\rep{\jdty{\Gamma}{A}{\rep{B}}}$,
                \item $\jdty{\Gamma}{v}{T_1[\rep{A/X}]}$,
                \item $\jdsub{\Gamma}{T_2[\rep{A/X}][v/x]}{T}$,
                \item $\jdsub{\Gamma, y: T_2[\rep{A/X}][v/x]}{C_{01}}{C_1[\rep{A/X}][v/x]}$, and
                \item $\jdsub{\Gamma}{C_2[\rep{A/X}][v/x]}{C_{02}}$~.
            \end{itemize}
        \item If $\jdty{\Gamma}{\explet{x}{c_1}{c_2}}{\tycomp{\Sigma}{T}{\square}}$, then
            there exists some $T_1$ such that
            \begin{itemize}
                \item $\jdty{\Gamma}{c_1}{\tycomp{\Sigma}{T_1}{\square}}$,
                \item $\jdty{\Gamma, x: T_1}{c_2}{\tycomp{\Sigma}{T}{\square}}$, and
                \item $x \notin \fv(T) \cup \fv(\Sigma)$~.
            \end{itemize}
        \item If $\jdty{\Gamma}{\explet{x}{c_1}{c_2}}{\tycomp{\Sigma}{T}{\tyctl{z}{C_1}{C_2}}}$, then
            there exist some $T_1$ and $C_0$ such that
            \begin{itemize}
                \item $\jdty{\Gamma}{c_1}{\tycomp{\Sigma}{T_1}{\tyctl{x}{C_0}{C_2}}}$,
                \item $\jdty{\Gamma, x: T_1}{c_2}{\tycomp{\Sigma}{T}{\tyctl{z}{C_1}{C_0}}}$, and
                \item $x \notin \fv(T) \cup \fv(\Sigma) \cup (\fv(C_1) \setminus \{z\})$~.
            \end{itemize}
    \end{enumerate}
\end{lemma}
\begin{proof}
    By induction on the derivations.
    \begin{enumit}
        \item Straightforward with Lemma \ref{lem:refl} and \ref{lem:trans}.
        \item Straightforward with Lemma \ref{lem:refl} and \ref{lem:trans}.
        \item Straightforward with Lemma \ref{lem:refl} and \ref{lem:trans}.
        \item 
        \def\currentprefix{inv:op}
        \begin{description}
            \item[Case \rulename{T-Op}:] Obvious with Lemma \ref{lem:refl}.
            \item[Case \rulename{T-CSub}:] We have
                \begin{enumrm}
                    \item\llabel{ty-op} $\jdty{\Gamma}{\op~v}{\tycomp{\Sigma'}{T'}{S'}}$,
                    \item\llabel{sub-C} $\jdsub{\Gamma}{\tycomp{\Sigma'}{T'}{S'}}{\tycomp{\Sigma}{T}{S}}$, and
                    \item $\jdwf{\Gamma}{\tycomp{\Sigma}{T}{S}}$
                \end{enumrm}
                for some $\Sigma', T'$, and $S'$.
                By the IH on \lref{ty-op}, we have
                \begin{enumrm}[resume]
                    \item\llabel{eq-S'} $S' = \tyctl{y}{C_{01}'}{C_{02}'}$,
                    \item\llabel{in-sig'} $\Sigma' \ni \op: \forall \rep{X: \rep{B}}. (x: T_1') \rarr ((y: T_2') \rarr C_1') \rarr C_2'$,
                    \item\llabel{wf-A} $\rep{\jdty{\Gamma}{A}{\rep{B}}}$,
                    \item\llabel{ty-v} $\jdty{\Gamma}{v}{T_1'[\rep{A/X}]}$,
                    \item\llabel{sub-T2'} $\jdsub{\Gamma}{T_2'[\rep{A/X}][v/x]}{T'}$,
                    \item\llabel{sub-C01'} $\jdsub{\Gamma, y: T_2'[\rep{A/X}][v/x]}{C_{01}'}{C_1'[\rep{A/X}][v/x]}$, and
                    \item\llabel{sub-C2'} $\jdsub{\Gamma}{C_2'[\rep{A/X}][v/x]}{C_{02}'}$
                \end{enumrm}
                for some $\rep{X}, \rep{\rep{B}}, \rep{A}, x, y, T_1', T_2', C_1', C_2', C_{01}'$, and $C_{02}'$.
                By inversion of \lref{sub-C}, we have
                \begin{enumrm}[resume]
                    \item\llabel{sub-sig} $\jdsub{\Gamma}{\Sigma}{\Sigma'}$,
                    \item\llabel{sub-T'} $\jdsub{\Gamma}{T'}{T}$, and
                    \item\llabel{sub-S'} $\jdsub{\Gamma \mid T'}{S'}{S}$~.
                \end{enumrm}
                By inversion of \lref{sub-S'} with \lref{eq-S'}, we have
                \begin{enumrm}[resume]
                    \item\llabel{eq-S} $S = \tyctl{y}{C_{01}}{C_{02}}$,
                    \item\llabel{sub-C01} $\jdsub{\Gamma, y: T'}{C_{01}}{C_{01}'}$, and
                    \item\llabel{sub-C02'} $\jdsub{\Gamma}{C_{02}'}{C_{02}}$
                \end{enumrm}
                for some $C_{01}$ and $C_{02}$.
                On the other hand,
                by inversion of \lref{sub-sig} with \lref{in-sig'} and Lemma~\ref{lem:rm-nonrfn}, we have
                \begin{enumrm}[resume]
                    \item\llabel{in-sig} $\Sigma \ni \op: \forall \rep{X: \rep{B}}. (x: T_1) \rarr ((y: T_2) \rarr C_1) \rarr C_2$,
                \end{enumrm}
                and
                \begin{itemize}
                    \item $\jdsub{\Gamma, \rep{X: \rep{B}}}{T_1'}{T_1}$,
                    \item $\jdsub{\Gamma, \rep{X: \rep{B}}, x: T_1'}{T_2}{T_2'}$,
                    \item $\jdsub{\Gamma, \rep{X: \rep{B}}, x: T_1', y: T_2}{C_1'}{C_1}$, and
                    \item $\jdsub{\Gamma, \rep{X: \rep{B}}, x: T_1'}{C_2}{C_2'}$
                \end{itemize}
                for some $T_1, T_2, C_1$ and $C_2$.
                Then, by Lemma~\ref{lem:subst-pred} with \lref{wf-A}
                and Lemma~\ref{lem:subst} with \lref{ty-v}, we have
                \begin{enumrm}[resume]
                    \item\llabel{sub-T1'} $\jdsub{\Gamma}{T_1'[\rep{A/X}]}{T_1[\rep{A/X}]}$,
                    \item\llabel{sub-T2} $\jdsub{\Gamma}{T_2[\rep{A/X}][v/x]}{T_2'[\rep{A/X}][v/x]}$,
                    \item\llabel{sub-C1'} $\jdsub{\Gamma, y: T_2[\rep{A/X}][v/x]}{C_1'[\rep{A/X}][v/x]}{C_1[\rep{A/X}][v/x]}$, and
                    \item\llabel{sub-C2} $\jdsub{\Gamma}{C_2[\rep{A/X}][v/x]}{C_2'[\rep{A/X}][v/x]}$~.
                \end{enumrm}
                By subsumption of \lref{ty-v} with \lref{sub-T1'},
                \begin{enumrm}[resume]
                    \item\llabel{ty-v-2} $\jdty{\Gamma}{v}{T_1[\rep{A/X}]}$~.
                \end{enumrm}
                By Lemma~\ref{lem:trans} with \lref{sub-T2}, \lref{sub-T2'} and \lref{sub-T'}, we have
                \begin{enumrm}[resume]
                    \item\llabel{sub-T2-T'} $\jdsub{\Gamma}{T_2[\rep{A/X}][v/x]}{T'}$ and
                    \item\llabel{sub-T2-T} $\jdsub{\Gamma}{T_2[\rep{A/X}][v/x]}{T}$~.
                \end{enumrm}
                By Lemma~\ref{lem:narrow} with ``\lref{sub-C01'} and \lref{sub-T2}'' and
                ``\lref{sub-C01} and \lref{sub-T2-T'}'' respectively, we have
                \begin{itemize}
                    \item $\jdsub{\Gamma, y: T_2[\rep{A/X}][v/x]}{C_{01}'}{C_1'[\rep{A/X}][v/x]}$ and
                    \item $\jdsub{\Gamma, y: T_2[\rep{A/X}][v/x]}{C_{01}}{C_{01}'}$~.
                \end{itemize}
                Then, by Lemma~\ref{lem:trans} with these two and \lref{sub-C1'}, we have
                \begin{enumrm}[resume]
                    \item\llabel{sub-C01-C1} $\jdsub{\Gamma, y: T_2[\rep{A/X}][v/x]}{C_{01}}{C_1[\rep{A/X}][v/x]}$~.
                \end{enumrm}
                Also, by Lemma~\ref{lem:trans} with \lref{sub-C2}, \lref{sub-C2'} and \lref{sub-C02'}, we have
                \begin{enumrm}[resume]
                    \item\llabel{sub-C2-C02} $\jdsub{\Gamma}{C_2[\rep{A/X}][v/x]}{C_{02}}$~.
                \end{enumrm}
                From \lref{in-sig}, \lref{wf-A}, \lref{ty-v-2}, \lref{sub-T2-T}, \lref{sub-C01-C1}
                and \lref{sub-C2-C02}, we have the conclusion.
        \end{description}
        \item
        \def\currentprefix{inv:letp}
        \begin{description}
            \item[Case \rulename{T-LetP}:] Obvious.
            \item[Case \rulename{T-LetIp}:] Contradictory.
            \item[Case \rulename{T-CSub}:] We have
                \begin{enumrm}
                    \item\llabel{ty-let} $\jdty{\Gamma}{\explet{x}{c_1}{c_2}}{\tycomp{\Sigma'}{T'}{S'}}$,
                    \item\llabel{sub-C'} $\jdsub{\Gamma}{\tycomp{\Sigma'}{T'}{S'}}{\tycomp{\Sigma}{T}{\square}}$, and
                    \item\llabel{wf-C} $\jdwf{\Gamma}{\tycomp{\Sigma}{T}{\square}}$
                \end{enumrm}
                for some $\Sigma', T'$, and $S'$.
                By inversion of \lref{sub-C'}, we have
                \begin{enumrm}[resume]
                    \item\llabel{eq-S'} $S = \square$,
                    \item\llabel{sub-sig'} $\jdsub{\Gamma}{\Sigma'}{\Sigma}$, and
                    \item\llabel{sub-T'} $\jdsub{\Gamma}{T'}{T}$~.
                \end{enumrm}
                Then, by the IH of \lref{ty-let}, we have
                \begin{enumrm}[resume]
                    \item\llabel{ty-c1} $\jdty{\Gamma}{c_1}{\tycomp{\Sigma'}{T_1}{\square}}$ and
                    \item\llabel{ty-c2} $\jdty{\Gamma, x: T_1}{c_2}{\tycomp{\Sigma'}{T'}{\square}}$
                \end{enumrm}
                for some $T_1$.
                By subsumption of \lref{ty-c1} with \lref{sub-sig'}, we have
                \begin{enumrm}[resume]
                    \item\llabel{ty-c1-2} $\jdty{\Gamma}{c_1}{\tycomp{\Sigma}{T_1}{\square}}$~.
                \end{enumrm}
                By Lemma~\ref{lem:wfg} with \lref{ty-c2}, we have
                \begin{enumrm}[resume]
                    \item\llabel{wf-Gx} $\jdwf{}{\Gamma, x: T_1}{}$~.
                \end{enumrm}
                Then it holds that $x \notin \dom(\Gamma)$ and hence from \lref{wf-C} we have
                \begin{enumrm}[resume]
                    \item\llabel{in-x} $x \notin \fv(T) \cup \fv(\Sigma)$~.
                \end{enumrm}
                Also, by Lemma~\ref{lem:weaken} with \lref{sub-C'} and \lref{wf-Gx}, we have
                \begin{itemize}
                    \item $\jdsub{\Gamma, x: T_1}{\tycomp{\Sigma'}{T'}{\square}}{\tycomp{\Sigma}{T}{\square}}$~.
                \end{itemize}
                Then by subsumption of \lref{ty-c2} we have
                \begin{enumrm}[resume]
                    \item\llabel{ty-c2-2} $\jdty{\Gamma, x: T_1}{c_2}{\tycomp{\Sigma}{T}{\square}}$~.
                \end{enumrm}
                Now we have the conclusion from \lref{ty-c1-2}, \lref{ty-c2-2} and \lref{in-x}.
        \end{description}
        \item
        \def\currentprefix{inv:leti}
        \begin{description}
            \item[Case \rulename{T-LetP}:] Contradictory.
            \item[Case \rulename{T-LetIp}:] Obvious.
            \item[Case \rulename{T-CSub}:] We have
                \begin{enumrm}
                    \item\llabel{ty-let} $\jdty{\Gamma}{\explet{x}{c_1}{c_2}}{\tycomp{\Sigma'}{T'}{S'}}$,
                    \item\llabel{sub-C'} $\jdsub{\Gamma}{\tycomp{\Sigma'}{T'}{S'}}{\tycomp{\Sigma}{T}{\tyctl{z}{C_1}{C_2}}}$, and
                    \item\llabel{wf-C} $\jdwf{\Gamma}{\tycomp{\Sigma}{T}{\tyctl{z}{C_1}{C_2}}}$
                \end{enumrm}
                for some $\Sigma', T'$, and $S'$.
                By inversion of \lref{sub-C'}, we have
                \begin{enumrm}[resume]
                    \item\llabel{sub-sig'} $\jdsub{\Gamma}{\Sigma'}{\Sigma}$,
                    \item\llabel{sub-T'} $\jdsub{\Gamma}{T'}{T}$, and
                    \item\llabel{sub-S'} $\jdsub{\Gamma \mid T'}{S'}{\tyctl{z}{C_1}{C_2}}$~.
                \end{enumrm}
                Case analysis on the derivation of \lref{sub-S'}.
                \begin{description}
                    \item[Case \rulename{S-Pure}:] Contradictory.
                    \item[Case \rulename{S-ATM}:] We have
                        \begin{enumrm}[resume]
                            \item\llabel{eq-S'} $S' = \tyctl{z}{C_1'}{C_2'}$,
                            \item\llabel{sub-C1} $\jdsub{\Gamma, z: T'}{C_1}{C_1'}$, and
                            \item\llabel{sub-C2'} $\jdsub{\Gamma}{C_2'}{C_2}$
                        \end{enumrm}
                        for some $C_1'$ and $C_2'$.
                        Then, by the IH of \lref{ty-let}, we have
                        \begin{enumrm}[resume]
                            \item\llabel{ty-c1} $\jdty{\Gamma}{c_1}{\tycomp{\Sigma'}{T_1}{\tyctl{x}{C_0}{C_2'}}}$ and
                            \item\llabel{ty-c2} $\jdty{\Gamma, x: T_1}{c_2}{\tycomp{\Sigma'}{T'}{\tyctl{z}{C_1'}{C_0}}}$
                        \end{enumrm}
                        for some $T_1$ and $C_0$.
                        By subsumption of \lref{ty-c1} with \lref{sub-sig'} and \lref{sub-C2'}, we have
                        \begin{enumrm}[resume]
                            \item\llabel{ty-c1-2} $\jdty{\Gamma}{c_1}{\tycomp{\Sigma}{T_1}{\tyctl{x}{C_0}{C_2}}}$~.
                        \end{enumrm}
                        By Lemma~\ref{lem:wfg} with \lref{ty-c2}, we have
                        \begin{enumrm}[resume]
                            \item\llabel{wf-Gx} $\jdwf{}{\Gamma, x: T_1}$~.
                        \end{enumrm}
                        Then it holds that $x \notin \dom(\Gamma)$ and hence from \lref{wf-C} we have
                        \begin{enumrm}[resume]
                            \item\llabel{in-x} $x \notin \fv(T) \cup \fv(\Sigma) \cup (\fv(C_1) \setminus \{z\})$~.
                        \end{enumrm}
                        Also, by Lemma~\ref{lem:weaken} with \lref{sub-sig'}, \lref{sub-T'}, \lref{sub-C1}
                        and \lref{wf-Gx}, we have
                        \begin{itemize}
                            \item $\jdsub{\Gamma, x: T_1}{\Sigma'}{\Sigma}$,
                            \item $\jdsub{\Gamma, x: T_1}{T'}{T}$, and
                            \item $\jdsub{\Gamma, x: T_1, z: T'}{C_1}{C_1'}$~.
                        \end{itemize}
                        Then by subsumption of \lref{ty-c2} we have
                        \begin{enumrm}[resume]
                            \item\llabel{ty-c2-2} $\jdty{\Gamma, x: T_1}{c_2}{\tycomp{\Sigma}{T}{\tyctl{z}{C_1}{C_0}}}$~.
                        \end{enumrm}
                        Now we have the conclusion from \lref{ty-c1-2}, \lref{ty-c2-2} and \lref{in-x}.
                    \item[Case \rulename{S-Embed}:] We have
                        \begin{enumrm}[resume]
                            \item\llabel{eq-S'-pure} $S' = \square$,
                            \item\llabel{sub-C1-C2} $\jdsub{\Gamma, z: T}{C_1}{C_2}$, and
                            \item\llabel{in-z} $z \notin \fv(C_2)$~.
                        \end{enumrm}
                        Then, by Lemma~\ref{lem:inv} with \lref{ty-let}, we have
                        \begin{enumrm}[resume]
                            \item\llabel{ty-c1-} $\jdty{\Gamma}{c_1}{\tycomp{\Sigma'}{T_1}{\square}}$ and
                            \item\llabel{ty-c2-} $\jdty{\Gamma, x: T_1}{c_2}{\tycomp{\Sigma'}{T'}{\square}}$
                        \end{enumrm}
                        for some $T_1$.
                        By Lemma~\ref{lem:wfg} with \lref{ty-c2-}, we have
                        \begin{enumrm}[resume]
                            \item\llabel{wf-Gx-} $\jdwf{}{\Gamma, x: T_1}$~.
                        \end{enumrm}
                        Then it holds that $x \notin \dom(\Gamma)$ and hence from \lref{wf-C} we have
                        \begin{enumrm}[resume]
                            \item\llabel{in-x-} $x \notin \fv(T) \cup \fv(\Sigma) \cup (\fv(C_1) \setminus \{z\})$ and
                            \item\llabel{in-x-2-} $x \notin \fv(C_2)$~.
                        \end{enumrm}
                        Also, by inversion of \lref{wf-C} we have $\jdwf{\Gamma}{C_2}$, and so we have
                        $\jdsub{\Gamma}{C_2}{C_2}$ by Lemma~\ref{lem:refl}.
                        Then, by Lemma~\ref{lem:weaken} with \lref{wf-Gx-} we have
                        $\jdsub{\Gamma, x: T_1}{C_2}{C_2}$.
                        And hence, by \rulename{S-Embed} with \lref{in-x-2-} we have
                        \begin{itemize}
                            \item $\jdsub{\Gamma \mid T_1}{\square}{\tyctl{x}{C_2}{C_2}}$~.
                        \end{itemize}
                        Therefore, by subsumption of \lref{ty-c1-} with \lref{sub-sig'}, we have
                        \begin{enumrm}[resume]
                            \item\llabel{ty-c1-2-} $\jdty{\Gamma}{c_1}{\tycomp{\Sigma}{T_1}{\tyctl{x}{C_2}{C_2}}}$~.
                        \end{enumrm}
                        Moreover, by Lemma~\ref{lem:weaken} with \lref{sub-C'} and \lref{wf-Gx-}, we have
                        \begin{itemize}
                            \item $\jdsub{\Gamma, x: T_1}{\tycomp{\Sigma'}{T'}{\square}}{\tycomp{\Sigma}{T}{\tyctl{z}{C_1}{C_2}}}$~.
                        \end{itemize}
                        Then by subsumption of \lref{ty-c2-} we have
                        \begin{enumrm}[resume]
                            \item\llabel{ty-c2-2-} $\jdty{\Gamma, x: T_1}{c_2}{\tycomp{\Sigma}{T}{\tyctl{z}{C_1}{C_2}}}$~.
                        \end{enumrm}
                        Now we have the conclusion from \lref{ty-c1-2-}, \lref{ty-c2-2-} and \lref{in-x-}.
                \end{description}
        \end{description}
    \end{enumit}
\end{proof}

\begin{lemma}[Inversion with pure evaluation contexts] \label{lem:inv-ctx}
    If $\jdty{\Gamma}{K[c]}{\tycomp{\Sigma}{T}{\tyctl{z}{C_1}{C_2}}}$, then
    there exist some $y, T_1$, and $C_0$ such that
    \begin{itemize}
        \item $\jdty{\Gamma}{c}{\tycomp{\Sigma}{T_1}{\tyctl{y}{C_0}{C_2}}}$ and
        \item $\jdty{\Gamma, y: T_1}{K[\expret{y}]}{\tycomp{\Sigma}{T}{\tyctl{z}{C_1}{C_0}}}$~.
    \end{itemize}
\end{lemma}
\begin{proof}
    By induction on the structure of $K$.
    \begin{description}
        \item[{Case $K = [\ ]$:}]
            \def\currentprefix{inv-ctx:empty}
            We have $\jdty{\Gamma}{c}{\tycomp{\Sigma}{T}{\tyctl{z}{C_1}{C_2}}}$.
            By $\alpha$-renaming, we have
            \begin{enumrm}
                \item\llabel{ty-c} $\jdty{\Gamma}{c}{\tycomp{\Sigma}{T}{\tyctl{y}{C_1[y/z]}{C_2}}}$~.
            \end{enumrm}
            Therefore, we have the first half of the conclusion with $T_1 = T$ and $C_0 = C_1[y/z]$.

            On the other hand, from \lref{ty-c}, it holds that
            \begin{enumrm}[resume]
                \item\llabel{wf-gy} $\jdwf{}{\Gamma, y: T}$
            \end{enumrm}
            by Lemma~\ref{lem:wft}, Lemma~\ref{lem:wfg}, and inversion.
            We show the second half of the conclusion by case analysis on $T$.
            \begin{description}
                \item[Case that $T$ is a refinement type $\tyrfn{z_0}{B}{\phi}$:]
                    By \rulename{T-CVar} and \rulename{T-Ret} with \lref{wf-gy}, it holds that
                    \begin{enumrm}[resume]
                        \item\llabel{ty-rety} $\jdty{\Gamma, y: T}{\expret{y}}{\tycomp{\emptyset}{\tyrfn{z_0}{B}{z_0 = y}}{\square}}$~.
                    \end{enumrm}
                    Also, we have the following subtyping with Lemma~\ref{lem:sub-eq}:
                    \[
                        \infer{\jdsub{\Gamma, y: T \mid \tyrfn{z_0}{B}{z_0 = y}}{\square}{\tyctl{z}{C_1}{C_1[y/z]}}}
                        {
                            \infer{\jdsub{\Gamma, y: T, z: \tyrfn{z_0}{B}{z_0 = y}}{C_1}{C_1[y/z]}}
                            {}
                        }
                    \]
                    Then it holds that
                    \begin{enumrm}[resume]
                        \item\llabel{sub-comp} $\jdsub{\Gamma, y: T}{\tycomp{\emptyset}{\tyrfn{z_0}{B}{z_0 = y}}{\square}}{\tycomp{\Sigma}{T}{\tyctl{z}{C_1}{C_1[y/z]}}}$
                    \end{enumrm}
                    by subtyping.
                    Therefore, by subsumption with \lref{ty-rety} and \lref{sub-comp}, we have the conclusion.
                \item[Case that $T$ is not a refinement type:]
                    By \rulename{T-Var} and \rulename{T-Ret} with \lref{wf-gy}, it holds that
                    \begin{enumrm}[resume]
                        \item\llabel{ty-rety2} $\jdty{\Gamma, y: T}{\expret{y}}{\tycomp{\emptyset}{T}{\square}}$~.
                    \end{enumrm}
                    Also, since $T$ is not a refinement type, by Lemma~\ref{lem:notin-nonrfn}
                    we have $z \notin C_1$ and so $C_1[y/z] = C_1$.
                    Then, we have the following subtyping with Lemma~\ref{lem:refl}:
                    \[
                        \infer{\jdsub{\Gamma, y: T \mid T}{\square}{\tyctl{z}{C_1}{C_1[y/z]}}}
                        {
                            \infer{\jdsub{\Gamma, y: T, z: T}{C_1}{C_1[y/z]}}
                            {}
                        }
                    \]
                    Then it holds that
                    \begin{enumrm}[resume]
                        \item\llabel{sub-comp2} $\jdsub{\Gamma, y: T}{\tycomp{\emptyset}{T}{\square}}{\tycomp{\Sigma}{T}{\tyctl{z}{C_1}{C_1[y/z]}}}$
                    \end{enumrm}
                    by subtyping.
                    Therefore, by subsumption with \lref{ty-rety2} and \lref{sub-comp2}, we have the conclusion.
            \end{description}
        \item[Case $K = \explet{x}{K_1}{c_2}$:]
            \def\currentprefix{inv-ctx:let}
            We have $\jdty{\Gamma}{\explet{x}{K_1[c]}{c_2}}{\tycomp{\Sigma}{T}{\tyctl{z}{C_1}{C_2}}}$.
            By Lemma~\ref{lem:inv}, we have
            \begin{enumrm}
                \item\llabel{ty-k1c} $\jdty{\Gamma}{K_1[c]}{\tycomp{\Sigma}{T'}{\tyctl{x}{C'}{C_2}}}$,
                \item\llabel{ty-c2} $\jdty{\Gamma, x: T'}{c_2}{\tycomp{\Sigma}{T}{\tyctl{z}{C_1}{C'}}}$, and
                \item\llabel{in-x} $x \notin \fv(T) \cup \fv(\Sigma) \cup (\fv(C_1) \setminus \{z\})$
            \end{enumrm}
            for some $T'$ and $C'$.
            By the IH of \lref{ty-k1c}, we have
            \begin{enumrm}[resume]
                \item\llabel{ty-c} $\jdty{\Gamma}{c}{\tycomp{\Sigma}{T_1}{\tyctl{y}{C_0}{C_2}}}$ and
                \item\llabel{ty-k1y} $\jdty{\Gamma, y: T_1}{K_1[\expret{y}]}{\tycomp{\Sigma}{T'}{\tyctl{x}{C'}{C_0}}}$
            \end{enumrm}
            for some $y, T_1$ and $C_0$.

            By Lemma~\ref{lem:wfg} with \lref{ty-c2}, we have $\jdwf{}{\Gamma, x: T'}$.
            By inversion, we have $x \notin \dom(\Gamma)$ and $\jdwf{\Gamma}{T'}$.
            Also, By Lemma~\ref{lem:wfg} with \lref{ty-k1y}, we have $\jdwf{}{\Gamma, y: T_1}$.
            Then, by Lemma~\ref{lem:weaken} we have $\jdwf{\Gamma, y: T_1}{T'}$.
            Moreover, w.l.o.g, we can assume $x \ne y$, and so
            $x \notin \dom(\Gamma) \cup \{y\} = \dom(\Gamma, y: T_1)$.
            Then we have $\jdwf{}{\Gamma, y: T_1, x: T'}$.

            Therefore, by Lemma~\ref{lem:weaken} with \lref{ty-c2}, we have
            \[
                \jdty{\Gamma, y: T_1, x: T'}{c_2}{\tycomp{\Sigma}{T}{\tyctl{z}{C_1}{C'}}}~.
            \]
            Then, by \rulename{T-LetIp} with \lref{ty-k1y} and \lref{in-x}, we have
            \[
                \jdty{\Gamma, y: T_1}{\explet{x}{K_1[\expret{y}]}{c_2}}{\tycomp{\Sigma}{T}{\tyctl{z}{C_1}{C_0}}}~,
            \]
            that is,
            \begin{enumrm}[resume]
                \item\llabel{ty-ky} $\jdty{\Gamma, y: T_1}{K[\expret{y}]}{\tycomp{\Sigma}{T}{\tyctl{z}{C_1}{C_0}}}$~.
            \end{enumrm}
            Therefore, from \lref{ty-c} and \lref{ty-ky} we have the conclusion.
    \end{description}
\end{proof}

\begin{theorem}[Subject reduction] \label{thm:subjred}
    If $\jdty{\emptyset}{c}{C}$ and $c \eval c'$, then $\jdty{\emptyset}{c'}{C}$.
\end{theorem}
\begin{proof}
    \def\currentprefix{subred}
    By induction on the typing derivation.
    \begin{description}
        \item[Case \rulename{T-Ret} and \rulename{T-Op}:]
            Contradictory because there is no evaluation rule for $c$.
        \item[Case \rulename{T-App}:] We have
            \def\currentprefix{subred:app}
            \begin{enumrm}
                \item\llabel{eq-c} $c = v_1~v_2$,
                \item\llabel{eq-C} $C = C_1[v_2/x]$,
                \item\llabel{ty-v1} $\jdty{}{v_1}{(x:T_1) \rarr C_1}$, and
                \item\llabel{ty-v2} $\jdty{}{v_2}{T_1}$
            \end{enumrm}
            for some $x, v_1, v_2, T_1$ and $C_1$.
            Case analysis on the evaluation derivation.
            \begin{description}
                \item[Case \rulename{E-App}:] We have
                    \def\currentprefix{subred:app:app}
                    \begin{enumrm}[resume]
                        \item\llabel{eq-v1} $v_1 = \exprec{f}{x}{c_1}$, and
                        \item\llabel{eq-c'} $c' = c_1[v_2/x][(\exprec{f}{x}{c_1})/f]$
                    \end{enumrm}
                    for some $f, x$ and $c_1$.
                    By Lemma \ref{lem:inv} with \lref[subred:app]{ty-v1}, we have
                    \begin{enumrm}[resume]
                        \item\llabel{ty-v1-2} $\jdty{}{v_1}{(x: T_0) \rarr C_0}$,
                        \item\llabel{sub-fun} $\jdsub{}{(x: T_0) \rarr C_0}{(x: T_1) \rarr C_1}$, and
                        \item\llabel{ty-c} $\jdty{f: (x: T_0) \rarr C_0, x: T_0}{c}{C_0}$
                    \end{enumrm}
                    for some $T_0$ and $C_0$.
                    By Lemma \ref{lem:wfg} with \lref{ty-c}, inversion, and Lemma \ref{lem:rm-nonrfn},
                    we have $\jdwf{}{T_0}$.
                    Also, by inversion of \lref{sub-fun}, we have $\jdsub{}{T_1}{T_0}$.
                    Then, By \rulename{T-VSub} with \lref[subred:app]{ty-v2},
                    we have $\jdty{}{v_2}{T_0}$.
                    Using this and \lref{ty-v1-2}, we have the conclusion
                    by Lemma \ref{lem:subst} with \lref{ty-c}.
                \item[Case \rulename{E-Prim}:] We have
                    \def\currentprefix{subred:app:prim}
                    \begin{enumrm}[resume]
                        \item\llabel{eq-v1} $v_1 = p$, and
                        \item\llabel{eq-c'} $c' = \zeta(p, v_2)$
                    \end{enumrm}
                    for some $p$.
                    By Lemma \ref{lem:inv} with \lref[subred:app]{ty-v1}, we have
                    \begin{enumrm}[resume]
                        \item\llabel{ty-p} $\jdty{}{p}{\ty(p)}$, and
                        \item\llabel{sub-typ} $\jdsub{}{\ty(p)}{(x:T_1) \rarr C_1}$~.
                    \end{enumrm}
                    By inversion of \lref{sub-typ}, we have
                    \begin{enumrm}[resume]
                        \item\llabel{eq-typ} $\ty(p) = (x:T_0) \rarr C_0$,
                        \item\llabel{sub-t1} $\jdsub{}{T_1}{T_0}$, and
                        \item\llabel{sub-c0} $\jdsub{x: T_1}{C_0}{C_1}$
                    \end{enumrm}
                    for some $T_0$ and $C_0$.
                    By Lemma \ref{lem:wfg} with \lref{ty-p} and \lref{eq-typ} and inversion,
                    we have $\jdwf{}{T_0}$.
                    Then, by \rulename{T-VSub} with \lref[subred:app]{ty-v2} and \lref{sub-t1},
                    we have $\jdty{}{v_2}{T_0}$.
                    Therefore, by Assumption \ref{asm:prim} with \lref{eq-typ}, we have
                    \begin{enumrm}[resume]
                        \item\llabel{ty-z} $\jdty{}{\zeta(p, v_2)}{C_0[v_2/x]}$~.
                    \end{enumrm}
                    Also, by Lemma \ref{lem:wft} with \lref[subred:app]{ty-v1} and inversion,
                    we have
                    \begin{enumrm}[resume]
                        \item\llabel{wf-C1} $\jdwf{x: T_1}{C_1}$~.
                    \end{enumrm}
                    Using \lref[subred:app]{ty-v2},
                    by Lemma \ref{lem:subst} with \lref{sub-c0} and \lref{wf-C1} respectively,
                    we have
                    \begin{itemize}
                        \item $\jdsub{}{C_0[v_2/x]}{C_1[v_2/x]}$ and
                        \item $\jdwf{}{C_1[v_2/x]}$~.
                    \end{itemize}
                    Therefore, by \rulename{T-CSub} with \lref{ty-z}, we have the conclusion.
            \end{description}
        \item[Case \rulename{T-If}:] We have
            \def\currentprefix{subred:if}
            \begin{enumrm}
                \item\llabel{eq-c} $c = \expif{v}{c_1}{c_2}$,
                \item\llabel{ty-v} $\jdty{}{v}{\tyrfn{x}{\tybool}{\phi}}$,
                \item\llabel{ty-c1} $\jdty{v = \exptrue}{c_1}{C}$, and
                \item\llabel{ty-c2} $\jdty{v = \expfalse}{c_2}{C}$
            \end{enumrm}
            for some $x, v, c_1, c_2$, and $\phi$.
            Case analysis on the evaluation derivation.
            \begin{description}
                \item[Case \rulename{E-IfT}: ] We have
                    \begin{enumrm}[resume]
                        \item\llabel{eq-v} $v = \exptrue$, and
                        \item\llabel{eq-c'} $c' = c_1$~.
                    \end{enumrm}
                    We have the conclusion by Lemma \ref{lem:rm-tauto} with \lref{ty-c1}.
                \item[Case \rulename{E-IfF}: ] Similar.
            \end{description}
        \item[Case \rulename{T-CSub}:] By the IH and \rulename{T-CSub}.
        \item[Case \rulename{T-LetP}:] We have
            \def\currentprefix{subred:letp}
            \begin{enumrm}
                \item\llabel{eq-c} $c = \explet{x}{c_1}{c_2}$,
                \item\llabel{eq-C} $C = \tycomp{\Sigma}{T_2}{\square}$,
                \item\llabel{ty-c1} $\jdty{}{c_1}{\tycomp{\Sigma}{T_1}{\square}}$,
                \item\llabel{ty-c2} $\jdty{x: T_1}{c_2}{\tycomp{\Sigma}{T_2}{\square}}$, and
                \item\llabel{in-x} $x \notin \fv(T_2) \cup \fv(\Sigma)$
            \end{enumrm}
            for some $x, c_1, c_2, \Sigma, T_1$ and $T_2$.
            Case analysis on the evaluation derivation.
            \begin{description}
                \item[Case \rulename{E-Let}:] 
                    By the IH and \rulename{T-LetP}.
                \item[Case \rulename{E-LetRet}:] We have
                    \def\currentprefix{subred:letp:ret}
                    \begin{enumrm}[resume]
                        \item\llabel{eq-c1} $c_1 = \expret{v}$, and
                        \item\llabel{eq-c'} $c' = c_2[v/x]$
                    \end{enumrm}
                    for some $v$.
                    By Lemma \ref{lem:inv} with \lref[subred:letp]{ty-c1}, we have
                    \begin{enumrm}[resume]
                        \item\llabel{sub-T0} $\jdsub{}{T_0}{T_1}$ and
                        \item\llabel{ty-v} $\jdty{}{v}{T_0}$
                    \end{enumrm}
                    for some $T_0$.
                    By Lemma \ref{lem:wft} with \lref[subred:letp]{ty-c1} and inversion,
                    we have $\jdwf{}{T_1}$.
                    Then, by \rulename{T-VSub} with \lref{ty-v} and \lref{sub-T0},
                    we have $\jdty{}{v}{T_1}$.
                    Therefore, by Lemma \ref{lem:subst} with \lref[subred:letp]{ty-c2},
                    we have
                    \[
                        \jdty{}{c_2[v/x]}{\tycomp{\Sigma}{T_2}{\square}}
                    \]
                    (Note that since \lref[subred:letp]{in-x},
                    it holds that $\Sigma[v/x] = \Sigma$ and $T_2[v/x] = T_2$.)
                    That is, we have the conclusion.
            \end{description}
        \item[Case \rulename{T-LetIp}:] We have
            \def\currentprefix{subred:leti}
            \begin{enumrm}
                \item\llabel{eq-c} $c = \explet{x}{c_1}{c_2}$,
                \item\llabel{eq-C} $C = \tycomp{\Sigma}{T_2}{\tyctl{z}{C_{21}}{C_{12}}}$,
                \item\llabel{ty-c1} $\jdty{}{c_1}{\tycomp{\Sigma}{T_1}{\tyctl{x}{C_0}{C_{12}}}}$,
                \item\llabel{ty-c2} $\jdty{x: T_1}{c_2}{\tycomp{\Sigma}{T_2}{\tyctl{z}{C_{21}}{C_0}}}$, and
                \item\llabel{in-x} $x \notin \fv(T_2) \cup \fv(\Sigma) \cup (\fv(C_{21}) \setminus \{z\})$
            \end{enumrm}
            for some $x, z, c_1, c_2, \Sigma, T_1, T_2, C_0, C_{12}$ and $C_{21}$.
            Case analysis on the evaluation derivation.
            \begin{description}
                \item[Case \rulename{E-Let}:] 
                    By the IH and \rulename{T-LetIp}.
                \item[Case \rulename{E-LetRet}:] We have
                    \def\currentprefix{subred:leti:ret}
                    \begin{enumrm}[resume]
                        \item\llabel{eq-c1} $c_1 = \expret{v}$, and
                        \item\llabel{eq-c'} $c' = c_2[v/x]$
                    \end{enumrm}
                    for some $v$.
                    By Lemma \ref{lem:inv} with \lref[subred:leti]{ty-c1}, we have
                    \begin{enumrm}[resume]
                        \item\llabel{sub-T0} $\jdsub{}{T_0}{T_1}$,
                        \item\llabel{ty-v} $\jdty{}{v}{T_0}$, and
                        \item\llabel{sub-S1} $\jdsub{ \mid T_0}{\square}{\tyctl{x}{C_0}{C_{12}}}$
                    \end{enumrm}
                    for some $T_0$.
                    By Lemma \ref{lem:wft} with \lref[subred:leti]{ty-c1} and inversion,
                    we have $\jdwf{}{T_1}$.
                    Then, by \rulename{T-VSub} with \lref{ty-v} and \lref{sub-T0},
                    we have $\jdty{}{v}{T_1}$.
                    Therefore, by Lemma \ref{lem:subst} with \lref[subred:leti]{ty-c2},
                    we have
                    \begin{enumrm}[resume]
                        \item\llabel{ty-c2-2} $\jdty{}{c_2[v/x]}{\tycomp{\Sigma}{T_2}{\tyctl{z}{C_{21}}{C_0[v/x]}}}$~.
                    \end{enumrm}
                    (Note that since \lref[subred:leti]{in-x},
                    it holds that $\Sigma[v/x] = \Sigma$, $T_2[v/x] = T_2$ and $C_{21}[v/x] = C_{21}$.)
                    By inversion of \lref{sub-S1}, we have
                    \begin{enumrm}[resume]
                        \item\llabel{sub-C0} $\jdsub{x:T_0}{C_0}{C_{12}}$~.
                    \end{enumrm}
                    By Lemma \ref{lem:wft} with \lref[subred:leti]{ty-c1} and inversion,
                    we have $\jdwf{}{C_{12}}$, which means $x \notin \fv(C_{12})$.
                    Therefore, by Lemma \ref{lem:subst} with \lref{sub-C0},
                    we have
                    \begin{enumrm}[resume]
                        \item\llabel{sub-C0-2} $\jdsub{}{C_0[v/x]}{C_{12}}$~.
                    \end{enumrm}
                    On the other hand, by Lemma \ref{lem:wft} with \lref[subred:leti]{ty-c2} and inversion,
                    we have $\jdwf{x:T_1, z:T_2}{C_{21}}$.
                    By Lemma \ref{lem:rm-unused} with \lref[subred:leti]{in-x},
                    we have $\jdwf{z:T_2}{C_{21}}$.
                    Then, by Lemma \ref{lem:refl} we have
                    \begin{enumrm}[resume]
                        \item\llabel{sub-C2} $\jdsub{z:T_2}{C_{21}}{C_{21}}$~.
                    \end{enumrm}
                    Hence, by \rulename{S-ATM} with \lref{sub-C0-2} and \lref{sub-C2},
                    we have
                    $\jdsub{\mid T_2}{\tyctl{z}{C_{21}}{C_0[v/x]}}{\tyctl{z}{C_{21}}{C_{12}}}$~.
                    Now we have the conclusion by subsumption of \lref{ty-c2-2}.
            \end{description}
        \item[Case \rulename{T-Hndl}:] We have
            \def\currentprefix{subred:hndl}
            \begin{enumrm}
                \item\llabel{eq-c} $c = \expwith{h}{c_0}$,
                \item\llabel{eq-h} $h = \{ \expret{x_r} \mapsto c_r, \repi{\op_i(x_i, k_i) \mapsto c_i} \}$,
                \item\llabel{ty-c0} $\jdty{}{c_0}{\tycomp{\Sigma_0}{T_0}{\tyctl{x_r}{C_1}{C}}}$,
                \item\llabel{ty-cr} $\jdty{x_r: T_0}{c_r}{C_1}$,
                \item\llabel{ty-ci} $\bigrepi{\jdty{\rep{X_i: \rep{B}_i}, x_i: T_{i1}, k_i: (y_i: T_{i2}) \rarr C_{i1}}{c_i}{C_{i2}}}$, and
                \item\llabel{eq-sig} $\Sigma_0 = \{ \repi{\op_i: \forall \rep{X_i: \rep{B}_i}. (x_i: T_{i1}) \rarr ((y_i: T_{i2}) \rarr C_{i1}) \rarr C_{2i}} \}$
            \end{enumrm}
            Case analysis on the evaluation derivation.
            \begin{description}
                \item[Case \rulename{E-Hndl}:]
                    By the IH and \rulename{T-Hndl}.
                \item[Case \rulename{E-HndlRet}:] We have
                    \def\currentprefix{subred:hndl:ret}
                    \begin{enumrm}[resume]
                        \item\llabel{eq-c0} $c_0 = \expret{v}$ and
                        \item\llabel{eq-c'} $c' = c_r[v/x_r]$
                    \end{enumrm}
                    for some $v$.
                    By Lemma \ref{lem:inv} with \lref[subred:hndl]{ty-c0}, we have
                    \begin{enumrm}[resume]
                        \item\llabel{sub-T0} $\jdsub{}{T_0'}{T_0}$,
                        \item\llabel{ty-v} $\jdty{}{v}{T_0'}$, and
                        \item\llabel{sub-S0} $\jdsub{ \mid T_0'}{\square}{\tyctl{x_r}{C_1}{C}}$
                    \end{enumrm}
                    for some $T_0'$.
                    By inversion of \lref{sub-S0}, we have
                    \begin{enumrm}[resume]
                        \item\llabel{sub-C1} $\jdsub{x_r: T_0'}{C_1}{C}$ and
                        \item\llabel{in-xr} $x_r \notin \fv(C)$~.
                    \end{enumrm}
                    By Lemma \ref{lem:narrow} with \lref[subred:hndl]{ty-cr} and \lref{sub-T0},
                    we have
                    \begin{enumrm}[resume]
                        \item\llabel{ty-cr-2} $\jdty{x_r: T_0'}{c_r}{C_1}$~.
                    \end{enumrm}
                    By Lemma \ref{lem:subst} with \lref{ty-v}
                    applied to \lref{sub-C1} and \lref{ty-cr-2}, we have
                    \begin{enumrm}[resume]
                        \item\llabel{sub-C1-2} $\jdsub{}{C_1[v/x_r]}{C}$ and
                        \item\llabel{ty-cr-3} $\jdty{}{c_r[v/x_r]}{C_1[v/x_r]}$
                    \end{enumrm}
                    respectively.
                    (Note that $C[v/x_r] = C$ since \lref{in-xr}.)
                    By Lemma \ref{lem:wft} with \lref[subred:hndl]{ty-c0} and inversion,
                    we have $\jdwf{}{C}$.
                    From this and \lref{sub-C1-2} and \lref{ty-cr-3},
                    we have the conclusion by \rulename{T-CSub}.
                \item[Case \rulename{E-HndlOp}:] We have
                    \def\currentprefix{subred:hndl:op}
                    \begin{enumrm}[resume]
                        \item\llabel{eq-c0} $c_0 = K[\op_i~v]$ and
                        \item\llabel{eq-c'} $c' = c_i[v/x_i][(\lambda y. \expwith{h}{K[\expret{y}]})/k_i]$
                    \end{enumrm}
                    for some $K$ and $v$.
                    W.l.o.g., we can assume that $y$ is disjoint from
                    any other existing variables.
                    By Lemma \ref{lem:inv-ctx} with \lref[subred:hndl]{ty-c0}, we have
                    \begin{enumrm}[resume]
                        \item\llabel{ty-opv} $\jdty{}{\op~v}{\tycomp{\Sigma_0}{T_1}{\tyctl{y}{C_0}{C}}}$ and
                        \item\llabel{ty-Ky} $\jdty{y: T_1}{K[\expret{y}]}{\tycomp{\Sigma_0}{T_0}{\tyctl{x_r}{C_1}{C_0}}}$
                    \end{enumrm}
                    for some $y, T_1$ and $C_0$.
                    By Lemma \ref{lem:inv} with \lref{ty-opv}, we have
                    \begin{enumrm}[resume]
                        \item\llabel{in-sig} $\Sigma_0 \ni \op_i: \forall \rep{X_i: \rep{B}_i}. (x_i: T_{i1}) \rarr ((y: T_{i2}) \rarr C_{i1}) \rarr C_{i2}$,
                        \item\llabel{wf-A} $\rep{\jdty{}{A}{\rep{B}_i}}$,
                        \item\llabel{ty-v} $\jdty{}{v}{T_{i1}[\rep{A/X_i}]}$,
                        \item\llabel{sub-Ti2} $\jdsub{}{T_{i2}[\rep{A/X_i}][v/x_i]}{T_1}$,
                        \item\llabel{sub-C0} $\jdsub{y: T_{i2}[\rep{A/X_i}][v/x_i]}{C_0}{C_{i1}[\rep{A/X_i}][v/x_i]}$, and
                        \item\llabel{sub-Ci2} $\jdsub{}{C_{i2}[\rep{A/X_i}][v/x_i]}{C}$
                    \end{enumrm}
                    for some $\rep{A}$.
                    Note that since \lref[subred:hndl]{eq-sig} holds, it holds that $y = y_i$ and we use
                    $\rep{X_i}, \rep{\rep{B}_i}, x_i, T_{i1}, T_{i2}, C_{i1}$, and $C_{i2}$ here
                    instead of introducing new ones.

                    Also, by Lemma~\ref{lem:narrow} with \lref{ty-Ky} and \lref{sub-Ti2}, we have
                    \[
                        \jdty{y: T_{i2}[\rep{A/X_i}][v/x_i]}{K[\expret{y}]}{\tycomp{\Sigma_0}{T_0}{\tyctl{x_r}{C_1}{C_0}}}~.
                    \]
                    Then, by subsumption with \lref{sub-C0}, we have
                    \begin{enumrm}[resume]
                        \item\llabel{ty-Ky-2} $\jdty{y: T_{i2}[\rep{A/X_i}][v/x_i]}{K[\expret{y}]}{\tycomp{\Sigma_0}{T_0}{\tyctl{x_r}{C_1}{C_{i1}[\rep{A/X_i}][v/x_i]}}}$~.
                    \end{enumrm}
                    On the other hand, by Lemma~\ref{lem:wfg} with \lref{ty-Ky-2} we have
                    $\jdwf{}{y: T_{i2}[\rep{A/X_i}][v/x_i]}$,
                    and hence by Lemma~\ref{lem:weaken} with \lref[subred:hndl]{ty-cr} and \lref[subred:hndl]{ty-ci}, we have
                    \begin{enumrm}[resume]
                        \item\llabel{ty-cr-2} $\jdty{y: T_{i2}[\rep{A/X_i}][v/x_i], x_r: T_0}{c_r}{C_1}$ and
                        \item\llabel{ty-ci-2} $\bigrepi{\jdty{y: T_{i2}[\rep{A/X_i}][v/x_i], \rep{X_i: \rep{B}_i}, x_i: T_{i1}, k_i: (y_i: T_{i2}) \rarr C_{i1}}{c_i}{C_{i2}}}$~.
                    \end{enumrm}
                    Therefore, by \rulename{T-Hndl} with
                    \lref[subred:hndl]{eq-h}, \lref[subred:hndl]{eq-sig},
                    \lref{ty-Ky-2}, \lref{ty-cr-2}, and \lref{ty-ci-2}, we have
                    \[
                        \jdty{y: T_{i2}[\rep{A/X_i}][v/x_i]}{\expwith{h}{K[\expret{y}]}}{C_{i1}[\rep{A/X_i}][v/x_i]}~.
                    \]
                    Then by \rulename{T-Fun} we have
                    \begin{enumrm}[resume]
                        \item\llabel{ty-yhKy} $\jdty{}{\lambda y. \expwith{h}{K[\expret{y}]}}{(y: y: T_{i2}[\rep{A/X_i}][v/x_i]) \rarr C_{i1}[\rep{A/X_i}][v/x_i]}$~.
                    \end{enumrm}

                    Now, by Lemma \ref{lem:subst-pred} with \lref{wf-A}
                    applied to \lref[subred:hndl]{ty-ci}, we have
                    \[
                        \jdty{x_i: T_{i1}[\rep{A/X_i}], k_i: (y_i: T_{i2}[\rep{A/X_i}]) \rarr C_{i1}[\rep{A/X_i}]}
                            {c_i}{C_{i2}[\rep{A/X_i}]}~.
                    \]
                    By applying \ref{lem:subst} twice with \lref{ty-v} and \lref{ty-yhKy} in a row, we have
                    \[
                        \jdty{}{c_i[v/x_i][(\lambda y. \expwith{h}{K[\expret{y}]})/k_i]}{C_{i2}[\rep{A/X_i}][v/x_i]}~.
                    \]
                    Note that
                    $C_{i2}[\rep{A/X_i}][v/x_i][(\lambda y. \expwith{h}{K[\expret{y}]})/k_i] = C_{i2}[\rep{A/X_i}][v/x_i]$
                    since $k_i \notin \fv(C_{i2}[\rep{A/X_i}][v/x_i])$ by Lemma \ref{lem:notin-nonrfn}.
                    Now we have the conclusion by subsumption with \lref{sub-Ci2}.
            \end{description}
    \end{description}
\end{proof}

\subsection{Type Safety}

\begin{theorem}[Type safety] \label{thm:safety}
    If $\jdty{\emptyset}{c}{\tycomp{\Sigma}{T}{S}}$ and $c \eval^* c'$, then either:
    \begin{itemize}
        \item $c' = \expret{v}$ for some $v$ such that $\jdty{\emptyset}{v}{T}$,
        \item $c' = K[\op~v]$ for some $K, \op$ and $v$ such that $\op \in \dom(\Sigma)$, or
        \item $c' \eval c''$ for some $c''$ such that $\jdty{\emptyset}{c''}{\tycomp{\Sigma}{T}{S}}$.
    \end{itemize}
\end{theorem}
\begin{proof}
    By induction on the length of $\eval^*$
    with Theorem~\ref{thm:progress} and Theorem~\ref{thm:subjred}.
\end{proof}

\section{Definitions for the CPS transformation}

\subsection{Evaluation rules for the target language of the CPS transformation}

\begin{align}
    \text{evaluation context} \quad
    E ::= [\ ] \mid E~v \mid E~\rep{A} \mid E~\tau
\end{align}
\fbox{$c \eval c'$}
\begin{gather}
    \infersc[Ec-Ctx]{E[c] \eval E[c']}
    {c \eval c'}
    \quad
    \infersc[Ec-IfT]{\expif{\exptrue}{c_1}{c_2} \eval c_1}
    {}
    \quad
    \infersc[Ec-IfF]{\expif{\expfalse}{c_1}{c_2} \eval c_2}
    {}
    \\
    \infersc[Ec-App]{(\exprec{f:\tau_1}{x:\tau_2}{c})~v \eval c[v/x][(\exprec{f:\tau_1}{x:\tau_2}{c})/f]}
    {}
    \\[1.5ex]
    \infersc[Ec-Prim]{p~v \eval \zeta_{\textit{cps}}(p, v)}
    {}
    \quad
    \infersc[Ec-PApp]{(\Lambda \rep{X: \rep{B}}. c)~\rep{A} \eval c[\rep{A/X}]}
    {}
    \\[1.5ex]
    \infersc[Ec-Proj]{\{ \repi{\op_i = v_i} \}\#\op_i \eval v_i}
    {}
    \quad
    \infersc[Ec-TApp]{(\Lambda \alpha. c)~\tau \eval c[\tau/\alpha]}
    {}
    \quad
    \infersc[Ec-Acsr]{(c: \tau) \eval c}
    {}
\end{gather}

\subsection{Syntax of typing contexts of the target language of the CPS transformation}

\begin{gather}
    \Gamma ::= \emptyset \mid \Gamma, x: \tau
        \mid \Gamma, X:\rep{B} \mid \Gamma, \alpha
\end{gather}

\subsection{Well-formedness rules of the target language of the CPS transformation}

\fbox{$\jdwf{}{\Gamma}$} \quad \fbox{$\jdwf{\Gamma}{\tau}$}
\begin{gather}
    \infersc[WEc-Empty]{\jdwf{}{\emptyset}}
    {}
    \quad
    \infersc[WEc-Var]{\jdwf{}{\Gamma, x: \tau}}
    {
        \jdwf{}{\Gamma} &
        x \notin \dom(\Gamma) &
        \jdwf{\Gamma}{\tau}
    }
    \\
    \infersc[WEc-PVar]{\jdwf{}{\Gamma, X: \rep{B}}}
    {
        \jdwf{}{\Gamma} &
        X \notin \dom(\Gamma)
    }
    \quad
    \infersc[WEc-TVar]{\jdwf{}{\Gamma, \alpha}}
    {
        \jdwf{}{\Gamma} &
        \alpha \notin \dom(\Gamma)
    }
    \\
    \infersc[WTc-Rfn]{\jdwf{\Gamma}{\tyrfn{x}{B}{\phi}}}
    {\Gamma, x: B \vdash \phi}
    \quad
    \infersc[WTc-Fun]{\jdwf{\Gamma}{(x: \tau_1) \rarr \tau_2}}
    {
        \jdwf{\Gamma, x: \tau_1}{\tau_2}
    }
    \quad
    \infersc[WTc-PPoly]{\jdwf{\Gamma}{\forall \rep{X: \rep{B}}. \tau}}
    {
        \jdwf{\Gamma, \rep{X: \rep{B}}}{\tau}
    }
    \\
    \infersc[WTc-Rcd]{\jdwf{\Gamma}{\{\repi{\op_i: \tau_i}\}}}
    {
        \bigrepi{\jdwf{\Gamma}{\tau_i}}
    }
    \quad
    \infersc[WTc-TVar]{\jdwf{\Gamma}{\alpha}}
    {
        \alpha \in \Gamma
    }
    \quad
    \infersc[WTc-TPoly]{\jdwf{\Gamma}{\forall \alpha. \tau}}
    {
        \jdwf{\Gamma, \alpha}{\tau}
    }
    \quad
\end{gather}

\subsection{Typing rules of the target language of the CPS transformation}

\fbox{$\jdty{\Gamma}{c}{\tau}$}
\begin{gather}
    \infersc[Tc-CVar]{\jdty{\Gamma}{x}{\tyrfn{y}{B}{x = y}}}
    {
        \jdwf{}{\Gamma} &
        \Gamma(x) = \tyrfn{y}{B}{\phi}
    }
    \quad
    \infersc[Tc-Var]{\jdty{\Gamma}{x}{\Gamma(x)}}
    {
        \jdwf{}{\Gamma} &
        \forall y, B, \phi. \Gamma(x) \neq \tyrfn{y}{B}{\phi}
    }
    \quad
    \infersc[Tc-Prim]{\jdty{\Gamma}{p}{\tycps(p)}}
    {\jdwf{}{\Gamma}}
    \\
    \infersc[Tc-Fun]{\jdty{\Gamma}{\exprec{f:(x: \tau_1) \rarr \tau_2}{x:\tau_1}{c}}{(x: \tau_1) \rarr \tau_2}}
    {
        \jdty{\Gamma, f: (x: \tau_1) \rarr \tau_2, x: \tau_1}{c}{\tau_2}
    }
    \quad
    \infersc[Tc-App]{\jdty{\Gamma}{c~v}{\tau_2[v/x]}}
    {
        \jdty{\Gamma}{c}{(x: \tau_1) \rarr \tau_2} &
        \jdty{\Gamma}{v}{\tau_1}
    }
    \\
    \infersc[Tc-TAbs]{\jdty{\Gamma}{\Lambda \alpha. c}{\forall \alpha. \tau}}
    {\jdty{\Gamma, \alpha}{c}{\tau}}
    \quad
    \infersc[Tc-TApp]{\jdty{\Gamma}{c~\tau}{\tau'[\tau/\alpha]}}
    {
        \jdty{\Gamma}{c}{\forall \alpha. \tau'} &
        \jdwf{\Gamma}{\tau}
    }
    \\
    \infersc[Tc-PAbs]{\jdty{\Gamma}{\Lambda \rep{X: \rep{B}}. c}{\forall \rep{X: \rep{B}}. \tau}}
    {
        \jdty{\Gamma, \rep{X: \rep{B}}}{c}{\tau}
    }
    \quad
    \infersc[Tc-PApp]{\jdty{\Gamma}{c~\rep{A}}{\tau[\rep{A/X}]}}
    {
        \jdty{\Gamma}{c}{\forall \rep{X: \rep{B}}. \tau} &
        \rep{\jdty{\Gamma}{A}{\rep{B}}}
    }
    \\
    \infersc[Tc-Rcd]{\jdty{\Gamma}{\{ \repi{\op_i = v_i} \}}{\{ \repi{\op_i : \tau_i} \}}}
    {
        \repi{\jdty{\Gamma}{v_i}{\tau_i}}
    }
    \quad
    \infersc[Tc-Proj]{\jdty{\Gamma}{v\#\op_i}{\tau_i}}
    {
        \jdty{\Gamma}{v}{\{ \repi{\op_i : \tau_i} \}}
    }
    \\
    \infersc[Tc-If]{\jdty{\Gamma}{\expif{v}{c_1}{c_2}}{\tau}}
    {
        \jdty{\Gamma}{v}{\tyrfn{x}{\tybool}{\phi}} &
        \jdty{\Gamma, v = \exptrue}{c_1}{\tau} &
        \jdty{\Gamma, v = \expfalse}{c_2}{\tau}
    }
    \\
    \infersc[Tc-Ascr]{\jdty{\Gamma}{(c : \tau)}{\tau}}
    {
        \jdty{\Gamma}{c}{\tau'} &
        \jdsub{\Gamma}{\tau'}{\tau} &
        \jdwf{\Gamma}{\tau}
    }
    \quad
    \infersc[Tc-Sub]{\jdty{\Gamma}{c}{\tau_2}}
    {
        \jdty{\Gamma}{c}{\tau_1} &
        \jdsub{\Gamma}{\tau_1}{\tau_2} &
        \jdwf{\Gamma}{\tau_2}
    }
\end{gather}

\subsection{Subtyping rules of the target language of the CPS transformation}

\fbox{$\jdsub{\Gamma}{\tau_1}{\tau_2}$}
\begin{gather}
    \infersc[Sc-Rfn]{\jdsub{\Gamma}{\tyrfn{x}{B}{\phi_1}}{\tyrfn{x}{B}{\phi_2}}}
    {\Gamma, x: B \vDash \phi_1 \implies \phi_2}
    \quad
    \infersc[Sc-Fun]{\jdsub{\Gamma}{(x: \tau_{11}) \rarr \tau_{12}}{(x: \tau_{21}) \rarr \tau_{22}}}
    {
        \jdsub{\Gamma}{\tau_{21}}{\tau_{11}} &
        \jdsub{\Gamma, x: \tau_{21}}{\tau_{12}}{\tau_{22}}
    }
    \\
    \infersc[Sc-PPoly]{\jdsub{\Gamma}{\forall \rep{X: \rep{B}}. \tau_1}{\forall \rep{X: \rep{B}}. \tau_2}}
    {
        \jdsub{\Gamma, \rep{X: \rep{B}}}{\tau_1}{\tau_2}
    }
    \quad
    \infersc[Sc-Rcd]{\jdsub{\Gamma}{\{ \repi{\op_i: \tau_{1i}}, \repi{\op'_i: \tau'_i} \}}{\{ \repi{\op_i:  \tau_{2i}} \}}}
    {\repi{\jdsub{\Gamma}{\tau_{1i}}{\tau_{2i}}}}
    \\
    \infersc[Sc-TVar]{\jdsub{\Gamma}{\alpha}{\alpha}}
    {\alpha \in \Gamma}
    \quad
    \infersc[Sc-Poly]{\jdsub{\Gamma}{\forall \alpha. \tau_1}{\forall \beta. \tau_2}}
    {
        \jdsub{\Gamma, \beta}{\tau_1[\tau/\alpha]}{\tau_2} &
        \jdwf{\Gamma, \beta}{\tau}&
        \beta \notin \fv(\forall \alpha. \tau_1)
    }
\end{gather}

\subsection{CPS transformation of expressions}

\begingroup
\allowdisplaybreaks
\begin{align}
    \cps{x} &\defeq x \\
    \cps{p} &\defeq \mathit{cps}(p) \\
    \cps{\exprec{f^{(x:T_1) \rarr C_1}}{x^{T_2}}{c}} &\defeq \exprec{f:\cps{(x:T_1) \rarr C_1}}{x:\cps{T_2}}{\cps{c}} \\
    \cps{\expret{v^T}} &\defeq \stLambda \alpha. \stlambda h:\{\}. \stlambda k:\cps{T} \rarr \alpha. k~\cps{v} \\
    \cps{\explet{x}{c_1^{\tycomp{\Sigma}{T_1}{\square}}}{c_2^{\tycomp{\Sigma}{T_2}{\square}}}}
        &\defeq \stLambda \alpha. \stlambda h: \cps{\Sigma}. \stlambda k:\cps{T_2} \rarr \alpha.
        \cps{c_1} \stapp \alpha \stapp h \stapp (\lambda x:\cps{T_1}. \cps{c_2} \stapp \alpha \stapp h \stapp k) \\
    & \hspace*{-120pt} \cps{\explet{x}{c_1^{\tycomp{\Sigma}{T_1}{\tyctl{x}{C_1}{C_2}}}}{c_2^{\tycomp{\Sigma}{T_2}{\tyctl{z}{C_0}{C_1}}}}} \defeq \\
        & \hspace*{-90pt} \stLambda \alpha. \stlambda h: \cps{\Sigma}. \stlambda k:(z:\cps{T_2}) \rarr \cps{C_0}.
        \cps{c_1} \stapp \cps{C_2} \stapp h \stapp (\lambda x:\cps{T_1}. \cps{c_2} \stapp \cps{C_1} \stapp h \stapp k) \\
    \cps{v_1~v_2} &\defeq \cps{v_1}~\cps{v_2} \\
    \cps{(\expif{v}{c_1}{c_2})^{C}} &\defeq (\expif{\cps{v}}{\cps{c_1}}{\cps{c_2}} : \cps{C}) \\
    \cps{(\op^{\rep{\mathit{A}}}~v)^{\tycomp{\Sigma}{T}{\tyctl{y}{C_1}{C_2}}}} &\defeq
        \stLambda \alpha. \stlambda h:\cps{\Sigma}. \stlambda k:(y: \cps{T} \rarr \cps{C_1}).
        h\#\op~\rep{A}~\cps{v}~(\lambda y': \cps{T}. k~y') \\
    \cps{(\expwith{h}{c})^C} &\defeq \cps{c} \stapp \cps{C} \stapp \cps{h^{\mathit{ops}}} \stapp \cps{h^{\mathit{ret}}} \\[-.5ex]
        \text{where} \hspace{-25pt} & \hspace{25pt} \left\{ \begin{aligned}
        h &= \{ \expret{x_r^{T_r}} \mapsto c_r, \repi{\op_i^{\rep{X_i: \rep{B_i}}}(x_i^{T_{x_i}}, k_i^{T_{k_i}}) \mapsto c_i} \} \\[-.5ex]
        \cps{h^{\mathit{ops}}} &\defeq
            \{ \repi{\op_i = \Lambda \rep{X_i: \rep{B_i}}. \lambda x_i:\cps{T_{x_i}}. \lambda k_i:\cps{T_{k_i}}. \cps{c_i}} \} \\[-.5ex]
        \cps{h^{\mathit{ret}}} &\defeq
            \lambda x_r:\cps{T_r}. \cps{c_r}
    \end{aligned} \right.
\end{align}
\endgroup

\subsection{CPS transformation of types and typing contexts}

\begin{align}
    \cps{\tyrfn{x}{B}{\phi}} &\defeq \tyrfn{x}{B}{\phi} \\
    \cps{(x: T) \rarr C} &\defeq (x: \cps{T}) \rarr \cps{C} \\
    \cps{\tycomp{\Sigma}{T}{\tyctl{x}{C_1}{C_2}}} &\defeq
        \forall \_. \cps{\Sigma} \rarr ((x: \cps{T}) \rarr \cps{C_1}) \rarr \cps{C_2} \\
    \cps{\tycomp{\Sigma}{T}{\square}} &\defeq
        \forall \alpha. \cps{\Sigma} \rarr (\cps{T} \rarr \alpha) \rarr \alpha \\
    \cps{\{ \repi{\op_i : \forall \rep{X_i: \rep{B}_i}. F_i} \}} &\defeq
        \{ \repi{\op_i : \forall \rep{X_i: \rep{B}_i}. \cps{F_i}^\mathcal{F}} \} \\
    \cps{(x: T_1) \rarr ((y: T_2) \rarr C_1) \rarr C_2}^\mathcal{F} &\defeq
        (x: \cps{T_1}) \rarr \cps{((y: T_2) \rarr C_1)} \rarr \cps{C_2}
\end{align}
\begin{align}
    \cps{\emptyset} &\defeq \emptyset \\
    \cps{\Gamma, x: T} &\defeq \cps{\Gamma}, x: \cps{T} \\
    \cps{\Gamma, X:\rep{B}} &\defeq \cps{\Gamma}, X:\rep{B}
\end{align}

\section{Proof of dynamic semantics preservation of the CPS transformation}

Regarding dynamic semantics,
we identify values and computations modulo types and predicates
since they are irrelevant to the dynamic semantics.
That is, the following equations hold, for example:
\begin{align}
    \lambda x: \tau_1. c &= \lambda x: \tau_2. c \\
    c[\tau/\alpha] &= c \\
    c[A/X] &= c
\end{align}
Also, We often omit type annotations when they are unnecessary.

Moreover, We also identify values and computations
modulo $\beta$ equivalence of the (static) meta language
(this is admissible because the meta language is pure).
Formally, we define a relation $\equiv_\beta$
as the smallest congruence relation over expressions in the target language
that satisfies the following equations:
\begin{align}
    (\stlambda x: \tau. c) \stapp v &\equiv_\beta c[v/x] \\
    (\stLambda \alpha. c) \stapp \tau &\equiv_\beta c[\tau/\alpha]
\end{align}
and we admit the $\equiv_\beta$-equivalence.

\newcommand{\stappBHK}{\stapp \tau \stapp v_h \stapp v_k}

\begin{assumption} \label{asm:cps:prim-dyn}
    \quad
    \begin{itemize}
        \item $\mathit{cps}(p)~\cps{v} \stappBHK \eval^* \cps{\zeta(p, v)} \stappBHK$
        \item If $\zeta(p, v)$ is undefined, then $\mathit{cps}(p)~\cps{v}$ gets stuck.
        \item $p = \exptrue \iff \mathit{cps}(p) = \exptrue$
        \item $p = \expfalse \iff \mathit{cps}(p) = \expfalse$
    \end{itemize}
\end{assumption}

\begin{lemma}[CPS transformation is homomorphic for substitution] \label{lem:cps:homo-subst-term}
    \quad
    \begin{itemize}
        \item $\cps{v[v_0/x]} = \cps{v}[\cps{v_0}/x]$
        \item $\cps{c[v_0/x]} = \cps{c}[\cps{v_0}/x]$
    \end{itemize}
\end{lemma}
\begin{proof}
    By simultaneous induction on the structure of $v$ and $c$.
\end{proof}

\begin{lemma}[Evaluation with pure evaluation context] \label{lem:cps:pure-eval-ctx}
    \begin{align}
        \cps{K[\op~v]} \stappBHK \eval^*
        v_h\#\op~\rep{A}~\cps{v}~(\lambda y. \cps{K[\expret{y}]} \stappBHK)~.
    \end{align}
\end{lemma}
\begin{proof}
    By induction on the structure of $K$.
    \begin{description}
        \item[{Case $K = [\ ]$:}]
            \begin{align}
                \text{LHS} &= \cps{\op~v} \stappBHK \\
                &= (\stLambda \alpha. \stlambda h. \stlambda k.
                    h\#\op~\rep{A}~\cps{v}~(\lambda y. k~y)) \stappBHK \\
                &\eval^* v_h\#\op~\rep{A}~\cps{v}~(\lambda y. v_k~y) \\
                &\equiv_\beta v_h\#\op~\rep{A}~\cps{v}~(\lambda y. (\stLambda \alpha. \stlambda h. \stlambda k. k~y) \stappBHK) \\
                &= v_h\#\op~\rep{A}~\cps{v}~(\lambda y. \cps{\expret{y}} \stappBHK) \\
                &= \text{RHS}
            \end{align}
        \item[Case $K = \explet{x}{K_1}{c_2}$:]
            \begingroup
            \allowdisplaybreaks
            \begin{align}
                \text{LHS} &= \cps{\explet{x}{K_1[\op~v]}{c_2}} \stappBHK \\
                &= (\stLambda \alpha. \stlambda h. \stlambda k.
                    \cps{K_1[\op~v]} \stapp \tau \stapp h \stapp (\lambda x. \cps{c_2} \stapp \tau \stapp h \stapp k))
                    \stappBHK \\
                &\eval^* \cps{K_1[\op~v]} \stapp \tau \stapp v_h \stapp (\lambda x. \cps{c_2} \stappBHK) \\
                &\text{(by the IH)} \\
                &\eval^* v_h\#\op~\rep{A}~\cps{v}
                    ~(\lambda y. \cps{K_1[\expret{y}]} \stapp \tau \stapp v_h \stapp (\lambda x. \cps{c_2} \stappBHK)) \\
                &\equiv_\beta v_h\#\op~\rep{A}~\cps{v}
                    ~(\lambda y. (\stLambda \alpha. \stlambda h. \stlambda k.
                    \cps{K_1[\expret{y}]} \stapp \tau \stapp h \stapp (\lambda x. \cps{c_2} \stapp \tau \stapp h \stapp k)
                    ) \stappBHK) \\
                &= v_h\#\op~\rep{A}~\cps{v}~(\lambda y. \cps{\explet{x}{K_1[\expret{y}]}{c_2}} \stappBHK) \\
                &= v_h\#\op~\rep{A}~\cps{v}~(\lambda y. \cps{K[\expret{y}]} \stappBHK) \\
                &= \text{RHS}
            \end{align}
            \endgroup
    \end{description}
\end{proof}

\begin{lemma}[One-step simulation] \label{lem:cps:sim-onestep}
    If $c \eval c'$, then
    $\cps{c} \stappBHK \eval^* \cps{c'} \stappBHK$~.
\end{lemma}
\begin{proof}
    By induction on the derivation of $c \eval c'$.
    In the following, we implicitly use Lemma~\ref{lem:cps:homo-subst-term} and
    the equality $c[\tau/\alpha] = c$ and $c[A/X] = c$
    (note that we identify computations modulo types and predicates regarding dynamic semantics).
    \begin{description}
        \item[Case \rulename{E-Let}:]
            \begin{align}
                \text{LHS} &= \cps{\explet{x}{c_1}{c_2}} \stappBHK \\
                &= (\stLambda \alpha. \stlambda h. \stlambda k.
                    \cps{c_1} \stapp \tau \stapp h \stapp (\lambda x. \cps{c_2} \stapp \tau \stapp h \stapp k))
                    \stappBHK \\
                &\eval^* \cps{c_1} \stapp \tau \stapp v_h \stapp (\lambda x. \cps{c_2} \stapp \tau \stapp v_h \stapp v_k) \\
                &\text{(by the IH)} \\
                &\eval^* \cps{c_1'} \stapp \tau \stapp v_h \stapp (\lambda x. \cps{c_2} \stapp \tau \stapp v_h \stapp v_k) \\
                &\equiv_\beta (\stLambda \alpha. \stlambda h. \stlambda k.
                    \cps{c_1'} \stapp \tau \stapp h \stapp (\lambda x. \cps{c_2} \stapp \tau \stapp h \stapp k))
                    \stappBHK \\
                &= \cps{\explet{x}{c_1'}{c_2}} \stappBHK \\
                &= \text{RHS}
            \end{align}
        \item[Case \rulename{E-LetRet}:]
            First, w.l.o.g., we can assume that $x \notin \fv(v_h) \cup \fv(v_k)$. Then,
            \begin{align}
                \text{LHS} &= \cps{\explet{x}{\expret{v}}{c_2}} \stappBHK \\
                &= (\stLambda \alpha. \stlambda h. \stlambda k.
                    \cps{\expret{v}} \stapp \tau \stapp h \stapp (\lambda x. \cps{c_2} \stapp \tau \stapp h \stapp k))
                    \stappBHK \\
                &\eval^* \cps{\expret{v}} \stapp \tau \stapp v_h \stapp (\lambda x. \cps{c_2} \stapp \tau \stapp v_h \stapp v_k) \\
                &= (\stLambda \alpha. \stlambda h. \stlambda k. k~\cps{v})
                    \stapp \tau \stapp v_h \stapp (\lambda x. \cps{c_2} \stapp \tau \stapp v_h \stapp v_k) \\
                &\eval^* (\lambda x. \cps{c_2} \stapp \tau \stapp v_h \stapp v_k)~\cps{v} \\
                &\eval (\cps{c_2} \stapp \tau \stapp v_h \stapp v_k)[\cps{v}/x] \\
                &= (\cps{c_2} \stapp \tau \stapp v_h \stapp v_k)[\cps{v}/x] \\
                &= \cps{c_2}[\cps{v}/x] \stapp \tau \stapp v_h \stapp v_k \\
                &= \cps{c_2[v/x]} \stapp \tau \stapp v_h \stapp v_k \\
                &= \text{RHS}
            \end{align}
        \item[Case \rulename{E-IfT}:]
            \begin{align}
                \text{LHS} &= \cps{(\expif{\exptrue}{c_1}{c_2})^C} \stappBHK \\
                &= (\expif{\exptrue}{\cps{c_1}}{\cps{c_2}} : \cps{C}) \stappBHK \\
                &\eval^* \cps{c_1} \stappBHK \\
                &= \text{RHS}
            \end{align}
        \item[Case \rulename{E-IfF}:] similar.
        \item[Case \rulename{E-App}:]
            \begin{align}
                \text{LHS} &= \cps{(\exprec{f}{x}{c})~v} \stappBHK \\
                &= (\exprec{f}{x}{\cps{c}})~\cps{v} \stappBHK \\
                &\eval \cps{c}[\exprec{f}{x}{\cps{c}}/f, \cps{v}/x] \stappBHK \\
                &= \cps{c[\exprec{f}{x}{c}/f, v/x]} \stappBHK \\
                &= \text{RHS}
            \end{align}
        \item[Case \rulename{E-Prim}:] By Assumption~\ref{asm:cps:prim-dyn}.
        \item[Case \rulename{E-Hndl}:]
            \begin{align}
                \text{LHS} &= \cps{\expwith{h}{c}} \stappBHK \\
                &= (\cps{c} \stapp \tau \stapp \cps{h^{\mathit{ops}}} \stapp \cps{h^{\mathit{ret}}})
                    \stappBHK \\
                &\text{(by the IH)} \\
                &= (\cps{c'} \stapp \tau \stapp \cps{h^{\mathit{ops}}} \stapp \cps{h^{\mathit{ret}}})
                    \stappBHK \\
                &= \cps{\expwith{h}{c'}} \stappBHK \\
                &= \text{RHS}
            \end{align}
        \item[Case \rulename{E-HndlRet}:]
            \begin{align}
                \text{LHS} &= \cps{\expwith{h}{\expret{v}}} \stappBHK \\
                &= (\cps{\expret{v}} \stapp \tau \stapp \cps{h^{\mathit{ops}}} \stapp \cps{h^{\mathit{ret}}})
                    \stappBHK \\
                &= (((\stLambda \alpha. \stlambda h. \stlambda k. k~\cps{v}))
                    \stapp \tau \stapp \cps{h^{\mathit{ops}}} \stapp \cps{h^{\mathit{ret}}})
                    \stappBHK \\
                &\eval^* (\cps{h^{\mathit{ret}}}~\cps{v}) \stappBHK \\
                &= ((\lambda x_r. \cps{c_r})~\cps{v}) \stappBHK \\
                &= \cps{c_r}[\cps{v}/x_r] \stappBHK \\
                &= \cps{c_r[v/x_r]} \stappBHK \\
                &= \text{RHS}
            \end{align}
        \item[Case \rulename{E-HndlOp}:]
            \begin{align}
                \text{LHS} &= \cps{\expwith{h}{K[\op_i~v]}} \stappBHK \\
                &= (\cps{K[\op_i~v]} \stapp \tau \stapp \cps{h^{\mathit{ops}}} \stapp \cps{h^{\mathit{ret}}})
                    \stappBHK \\
                &\text{(by Lemma~\ref{lem:cps:pure-eval-ctx})} \\
                &\eval^* \cps{h^{\mathit{ops}}}\#\op_i~\rep{A}~\cps{v}
                    ~(\lambda y. \cps{K[\expret{y}]} \stapp \tau \stapp \cps{h^{\mathit{ops}}} \stapp \cps{h^{\mathit{ret}}}) \\
                &= \cps{h^{\mathit{ops}}}\#\op_i~\rep{A}~\cps{v}~(\lambda y. \cps{\expwith{h}{K[\expret{y}]}}) \\
                &\eval (\Lambda \rep{X_i}. \lambda x_i. \lambda k_i. \cps{c_i})
                    ~\rep{A}~\cps{v}~(\lambda y. \cps{\expwith{h}{K[\expret{y}]}}) \\
                &\eval^* \cps{c_i}[\cps{v}/x_i][\lambda y. \cps{\expwith{h}{K[\expret{y}]}}/k_i] \\
                &= \cps{c_i[v/x_i][\lambda y. \expwith{h}{K[\expret{y}]}/k_i]} \\
                &= \text{RHS}
            \end{align}
    \end{description}
\end{proof}

\newcommand{\stappTop}{\stapp \tau \stapp \{\} \stapp (\lambda x: \tau. x)}

\begin{theorem}[Forward (multi-step) simulation] \label{thm:cps:sim-forward}
    If $c \eval^* \expret{v}$, then
    $\cps{c} \stappTop \eval^+ \cps{v}$~.
\end{theorem}
\begin{proof}
    By applying Lemma~\ref{lem:cps:sim-onestep} repeatedly, we have
    \begin{align}
        \cps{c} \stappTop \eval^* \cps{\expret{v}} \stappTop~.
    \end{align}
    Then,
    \begin{align}
        &\cps{\expret{v}} \stappTop \\
        &= (\stLambda \alpha. \stlambda h. \stlambda k. k~\cps{v}) \stappTop \\
        &\eval^* (\lambda x. x)~\cps{v} \\
        &\eval \cps{v}
    \end{align}
    and therefore we have the conclusion.
\end{proof}

\begin{definition}
    We define evaluation contexts $E$ as follows:
    \begin{align}
        E ::= [\ ] \mid \explet{x}{E}{c} \mid \expwith{h}{E}
    \end{align}
\end{definition}

\newcommand{\bop}{\mathit{bop}}

\begin{definition}
    We define a function $\bop$ as follows:
    \begin{align}
        \bop([\ ]) &\defeq \emptyset \\
        \bop(\explet{x}{E}{c}) &\defeq \bop(E) \\
        \bop(\expwith{h}{E}) &\defeq \dom(h) \cup \bop(E)
    \end{align}
    That is, $\bop(E)$ is a set of operations that are handled by a handler in $E$.
\end{definition}

We say \emph{$c$ is stuck} if $c$ is irreducible and $c \ne \expret{v}$.
We proceed the proof of the backward simulation following \citet{Danvy03}.

\newcommand{\hopsz}{h_0^{\mathtt{ops}}}
\newcommand{\stappBZK}{\stapp \tau \stapp \cps{\hopsz} \stapp v_k}

\begin{lemma}[Preservation of the specific forms of stuck computations] \label{lem:cps:sim-stuck-cases}
    \quad
    \begin{enumerate}
        \item If $c = E[\expif{v}{c_1}{c_2}]$
            where $v$ is not $\exptrue$ nor $\expfalse$,
            then $\cps{c} \stappBHK$ gets stuck.
        \item If $c = E[v_1~v_2]$
            where $v_1$ is not $\exprec{f}{x}{c}$ nor $p$ such that $\zeta(p, v_2)$ is defined,
            then $\cps{c} \stappBHK$ gets stuck.
        \item Let $h_0$ be a handler.
            If $c = E[\op~v]$
            where $\op \notin \bop(E) \cup \dom(h_0)$,
            then $\cps{c} \stapp \tau \stapp \cps{h_0^{\mathtt{ops}}} \stapp v_k$ gets stuck.
    \end{enumerate}
\end{lemma}
\begin{proof}
    \quad
    \begin{enumerate}
        \item \label{enum:stuck-1} By induction on the structure of $E$.
            \begin{description}
                \item[{Case $E = [\ ]$:}]
                    \begin{align}
                        \cps{c} \stappBHK &= \cps{\expif{v}{c_1}{c_2}} \stappBHK \\
                        &= \expif{\cps{v}}{\cps{c_1}}{\cps{c_2}} \stappBHK \\
                    \end{align}
                    From Assumption~\ref{asm:cps:prim-dyn},
                    $\cps{v}$ is neither $\exptrue$ nor $\expfalse$.
                    Therefore, there is no applicable evaluation rule,
                    and hence this computation is stuck.
                \item[Case $E = \explet{x}{E_1}{c}$:]
                    \begin{align}
                        \cps{c} \stappBHK &= \cps{\explet{x}{E_1[\expif{v}{c_1}{c_2}]}{c}} \stappBHK \\
                        &= (\stLambda \alpha. \stlambda h. \stlambda k.
                            \cps{E_1[\expif{v}{c_1}{c_2}]} \stapp \tau \stapp h \stapp (\lambda x. \cps{c} \stapp \tau \stapp h \stapp k)) \stappBHK \hspace*{-2ex} \\
                        &\eval^* \cps{E_1[\expif{v}{c_1}{c_2}]} \stapp \tau \stapp v_h \stapp (\lambda x. \cps{c} \stappBHK)
                    \end{align}
                    By the IH, this computation gets stuck.
                \item[Case $E = \expwith{h}{E_1}$:]
                    \begin{align}
                        \cps{c} \stappBHK &= \cps{\expwith{h}{E_1[\expif{v}{c_1}{c_2}]}} \stappBHK \\
                        &= (\cps{E_1[\expif{v}{c_1}{c_2}]} \stapp \tau \stapp \cps{h^{\mathit{ops}}} \stapp \cps{h^{\mathit{ret}}}) \stappBHK
                    \end{align}
                    By the IH, this computation gets stuck.
            \end{description}
        \item Similar to the case \ref{enum:stuck-1}.
        \item By induction on the structure of $E$.
            \begin{description}
                \item[{Case $E = [\ ]$:}]
                    \begin{align}
                        \cps{c} \stappBZK &= \cps{\op~v} \stappBZK \\
                        &= (\stLambda \alpha. \stlambda h. \stlambda k.
                            h\#\op~\rep{A}~\cps{v}~(\lambda y. \cps{\expret{y}} \stapp \tau \stapp h \stapp k))
                            \stappBZK \\
                        &\eval^* \hopsz\#\op~\rep{A}~\cps{v}~(\lambda y. \cps{\expret{y}} \stappBZK)
                    \end{align}
                    Here, $\hopsz$ does not have a field with $\op$ since $\op \notin \dom(h_0)$.
                    Therefore, there is no applicable evaluation rule,
                    and hence this computation is stuck.
                \item[Case $E = \explet{x}{E_1}{c}$:]
                    \begin{align}
                        \cps{c} \stappBZK &= \cps{\explet{x}{E_1[\op~v]}{c}} \stappBZK \\
                        &= (\stLambda \alpha. \stlambda h. \stlambda k.
                            \cps{E_1[\op~v]} \stapp \tau \stapp h \stapp (\lambda x. \cps{c} \stapp \tau \stapp h \stapp k)) \stappBZK \hspace*{-2ex} \\
                        &\eval^* \cps{E_1[\op~v]} \stapp \tau \stapp \cps{\hopsz} \stapp (\lambda x. \cps{c} \stappBZK)
                    \end{align}
                    Since $\op \notin \bop(E) \cup \dom(h_0)$ and $\bop(E) = \bop(\explet{x}{E_1}{c}) = \bop(E_1)$,
                    it holds that $\op \notin \bop(E_1) \cup \dom(h_0)$.
                    Then, by the IH, this computation gets stuck.
                \item[Case $E = \expwith{h}{E_1}$:]
                    \begin{align}
                        \cps{c} \stappBZK &= \cps{\expwith{h}{E_1[\op~v]}} \stappBZK \\
                        &= (\cps{E_1[\op~v]} \stapp \tau \stapp \cps{h^{\mathit{ops}}} \stapp \cps{h^{\mathit{ret}}}) \stappBZK
                    \end{align}
                    Here, $\op \notin \bop(E) = \bop(\expwith{h}{E_1}) = \bop(E_1) \cup \dom(h)$.
                    Therefore, by the IH, this computation gets stuck.
            \end{description}
    \end{enumerate}
\end{proof}

\begin{lemma}[Preservation of stuck computations] \label{lem:cps:sim-stuck}
    If $c$ is a stuck computation,
    then $\cps{c} \stappTop$ also gets stuck.
\end{lemma}
\begin{proof}
    A stuck computation $c$ is either:
    \begin{itemize}
        \item $E[\expif{v}{c_1}{c_2}]$ where $v$ is not $\exptrue$ nor $\expfalse$,
        \item $E[v_1~v_2]$ where $v_1$ is not $\exprec{f}{x}{c}$ nor $p$ such that $\zeta(p, v_2)$ is defined, or
        \item $E[\op~v]$ where $\op \notin \bop(E)$.
    \end{itemize}
    Therefore, it is immediate from Lemma~\ref{lem:cps:sim-stuck-cases}.
\end{proof}

\begin{theorem}[Backward simulation] \label{thm:cps:sim-backward}
    If $\cps{c} \stappTop \eval^+ v'$, then
    $c \eval^* \expret{v}$ and $\cps{v} = v'$~.
\end{theorem}
\begin{proof}
    We show this theorem by proving its contraposition:
    If ``$c \eval^* \expret{v}$ and $\cps{v} = v'$'' does not hold, then
    $\cps{c} \stappTop \eval^+ v'$
    also does not hold.
    We can divide the situation into two cases:
    \begin{description}
        \item[Case that $c \eval^* \expret{v}$ does not hold:]
            There are two possibilities where $c$ does not evaluate to a value-return.
            \begin{description}
                \item[Case that $c$ diverges:]
                    Since $c$ diverges, for all natural numbers $n$,
                    there exists a sequence
                    \begin{align}
                        c \eval c_1 \eval \cdots \eval c_n~.
                    \end{align}
                    Then, by Lemma~\ref{lem:cps:sim-onestep},
                    we have a sequence
                    \begin{align}
                        \cps{c} \stappTop \eval^+ \cps{c_1} \stappTop \eval^+ \cdots \eval^+ \cps{c_n} \stappTop
                    \end{align}
                    for all $n$.
                    The length of the sequence is at least $n$,
                    and therefore $\cps{c} \stappTop$ has evaluation sequences of arbitrary length,
                    which means it cannot be evaluated to a value.
                \item[Case that $c \eval^* c'$ and $c'$ is stuck:]
                    By applying Lemma~\ref{lem:cps:sim-onestep} repeatedly,
                    we have
                    \begin{align}
                        \cps{c} \stappTop \eval^* \cps{c'} \stappTop~.
                    \end{align}
                    Also, by Lemma~\ref{lem:cps:sim-stuck},
                    it holds that $\cps{c'} \stappTop$ gets stuck.
                    Therefore, $\cps{c} \stappTop$ cannot be evaluated to a value.
            \end{description}
        \item[Case that $c \eval^* \expret{v}$ holds but $\cps{v} = v'$ does not:]
            By Theorem~\ref{thm:cps:sim-forward}, we have
            \begin{align}
                \cps{c} \stappTop \eval^+ \cps{v}~.
            \end{align}
            Then, from the premise $\cps{v} \ne v'$
            and the fact that the evaluation of the target language is deterministic,
            it cannot be the case that $\cps{c} \stappTop \eval^+ v'$~.
    \end{description}
\end{proof}

\begin{corollary}[Simulation] \label{cor:cps:sim}
    If $c \eval^* \expret{v}$, then
    $\cps{c} \stappTop \eval^+ \cps{v}$~.
    Also, if $\cps{c} \stappTop \eval^+ v'$, then
    $c \eval^* \expret{v}$ and $\cps{v} = v'$~.
\end{corollary}
\begin{proof}
    Immediate from Theorem~\ref{thm:cps:sim-forward} and \ref{thm:cps:sim-backward}.
\end{proof}

\section{Proof of type preservation of the CPS transformation}

In the following, we consider static expressions and dynamic ones as identical
since the distinction is irrelevant to the discussion on the type preservation.
In other words, we write $\cps{c} \stappBHK$ as $\cps{c}~\tau~v_h~v_k$ below, for example.

\subsection{Basic properties for the target language of the CPS transformation}

\begin{assumption} \label{asm:cps:formula} \quad
    \begin{itemize}
        \item If $\jdwf{}{\Gamma}$ and $\dom(\Gamma) \supseteq \fv(\phi)$, then $\jdwf{\Gamma}{\phi}$.
        \item If $\jdwf{\Gamma}{\phi}$, then $\jdwf{}{\Gamma}$.
        \item If $\jdwf{\Gamma}{\phi}$, then $\Gamma \vDash \phi \Rarr \phi$.
        \item If $\Gamma \vDash \phi_1 \Rarr \phi_2$ and $\Gamma \vDash \phi_2 \Rarr \phi_3$, then $\Gamma \vDash \phi_1 \Rarr \phi_3$.
        \item If $\jdty{\Gamma}{v}{\tau}$ and $\jdty{\Gamma, x: \tau, \Gamma'}{A}{\rep{B}}$, then $\jdty{\Gamma, \Gamma'[v/x]}{A[v/x]}{\rep{B}}$.
        \item If $\jdty{\Gamma}{v}{\tau}$ and $\jdwf{\Gamma, x: \tau, \Gamma'}{\phi}$, then $\jdwf{\Gamma, \Gamma'[v/x]}{\phi[v/x]}$.
        \item If $\jdty{\Gamma}{v}{\tau}$ and $\valid{\Gamma, x: \tau, \Gamma'}{\phi}$, then $\valid{\Gamma, \Gamma'[v/x]}{\phi[v/x]}$.
        \item If $\jdty{\Gamma}{A}{\rep{B}}$ and $\jdty{\Gamma, X: \rep{B}, \Gamma'}{A'}{\rep{B'}}$, then $\jdty{\Gamma, \Gamma'[A/X]}{A'[A/X]}{\rep{B'}}$.
        \item If $\jdty{\Gamma}{A}{\rep{B}}$ and $\jdwf{\Gamma, X: \rep{B}, \Gamma'}{\phi}$, then $\jdwf{\Gamma, \Gamma'[A/X]}{\phi[A/X]}$.
        \item If $\jdty{\Gamma}{A}{\rep{B}}$ and $\valid{\Gamma, X: \rep{B}, \Gamma'}{\phi}$, then $\valid{\Gamma, \Gamma'[A/X]}{\phi[A/X]}$.
        \item If $\jdwf{}{\Gamma_1, \Gamma_2, \Gamma_3}$ and $\jdwf{\Gamma_1, \Gamma_2}{\phi}$, then $\jdwf{\Gamma_1, \Gamma_2, \Gamma_3}{\phi}$.
        \item If $\jdwf{}{\Gamma_1, \Gamma_2, \Gamma_3}$ and $\jdty{\Gamma_1, \Gamma_2}{A}{\rep{B}}$, then $\jdty{\Gamma_1, \Gamma_2, \Gamma_3}{A}{\rep{B}}$.
        \item If $\jdwf{}{\Gamma_1, \Gamma_2, \Gamma_3}$ and $\Gamma_1, \Gamma_2 \vDash \phi$, then $\Gamma_1, \Gamma_2, \Gamma_3 \vDash \phi$.
        \item If $\jdsub{\Gamma}{\tau_1}{\tau_2}$, $\jdwf{}{\Gamma, x:\tau_1, \Gamma'}$ and $\jdty{\Gamma, x:\tau_2, \Gamma'}{A}{\rep{B}}$, then $\jdty{\Gamma, x:\tau_1, \Gamma'}{A}{\rep{B}}$.
        \item If $\jdsub{\Gamma}{\tau_1}{\tau_2}$, $\jdwf{}{\Gamma, x:\tau_1, \Gamma'}$ and $\jdwf{\Gamma, x:\tau_2, \Gamma'}{\phi}$, then $\jdwf{\Gamma, x:\tau_1, \Gamma'}{\phi}$.
        \item If $\jdsub{\Gamma}{\tau_1}{\tau_2}$ and $\valid{\Gamma, x:\tau_2, \Gamma'}{\phi}$, then $\valid{\Gamma, x:\tau_1, \Gamma'}{\phi}$.
        \item If $x \notin \fv(\Gamma', \phi)$ and $\jdwf{\Gamma, x: \tau_0, \Gamma'}{\phi}$, then $\jdwf{\Gamma, \Gamma'}{\phi}$.
        \item If $x \notin \fv(\Gamma', A)$ and $\jdty{\Gamma, x: \tau_0, \Gamma'}{A}{\rep{B}}$, then $\jdty{\Gamma, \Gamma'}{A}{\rep{B}}$.
        \item If $\jdwf{\Gamma, x: \tau, \Gamma'}{\phi}$ and $\tau$ is not a refinement type, then $x \notin \fv(\Gamma', \phi)$.
        \item If $\jdty{\Gamma, x: \tau, \Gamma'}{A}{\rep{B}}$ and $\tau$ is not a refinement type, then $x \notin \fv(\Gamma', A)$.
        \item If $\Gamma, x: \tau, \Gamma' \vDash \phi$ and $\tau$ is not a refinement type, then $x \notin \fv(\Gamma', \phi)$ and $\Gamma, \Gamma' \vDash \phi$.
        \item If $\alpha \notin \fv(\Gamma', \phi)$ and $\jdwf{\Gamma, \alpha, \Gamma'}{\phi}$, then $\jdwf{\Gamma, \Gamma'}{\phi}$.
        \item If $\alpha \notin \fv(\Gamma', \phi)$ and $\Gamma, \alpha, \Gamma' \vDash \phi$, then $\Gamma, \Gamma' \vDash \phi$.
    \end{itemize}
\end{assumption}

\begin{assumption} \label{asm:cps:prim} \quad
    \begin{itemize}
        \item $\jdwf{}{\tycps(p)}$ for all $p$.
    \end{itemize}
\end{assumption}

\begin{lemma}[Weakening] \label{lem:cps:weaken} \quad
    Assume that $\jdwf{}{\Gamma_1, \Gamma_2, \Gamma_3}$.
    \begin{itemize}
        \item If $\jdwf{\Gamma_1, \Gamma_3}{\tau}$, then $\jdwf{\Gamma_1, \Gamma_2, \Gamma_3}{\tau}$.
        \item If $\jdty{\Gamma_1, \Gamma_3}{c}{\tau}$, then $\jdty{\Gamma_1, \Gamma_2, \Gamma_3}{c}{\tau}$.
        \item If $\jdsub{\Gamma_1, \Gamma_3}{\tau_1}{\tau_2}$, then $\jdsub{\Gamma_1, \Gamma_2, \Gamma_3}{\tau_1}{\tau_2}$.
    \end{itemize}
\end{lemma}
\begin{proof}
    By induction on the derivation. Assumption \ref{asm:cps:formula} is used.
\end{proof}

\begin{lemma}[Narrowing] \label{lem:cps:narrow}
    Assume that $\jdsub{\Gamma}{\tau_1}{\tau_2}$.
    \begin{itemize}
        \item If $\jdwf{}{\Gamma, x: \tau_1, \Gamma'}$ and $\jdwf{\Gamma, x: \tau_2, \Gamma'}{\tau}$,
            then $\jdwf{\Gamma, x: \tau_1, \Gamma'}{\tau}$.
        \item If $\jdwf{}{\Gamma, x: \tau_1, \Gamma'}$ and $\jdty{\Gamma, x: \tau_2, \Gamma'}{c}{\tau}$,
            then $\jdty{\Gamma, x: \tau_1, \Gamma'}{c}{\tau}$.
        \item If $\jdsub{\Gamma, x: \tau_2, \Gamma'}{\tau_1}{\tau_2}$,
            then $\jdsub{\Gamma, x: \tau_1, \Gamma'}{\tau_1}{\tau_2}$.
    \end{itemize}
\end{lemma}
\begin{proof}
    By induction on the derivation. Assumption \ref{asm:cps:formula} is used.
\end{proof}

\begin{lemma}[Remove unused type bindings] \label{lem:cps:rm-unused} \quad
    \begin{itemize}
        \item If $x \notin \fv(\Gamma')$ and $\jdwf{}{\Gamma, x: \tau_0, \Gamma'}$,
            then $\jdwf{}{\Gamma, \Gamma'}$.
        \item If $x \notin \fv(\Gamma', \tau)$ and $\jdwf{\Gamma, x: \tau_0, \Gamma'}{\tau}$,
            then $\jdwf{\Gamma, \Gamma'}{\tau}$.
    \end{itemize}
\end{lemma}
\begin{proof}
    By induction on the derivation.
    The case for \rulename{WTc-Rfn} uses Assumption \ref{asm:cps:formula}.
\end{proof}

\begin{lemma}[Variables of non-refinement types do not apper in types] \label{lem:cps:notin-nonrfn}
    Assume that $\tau_0$ is not a refinement type.
    \begin{itemize}
        \item If $\jdwf{}{\Gamma, x: \tau_0, \Gamma'}$, then  $x \notin \fv(\Gamma')$.
        \item If $\jdwf{\Gamma, x: \tau_0, \Gamma'}{\tau}$, then $x \notin \fv(\Gamma', \tau)$.
    \end{itemize}
\end{lemma}
\begin{proof}
    By induction on the derivation.
    The case for \rulename{WTc-Rfn} uses Assumption \ref{asm:cps:formula}.
\end{proof}

\begin{lemma}[Remove non-refinement type bindings] \label{lem:cps:rm-nonrfn}
    Assume that $\tau_0$ is not a refinement type.
    \begin{enumerate}
        \item If $\jdwf{}{\Gamma, x: \tau_0, \Gamma'}$,
            then $\jdwf{}{\Gamma, \Gamma'}$.
        \item If $\jdwf{\Gamma, x: \tau_0, \Gamma'}{\tau}$,
            then $\jdwf{\Gamma, \Gamma'}{\tau}$.
        \item If $x \notin \fv(c)$ and $\jdty{\Gamma, x: \tau_0, \Gamma'}{c}{\tau}$,
            then $\jdty{\Gamma, \Gamma'}{c}{\tau}$.
        \item If $\jdsub{\Gamma, x: \tau_0, \Gamma'}{\tau_1}{\tau_2}$,
            then $\jdsub{\Gamma, \Gamma'}{\tau_1}{\tau_2}$.
    \end{enumerate}
\end{lemma}
\begin{proof} \quad
    \begin{enumit}
        \item Immediate by Lemma \ref{lem:cps:notin-nonrfn} and \ref{lem:cps:rm-unused}.
        \item Immediate by Lemma \ref{lem:cps:notin-nonrfn} and \ref{lem:cps:rm-unused}.
        \item By induction on the derivation.
            The case for \rulename{Tc-PApp} uses Assumption \ref{asm:cps:formula}.
        \item By induction on the derivation.
            The case for \rulename{Sc-Rfn} uses Assumption \ref{asm:cps:formula}.
        \end{enumit}
\end{proof}

\begin{lemma}[Remove unused type variable bindings] \label{lem:cps:rm-unused-tvar} \quad
    \begin{itemize}
        \item If $\alpha \notin \fv(\Gamma')$
            and $\jdwf{}{\Gamma, \alpha, \Gamma'}$,
            then $\jdwf{}{\Gamma, \Gamma'}$.
        \item If $\alpha \notin \fv(\Gamma', \tau)$
            and $\jdwf{\Gamma, \alpha, \Gamma'}{\tau}$,
            then $\jdwf{\Gamma, \Gamma'}{\tau}$.
        \item If $\alpha \notin \fv(\Gamma', \tau_1, \tau_2)$
            and $\jdsub{\Gamma, \alpha, \Gamma'}{\tau_1}{\tau_2}$,
            then $\jdsub{\Gamma, \Gamma'}{\tau_1}{\tau_2}$.
    \end{itemize}
\end{lemma}
\begin{proof}
    By induction on the derivation.
    The case for \rulename{WTSc-Rfn} and \rulename{Sc-Rfn} uses Assumption \ref{asm:cps:formula}.
\end{proof}

\begin{lemma}[Substitution] \label{lem:cps:subst}
    Assume that $\jdty{\Gamma}{v}{\tau_0}$.
    \begin{itemize}
        \item If $\jdwf{}{\Gamma, x: \tau_0, \Gamma'}$,
            then $\jdwf{}{\Gamma, \Gamma'[v/x]}$.
        \item If $\jdwf{\Gamma, x: \tau_0, \Gamma'}{\tau}$,
            then $\jdwf{\Gamma, \Gamma'[v/x]}{\tau[v/x]}$.
        \item If $\jdty{\Gamma, x: \tau_0, \Gamma'}{c}{\tau}$,
            then $\jdty{\Gamma, \Gamma'[v/x]}{c[v/x]}{\tau[v/x]}$.
        \item If $\jdsub{\Gamma, x: \tau_0, \Gamma'}{\tau_1}{\tau_2}$,
            then $\jdsub{\Gamma, \Gamma'[v/x]}{\tau_1[v/x]}{\tau_2[v/x]}$.
    \end{itemize}
\end{lemma}
\begin{proof}
    By induction on the derivation. Assumption \ref{asm:cps:formula} is used.
\end{proof}

\begin{lemma}[Predicate substitution] \label{lem:cps:subst-pred}
    Assume that $\jdty{\Gamma}{A}{\rep{B}}$.
    \begin{itemize}
        \item If $\jdwf{}{\Gamma, X: \rep{B}, \Gamma'}$,
            then $\jdwf{}{\Gamma, \Gamma'[A/X]}$.
        \item If $\jdwf{\Gamma, X: \rep{B}, \Gamma'}{\tau}$,
            then $\jdwf{\Gamma, \Gamma'[A/X]}{\tau[A/X]}$.
        \item If $\jdty{\Gamma, X: \rep{B}, \Gamma'}{c}{\tau}$,
            then $\jdty{\Gamma, \Gamma'[A/X]}{c[A/X]}{\tau[A/X]}$.
        \item If $\jdsub{\Gamma, X: \rep{B}, \Gamma'}{\tau_1}{\tau_2}$,
            then $\jdsub{\Gamma, \Gamma'[A/X]}{\tau_1[A/X]}{\tau_2[A/X]}$.
    \end{itemize}
\end{lemma}
\begin{proof}
    By induction on the derivation. Assumption \ref{asm:cps:formula} is used.
\end{proof}

\begin{lemma}[Type substitution] \label{lem:cps:subst-type}
    Assume that $\jdwf{\Gamma}{\tau_0}$.
    \begin{itemize}
        \item If $\jdwf{}{\Gamma, \alpha, \Gamma'}$,
            then $\jdwf{}{\Gamma, \Gamma'[\tau_0/\alpha]}$.
        \item If $\jdwf{\Gamma, \alpha, \Gamma'}{\tau}$,
            then $\jdwf{\Gamma, \Gamma'[\tau_0/\alpha]}{\tau[\tau_0/\alpha]}$.
        \item If $\jdty{\Gamma, \alpha, \Gamma'}{c}{\tau}$,
            then $\jdty{\Gamma, \Gamma'[\tau_0/\alpha]}{c[\tau_0/\alpha]}{\tau[\tau_0/\alpha]}$.
        \item If $\jdsub{\Gamma, \alpha, \Gamma'}{\tau_1}{\tau_2}$,
            then $\jdsub{\Gamma, \Gamma'[\tau_0/\alpha]}{\tau_1[\tau_0/\alpha]}{\tau_2[\tau_0/\alpha]}$.
    \end{itemize}
\end{lemma}
\begin{proof}
    By induction on the derivation. Assumption \ref{asm:cps:formula} is used.
\end{proof}

\begin{lemma}[Well-formedness of typing contexts from that of types] \label{lem:cps:wfg}
    If $\jdwf{\Gamma}{\tau}$, then $\jdwf{}{\Gamma}$.
\end{lemma}
\begin{proof}
    By induction on the derivation.
    The case for \rulename{WTc-Rfn} uses Assumption \ref{asm:cps:formula}.
\end{proof}

\begin{lemma}[Well-formedness of types from typings] \label{lem:cps:wft}
    If $\jdty{\Gamma}{c}{\tau}$, then $\jdwf{\Gamma}{\tau}$.
\end{lemma}
\begin{proof}
    By induction on the derivation.
    \begin{description}
        \item[Case \rulename{Tc-CVar}:] By Assumption \ref{asm:cps:formula}.
        \item[Case \rulename{Tc-Var}:] By Lemma \ref{lem:cps:weaken}.
        \item[Case \rulename{Tc-Prim}:]
            By Assumption \ref{asm:cps:prim} and Lemma \ref{lem:cps:weaken}.
        \item[Case \rulename{Tc-Fun}:]
            By the IH, Lemma \ref{lem:cps:rm-nonrfn}, and \rulename{WTc-Fun}.
        \item[Case \rulename{Tc-App}:]
            By the IH, inversion, and Lemma \ref{lem:cps:subst}.
        \item[Case \rulename{Tc-TAbs}:]
            By the IH and \rulename{WTc-TPoly}.
        \item[Case \rulename{Tc-TApp}:]
            By the IH, inversion, and Lemma \ref{lem:cps:subst-type}.
        \item[Case \rulename{Tc-PAbs}:]
            By the IH and \rulename{WTc-PPoly}.
        \item[Case \rulename{Tc-PApp}:]
            By the IH, inversion, and Lemma \ref{lem:cps:subst-pred}.
        \item[Case \rulename{Tc-If}:]
            By the IH and Lemma \ref{lem:cps:rm-unused}.
        \item[Case \rulename{Tc-Ascr} and \rulename{Tc-Sub}:]
            Immediate.
    \end{description}
\end{proof}

\begin{lemma}[Reflexivity] \label{lem:cps:refl}
    If $\jdwf{\Gamma}{\tau}$, then $\jdsub{\Gamma}{\tau}{\tau}$.
\end{lemma}
\begin{proof}
    By induction on the derivation.
    The case for \rulename{WTc-Rfn} uses Assumption \ref{asm:cps:formula}.
\end{proof}

\begin{lemma}[Transitivity] \label{lem:cps:trans}
    If $\jdsub{\Gamma}{\tau_1}{\tau_2}$ and $\jdsub{\Gamma}{\tau_2}{\tau_3}$, then $\jdsub{\Gamma}{\tau_1}{\tau_3}$.
\end{lemma}
\begin{proof}
    By induction on the structure of $\tau_2$.
    Assumption~\ref{asm:cps:formula}, Lemma~\ref{lem:cps:narrow}, and \ref{lem:cps:weaken} are used.
\end{proof}

\begin{lemma}[Inversion] \label{lem:cps:inv} \quad
    \begin{itemize}
        \item If $\jdty{\Gamma}{x}{\tau}$, then either
            \begin{itemize}
                \item $\jdwf{}{\Gamma}$ and  $\jdsub{\Gamma}{\tyrfn{z}{B}{z = x}}{\tau}$
                    (if $\Gamma(x) = \tyrfn{z}{B}{\phi}$ for some $z, B$ and $\phi$)
                \item $\jdwf{}{\Gamma}$ and $\jdsub{\Gamma}{\Gamma(x)}{\tau}$ (otherwise)
            \end{itemize}
        \item If $\jdty{\Gamma}{p}{\tau}$, then
            $\jdwf{}{\Gamma}$ and  $\jdsub{\Gamma}{\tycps(p)}{\tau}$.
        \item If $\jdty{\Gamma}{\exprec{f:(x:\tau_1) \rarr \tau_2}{x:\tau_1}{c}}{\tau}$, then
            $\jdty{\Gamma, f:(x:\tau_1) \rarr \tau_2, x:\tau_1}{c}{\tau_2}$ and
            $\jdsub{\Gamma}{(x:\tau_1) \rarr \tau_2}{\tau}$.
        \item If $\jdty{\Gamma}{\Lambda \alpha. c}{\tau}$, then
            $\jdty{\Gamma, \alpha}{c}{\tau'}$ and $\jdsub{\Gamma}{\forall \alpha. \tau'}{\tau}$
            for some $\tau'$.
        \item If $\jdty{\Gamma}{\{\repi{\op_i = v_i}\}}{\tau}$, then
            $\bigrepi{\jdty{\Gamma}{v_i}{\tau_i}}$ and $\jdsub{\Gamma}{\{\op_i: \tau_i\}}{\tau}$
            for some $\repi{\tau_i}$.
        \item If $\jdty{\Gamma}{c~v}{\tau}$, then
            $\jdty{\Gamma}{c}{(x:\tau_1) \rarr \tau_2}$, $\jdty{\Gamma}{v}{\tau_1}$
            and $\jdsub{\Gamma}{\tau_2[v/x]}{\tau}$
            for some $x, \tau_1$ and $\tau_2$.
        \item If $\jdty{\Gamma}{c~\rep{A}}{\tau}$, then
            $\jdty{\Gamma}{c}{\forall \rep{X:\rep{B}}. \tau'}$, $\rep{\jdty{\Gamma}{A}{\rep{B}}}$
            and $\jdsub{\Gamma}{\tau'[\rep{A/X}]}{\tau}$
            for some $\rep{X}, \rep{\rep{B}}$ and $\tau'$.
        \item If $\jdty{\Gamma}{c~\tau'}{\tau}$, then
            $\jdty{\Gamma}{c}{\forall \alpha. \tau_1}$, $\jdwf{\Gamma}{\tau'}$
            and $\jdsub{\Gamma}{\tau_1[\tau'/\alpha]}{\tau}$
            for some $\alpha$ and $\tau_1$.
        \item If $\jdty{\Gamma}{v\#\op}{\tau}$, then
            $\jdty{\Gamma}{v}{\{\ldots, \op:\tau, \ldots\}}$.
        \item If $\jdty{\Gamma}{(c : \tau')}{\tau}$, then
            $\jdty{\Gamma}{c}{\tau'}$ and $\jdsub{\Gamma}{\tau'}{\tau}$.
        \item If $\jdty{\Gamma}{\expif{v}{c_1}{c_2}}{\tau}$, then
            $\jdty{\Gamma}{v}{\tyrfn{z}{\tybool}{\phi}}$,
            $\jdty{\Gamma, v = \exptrue}{c_1}{\tau'}$, $\jdty{\Gamma, v = \expfalse}{c_2}{\tau'}$, 
            and $\jdsub{\Gamma}{\tau'}{\tau}$
            for some $z, \phi$ and $\tau'$.
    \end{itemize}
\end{lemma}
\begin{proof}
    By induction on the derivation. Lemma \ref{lem:cps:refl} and \ref{lem:cps:trans} are used.
\end{proof}

\begin{lemma}[Inversion for CPS-transformed computations] \label{lem:cps:inv-c}
    If $\jdty{\Gamma}{\Lambda \alpha. \lambda h:\tau_h. \lambda k:\tau_k. c}{\tau}$
    and neither $\tau_h$ nor $\tau_k$ is a refinement type,
    then there exists some $\tau'$ such that
    \begin{itemize}
        \item $\jdty{\Gamma, \alpha, h:\tau_h, k:\tau_k}{c}{\tau'}$ and
        \item $\jdsub{\Gamma}{\forall \alpha. \tau_h \rarr \tau_k \rarr \tau'}{\tau}$~.
    \end{itemize}
\end{lemma}
\begin{proof}
    By Lemma \ref{lem:cps:inv}, \rulename{Sc-Poly}, \rulename{Sc-Fun}, and Lemma \ref{lem:cps:trans}.
\end{proof}

\begin{lemma}[Inversion for the specific form of application] \label{lem:cps:inv-c-app}
    If $\jdty{\Gamma}{c~\tau_0~v_1~v_2}{\tau}$, then
    there exist some $\tau', \tau_1$, and $\tau_2$ such that
    \begin{itemize}
        \item $\jdty{\Gamma}{c}{\tau'}$,
        \item $\jdty{\Gamma}{v_1}{\tau_1}$, and
        \item $\jdty{\Gamma}{v_2}{\tau_2}$~.
    \end{itemize}
    In addition, if $\jdsub{\Gamma}{\tau_1'}{\tau_1}$ and $\jdsub{\Gamma}{\tau_2'}{\tau_2}$
    for some $\tau_1'$ and $\tau_2'$
    and neither $\tau_1'$ nor $\tau_2'$ is a refinement type,
    then $\jdsub{\Gamma}{\tau'}{\forall \alpha. \tau_1' \rarr \tau_2' \rarr \tau}$
    where $\alpha$ is fresh.
\end{lemma}
\begin{proof}
    The first half is by Lemma \ref{lem:cps:inv}.
    The second half is by Lemma \ref{lem:cps:wft}, \ref{lem:cps:weaken} and \ref{lem:cps:trans}
    with the results of the first half.
\end{proof}

\subsection{Forward type preservation}

\begin{assumption} \label{asm:cps:formula-cps} \quad
    \begin{itemize}
        \item If $\jdwf{\Gamma}{\phi}$, then $\jdwf{\cps{\Gamma}}{\phi}$.
        \item If $\jdty{\Gamma}{A}{\rep{B}}$, then $\jdty{\cps{\Gamma}}{A}{\rep{B}}$.
        \item If $\Gamma \vDash \phi$, then $\cps{\Gamma} \vDash \phi$.
    \end{itemize}
\end{assumption}

\begin{assumption} \label{asm:cps:prim-cps} \quad
    \begin{itemize}
        \item $\cps{\ty(p)} = \tycps(\cps{p})$.
        \item If $\ty(p) = \tyrfn{x}{B}{\phi}$ for some $x, B$ and $\phi$, then $\cps{p} = p$.
    \end{itemize}
\end{assumption}

\begin{lemma}[CPS transformation preserves free variables in types] \label{lem:cps:presv-fv} \quad
    \begin{itemize}
        \item $\fv(\cps{T}) = \fv(T)$.
        \item $\fv(\cps{C}) = \fv(C)$.
        \item $\fv(\cps{\Sigma}) = \fv(\Sigma)$.
    \end{itemize}
\end{lemma}
\begin{proof}
    By simultaneous induction on the structure of types.
\end{proof}

\begin{lemma}[CPS transformation is homomorphic for substitution] \label{lem:cps:homo-subst} \quad
    \begin{itemize}
        \item $\cps{T[v/x]} = \cps{T}[\cps{v}/x]$.
        \item $\cps{C[v/x]} = \cps{C}[\cps{v}/x]$.
        \item $\cps{\Sigma[v/x]} = \cps{\Sigma}[\cps{v}/x]$.
        \item $\cps{T[A/X]} = \cps{T}[A/X]$.
        \item $\cps{C[A/X]} = \cps{C}[A/X]$.
        \item $\cps{\Sigma[A/X]} = \cps{\Sigma}[A/X]$.
    \end{itemize}
\end{lemma}
\begin{proof}
    By simultaneous induction on the structure of types.
    The case for $T = \tyrfn{x}{B}{\phi}$ uses Assumption \ref{asm:cps:prim-cps}.
\end{proof}

\begin{lemma}[CPS transformation preserves well-formedness] \label{lem:cps:presv-wf} \quad
    \begin{itemize}
        \item If $\jdwf{}{\Gamma}$, then $\jdwf{}{\cps{\Gamma}}$.
        \item If $\jdwf{\Gamma}{T}$, then $\jdwf{\cps{\Gamma}}{\cps{T}}$.
        \item If $\jdwf{\Gamma}{C}$, then $\jdwf{\cps{\Gamma}}{\cps{C}}$.
        \item If $\jdwf{\Gamma}{\Sigma}$, then $\jdwf{\cps{\Gamma}}{\cps{\Sigma}}$.
    \end{itemize}
\end{lemma}
\begin{proof}
    By simultaneous induction on the derivations. Lemma \ref{lem:cps:weaken} is used.
    The case for \rulename{WT-Rfn} uses Assumption \ref{asm:cps:formula-cps}.
\end{proof}

\begin{lemma}[CPS transformation preserves subtyping] \label{lem:cps:presv-sub} \quad
    \begin{itemize}
        \item If $\jdsub{\Gamma}{T_1}{T_2}$, then $\jdsub{\cps{\Gamma}}{\cps{T_1}}{\cps{T_2}}$.
        \item If $\jdsub{\Gamma}{C_1}{C_2}$, then $\jdsub{\cps{\Gamma}}{\cps{C_1}}{\cps{C_2}}$.
        \item If $\jdsub{\Gamma}{\Sigma_1}{\Sigma_2}$, then $\jdsub{\cps{\Gamma}}{\cps{\Sigma_1}}{\cps{\Sigma_2}}$.
    \end{itemize}
\end{lemma}
\begin{proof}
    By simultaneous induction on the derivations. Lemma \ref{lem:cps:weaken} is used.
    The case for \rulename{S-Rfn} uses Assumption \ref{asm:cps:formula-cps}.
\end{proof}

\begin{theorem}[Forward type preservation] \quad
    \begin{enumerate}
        \item If\, $\jdty{\Gamma}{v}{T}$, then $\jdty{\cps{\Gamma}}{\cps{v}}{\cps{T}}$.
        \item If\, $\jdty{\Gamma}{c}{C}$, then $\jdty{\cps{\Gamma}}{\cps{c}}{\cps{C}}$.
    \end{enumerate}
\end{theorem}
\begin{proof}
    By simultaneous induction on the typing derivation of the source language.
    \begin{enumit}
        \item 
        \begin{description}
            \item[Case \rulename{T-CVar}:] By Lemma \ref{lem:cps:presv-wf},
                definition of CPS transformation of typing contexts, and \rulename{Tc-CVar}.
            \item[Case \rulename{T-Var}:] By Lemma \ref{lem:cps:presv-wf},
                definition of CPS transformation of typing contexts, and \rulename{Tc-Var}.
            \item[Case \rulename{T-Prim}:] By Lemma \ref{lem:cps:presv-wf},
                Assumption \ref{asm:cps:prim-cps}, and \rulename{Tc-Prim}.
            \item[Case \rulename{T-Fun}:] By the IH and \rulename{Tc-Fun}.
            \item[Case \rulename{T-VSub}:] By the IH, Lemma \ref{lem:cps:presv-sub},
                Lemma \ref{lem:cps:presv-wf} and \rulename{Tc-Sub}.
        \end{description}
        \item 
        \begin{description}
            \item[Case \rulename{T-Ret}:] we have
                \begin{itemize}
                    \item $c = \expret{v}$,
                    \item $C = \tycomp{\{\}}{T}{\square}$, and
                    \item $\jdty{\Gamma}{v}{T}$
                \end{itemize}
                for some $v$ and $T$.
                Then, we have
                \begin{itemize}
                    \item $\cps{c} = \Lambda \alpha. \lambda h:\{\}. \lambda k:\cps{T} \rarr \alpha. k~\cps{v}$ and
                    \item $\cps{C} = \forall \alpha. \{\} \rarr (\cps{T} \rarr \alpha) \rarr \alpha$~.
                \end{itemize}
                By the IH, we have
                \begin{itemize}
                    \item $\jdty{\cps{\Gamma}}{\cps{v}}{\cps{T}}$~.
                \end{itemize}
                We have the conclusion by the following derivation with Lemma \ref{lem:cps:weaken}:
                \[
                    \infersc[Tc-TAbs]{\jdty{\cps{\Gamma}}{\Lambda \alpha. \lambda h:\{\}. \lambda k:\cps{T} \rarr \alpha. k~\cps{v}}
                        {\forall \alpha. \{\} \rarr (\cps{T} \rarr \alpha) \rarr \alpha}}
                    {
                        \infersc[Tc-Lam]{\jdty{\cps{\Gamma}, \alpha}{\lambda h:\{\}. \lambda k:\cps{T} \rarr \alpha. k~\cps{v}}
                            {\{\} \rarr (\cps{T} \rarr \alpha) \rarr \alpha}}
                        {
                            \infersc[Tc-Lam]{\jdty{\cps{\Gamma}, \alpha, h: \{\}}{\lambda k\cps{T} \rarr \alpha. k~\cps{v}}
                                {(\cps{T} \rarr \alpha) \rarr \alpha}}
                            {
                                \infersc[Tc-App]{\jdty{\cps{\Gamma}, \alpha, h: \{\}, k:\cps{T} \rarr \alpha}
                                    {k~\cps{v}}{\alpha}}
                                {
                                    \infersc[Tc-Var]{\jdty{\Gamma_{\alpha,h,k}}
                                        {k}{\cps{T} \rarr \alpha}}
                                    {}
                                    &
                                    \jdty{\Gamma_{\alpha,h,k}}
                                        {\cps{v}}{\cps{T}}
                                }
                            }
                        }
                    }
                \]
                where $\Gamma_{\alpha,h,k} \defeq \cps{\Gamma}, \alpha, h: \{\}, k:\cps{T} \rarr \alpha$~.
            \item[Case \rulename{T-App}:] By the IH, Lemma \ref{lem:cps:homo-subst}
                and \rulename{Tc-App}.
            \item[Case \rulename{T-If}:] By the IH, \rulename{Tc-If} and \rulename{Tc-Ascr}.
            \item[Case \rulename{T-CSub}:] similar to the case for \rulename{T-VSub}.
            \item[Case \rulename{T-LetP}:] We have
                \begin{itemize}
                    \item $c = \explet{x}{c_1}{c_2}$,
                    \item $C = \tycomp{\Sigma}{T_2}{\square}$,
                    \item $\jdty{\Gamma}{c_1}{\tycomp{\Sigma}{T_1}{\square}}$,
                    \item $\jdty{\Gamma, x: T_1}{c_2}{\tycomp{\Sigma}{T_2}{\square}}$, and
                    \item $x \notin \fv(T_2) \cup \fv(\Sigma)$
                \end{itemize}
                for some $x, c_1, c_2, \Sigma, T_1$ and $T_2$.
                Then we have
                \begin{itemize}
                    \item $\cps{c} = \Lambda \alpha. \lambda h: \cps{\Sigma}. \lambda k:\cps{T_2} \rarr \alpha.
                        \cps{c_1}~\alpha~h~(\lambda x:\cps{T_1}. \cps{c_2}~\alpha~h~k)$ and
                    \item $\cps{C} = \forall \alpha. \cps{\Sigma} \rarr (\cps{T_2} \rarr \alpha) \rarr \alpha$~.
                \end{itemize}
                By Lemma \ref{lem:cps:presv-fv}, we have
                \begin{itemize}
                    \item $x \notin \fv(\cps{T_2}) \cup \fv(\cps{\Sigma})$~.
                \end{itemize}
                Also, by the IHs, we have
                \begin{itemize}
                    \item $\jdty{\cps{\Gamma}}{\cps{c_1}}
                        {\forall \beta. \cps{\Sigma} \rarr (\cps{T_1} \rarr \beta) \rarr \beta}$ and
                    \item $\jdty{\cps{\Gamma}, x:\cps{T_1}}{\cps{c_2}}
                        {\forall \gamma. \cps{\Sigma} \rarr (\cps{T_2} \rarr \gamma) \rarr \gamma}$~.
                \end{itemize}
                We have the conclusion by the following derivations with Lemma \ref{lem:cps:weaken}:
                \[  (A):
                    \infersc[Tc-App]{
                        \jdty{\cps{\Gamma}_{\alpha, h, k}}
                            {\cps{c_1}~\alpha~h}{(\cps{T_1} \rarr \alpha) \rarr \alpha}
                    }{
                        \infersc[Tc-App]{
                            \jdty{\Gamma_{\alpha, h, k}}
                                {\cps{c_1}~\alpha~h}{(\cps{T_1} \rarr \alpha) \rarr \alpha}
                        }{
                            \infersc[Tc-TApp]{
                                \jdty{\Gamma_{\alpha, h, k}}
                                    {\cps{c_1}~\alpha}{\cps{\Sigma} \rarr (\cps{T_1} \rarr \alpha) \rarr \alpha}
                            }{
                                \jdty{\Gamma_{\alpha, h, k}}
                                    {\cps{c_1}}{\forall \beta. \cps{\Sigma} \rarr (\cps{T_1} \rarr \beta) \rarr \beta}
                                &
                                \jdwf{\Gamma_{\alpha, h, k}}{\alpha}
                            }
                            &
                            \infersc[Tc-Var]{
                                \jdty{\Gamma_{\alpha, h, k}}
                                    {h}{\cps{\Sigma}}
                            }{}
                        }
                    }
                \]
                \[  (B):
                    \infersc[Tc-App]{
                        \jdty{\Gamma_{\alpha, h, k, x}}
                            {\cps{c_2}~\alpha~h}{(\cps{T_2} \rarr \alpha) \rarr \alpha}
                    }{
                        \infersc[Tc-TApp]{
                            \jdty{\Gamma_{\alpha, h, k, x}}
                                {\cps{c_2}~\alpha}{\cps{\Sigma} \rarr (\cps{T_2} \rarr \alpha) \rarr \alpha}
                        }{
                            \jdty{\Gamma_{\alpha, h, k, x}}
                                {\cps{c_2}}{\forall \gamma. \cps{\Sigma} \rarr (\cps{T_2} \rarr \gamma) \rarr \gamma}
                            &
                            \jdwf{\Gamma_{\alpha, h, k, x}}{\alpha}
                        }
                        &
                        \infersc[Tc-Var]{
                            \jdty{\Gamma_{\alpha, h, k, x}}
                                {h}{\cps{\Sigma}}
                        }{}
                    }
                \]
                \[
                    \infersc[Tc-TAbs]{
                        \jdty{\cps{\Gamma}}{\Lambda \alpha. \lambda h: \cps{\Sigma}. \lambda k:\cps{T_2} \rarr \alpha.
                            \cps{c_1}~\alpha~h~(\lambda x:\cps{T_1}. \cps{c_2}~\alpha~h~k)}
                            {\forall \alpha. \cps{\Sigma} \rarr (\cps{T_2} \rarr \alpha) \rarr \alpha}
                    }{
                        \infersc[Tc-Fun]{
                            \jdty{\cps{\Gamma}, \alpha}{\lambda h: \cps{\Sigma}. \lambda k:\cps{T_2} \rarr \alpha.
                                \cps{c_1}~\alpha~h~(\lambda x:\cps{T_1}. \cps{c_2}~\alpha~h~k)}
                                {\cps{\Sigma} \rarr (\cps{T_2} \rarr \alpha) \rarr \alpha}
                        }{
                            \infersc[Tc-Fun]{
                                \jdty{\cps{\Gamma}, \alpha, h: \cps{\Sigma}}{\lambda k:\cps{T_2} \rarr \alpha.
                                    \cps{c_1}~\alpha~h~(\lambda x:\cps{T_1}. \cps{c_2}~\alpha~h~k)}
                                    {(\cps{T_2} \rarr \alpha) \rarr \alpha}
                            }{
                                \infersc[Tc-App]{
                                    \jdty{\cps{\Gamma}, \alpha, h: \cps{\Sigma}, k:\cps{T_2} \rarr \alpha}{
                                        \cps{c_1}~\alpha~h~(\lambda x:\cps{T_1}. \cps{c_2}~\alpha~h~k)}
                                        {\alpha}
                                }{
                                    (A)
                                    &
                                    \infersc[Tc-Fun]{
                                        \jdty{\Gamma_{\alpha, h, k}}
                                            {\lambda x:\cps{T_1}. \cps{c_2}~\alpha~h~k}{\cps{T_1} \rarr \alpha}
                                    }{
                                        \infersc[Tc-App]{
                                            \jdty{\Gamma_{\alpha, h, k}, x:\cps{T_1}}
                                                {\cps{c_2}~\alpha~h~k}{\alpha}
                                        }{
                                            (B)
                                            &
                                            \infersc[Tc-Var]{
                                                \jdty{\Gamma_{\alpha, h, k, x}}
                                                    {k}{\cps{T_2} \rarr \alpha}
                                            }{}
                                        }
                                    }
                                }
                            }
                        }
                    }
                \]
                where $\Gamma_{\alpha, h, k} \defeq \cps{\Gamma}, \alpha, h: \cps{\Sigma}, k:\cps{T_2} \rarr \alpha$
                and $\Gamma_{\alpha, h, k, x} \defeq \Gamma_{\alpha, h, k}, x:\cps{T_1}$~.
            \item[Case \rulename{T-LetIp}:] We have
                \begin{itemize}
                    \item $c = \explet{x}{c_1}{c_2}$,
                    \item $C = \tycomp{\Sigma}{T_2}{\tyctl{z}{C_0}{C_2}}$,
                    \item $\jdty{\Gamma}{c_1}{\tycomp{\Sigma}{T_1}{\tyctl{x}{C_1}{C_2}}}$,
                    \item $\jdty{\Gamma, x: T_1}{c_2}{\tycomp{\Sigma}{T_2}{\tyctl{z}{C_0}{C_1}}}$, and
                    \item $x \notin \fv(T_2) \cup \fv(\Sigma) \cup (\fv(C_0) \setminus \{z\})$
                \end{itemize}
                for some $x, z, c_1, c_2, \Sigma, T_1, T_2, C_0, C_1$ and $C_2$.
                Then we have
                \begin{itemize}
                    \item $\cps{c} = \Lambda \alpha. \lambda h: \cps{\Sigma}. \lambda k:(z:\cps{T_2}) \rarr \cps{C_0}.
                        \cps{c_1}~\cps{C_2}~h~(\lambda x:\cps{T_1}. \cps{c_2}~\cps{C_1}~h~k)$ and
                    \item $\cps{C} = \forall \alpha. \cps{\Sigma} \rarr ((z:\cps{T_2}) \rarr \cps{C_0}) \rarr \cps{C_2}$~.
                \end{itemize}
                By Lemma \ref{lem:cps:presv-fv}, we have
                \begin{itemize}
                    \item $x \notin \fv(\cps{T_2}) \cup \fv(\cps{\Sigma}) \cup (\fv(\cps{C_0}) \setminus \{z\})$~.
                \end{itemize}
                Also, by the IHs, we have
                \begin{itemize}
                    \item $\jdty{\cps{\Gamma}}{\cps{c_1}}
                        {\forall \alpha. \cps{\Sigma} \rarr ((x:\cps{T_1}) \rarr \cps{C_1}) \rarr \cps{C_2}}$ and
                    \item $\jdty{\cps{\Gamma}, x:\cps{T_1}}{\cps{c_2}}
                        {\forall \alpha. \cps{\Sigma} \rarr ((z:\cps{T_2}) \rarr \cps{C_0}) \rarr \cps{C_1}}$~.
                \end{itemize}
                We have the conclusion by a straightforward derivation
                like the case for \rulename{T-LetP}
                using those judgements shown so far and Lemma \ref{lem:cps:weaken}.
            \item[Case \rulename{T-Op}:]
                (In this case, we use Lemma \ref{lem:cps:homo-subst} frequently and implicitly.) \\
                We have
                \begin{itemize}
                    \item $c = \op~v$,
                    \item $C = \tycomp{\Sigma}{T_2[\rep{A/X}][v/x]}{\tyctl{y}{C_1[\rep{A/X}][v/x]}{C_2[\rep{A/X}][v/x]}}$,
                    \item $\Sigma \ni \op: \forall \rep{X: \rep{B}}. (x: T_1) \rarr ((y: T_2) \rarr C_1) \rarr C_2$,
                    \item $\jdwf{\Gamma}{\Sigma}$,
                    \item $\rep{\jdty{\Gamma}{A}{\rep{B}}}$, and
                    \item $\jdty{\Gamma}{v}{T_1[\rep{A/X}]}$
                \end{itemize}
                for some $x, y, v, \rep{X}, \rep{A}, \rep{\rep{B}}, \Sigma, T_1, T_2, C_1$ and $C_2$.
                Then, we have
                \begin{itemize}
                    \item $\cps{c} = \Lambda \alpha. \lambda h:\cps{\Sigma}. \lambda k:(y: \cps{T_2}[\rep{A/X}][\cps{v}/x] \rarr \cps{C_1}[\rep{A/X}][\cps{v}/x]).$\\
                        \hspace*{180pt} $h\#\op~\rep{A}~\cps{v}~(\lambda y':\cps{T_2}[\rep{A/X}][\cps{v}/x]. k~y')$,
                    \item $\cps{C} = \forall \alpha. \cps{\Sigma} \rarr ((y: \cps{T_2}[\rep{A/X}][\cps{v}/x]) \rarr \cps{C_1}[\rep{A/X}][\cps{v}/x]) \rarr \cps{C_2}[\rep{A/X}][\cps{v}/x]$, and
                    \item $\cps{\Sigma} \ni \op: \forall \rep{X: \rep{B}}. (x: \cps{T_1}) \rarr ((y: \cps{T_2}) \rarr \cps{C_1}) \rarr \cps{C_2}$~.
                \end{itemize}
                Also, by the IHs, we have
                \begin{itemize}
                    \item $\jdty{\cps{\Gamma}}{\cps{v}}{\cps{T_1}[\rep{A/X}]}$~.
                \end{itemize}
                By Assumption~\ref{asm:cps:formula-cps}, we have
                \begin{itemize}
                    \item $\rep{\jdty{\cps{\Gamma}}{A}{\rep{B}}}$~.
                \end{itemize}
                We have the conclusion by a straightforward derivation
                like the cases for \rulename{T-Ret} and \rulename{T-LetP}
                using those judgements shown so far and Lemma \ref{lem:cps:weaken}.
            \item[Case \rulename{T-Hndl}:] We have
                \begin{itemize}
                    \item $c = \expwith{h}{c_0}$,
                    \item $h = \{ \expret{x_r} \mapsto c_r, \repi{\op_i(x_i, k_i) \mapsto c_i} \}$,
                    \item $\jdty{\Gamma}{c_0}{\tycomp{\Sigma_0}{T_r}{\tyctl{x_r}{C_1}{C}}}$,
                    \item $\jdty{\Gamma, x_r: T_r}{c_r}{C_1}$,
                    \item $\bigrepi{\jdty{\Gamma, \rep{X_i: \rep{B}_i}, x_i: T_{i1}, k_i: (y_i: T_{i2}) \rarr C_{i1}}{c_i}{C_{i2}}}$, and
                    \item $\Sigma_0 = \{ \repi{\op_i: \forall \rep{X_i: \rep{B}_i}. (x_i: T_{i1}) \rarr ((y_i: T_{i2}) \rarr C_{i1}) \rarr C_{2i}} \}$
                \end{itemize}
                Then, we have
                \begin{itemize}
                    \item $\cps{c} = \cps{c_0}~\cps{C}~\cps{h^{\mathit{ops}}}~\cps{h^{\mathit{ret}}}$,
                    \item $\cps{h^{\mathit{ret}}} = \lambda x_r:\cps{T_r}. \cps{c_r}$,
                    \item $\cps{h^{\mathit{ops}}} = \{ \repi{\op_i = \Lambda \rep{X_i: \rep{B_i}}. \lambda x_i:\cps{T_{i1}}. \lambda k_i:(y_i:\cps{T_{i2}}) \rarr \cps{C_{i1}}. \cps{c_i}} \}$, and
                    \item $\cps{\Sigma_0} = \{ \repi{\op_i: \forall \rep{X_i: \rep{B}_i}. (x_i: \cps{T_{i1}}) \rarr ((y_i: \cps{T_{i2}}) \rarr \cps{C_{i1}}) \rarr \cps{C_{i2}}} \}$~.
                \end{itemize}
                Also, by the IHs, we have
                \begin{itemize}
                    \item $\jdty{\cps{\Gamma}}{\cps{c_0}}{\forall \alpha. \cps{\Sigma_0} \rarr ((x_r: \cps{T_r}) \rarr \cps{C_1}) \rarr \cps{C}}$,
                    \item $\jdty{\cps{\Gamma}, x_r: \cps{T_r}}{\cps{c_r}}{\cps{C_1}}$, and
                    \item $\bigrepi{\jdty{\cps{\Gamma}, \rep{X_i: \rep{B}_i}, x_i: \cps{T_{i1}}, k_i: (y_i: \cps{T_{i2}}) \rarr \cps{C_{i1}}}{\cps{c_i}}{\cps{C_{i2}}}}$~.
                \end{itemize}
                We have the conclusion by a straightforward derivation
                like the cases for \rulename{T-Ret} and \rulename{T-LetP}
                using those judgements shown so far and Lemma \ref{lem:cps:weaken}.
        \end{description}
    \end{enumit}
\end{proof}

\subsection{Backward type preservation}

For the backward direction, we define some notations.
\begin{definition}
    $\Gamma$ is \emph{cps-wellformed}
    if for all $(x:\tau) \in \Gamma$, it holds that $\tau = \cps{T}$ for some $T$.
\end{definition}
\begin{definition}
    $\rmtv$ is a function which removes all bindings of type variables
    from a typing context. Formally, it is defined as follows:
    \begin{align}
        \rmtv(\emptyset) &\defeq \emptyset &
        \rmtv(\Gamma, x: \tau) &\defeq \rmtv(\Gamma), x: \tau \\
        \rmtv(\Gamma, X: \rep{B}) &\defeq \rmtv(\Gamma), X: \rep{B} &
        \rmtv(\Gamma, \alpha) &\defeq \rmtv(\Gamma)
    \end{align}
\end{definition}

\begin{lemma}[CPS-wellformed target typing contexts have corresponding source ones] \label{lem:cps:cpswf-rmtv}
    If $\Gamma$ is cps-wellformed,
    then there exists some $\Gamma'$ such that $\cps{\Gamma'} = \rmtv(\Gamma)$.
\end{lemma}
\begin{proof}
    By induction on the structure of $\Gamma$.
\end{proof}

Since the CPS transformation is injective,
there is only one $\Gamma'$ which satisfies the equation in Lemma \ref{lem:cps:cpswf-rmtv}.
Therefore, we define a function $\cpsinv{-}$ that maps $\Gamma$ to $\Gamma'$:
\begin{definition}
    Let $\Gamma$ be a cps-wellformed typing context in the target language.
    We define $\cpsinv{\Gamma}$ to be the typing context in the source language
    such that $\cps{\cpsinv{\Gamma}} = \rmtv(\Gamma)$.
\end{definition}

\begin{assumption} \label{asm:cps:formula-cpsinv}
    Assume that $\Gamma$ is cps-wellformed.
    \begin{itemize}
        \item If $\jdwf{\Gamma}{\phi}$, then $\jdwf{\cpsinv{\Gamma}}{\phi}$.
        \item If $\jdty{\Gamma}{A}{\rep{B}}$, then $\jdty{\cpsinv{\Gamma}}{A}{\rep{B}}$.
        \item If $\Gamma \vDash \phi$, then $\cpsinv{\Gamma} \vDash \phi$.
    \end{itemize}
\end{assumption}

\begin{lemma}[Computation types in the specific form of subtyping are pure] \label{lem:cps:sub-pure}
    If $\jdsub{\Gamma}{\cps{C}}{\forall \alpha. \tau_1 \rarr (\tau_2 \rarr \beta) \rarr \tau_4}$
    and $\beta \in \Gamma$,
    then $C = \tycomp{\Sigma}{T}{\square}$ (for some $\Sigma$ and $T$),
    and $\tau_4 = \beta$.
\end{lemma}
\begin{proof}
    Assume that $C = \tycomp{\Sigma}{T}{\tyctl{x}{C_1}{C_2}}$ for some $\Sigma, T, x, C_1$ and $C_2$.
    Then, we have
    \[
        \jdsub{\Gamma}
            {\forall \gamma. \cps{\Sigma} \rarr ((x: \cps{T}) \rarr \cps{C_1}) \rarr \cps{C_2}}
            {\forall \alpha. \tau_1 \rarr (\tau_2 \rarr \beta) \rarr \tau_4}
    \]
    where $\gamma$ is fresh.
    By inversion, we have $\jdsub{\Gamma, \alpha, h:\tau_1, x:\cps{T}}{\beta}{\cps{C_1}}$,
    that is, $\jdsub{\Gamma, \alpha, h:\tau_1, x:\cps{T}}{\beta}{\forall \delta. \tau_5}$
    for some $\tau_5$ and $\delta$.
    This is contradictory since there is no subtyping rule for such a judgment.

    Therefore, $C = \tycomp{\Sigma}{T}{\square}$ for some $\Sigma$ and $T$.
    In this case, we have
    \[
        \jdsub{\Gamma}
            {\forall \gamma. \cps{\Sigma} \rarr (\cps{T} \rarr \gamma) \rarr \gamma}
            {\forall \alpha. \tau_1 \rarr (\tau_2 \rarr \beta) \rarr \tau_4}
    \]
    where $\gamma$ is fresh.
    By inversion, we have
    \begin{itemize}
        \item $\jdwf{\Gamma, \alpha}{\tau_6}$,
        \item $\jdsub{\Gamma, \alpha, h:\tau_1, x:\cps{T}[\tau_6/\gamma]}{\beta}{\gamma[\tau_6/\gamma]}$, and
        \item $\jdsub{\Gamma, \alpha, h:\tau_1, x:\cps{T}[\tau_6/\gamma]}{\gamma[\tau_6/\gamma]}{\tau_4}$
    \end{itemize}
    for some $\tau_6$.
    The second judgment can be derived by only \rulename{Sc-TVar} where $\gamma[\tau_6/\gamma] = \beta$.
    Therefore, the third judgment becomes
    $\jdsub{\Gamma, \alpha, h:\tau_1, x:\cps{T}[\tau_6/\gamma]}{\beta}{\tau_4}$,
    which can be derived by only \rulename{Sc-TVar} where $\tau_4 = \beta$.
\end{proof}

\begin{lemma}[Computation types can be assumed to be impure] \label{lem:assume-atm}
    If $\jdty{\Gamma}{c}{C}$, then w.l.o.g.,
    we can assume that $C = \tycomp{\Sigma}{T}{\tyctl{x}{C_1}{C_2}}$
    for some $\Sigma, T, x, C_1$ and $C_2$.
\end{lemma}
\begin{proof} \quad
    \begin{description}
        \item[Case $C = \tycomp{\Sigma}{T}{\tyctl{x}{C_1}{C_2}}$:] Immediate.
        \item[Case $C = \tycomp{\Sigma}{T}{\square}$:]
            It holds that
            $\jdsub{\Gamma}{\tycomp{\Sigma}{T}{\square}}{\tycomp{\Sigma}{T}{\tyctl{x}{C_0}{C_0}}}$
            for any $C_0$ such that $\jdwf{\Gamma}{C_0}$.
            Therefore, by subsumption
            we have $\jdty{\Gamma}{c}{\tycomp{\Sigma}{T}{\tyctl{x}{C_0}{C_0}}}$~.
    \end{description}
\end{proof}

\begin{lemma}[Backward preservation on well-formedness] \label{lem:cps:presv-b-wf}
    Assume that $\Gamma$ is cps-wellformed.
    \begin{enumerate}
        \item If $\jdwf{}{\Gamma}$, then $\jdwf{}{\cpsinv{\Gamma}}$.
        \item If $\jdwf{\Gamma}{\cps{T}}$, then $\jdwf{\cpsinv{\Gamma}}{T}$.
        \item If $\jdwf{\Gamma}{\cps{C}}$, then $\jdwf{\cpsinv{\Gamma}}{C}$.
        \item If $\jdwf{\Gamma}{\cps{\Sigma}}$, then $\jdwf{\cpsinv{\Gamma}}{\Sigma}$.
    \end{enumerate}
\end{lemma}

\begin{proof}
    By simultaneous induction on the derivation.
    \begin{enumit}
        \item
        \begin{description}
            \item[Case \rulename{WEc-Empty}:] Obvious since $\cpsinv{\emptyset} = \emptyset$.
            \item[Case \rulename{WEc-Var}:] We have
                \begin{itemize}
                    \item $\Gamma = \Gamma', x:\tau$,
                    \item $\jdwf{}{\Gamma'}$,
                    \item $x \notin \dom(\Gamma')$, and
                    \item $\jdwf{\Gamma'}{\tau}$
                \end{itemize}
                for some $\Gamma', x$, and $\tau$.
                By the IH, we have $\jdwf{}{\cpsinv{\Gamma'}}$.
                Also, we have $x \notin \dom(\cpsinv{\Gamma'})$
                since $\dom(\Gamma') \supseteq \dom(\cpsinv{\Gamma'})$.
                Moreover, since $\Gamma$ is cps-wellformed, $\tau = \cps{T}$ for some $T$.
                Then, by the IH, we have $\jdwf{\cpsinv{\Gamma'}}{T}$.
                We have the conclusion by \rulename{WE-Var}.
                (Note that $\cpsinv{\Gamma} = \cpsinv{\Gamma', x:\cps{T}} = \cpsinv{\Gamma'}, x: T$.)
            \item[Case \rulename{WEc-BVar}:] By the IH and \rulename{WE-BVar}.
            \item[Case \rulename{WEc-PVar}:] By the IH and \rulename{WE-PVar}.
            \item[Case \rulename{WEc-TVar}:] By the IH.
                Note that $\cpsinv{\Gamma', \alpha} = \cpsinv{\Gamma'}$.
        \end{description}
        \item Case analysis on $T$.
        \begin{description}
            \item[Case $T = \tyrfn{z}{B}{\phi}$:]
                By Assumption \ref{asm:cps:formula-cpsinv} and \rulename{WT-Rfn}.
            \item[Case $T = (x:T_1) \rarr C_1$:] By the IHs and \rulename{WT-Fun}.
        \end{description}
        \item Case analysis on $C$.
        \begin{description}
            \item[Case $C = \tycomp{\Sigma}{T}{\square}$:] We have
                $\cps{C} = \forall \alpha. \cps{\Sigma} \rarr (\cps{T} \rarr \alpha) \rarr \alpha$
                for some $\alpha$.
                By inversion, we have
                \begin{itemize}
                    \item $\jdwf{\Gamma, \alpha}{\cps{\Sigma}}$ and
                    \item $\jdwf{\Gamma, \alpha, h: \cps{\Sigma}}{\cps{T}}$~.
                \end{itemize}
                By \ref{lem:rm-nonrfn}, we have
                \begin{itemize}
                    \item $\jdwf{\Gamma, \alpha}{\cps{T}}$~.
                \end{itemize}
                By the IHs, we have
                \begin{itemize}
                    \item $\jdwf{\cpsinv{\Gamma}}{\Sigma}$ and
                    \item $\jdwf{\cpsinv{\Gamma}}{T}$~.
                \end{itemize}
                Also, by \rulename{WT-Pure}, we have $\jdwf{\cpsinv{\Gamma} \mid T}{\square}$.
                Then we have the conclusion by \rulename{WT-Comp}.
            \item[Case $C = \tycomp{\Sigma}{T}{\tyctl{x}{C_1}{C_2}}$:] We have
                $\cps{C} = \forall \alpha. \cps{\Sigma} \rarr ((x:\cps{T}) \rarr \cps{C_1}) \rarr \cps{C_2}$~.
                By inversion, we have
                \begin{itemize}
                    \item $\jdwf{\Gamma, \alpha}{\cps{\Sigma}}$,
                    \item $\jdwf{\Gamma, \alpha, h: \cps{\Sigma}}{\cps{T}}$,
                    \item $\jdwf{\Gamma, \alpha, h: \cps{\Sigma}, x:\cps{T}}{\cps{C_1}}$, and
                    \item $\jdwf{\Gamma, \alpha, h: \cps{\Sigma}, k: (x:\cps{T}) \rarr \cps{C_1}}{\cps{C_2}}$~.
                \end{itemize}
                By \ref{lem:rm-nonrfn}, we have
                \begin{itemize}
                    \item $\jdwf{\Gamma, \alpha}{\cps{T}}$,
                    \item $\jdwf{\Gamma, \alpha, x:\cps{T}}{\cps{C_1}}$, and
                    \item $\jdwf{\Gamma, \alpha}{\cps{C_2}}$~.
                \end{itemize}
                By the IHs, we have
                \begin{itemize}
                    \item $\jdwf{\cpsinv{\Gamma}}{\Sigma}$,
                    \item $\jdwf{\cpsinv{\Gamma}}{T}$,
                    \item $\jdwf{\cpsinv{\Gamma}, x:T}{C_1}$, and
                    \item $\jdwf{\cpsinv{\Gamma}}{C_2}$~.
                \end{itemize}
                Then we have the conclusion by \rulename{WT-ATM} and \rulename{WT-Comp}.
        \end{description}
        \item By the IHs and \rulename{WT-Sig}.
    \end{enumit}
\end{proof}

\begin{lemma}[Backward preservation on subtyping] \label{lem:cps:cpsinv-sub}
    Assume that $\Gamma$ is cps-wellformed.
    \begin{enumerate}
        \item If $\jdsub{\Gamma}{\cps{T_1}}{\cps{T_2}}$, then $\jdsub{\cpsinv{\Gamma}}{T_1}{T_2}$.
        \item If $\jdsub{\Gamma}{\cps{C_1}}{\cps{C_2}}$, then $\jdsub{\cpsinv{\Gamma}}{C_1}{C_2}$.
        \item If $\jdsub{\Gamma}{\cps{\Sigma_1}}{\cps{\Sigma_2}}$, then $\jdsub{\cpsinv{\Gamma}}{\Sigma_1}{\Sigma_2}$.
    \end{enumerate}
\end{lemma}

\begin{proof}
    By simultaneous induction on the derivation.
    \begin{enumit}
        \item Case analysis on $T_1$ and $T_2$.
        \begin{description}
            \item[Case $T_1 = \tyrfn{z}{B}{\phi_1}$ and $T_2 = \tyrfn{z}{B}{\phi_2}$:] We have
                \begin{itemize}
                    \item $\cps{T_1} = \tyrfn{z}{B}{\phi_1}$ and
                    \item $\cps{T_2} = \tyrfn{z}{B}{\phi_2}$~.
                \end{itemize}
                We have the conclusion
                by Assumption \ref{asm:cps:formula-cpsinv} and \rulename{S-Rfn}.
            \item[Case $T_1 = (x:T_{10}) \rarr C_1$ and $T_1 = (x:T_{10}) \rarr C_1$:] We have
                \begin{itemize}
                    \item $\cps{T_1} = (x:\cps{T_{10}}) \rarr \cps{C_1}$ and
                    \item $\cps{T_2} = (x:\cps{T_{20}}) \rarr \cps{C_2}$.
                \end{itemize}
                We have the conclusion by the IHs and \rulename{S-Fun}.
            \item[Otherwise:] Contradictory since there is no applicable rule.
        \end{description}
        \item Case analysis on $C_1$ and $C_2$.
        \begin{description}
            \item[Case $C_1 = \tycomp{\Sigma_1}{T_1}{\square}$
                and $C_2 = \tycomp{\Sigma_2}{T_2}{\square}$:] We have
                \begin{itemize}
                    \item $\cps{C_1} = \forall \alpha. \cps{\Sigma_1} \rarr (\cps{T_1} \rarr \alpha) \rarr \alpha$ and
                    \item $\cps{C_2} = \forall \beta. \cps{\Sigma_2} \rarr (\cps{T_2} \rarr \beta) \rarr \beta$
                \end{itemize}
                for some $\alpha$ and $\beta$.
                By inversion, we have
                \begin{itemize}
                    \item $\jdwf{\Gamma, \beta}{\tau}$,
                    \item $\jdsub{\Gamma, \beta}{\cps{\Sigma_2}}{\cps{\Sigma_1}[\tau/\alpha]}$ and
                    \item $\jdsub{\Gamma, \beta, h: \cps{\Sigma_2}}{\cps{T_1}}{{\cps{T_2}[\tau/\alpha]}}$
                \end{itemize}
                for some $\tau$.
                Since CPS-transformed types do not contain type variables, we have
                \begin{itemize}
                    \item $\jdsub{\Gamma, \beta}{\cps{\Sigma_2}}{\cps{\Sigma_1}}$ and
                    \item $\jdsub{\Gamma, \beta, h: \cps{\Sigma_2}}{\cps{T_1}}{{\cps{T_2}}}$~.
                \end{itemize}
                By \ref{lem:rm-nonrfn}, we have
                \begin{itemize}
                    \item $\jdsub{\Gamma, \beta}{\cps{\Sigma_2}}{\cps{\Sigma_1}}$ and
                    \item $\jdsub{\Gamma, \beta}{\cps{T_1}}{{\cps{T_2}}}$~.
                \end{itemize}
                By the IHs, we have
                \begin{itemize}
                    \item $\jdsub{\cpsinv{\Gamma}}{\Sigma_2}{\Sigma_1}$ and
                    \item $\jdsub{\cpsinv{\Gamma}}{T_1}{T_2}$~.
                \end{itemize}
                Also, by \rulename{S-Pure}, we have $\jdsub{\cpsinv{\Gamma} \mid T_1}{\square}{\square}$.
                Then we have the conclusion by \rulename{S-Comp}.
            \item[Case $C_1 = \tycomp{\Sigma_1}{T_1}{\tyctl{x}{C_{11}}{C_{12}}}$
                and $C_2 = \tycomp{\Sigma_2}{T_2}{\tyctl{x}{C_{21}}{C_{22}}}$:] We have
                \begin{itemize}
                    \item $\cps{C_1} = \forall \alpha. \cps{\Sigma_1} \rarr ((x:\cps{T_1}) \rarr \cps{C_{11}}) \rarr \cps{C_{12}}$ and
                    \item $\cps{C_2} = \forall \beta. \cps{\Sigma_2} \rarr ((x:\cps{T_2}) \rarr \cps{C_{21}}) \rarr \cps{C_{22}}$~.
                \end{itemize}
                By inversion, we have
                \begin{itemize}
                    \item $\jdwf{\Gamma, \beta}{\tau}$,
                    \item $\jdsub{\Gamma, \beta}{\cps{\Sigma_2}}{\cps{\Sigma_1}[\tau/\alpha]}$,
                    \item $\jdsub{\Gamma, \beta, h: \cps{\Sigma_2}}{\cps{T_1}[\tau/\alpha]}{{\cps{T_2}}}$,
                    \item $\jdsub{\Gamma, \beta, h: \cps{\Sigma_2}, x:\cps{T_1}[\tau/\alpha]}{\cps{C_{21}}}{\cps{C_{11}}[\tau/\alpha]}$, and
                    \item $\jdsub{\Gamma, \beta, h: \cps{\Sigma_2}, k: (x:\cps{T_2}) \rarr \cps{C_{21}}}{\cps{C_{12}}[\tau/\alpha]}{\cps{C_22}}$~.
                \end{itemize}
                for some $\tau$.
                Since CPS-transformed types do not contain type variables, we have
                \begin{itemize}
                    \item $\jdsub{\Gamma, \beta}{\cps{\Sigma_2}}{\cps{\Sigma_1}}$,
                    \item $\jdsub{\Gamma, \beta, h: \cps{\Sigma_2}}{\cps{T_1}}{{\cps{T_2}}}$,
                    \item $\jdsub{\Gamma, \beta, h: \cps{\Sigma_2}, x:\cps{T_1}}{\cps{C_{21}}}{\cps{C_{11}}}$, and
                    \item $\jdsub{\Gamma, \beta, h: \cps{\Sigma_2}, k: (x:\cps{T_2}) \rarr \cps{C_{21}}}{\cps{C_{12}}}{\cps{C_22}}$~.
                \end{itemize}
                By \ref{lem:rm-nonrfn}, we have
                \begin{itemize}
                    \item $\jdsub{\Gamma, \beta}{\cps{\Sigma_2}}{\cps{\Sigma_1}}$,
                    \item $\jdsub{\Gamma, \beta}{\cps{T_1}}{{\cps{T_2}}}$,
                    \item $\jdsub{\Gamma, \beta, x:\cps{T_1}}{\cps{C_{21}}}{\cps{C_{11}}}$, and
                    \item $\jdsub{\Gamma, \beta}{\cps{C_{12}}}{\cps{C_22}}$~.
                \end{itemize}
                By the IHs, we have
                \begin{itemize}
                    \item $\jdsub{\cpsinv{\Gamma}}{\Sigma_2}{\Sigma_1}$,
                    \item $\jdsub{\cpsinv{\Gamma}}{T_1}{T_2}$,
                    \item $\jdsub{\cpsinv{\Gamma}, x:T_1}{C_{21}}{C_{11}}$, and
                    \item $\jdsub{\cpsinv{\Gamma}}{C_{12}}{C_{22}}$~.
                \end{itemize}
                Then we have the conclusion by \rulename{S-ATM} and \rulename{S-Comp}.
            \item[Case $C_1 = \tycomp{\Sigma_1}{T_1}{\square}$
                and $C_2 = \tycomp{\Sigma_2}{T_2}{\tyctl{x}{C_{21}}{C_{22}}}$:] We have
                \begin{itemize}
                    \item $\cps{C_1} = \forall \alpha. \cps{\Sigma_1} \rarr (\cps{T_1} \rarr \alpha) \rarr \alpha$ and
                    \item $\cps{C_2} = \forall \beta. \cps{\Sigma_2} \rarr ((x:\cps{T_2}) \rarr \cps{C_{21}}) \rarr \cps{C_{22}}$~.
                \end{itemize}
                W.l.o.g., we can assume that $x \notin \cps{C_{22}}$.
                By inversion, we have
                \begin{itemize}
                    \item $\jdwf{\Gamma, \beta}{\tau}$,
                    \item $\jdsub{\Gamma, \beta}{\cps{\Sigma_2}}{\cps{\Sigma_1}[\tau/\alpha]}$,
                    \item $\jdsub{\Gamma, \beta, h: \cps{\Sigma_2}}{\cps{T_1}[\tau/\alpha]}{{\cps{T_2}}}$,
                    \item $\jdsub{\Gamma, \beta, h: \cps{\Sigma_2}, x:\cps{T_1}[\tau/\alpha]}{\cps{C_{21}}}{\alpha[\tau/\alpha]}$, and
                    \item $\jdsub{\Gamma, \beta, h: \cps{\Sigma_2}, k: (x:\cps{T_2}) \rarr \cps{C_{21}}}{\alpha[\tau/\alpha]}{\cps{C_22}}$~.
                \end{itemize}
                for some $\tau$.
                Since CPS-transformed types do not contain type variables, we have
                \begin{itemize}
                    \item $\jdsub{\Gamma, \beta}{\cps{\Sigma_2}}{\cps{\Sigma_1}}$,
                    \item $\jdsub{\Gamma, \beta, h: \cps{\Sigma_2}}{\cps{T_1}}{{\cps{T_2}}}$,
                    \item $\jdsub{\Gamma, \beta, h: \cps{\Sigma_2}, x:\cps{T_1}}{\cps{C_{21}}}{\tau}$, and
                    \item $\jdsub{\Gamma, \beta, h: \cps{\Sigma_2}, k: (x:\cps{T_2}) \rarr \cps{C_{21}}}{\tau}{\cps{C_22}}$~.
                \end{itemize}
                By \ref{lem:rm-nonrfn}, we have
                \begin{itemize}
                    \item $\jdsub{\Gamma, \beta}{\cps{\Sigma_2}}{\cps{\Sigma_1}}$,
                    \item $\jdsub{\Gamma, \beta}{\cps{T_1}}{{\cps{T_2}}}$,
                    \item $\jdsub{\Gamma, \beta, x:\cps{T_1}}{\cps{C_{21}}}{\tau}$, and
                    \item $\jdsub{\Gamma, \beta}{\tau}{\cps{C_22}}$~.
                \end{itemize}
                By Lemma \ref{lem:cps:weaken} and \ref{lem:cps:trans}, we have
                \begin{itemize}
                    \item $\jdsub{\Gamma, \beta}{\cps{\Sigma_2}}{\cps{\Sigma_1}}$,
                    \item $\jdsub{\Gamma, \beta}{\cps{T_1}}{{\cps{T_2}}}$, and
                    \item $\jdsub{\Gamma, \beta, x:\cps{T_1}}{\cps{C_{21}}}{\cps{C_22}}$~.
                \end{itemize}
                By the IHs, we have
                \begin{itemize}
                    \item $\jdsub{\cpsinv{\Gamma}}{\Sigma_2}{\Sigma_1}$,
                    \item $\jdsub{\cpsinv{\Gamma}}{T_1}{T_2}$, and
                    \item $\jdsub{\cpsinv{\Gamma}, x:T_1}{C_{21}}{C_{22}}$~.
                \end{itemize}
                Then we have the conclusion by \rulename{S-Embed} and \rulename{S-Comp}.
            \item[Case $C_1 = \tycomp{\Sigma_1}{T_1}{\tyctl{x}{C_{11}}{C_{12}}}$
                and $C_2 = \tycomp{\Sigma_2}{T_2}{\square}$:] We have
                \begin{itemize}
                    \item $\cps{C_1} = \forall \alpha. \cps{\Sigma_1} \rarr ((x:\cps{T_1}) \rarr \cps{C_{11}}) \rarr \cps{C_{12}}$ and
                    \item $\cps{C_2} = \forall \beta. \cps{\Sigma_2} \rarr (\cps{T_2} \rarr \beta) \rarr \beta$~.
                \end{itemize}
                By inversion, we have
                \begin{itemize}
                    \item $\jdwf{\Gamma, \beta}{\tau}$, and
                    \item $\jdsub{\Gamma, \beta, h: \cps{\Sigma_2}, k: (x:\cps{T_2}) \rarr \beta}{\cps{C_{12}}[\tau/\alpha]}{\beta}$~.
                \end{itemize}
                for some $\tau$.
                Since CPS-transformed types do not contain type variables, we have
                \begin{itemize}
                    \item $\jdsub{\Gamma, \beta, h: \cps{\Sigma_2}, k: (x:\cps{T_2}) \rarr \beta}{\cps{C_{12}}}{\beta}$~.
                \end{itemize}
                This is contradictory since $\cps{C_{12}}$ cannot be a type variable
                and thus there is no applicable rule.
        \end{description}
        \item By the IHs, and \rulename{S-Sig}.
    \end{enumit}
\end{proof}

\begin{theorem}[Backward type preservation (for open expressions)] \label{thm:cps-backward}
    Assume that $\Gamma$ is cps-wellformed.
    \begin{enumerate}
        \item If $\jdty{\Gamma}{\cps{v}}{\tau}$, then
            there exists $T$ such that
            $\jdty{\cpsinv{\Gamma}}{v}{T}$ and $\jdsub{\Gamma}{\cps{T}}{\tau}$.
        \item If $\jdty{\Gamma}{\cps{c}}{\tau}$, then
            there exists $C$ such that
            $\jdty{\cpsinv{\Gamma}}{c}{C}$ and $\jdsub{\Gamma}{\cps{C}}{\tau}$.
    \end{enumerate}
\end{theorem}
\begin{proof}
    By simultaneous induction on the structure of $v$ and $c$.
    \begin{enumit}
        \item
        \begin{description}
            \item[Case $v = x$:] We have $\cps{v} = x$.
            By Lemma \ref{lem:cps:inv}, we have \emph{either}
            \begin{enumerate}
                \item $\jdwf{}{\Gamma}$ and  $\jdsub{\Gamma}{\tyrfn{z}{B}{z = x}}{\tau}$
                    (if $\Gamma(x) = \tyrfn{z}{B}{\phi}$ for some $z, B$ and $\phi$)
                \item $\jdwf{}{\Gamma}$ and $\jdsub{\Gamma}{\Gamma(x)}{\tau}$ (otherwise)
            \end{enumerate}
            \begin{description}
                \item[Case 1:]
                    By Lemma \ref{lem:cps:presv-b-wf}, we have $\jdwf{}{\cpsinv{\Gamma}}$.
                    Also, since $\cps{\tyrfn{z}{B}{\phi'}} = \tyrfn{z}{B}{\phi'}$ for any $\phi'$, we have
                    \begin{itemize}
                        \item $\cpsinv{\Gamma}(x) = \tyrfn{z}{B}{\phi}$ and
                        \item $\jdsub{\Gamma}{\cps{\tyrfn{z}{B}{z = x}}}{\tau}$~.
                    \end{itemize}
                    Then, by \rulename{T-CVar}, we have $\jdty{\cpsinv{\Gamma}}{x}{\tyrfn{z}{B}{z = x}}$.
                    Now we have the conclusion with $T = \tyrfn{z}{B}{z = x}$.
                \item[Case 2:]
                    By Lemma \ref{lem:cps:presv-b-wf}, we have $\jdwf{}{\cpsinv{\Gamma}}$.
                    Then, since
                    $\Gamma(x) = \rmtv(\Gamma)(x) = \cps{\cpsinv{\Gamma}}(x) = \cps{\cpsinv{\Gamma}(x)}$
                    holds by Lemma \ref{lem:cps:cpswf-rmtv},
                    we have $\jdsub{\Gamma}{\cps{\cpsinv{\Gamma}(x)}}{\tau}$.
                    Also, by \rulename{T-Var}, we have $\jdty{\cpsinv{\Gamma}}{x}{\cpsinv{\Gamma}(x)}$.
                    Now we have the conclusion with $T = \cpsinv{\Gamma}(x)$.
            \end{description}
            \item[Case $v = p$:] We have $\cps{v} = \mathit{cps}(p)$.
                By Lemma \ref{lem:cps:inv}, we have
                \begin{itemize}
                    \item $\jdwf{}{\Gamma}$ and
                    \item $\jdsub{\Gamma}{\tycps(\cps{p})}{\tau}$.
                \end{itemize}
                By Lemma \ref{lem:cps:presv-b-wf}, we have $\jdwf{}{\cpsinv{\Gamma}}$.
                Then, by \rulename{T-Prim}, we have $\jdty{\cpsinv{\Gamma}}{p}{\ty(p)}$.
                Also, by Assumption \ref{asm:cps:prim-cps},
                we have $\jdsub{\Gamma}{\cps{\ty(p)}}{\tau}$.
                Now we have the conclusion with $T = \ty(p)$.
            \item[Case $v = \exprec{f^{(x:T_1) \rarr C_1}}{x^{T_2}}{c}$:]
                We have $\cps{v} = \exprec{f:(x:\cps{T_1}) \rarr \cps{C_1}}{x:\cps{T_2}}{\cps{c}}$.
                By Lemma \ref{lem:cps:inv}, we have
                \begin{itemize}
                    \item $\jdty{\Gamma, f:(x:\cps{T_1}) \rarr \cps{C_1}, x: \cps{T_1}}{\cps{c}}{\cps{C_1}}$ and
                    \item $\jdsub{\Gamma}{(x:\cps{T_1}) \rarr \cps{C_1}}{\tau}$~.
                \end{itemize}
                By the IH, we have
                \begin{itemize}
                    \item $\jdty{\cpsinv{\Gamma}, f:(x:T_1) \rarr C_1, x: T_1}{c}{C_1'}$ and
                    \item $\jdsub{\Gamma, f:(x:\cps{T_1}) \rarr \cps{C_1}, x: \cps{T_1}}{\cps{C_1'}}{\cps{C_1}}$
                \end{itemize}
                for some $C_1'$.
                By Lemma \ref{lem:cps:cpsinv-sub}, we have
                \[
                    \jdsub{\cpsinv{\Gamma}, f:(x:T_1) \rarr C_1, x: T_1}{C_1'}{C_1}~.
                \]
                Then, by \rulename{T-CSub} and \rulename{T-Fun}, we have
                \[
                    \jdty{\Gamma}{\exprec{f^{(x:T_1) \rarr C_1}}{x^{T_2}}{c}}{(x:T_1) \rarr C_1}~.
                \]
                Now we have the conclusion with $T = (x:T_1) \rarr C_1$.
        \end{description}
        \item
        \begin{description}
            \item[Case $c = \expret{v^T}$:]
                We have $\cps{c} = \Lambda \alpha. \lambda h:\{\}. \lambda k:\cps{T} \rarr \alpha. k~\cps{v}$~.
                By Lemma \ref{lem:cps:inv-c}, we have
                \begin{itemize}
                    \item $\jdty{\Gamma, \alpha, h:\{\}, k:\cps{T} \rarr \alpha}{k~\cps{v}}{\tau'}$ and
                    \item $\jdsub{\Gamma}{\forall \alpha. \{\} \rarr (\cps{T} \rarr \alpha) \rarr \tau'}{\tau}$
                \end{itemize}
                for some $\tau'$.
                By Lemma \ref{lem:cps:inv}, we have
                \begin{itemize}
                    \item $\jdty{\Gamma, \alpha, h:\{\}, k:\cps{T} \rarr \alpha}{k}{(y:\tau_1) \rarr \tau_2}$,
                    \item $\jdty{\Gamma, \alpha, h:\{\}, k:\cps{T} \rarr \alpha}{\cps{v}}{\tau_1}$, and
                    \item $\jdsub{\Gamma, \alpha, h:\{\}, k:\cps{T} \rarr \alpha}{\tau_2[\cps{v}/y]}{\tau'}$
                \end{itemize}
                for some $y, \tau_1$, and $\tau_2$.
                By Lemma \ref{lem:cps:rm-nonrfn}, we have
                \begin{itemize}
                    \item $\jdty{\Gamma, \alpha}{\cps{v}}{\tau_1}$~.
                \end{itemize}
                Then, by the IH, we have
                \begin{itemize}
                    \item $\jdty{\cpsinv{\Gamma}}{v}{T}$ and
                    \item $\jdsub{\Gamma}{\cps{T}}{\tau_1}$~.
                \end{itemize}
                Therefore, by \rulename{T-Ret}, we have
                \begin{itemize}
                    \item $\jdty{\cpsinv{\Gamma}}{\expret{v}}{\tycomp{\{\}}{T}{\square}}$~.
                \end{itemize}
                On the other hand, by Lemma \ref{lem:cps:inv}, we have
                \begin{itemize}
                    \item $\jdsub{\Gamma, \alpha, h:\{\}, k:\cps{T} \rarr \alpha}{\cps{T} \rarr \alpha}{(y:\tau_1) \rarr \tau_2}$~.
                \end{itemize}
                Then, By inversion, we have $\tau_2 = \alpha$.
                By inversion again, we have $\tau' = \alpha$.
                Therefore, we have
                \begin{itemize}
                    \item $\jdsub{\Gamma}{\forall \alpha. \{\} \rarr (\cps{T} \rarr \alpha) \rarr \alpha}{\tau}$,
                \end{itemize}
                that is,
                \begin{itemize}
                    \item $\jdsub{\Gamma}{\cps{\tycomp{\{\}}{T}{\square}}}{\tau}$~.
                \end{itemize}
                Now we have the conclusion with $C = \tycomp{\{\}}{T}{\square}$.
            \item[Case $c = \explet{x}{c_1^{\tycomp{\Sigma}{T_1}{\square}}}{c_2^{\tycomp{\Sigma}{T_2}{\square}}}$:]
                \def\currentprefix{cps-bw:let-pure}
                We have $\cps{c} = \Lambda \alpha. \lambda h: \cps{\Sigma}. \lambda k:\cps{T_2} \rarr \alpha.
                \cps{c_1}~\alpha~h~(\lambda x:\cps{T_1}. \cps{c_2}~\alpha~h~k)$~.
                By Lemma \ref{lem:cps:inv-c}, we have
                \begin{enumrm}
                    \item\llabel{ty-body} $\jdty{\Gamma, \alpha, h:\cps{\Sigma}, k:\cps{T_2} \rarr \alpha}
                        {\cps{c_1}~\alpha~h~(\lambda x:\cps{T_1}. \cps{c_2}~\alpha~h~k)}{\tau'}$ and
                    \item\llabel{sub-t} $\jdsub{\Gamma}{\forall \alpha. \cps{\Sigma} \rarr (\cps{T_2} \rarr \alpha) \rarr \tau'}{\tau}$
                \end{enumrm}
                for some $\tau'$.
                By Lemma \ref{lem:cps:inv-c-app} with \lref{ty-body}, we have
                \begin{enumrm}[resume]
                    \item\llabel{ty-cpsc1} $\jdty{\Gamma, \alpha, h:\cps{\Sigma}, k:\cps{T_2} \rarr \alpha}
                        {\cps{c_1}}{\tau''}$,
                    \item\llabel{ty-h} $\jdty{\Gamma, \alpha, h:\cps{\Sigma}, k:\cps{T_2} \rarr \alpha}{h}{\tau_1}$, and
                    \item\llabel{ty-fun} $\jdty{\Gamma, \alpha, h:\cps{\Sigma}, k:\cps{T_2} \rarr \alpha}
                        {\lambda x:\cps{T_1}. \cps{c_2}~\alpha~h~k}{\tau_2}$
                \end{enumrm}
                for some $\tau'', \tau_1$ and $\tau_2$.
                By Lemma \ref{lem:cps:inv} with \lref{ty-h} and \lref{ty-fun} respectively, we have
                \begin{enumrm}[resume]
                    \item $\jdsub{\Gamma, \alpha, h:\cps{\Sigma}, k:\cps{T_2} \rarr \alpha}{\cps{\Sigma}}{\tau_1}$,
                    \item\llabel{ty-body-2} $\jdty{\Gamma, \alpha, h:\cps{\Sigma}, k:\cps{T_2} \rarr \alpha, x:\cps{T_1}}
                        {\cps{c_2}~\alpha~h~k}{\tau_3}$, and
                    \item $\jdsub{\Gamma, \alpha, h:\cps{\Sigma}, k:\cps{T_2} \rarr \alpha}
                        {(x:\cps{T_1}) \rarr \tau_3}{\tau_2}$
                \end{enumrm}
                for some $\tau_3$.
                Then, by the second half of Lemma \ref{lem:cps:inv-c-app}, we have
                \begin{enumrm}[resume]
                    \item\llabel{sub-t''} $\jdsub{\Gamma, \alpha, h:\cps{\Sigma}, k:\cps{T_2} \rarr \alpha}
                        {\tau''}{\forall \beta. \cps{\Sigma} \rarr ((x:\cps{T_1}) \rarr \tau_3) \rarr \tau'}$
                \end{enumrm}
                where $\beta$ is fresh.

                On the other hand, by Lemma \ref{lem:cps:inv-c-app} with \lref{ty-body-2}, we have
                \begin{enumrm}[resume]
                    \item\llabel{ty-cpsc2} $\jdty{\Gamma, \alpha, h:\cps{\Sigma}, k:\cps{T_2} \rarr \alpha, x:\cps{T_1}}
                        {\cps{c_2}}{\tau_3'}$,
                    \item\llabel{ty-h-2} $\jdty{\Gamma, \alpha, h:\cps{\Sigma}, k:\cps{T_2} \rarr \alpha, x:\cps{T_1}}
                        {h}{\tau_4}$, and
                    \item\llabel{ty-k} $\jdty{\Gamma, \alpha, h:\cps{\Sigma}, k:\cps{T_2} \rarr \alpha, x:\cps{T_1}}
                        {k}{\tau_5}$
                \end{enumrm}
                for some $\tau_3', \tau_4$ and $\tau_5$.
                By Lemma \ref{lem:cps:inv} with \lref{ty-h-2} and \lref{ty-k} respectively, we have
                \begin{itemize}
                    \item $\jdsub{\Gamma, \alpha, h:\cps{\Sigma}, k:\cps{T_2} \rarr \alpha, x:\cps{T_1}}
                        {\cps{\Sigma}}{\tau_4}$ and
                    \item $\jdsub{\Gamma, \alpha, h:\cps{\Sigma}, k:\cps{T_2} \rarr \alpha, x:\cps{T_1}}
                        {\cps{T_2} \rarr \alpha}{\tau_5}$~.
                \end{itemize}
                Then, by the second half of Lemma \ref{lem:cps:inv-c-app}, we have
                \begin{enumrm}[resume]
                    \item\llabel{sub-t3'} $\jdsub{\Gamma, \alpha, h:\cps{\Sigma}, k:\cps{T_2} \rarr \alpha, x:\cps{T_1}}
                        {\tau_3'}{\forall \gamma. \cps{\Sigma} \rarr (\cps{T_2} \rarr \alpha) \rarr \tau_3}$
                \end{enumrm}
                where $\gamma$ is fresh.

                By Lemma \ref{lem:cps:rm-nonrfn}
                with \lref{ty-cpsc1}, \lref{sub-t''}, \lref{ty-cpsc2} and \lref{sub-t3'}, we have
                \begin{enumrm}[resume]
                    \item\llabel{ty-cpsc1-2} $\jdty{\Gamma, \alpha}{\cps{c_1}}{\tau''}$,
                    \item\llabel{sub-t''-2} $\jdsub{\Gamma, \alpha}
                        {\tau''}{\forall \beta. \cps{\Sigma} \rarr ((x:\cps{T_1}) \rarr \tau_3) \rarr \tau'}$
                    \item\llabel{ty-cpsc2-2} $\jdty{\Gamma, \alpha, x:\cps{T_1}}{\cps{c_2}}{\tau_3'}$
                    \item\llabel{sub-t3'-2} $\jdsub{\Gamma, \alpha, x:\cps{T_1}}
                        {\tau_3'}{\forall \gamma. \cps{\Sigma} \rarr (\cps{T_2} \rarr \alpha) \rarr \tau_3}$~.
                \end{enumrm}
                Then, by the IHs of \lref{ty-cpsc1-2} and \lref{ty-cpsc2-2} respectively, we have
                \begin{enumrm}[resume]
                    \item\llabel{ty-c1} $\jdty{\cpsinv{\Gamma}}{c_1}{C_1}$,
                    \item\llabel{sub-cpsC1} $\jdsub{\Gamma, \alpha}{\cps{C_1}}{\tau''}$,
                    \item\llabel{ty-c2} $\jdty{\cpsinv{\Gamma}, x:T_1}{c_2}{C_2}$, and
                    \item\llabel{sub-cpsC2} $\jdsub{\Gamma, \alpha, x:\cps{T_1}}{\cps{C_2}}{\tau_3'}$
                \end{enumrm}
                for some $C_1$ and $C_2$.
                By Lemma \ref{lem:cps:trans} with ``\lref{sub-t''-2} and \lref{sub-cpsC1}''
                and ``\lref{sub-t3'-2} and \lref{sub-cpsC2}'' respectively, we have
                \begin{enumrm}[resume]
                    \item\llabel{sub-cpsC1-2} $\jdsub{\Gamma, \alpha}
                        {\cps{C_1}}{\forall \beta. \cps{\Sigma} \rarr ((x:\cps{T_1}) \rarr \tau_3) \rarr \tau'}$ and
                    \item\llabel{sub-cpsC2-2} $\jdsub{\Gamma, \alpha, x:\cps{T_1}}
                        {\cps{C_2}}{\forall \gamma. \cps{\Sigma} \rarr (\cps{T_2} \rarr \alpha) \rarr \tau_3}$~.
                \end{enumrm}
                By Lemma \ref{lem:cps:sub-pure} with \lref{sub-cpsC2-2}, we have
                \begin{itemize}
                    \item $C_1 = \tycomp{\Sigma_{11}}{T_{11}}{\square}$ and
                    \item $\tau_3 = \alpha$
                \end{itemize}
                for some $\Sigma_{11}$ and $T_{11}$.
                Then, by Lemma \ref{lem:cps:sub-pure} again with \lref{sub-cpsC1-2}, we have
                \begin{itemize}
                    \item $C_2 = \tycomp{\Sigma_{22}}{T_{22}}{\square}$ and
                    \item $\tau' = \alpha$
                \end{itemize}
                for some $\Sigma_{22}$ and $T_{22}$.
                By inversion of \lref{sub-cpsC1-2}, we have
                \begin{itemize}
                    \item $\jdsub{\Gamma, \alpha, \beta}{\cps{\Sigma}}{\cps{\Sigma_{11}}}$ and
                    \item $\jdsub{\Gamma, \alpha, h:\cps{\Sigma}, \beta}{\cps{T_{11}}}{\cps{T_1}}$~.
                \end{itemize}
                By Lemma \ref{lem:cps:rm-nonrfn}, \ref{lem:cps:rm-unused-tvar} and \ref{lem:cps:cpsinv-sub}, we have
                \begin{itemize}
                    \item $\jdsub{\cpsinv{\Gamma}}{\Sigma}{\Sigma_{11}}$ and
                    \item $\jdsub{\cpsinv{\Gamma}}{T_{11}}{T_1}$~.
                \end{itemize}
                Then, by subsumption on \lref{ty-c1}, we have
                \begin{enumrm}[resume]
                    \item\llabel{ty-c1-2} $\jdty{\cpsinv{\Gamma}}{c_1}{\tycomp{\Sigma}{T_1}{\square}}$~.
                \end{enumrm}
                In the same way, from \lref{sub-cpsC2-2}, we have
                \begin{enumrm}[resume]
                    \item\llabel{ty-c2-2} $\jdty{\cpsinv{\Gamma}, x:T_1}{c_2}{\tycomp{\Sigma}{T_2}{\square}}$~.
                \end{enumrm}
                Therefore, by \rulename{T-LetP} with \lref{ty-c1-2} and \lref{ty-c2-2}, we have
                \[
                    \jdty{\cpsinv{\Gamma}}{\explet{x}{c_1}{c_2}}{\tycomp{\Sigma}{T_2}{\square}}~.
                \]
                Also, since $\tau' = \alpha$, \lref{sub-t} implies
                \[
                    \jdsub{\Gamma}{\cps{\tycomp{\Sigma}{T_2}{\square}}}{\tau}~.
                \]
                Now we have the conclusion with $C = \tycomp{\Sigma}{T_2}{\square}$.
            \item[Case $c = \explet{x}{c_1^{\tycomp{\Sigma}{T_1}{\tyctl{x}{C_1}{C_2}}}}{c_2^{\tycomp{\Sigma}{T_2}{\tyctl{z}{C_0}{C_1}}}}$:]
                \def\currentprefix{cps-bw:let}
                We have $\cps{c} = \Lambda \alpha. \lambda h: \cps{\Sigma}. \lambda k:(z:\cps{T_2}) \rarr \cps{C_0}.
                    \cps{c_1}~\cps{C_2}~h~(\lambda x:\cps{T_1}. \cps{c_2}~\cps{C_1}~h~k)$~.
                In the similar way to the previous case, we have
                \begin{enumrm}
                    \item\llabel{sub-t} $\jdsub{\Gamma}
                        {\forall \alpha. \cps{\Sigma} \rarr ((z:\cps{T_2}) \rarr \cps{C_0}) \rarr \tau'}{\tau}$,
                    \item\llabel{ty-c1} $\jdty{\cpsinv{\Gamma}}{c_1}{C_1}$,
                    \item\llabel{sub-cpsC1} $\jdsub{\Gamma, \alpha}{\cps{C_1}}
                        {\forall \beta. \cps{\Sigma} \rarr ((x:\cps{T_1}) \rarr \tau_3) \rarr \tau'}$,
                    \item\llabel{ty-c2} $\jdty{\cpsinv{\Gamma}, x:T_1}{c_2}{C_2}$, and
                    \item\llabel{sub-cpsC2} $\jdsub{\Gamma, \alpha, x:\cps{T_1}}{\cps{C_2}}
                        {\forall \gamma. \cps{\Sigma} \rarr ((z:\cps{T_2}) \rarr \cps{C_0}) \rarr \tau_3}$
                \end{enumrm}
                for some $\tau', \tau_3, C_1$ and $C_2$.
                By Lemma \ref{lem:assume-atm}, we can assume that
                \begin{itemize}
                    \item $C_1 = \tycomp{\Sigma_1}{T_{10}}{\tyctl{x_1}{C_{11}}{C_{12}}}$ and
                    \item $C_2 = \tycomp{\Sigma_2}{T_{20}}{\tyctl{x_2}{C_{21}}{C_{22}}}$
                \end{itemize}
                for some $\Sigma_1, T_{10}, x_1, C_{11}, C_{12}, \Sigma_2, T_{20}, x_2, C_{21}$ and $C_{22}$.
                Then, by inversion of \lref{sub-cpsC1}, we have
                \begin{enumrm}[resume]
                    \item $x_1 = x$
                    \item\llabel{sub-cpssig} $\jdsub{\Gamma, \alpha, \beta}{\cps{\Sigma}}{\cps{\Sigma_1}}$,
                    \item\llabel{sub-cpsT10} $\jdsub{\Gamma, \alpha, \beta, h:\cps{\Sigma}}
                        {\cps{T_{10}}}{\cps{T_1}}$,
                    \item\llabel{sub-t3} $\jdsub{\Gamma, \alpha, \beta, h:\cps{\Sigma}, x:\cps{T_{10}}}
                        {\tau_3}{\cps{C_{11}}}$, and
                    \item\llabel{sub-cpsC12} $\jdsub{\Gamma, \alpha, \beta, h:\cps{\Sigma}, k:(x:\cps{T_1}) \rarr \tau_3}
                        {\cps{C_{12}}}{\tau'}$~.
                \end{enumrm}
                By Lemma \ref{lem:cps:rm-nonrfn}, \ref{lem:cps:rm-unused-tvar} and \ref{lem:cps:cpsinv-sub}
                with \lref{sub-cpssig} and \lref{sub-cpsT10} respectively, we have
                \begin{enumrm}[resume]
                    \item\llabel{sub-sig} $\jdsub{\cpsinv{\Gamma}}{\Sigma}{\Sigma_1}$ and
                    \item\llabel{sub-T10} $\jdsub{\cpsinv{\Gamma}}{T_{10}}{T_1}$~.
                \end{enumrm}
                By subsumption on \lref{ty-c1} with \lref{sub-sig}, we have
                \begin{enumrm}[resume]
                    \item\llabel{ty-c1-2} $\jdty{\cpsinv{\Gamma}}{c_1}{\tycomp{\Sigma}{T_{10}}{\tyctl{x_1}{C_{11}}{C_{12}}}}$~.
                \end{enumrm}
                On the other hand, by inversion of \lref{sub-cpsC2}, we have
                \begin{itemize}
                    \item $x_2 = z$
                    \item $\jdsub{\Gamma, \alpha, x: \cps{T_1}, \gamma}
                        {\cps{\Sigma}}{\cps{\Sigma_2}}$,
                    \item $\jdsub{\Gamma, \alpha, x: \cps{T_1}, \gamma, h:\cps{\Sigma}}
                        {\cps{T_{20}}}{\cps{T_2}}$,
                    \item $\jdsub{\Gamma, \alpha, x: \cps{T_1}, \gamma, h:\cps{\Sigma}, z:\cps{T_{20}}}
                        {\cps{C_0}}{\cps{C_{21}}}$, and
                    \item $\jdsub{\Gamma, \alpha, x: \cps{T_1}, \gamma, h:\cps{\Sigma}, k:(z:\cps{T_2}) \rarr \cps{C_0}}
                        {\cps{C_{22}}}{\tau_3}$~.
                \end{itemize}
                By Lemma \ref{lem:cps:narrow} with \lref{sub-cpsT10}, we have
                \begin{enumrm}[resume]
                    \item\llabel{sub-cpssig-2} $\jdsub{\Gamma, \alpha, x: \cps{T_{10}}, \gamma}
                        {\cps{\Sigma}}{\cps{\Sigma_2}}$,
                    \item\llabel{sub-cpsT20} $\jdsub{\Gamma, \alpha, x: \cps{T_{10}}, \gamma, h:\cps{\Sigma}}
                        {\cps{T_{20}}}{\cps{T_2}}$,
                    \item\llabel{sub-cpsC0} $\jdsub{\Gamma, \alpha, x: \cps{T_{10}}, \gamma, h:\cps{\Sigma}, z:\cps{T_{20}}}
                        {\cps{C_0}}{\cps{C_{21}}}$, and
                    \item\llabel{sub-cpsC22} $\jdsub{\Gamma, \alpha, x: \cps{T_{10}}, \gamma, h:\cps{\Sigma}, k:(z:\cps{T_2}) \rarr \cps{C_0}}
                        {\cps{C_{22}}}{\tau_3}$~.
                \end{enumrm}
                By Lemma \ref{lem:cps:rm-nonrfn}, \ref{lem:cps:rm-unused-tvar} and \ref{lem:cps:trans}
                with \lref{sub-t3} and \lref{sub-cpsC22}, we have
                \begin{enumrm}[resume]
                    \item\llabel{sub-cpsC22-2} $\jdsub{\Gamma, \alpha, x: \cps{T_{10}}}{\cps{C_{22}}}{\cps{C_{11}}}$~.
                \end{enumrm}
                By Lemma \ref{lem:cps:rm-nonrfn}, \ref{lem:cps:rm-unused-tvar} and \ref{lem:cps:cpsinv-sub}
                with \lref{sub-cpssig-2}, \lref{sub-cpsT20}, \lref{sub-cpsC0} and \lref{sub-cpsC22-2},
                we have
                \begin{itemize}
                    \item $\jdsub{\cpsinv{\Gamma}, x: T_{10}}{\Sigma}{\Sigma_2}$,
                    \item $\jdsub{\cpsinv{\Gamma}, x: T_{10}}{T_{20}}{T_2}$,
                    \item $\jdsub{\cpsinv{\Gamma}, x: T_{10}, z:T_{20}}{C_0}{C_{21}}$, and
                    \item $\jdsub{\cpsinv{\Gamma}, x: T_{10}}{C_{22}}{C_{11}}$~.
                \end{itemize}
                Then, by Lemma \ref{lem:cps:narrow} and subsumption on \lref{ty-c2}, we have
                \begin{enumrm}[resume]
                    \item\llabel{ty-c2-2} $\jdty{\cpsinv{\Gamma}, x:T_{10}}{c_2}
                        {\tycomp{\Sigma}{T_2}{\tyctl{z}{C_0}{C_{11}}}}$~.
                \end{enumrm}
                Therefore, by \rulename{T-LetIp}, we have
                \[
                    \jdty{\cpsinv{\Gamma}}{\explet{x}{c_1}{c_2}}{\tycomp{\Sigma}{T_2}{\tyctl{z}{C_0}{C_{12}}}}~.
                \]

                Also, by Lemma \ref{lem:cps:rm-nonrfn}, \ref{lem:cps:rm-unused-tvar}, \ref{lem:cps:weaken}, and \ref{lem:cps:trans}
                with \lref{sub-t} and \lref{sub-cpsC12}, we have
                \begin{itemize}
                    \item $\jdsub{\Gamma}
                    {\forall \alpha. \cps{\Sigma} \rarr ((z:\cps{T_2}) \rarr \cps{C_0}) \rarr \cps{C_{12}}}{\tau}$,
                \end{itemize}
                that is,
                \[
                    \jdsub{\Gamma}
                    {\cps{\tycomp{\Sigma}{T_2}{\tyctl{z}{C_0}{C_{12}}}}}{\tau}~.
                \]
                Now we have the conclusion with $C = \tycomp{\Sigma}{T_2}{\tyctl{z}{C_0}{C_{12}}}$~.
            \item[Case $c = v_1~v_2$:]
                We have $\cps{c} = \cps{v_1}~\cps{v_2}$~.
                By Lemma \ref{lem:cps:inv}, we have
                \begin{itemize}
                    \item $\jdty{\Gamma}{\cps{v_1}}{(x: \tau_1) \rarr \tau_2}$,
                    \item $\jdty{\Gamma}{\cps{v_2}}{\tau_1}$, and
                    \item $\jdsub{\Gamma}{\tau_2[\cps{v_2}/x]}{\tau}$
                \end{itemize}
                for some $x, \tau_1$ and $\tau_2$.
                By the IHs, we have
                \begin{itemize}
                    \item $\jdty{\cpsinv{\Gamma}}{v_1}{T_1}$,
                    \item $\jdty{\cpsinv{\Gamma}}{v_2}{T_2}$,
                    \item $\jdsub{\Gamma}{\cps{T_1}}{(x: \tau_1) \rarr \tau_2}$, and
                    \item $\jdsub{\Gamma}{\cps{T_2}}{\tau_1}$
                \end{itemize}
                for some $T_1$ and $T_2$.
                By inversion, we have
                \begin{itemize}
                    \item $T_1 = (x: T_{11}) \rarr C_{12}$,
                    \item $\jdsub{\Gamma}{\tau_1}{\cps{T_{11}}}$, and
                    \item $\jdsub{\Gamma, x: \tau_1}{\cps{C_{12}}}{\tau_2}$
                \end{itemize}
                for some $T_{11}$ and $C_{12}$.
                By Lemma \ref{lem:cps:trans}, we have $\jdsub{\Gamma}{\cps{T_2}}{\cps{T_{11}}}$~.
                Then, by Lemma \ref{lem:cps:cpsinv-sub},
                we have $\jdsub{\cpsinv{\Gamma}}{T_2}{T_{11}}$,
                and hence by \rulename{T-VSub} we have $\jdty{\cpsinv{\Gamma}}{v_2}{T_{11}}$~.
                Therefore, by \rulename{T-App},
                we have $\jdty{\cpsinv{\Gamma}}{v_1~v2}{C_{12}[v_2/x]}$~.
                
                On the other hand, by Lemma \ref{lem:cps:subst},
                we have $\jdsub{\Gamma}{\cps{C_{12}}[\cps{v_2}/x]}{\tau_2[\cps{v_2}/x]}$~.
                Then, by Lemma \ref{lem:cps:trans},
                we have $\jdsub{\Gamma}{\cps{C_{12}}[\cps{v_2}/x]}{\tau}$~.

                Now we have the conclusion with $C = C_{12}[v_2/x]$~.
            \item[Case $c = (\expif{v}{c_1}{c_2})^{C'}$:]
                We have $\cps{c} = (\expif{\cps{v}}{\cps{c_1}}{\cps{c_2}} : \cps{C'})$~.
                By Lemma \ref{lem:cps:inv}, we have
                \begin{itemize}
                    \item $\jdty{\Gamma}{\expif{\cps{v}}{\cps{c_1}}{\cps{c_2}}}{\cps{C'}}$ and
                    \item $\jdsub{\Gamma}{\cps{C'}}{\tau}$~.
                \end{itemize}
                By Lemma \ref{lem:cps:inv} again, we have
                \begin{itemize}
                    \item $\jdty{\Gamma}{\cps{v}}{\tyrfn{z}{\tybool}{\phi}}$,
                    \item $\jdty{\Gamma, \cps{v} = \exptrue}{\cps{c_1}}{\tau'}$,
                    \item $\jdty{\Gamma, \cps{v} = \expfalse}{\cps{c_2}}{\tau'}$, and
                    \item $\jdsub{\Gamma}{\tau'}{\cps{C'}}$
                \end{itemize}
                for some $z, \phi$ and $\tau'$.
                By the IHs, we have
                \begin{itemize}
                    \item $\jdty{\cpsinv{\Gamma}}{v}{\tyrfn{z}{\tybool}{\phi}}$,
                    \item $\jdty{\cpsinv{\Gamma}, v = \exptrue}{c_1}{C_1}$,
                    \item $\jdty{\cpsinv{\Gamma}, v = \expfalse}{c_2}{C_2}$,
                    \item $\jdsub{\Gamma, \cps{v} = \exptrue}{\cps{C_1}}{\tau'}$, and
                    \item $\jdsub{\Gamma, \cps{v} = \expfalse}{\cps{C_2}}{\tau'}$
                \end{itemize}
                for some $C_1$ and $C_2$.
                (Note that since $v$ is of a refinement type, it holds that $\cps{v} = v$.)
                By Lemma \ref{lem:cps:weaken} and \ref{lem:cps:trans}, we have
                \begin{itemize}
                    \item $\jdsub{\Gamma, \cps{v} = \exptrue}{\cps{C_1}}{\cps{C'}}$ and
                    \item $\jdsub{\Gamma, \cps{v} = \expfalse}{\cps{C_2}}{\cps{C'}}$~.
                \end{itemize}
                By Lemma \ref{lem:cps:cpsinv-sub}, we have
                \begin{itemize}
                    \item $\jdsub{\cpsinv{\Gamma}, v = \exptrue}{C_1}{C'}$ and
                    \item $\jdsub{\cpsinv{\Gamma}, v = \expfalse}{C_2}{C'}$~.
                \end{itemize}
                Then, by \rulename{T-CSub}, we have
                \begin{itemize}
                    \item $\jdty{\cpsinv{\Gamma}, v = \exptrue}{c_1}{C'}$ and
                    \item $\jdty{\cpsinv{\Gamma}, v = \expfalse}{c_2}{C'}$~.
                \end{itemize}
                Therefore by \rulename{T-If}, we have
                $\jdty{\cpsinv{\Gamma}}{\expif{v}{c_1}{c_2}}{C'}$~.
                Now we have the conclusion with $C = C'$~.
            \item[Case $c = (\op^{\rep{\mathit{A}}}~v)^{\tycomp{\Sigma}{T}{\tyctl{y}{C_1}{C_2}}}$:]
                \def\currentprefix{cpsbw:op}
                We have $\cps{c} = \Lambda \alpha. \lambda h:\cps{\Sigma}. \lambda k:(y: \cps{T}) \rarr \cps{C_1}.
                    h\#\op~\rep{A}~\cps{v}~(\lambda y':\cps{T}. k~y')$~.
                By Lemma \ref{lem:cps:inv-c}, we have
                \begin{enumrm}
                    \item\llabel{ty-body} $\jdty{\Gamma, \alpha, h:\cps{\Sigma}, k:(y: \cps{T}) \rarr \cps{C_1}}
                        {h\#\op~\rep{A}~\cps{v}~(\lambda y':\cps{T}. k~y')}{\tau'}$ and
                    \item\llabel{sub-t} $\jdsub{\Gamma}{\forall \alpha. \cps{\Sigma} \rarr ((y: \cps{T}) \rarr \cps{C_1}) \rarr \tau'}{\tau}$
                \end{enumrm}
                for some $\tau'$.
                (Below, we write $\Gamma_{\alpha,h,k}$ for $\Gamma, \alpha, h:\cps{\Sigma}, k:(y: \cps{T}) \rarr \cps{C_1}$~.)

                By Lemma \ref{lem:cps:inv} with \lref{ty-body}, we have
                \begin{enumrm}[resume]
                    \item\llabel{ty-fun} $\jdty{\Gamma_{\alpha,h,k}}{\lambda y':\cps{T}. k~y'}{\tau_1}$,
                    \item\llabel{ty-cpsv} $\jdty{\Gamma_{\alpha,h,k}}{\cps{v}}{\tau_3}$,
                    \item\llabel{wf-A} $\rep{\jdty{\Gamma_{\alpha,h,k}}{A}{\rep{B}}}$,
                    \item\llabel{sub-cpssig} $\jdsub{\Gamma_{\alpha,h,k}}{\cps{\Sigma}}{\{\ldots, \forall \rep{X:\rep{B}}. \tau_5, \ldots\}}$,
                    \item\llabel{sub-t5} $\jdsub{\Gamma_{\alpha,h,k}}{\tau_5[\rep{A/X}]}{(x: \tau_3) \rarr \tau_4}$,
                    \item\llabel{sub-t4} $\jdsub{\Gamma_{\alpha,h,k}}{\tau_4[\cps{v}/x]}{\tau_1 \rarr \tau_2}$, and
                    \item\llabel{sub-t2} $\jdsub{\Gamma_{\alpha,h,k}}{\tau_2}{\tau'}$~.
                \end{enumrm}
                By Assumption \ref{asm:cps:formula} and \ref{asm:cps:formula-cpsinv} with \lref{wf-A}, we have
                \begin{itemize}
                    \item $\rep{\jdty{\cpsinv{\Gamma}}{A}{\rep{B}}}$~.
                \end{itemize}

                By inversion of \lref{sub-cpssig}, we have
                \begin{itemize}
                    \item $\Sigma = \{\ldots, \op: \forall \rep{X:\rep{B}}.
                        (x_{\op}: T_{\op 1}) \rarr ((y_{\op}: T_{\op 2}) \rarr C_{\op 1}) \rarr C_{\op 2}, \ldots\}$ and
                    \item $\jdsub{\Gamma_{\alpha,h,k}, \rep{X:\rep{B}}}
                        {(x_{\op}: \cps{T_{\op 1}}) \rarr ((y_{\op}: \cps{T_{\op 2}}) \rarr \cps{C_{\op 1}}) \rarr \cps{C_{\op 2}}}{\tau_5}$~.
                \end{itemize}
                By repeatedly inverting this subtyping judgment
                with applying Lemma \ref{lem:cps:subst-pred} with \lref{wf-A},
                Lemma \ref{lem:cps:subst} with \lref{ty-cpsv},
                and Lemma \ref{lem:cps:trans} with \lref{sub-t5}, \lref{sub-t4} and \lref{sub-t2},
                we have
                \begin{enumrm}[resume]
                    \item $x = x_{\op}$,
                    \item\llabel{sub-t3} $\jdsub{\Gamma_{\alpha,h,k}}{\tau_3}{\cps{T_{\op 1}}[\rep{A/X}]}$,
                    \item\llabel{sub-t1} $\jdsub{\Gamma_{\alpha,h,k}}
                        {\tau_1}{(y_{\op}:\cps{T_{\op 2}}[\rep{A/X}][\cps{v}/x]) \rarr \cps{C_{\op 1}}[\rep{A/X}][\cps{v}/x]}$, and
                    \item\llabel{sub-cpsCop2} $\jdsub{\Gamma_{\alpha,h,k}}
                        {\cps{C_{\op 2}}[\rep{A/X}][\cps{v}/x]}{\tau'}$~.
                \end{enumrm}

                By Lemma \ref{lem:cps:rm-nonrfn} with \lref{ty-cpsv}, we have
                \begin{itemize}
                    \item $\jdty{\Gamma, \alpha}{\cps{v}}{\tau_3}$~.
                \end{itemize}
                Then, by the IH, we have
                \begin{enumrm}[resume]
                    \item\llabel{ty-v} $\jdty{\cpsinv{\Gamma}}{v}{T_v}$ and
                    \item\llabel{sub-cpsTv} $\jdsub{\Gamma, \alpha}{\cps{T_v}}{\tau_3}$
                \end{enumrm}
                for some $T_v$.
                By Lemma \ref{lem:cps:trans} with \lref{sub-t3} and \lref{sub-cpsTv}
                (using Lemma \ref{lem:cps:rm-nonrfn}), we have
                \begin{itemize}
                    \item $\jdsub{\Gamma, \alpha}{\cps{T_v}}{\cps{T_{\op 1}}[\rep{A/X}]}$~.
                \end{itemize}
                Then, by Lemma \ref{lem:cps:cpsinv-sub}, we have
                \begin{itemize}
                    \item $\jdsub{\cpsinv{\Gamma}}{T_v}{T_{\op 1}[\rep{A/X}]}$
                \end{itemize}
                and hence, by \rulename{T-VSub} with \lref{ty-v}, we have
                \begin{itemize}
                    \item $\jdty{\cpsinv{\Gamma}}{v}{T_{\op 1}[\rep{A/X}]}$~.
                \end{itemize}

                Also, by Lemma~\ref{lem:cps:wfg} and inversion, we have
                $\jdwf{\Gamma, \alpha}{\cps{\Sigma}}$.
                Then by Lemma~\ref{lem:cps:presv-b-wf}, we have
                $\jdwf{\cpsinv{\Gamma}}{\Sigma}$.

                Therefore, by \rulename{T-Op}, we have
                \[
                    \jdty{\cpsinv{\Gamma}}{\op~v}{\tycomp{\Sigma}{T_{\op 2}[\rep{A/X}][v/x]}{\tyctl{y_{\op}}{C_{\op 1}[\rep{A/X}][v/x]}{C_{\op 2}[\rep{A/X}][v/x]}}}~.
                \]

                On the other hand, by Lemma~\ref{lem:cps:inv} with \lref{ty-fun}, we have
                \begin{enumrm}[resume]
                    \item\llabel{sub-cpsT-t6} $\jdsub{\Gamma_{\alpha,h,k}}{(y':\cps{T}) \rarr \tau_6}{\tau_1}$,
                    \item\llabel{sub-t8} $\jdsub{\Gamma_{\alpha,h,k}, y':\cps{T}}{\tau_8[y'/y_0]}{\tau_6}$,
                    \item\llabel{sub-cpsT-cpsC1} $\jdsub{\Gamma_{\alpha,h,k}, y':\cps{T}}{(y:\cps{T}) \rarr \cps{C_1}}{(y_0:\tau_7) \rarr \tau_8}$, and
                    \item\llabel{ty-y'} $\jdty{\Gamma_{\alpha,h,k}, y':\cps{T}}{y'}{\tau_7}$~.
                \end{enumrm}
                By inversion of \lref{sub-cpsT-cpsC1}, we have
                \begin{itemize}
                    \item $y = y_0$ and
                    \item $\jdsub{\Gamma_{\alpha,h,k}, y':\cps{T}, y:\tau_7}{\cps{C_1}}{\tau_8}$~.
                \end{itemize}
                Then, by Lemma~\ref{lem:cps:subst} with \lref{ty-y'}, we have
                \begin{itemize}
                    \item $\jdsub{\Gamma_{\alpha,h,k}, y':\cps{T}}{\cps{C_1}[y'/y]}{\tau_8[y'/y]}$~.
                \end{itemize}
                Then, by Lemma~\ref{lem:cps:trans} with \lref{sub-t8}, we have
                \begin{itemize}
                    \item $\jdsub{\Gamma_{\alpha,h,k}, y':\cps{T}}{\cps{C_1}[y'/y]}{\tau_6}$
                \end{itemize}
                (Note that $y = y_0$).
                Then by \rulename{Sc-Fun}, we have
                \begin{itemize}
                    \item $\jdsub{\Gamma_{\alpha,h,k}}{(y':\cps{T}) \rarr \cps{C_1}[y'/y]}{(y':\cps{T}) \rarr \tau_6}$
                \end{itemize}
                and by $\alpha$-renaming we have
                \begin{itemize}
                    \item $\jdsub{\Gamma_{\alpha,h,k}}{(y:\cps{T}) \rarr \cps{C_1}}{(y':\cps{T}) \rarr \tau_6}$~.
                \end{itemize}
                Then, by Lemma~\ref{lem:cps:trans} with \lref{sub-cpsT-t6} and \lref{sub-t1}, we have
                \begin{enumrm}[resume]
                    \item\llabel{sub-cpsT-cpsC1-2} $\jdsub{\Gamma_{\alpha,h,k}}{(y:\cps{T}) \rarr \cps{C_1}}{(y_{\op}:\cps{T_{\op 2}}[\rep{A/X}][\cps{v}/x]) \rarr \cps{C_{\op 1}}[\rep{A/X}][\cps{v}/x]}$~.
                \end{enumrm}

                Therefore, by some subtyping rules with \lref{sub-cpsT-cpsC1-2} and \lref{sub-cpsCop2}, we have
                \begin{itemize}
                    \item $\jdsub{\Gamma}
                        {\forall \alpha. \cps{\Sigma} \rarr ((y_{\op}:\cps{T_{\op 2}}[\rep{A/X}][\cps{v}/x]) \rarr \cps{C_{\op 1}}[\rep{A/X}][\cps{v}/x]) \rarr \cps{C_{\op 2}}[\rep{A/X}][\cps{v}/x]}
                        {\forall \alpha. \cps{\Sigma} \rarr ((y: \cps{T}) \rarr \cps{C_1}) \rarr \tau'}$~.
                \end{itemize}
                Then by Lemma \ref{lem:cps:trans} with \lref{sub-t}, we have
                \begin{itemize}
                    \item $\jdsub{\Gamma}
                        {\forall \alpha. \cps{\Sigma} \rarr ((y_{\op}:\cps{T_{\op 2}}[\rep{A/X}][\cps{v}/x]) \rarr \cps{C_{\op 1}}[\rep{A/X}][\cps{v}/x]) \rarr \cps{C_{\op 2}}[\rep{A/X}][\cps{v}/x]}
                        {\tau}$~,
                \end{itemize}
                that is,
                \[
                    \jdsub{\Gamma}{\cps{\tycomp{\Sigma}{T_{\op 2}[\rep{A/X}][v/x]}{\tyctl{y_{\op}}{C_{\op 1}[\rep{A/X}][v/x]}{C_{\op 2}[\rep{A/X}][v/x]}}}}{\tau}~.
                \]
                Now we have the conclusion with $C = \tycomp{\Sigma}{T_{\op 2}[\rep{A/X}][v/x]}{\tyctl{y_{\op}}{C_{\op 1}[\rep{A/X}][v/x]}{C_{\op 2}[\rep{A/X}][v/x]}}$~.
            \item[Case $c = (\expwith{h}{c})^C$:]
                \def\currentprefix{cpsbw:hndl}
                We have $\cps{c} = \cps{c}~\cps{C}~\cps{h^{\mathit{ops}}}~\cps{h^{\mathit{ret}}}$
                where
                \[
                    \left\{ \begin{aligned}
                        h &= \{ \expret{x_r^{T_r}} \mapsto c_r, \repi{\op_i^{\rep{X_i: \rep{B_i}}}(x_i^{T_{i1}}, k_i^{(y_i:T_{i2}) \rarr C_{i1}}) \mapsto c_i} \} \\
                        \cps{h^{\mathit{ret}}} &= \lambda x_r:\cps{T_r}. \cps{c_r} \\
                        \cps{h^{\mathit{ops}}} &=
                            \{ \repi{\op_i = \Lambda \rep{X_i: \rep{B_i}}. \lambda x_i:\cps{T_{i1}}. \lambda k_i:(y_i:\cps{T_{i2}}) \rarr \cps{C_{i1}}. \cps{c_i}} \}
                    \end{aligned} \right.
                \]
                By Lemma \ref{lem:cps:inv-c-app}, we have
                \begin{enumrm}
                    \item\llabel{ty-cpsc} $\jdty{\Gamma}{\cps{c}}{\tau'}$,
                    \item\llabel{ty-cpshops} $\jdty{\Gamma}{\cps{h^{\mathit{ops}}}}{\tau_1}$, and
                    \item\llabel{ty-cpshret} $\jdty{\Gamma}{\cps{h^{\mathit{ret}}}}{\tau_2}$
                \end{enumrm}
                for some $\tau', \tau_1$ and $\tau_2$.

                By Lemma \ref{lem:cps:inv} with \lref{ty-cpshops} and \lref{ty-cpshret}, we have
                \begin{enumrm}[resume]
                    \item\llabel{ty-cpsopc} $\bigrepi{\jdty{\Gamma}
                        {\Lambda \rep{X_i: \rep{B_i}}. \lambda x_i:\cps{T_{i1}}. \lambda k_i:(y_i:\cps{T_{i2}}) \rarr \cps{C_{i1}}. \cps{c_i}}{\tau_i}}$,
                    \item\llabel{sub-t1} $\jdsub{\Gamma}{\{\repi{\op_i: \tau_i}\}}{\tau_1}$,
                    \item\llabel{ty-cpscr} $\jdty{\Gamma, x_r:\cps{T_r}}{\cps{c_r}}{\tau_3}$, and
                    \item\llabel{sub-t2} $\jdsub{\Gamma}{(x_r:\cps{T_r}) \rarr \tau_3}{\tau_2}$~.
                \end{enumrm}
                Then, by the second half of Lemma \ref{lem:cps:inv-c-app}, we have
                \begin{enumrm}[resume]
                    \item\llabel{sub-t'} $\jdsub{\Gamma}{\tau'}
                        {\forall \alpha. \{\repi{\op_i: \tau_i}\} \rarr ((x_r:\cps{T_r}) \rarr \tau_3) \rarr \tau}$
                \end{enumrm}
                where $\alpha$ is fresh.

                By the IH of \lref{ty-cpscr}, we have
                \begin{enumrm}[resume]
                    \item\llabel{ty-cr} $\jdty{\cpsinv{\Gamma}, x_r:T_r}{c_r}{C_r}$ and
                    \item\llabel{sub-cpsCr} $\jdsub{\Gamma, x_r:\cps{T_r}}{\cps{C_r}}{\tau_3}$
                \end{enumrm}
                for some $C_r$.

                By repeatedly inverting \lref{ty-cpsopc} and by Lemma \ref{lem:cps:trans}, we have
                \begin{enumrm}[resume]
                    \item\llabel{ty-cpsci} $\bigrepi{\jdty{\Gamma, \rep{X_i: \rep{B_i}}, x_i:\cps{T_{i1}}, k_i:(y_i:\cps{T_{i2}}) \rarr \cps{C_{i1}}}
                        {\cps{c_i}}{\tau_i'}}$ and
                    \item\llabel{sub-ti} $\bigrepi{\jdsub{\Gamma}
                        {\forall \rep{X_i: \rep{B_i}}. (x_i:\cps{T_{i1}}) \rarr ((y_i:\cps{T_{i2}}) \rarr \cps{C_{i1}}) \rarr \tau_i'}{\tau_i}}$
                \end{enumrm}
                for some $\tau_i'$.
                By the IH of \lref{ty-cpsci}, we have
                \begin{enumrm}[resume]
                    \item\llabel{ty-ci} $\bigrepi{\jdty{\cpsinv{\Gamma}, \rep{X_i: \rep{B_i}}, x_i:T_{i1}, k_i:(y_i:T_{i2}) \rarr C_{i1}}
                    {c_i}{C_i}}$ and
                    \item\llabel{sub-Ci} $\bigrepi{\jdsub{\Gamma, \rep{X_i: \rep{B_i}}, x_i:\cps{T_{i1}}, k_i:(y_i:\cps{T_{i2}}) \rarr \cps{C_{i1}}}
                        {\cps{C_i}}{\tau_i'}}$
                \end{enumrm}
                for some $C_i$'s.
                By \rulename{Sc-Fun} and \rulename{Sc-PPoly} with \lref{sub-Ci}, we have
                \begin{itemize}
                    \item {\small $\bigrepi{\jdsub{\Gamma}
                        {\forall \rep{X_i: \rep{B_i}}. (x_i:\cps{T_{i1}}) \rarr ((y_i:\cps{T_{i2}}) \rarr \cps{C_{i1}}) \rarr \cps{C_i}}
                        {\forall \rep{X_i: \rep{B_i}}. (x_i:\cps{T_{i1}}) \rarr ((y_i:\cps{T_{i2}}) \rarr \cps{C_{i1}}) \rarr \tau_i'}}$~.
                    }
                \end{itemize}
                Then, by Lemma \ref{lem:cps:trans} with \lref{sub-ti}, we have
                \begin{enumrm}[resume]
                    \item\llabel{sub-ti-2} $\bigrepi{\jdsub{\Gamma}
                        {\forall \rep{X_i: \rep{B_i}}. (x_i:\cps{T_{i1}}) \rarr ((y_i:\cps{T_{i2}}) \rarr \cps{C_{i1}}) \rarr \cps{C_i}}
                        {\tau_i}}$~.
                \end{enumrm}
                Thus, by Lemma \ref{lem:cps:trans}
                and subtyping with \lref{sub-t'}, \lref{sub-cpsCr} and \lref{sub-ti-2}, we have
                \begin{itemize}
                    \item $\jdsub{\Gamma}{\tau'}
                    {\forall \alpha. \tau_s \rarr ((x_r:\cps{T_r}) \rarr \cps{C_r}) \rarr \tau}$
                \end{itemize}
                where $\tau_s \defeq \{\repi{\op_i: \forall \rep{X_i: \rep{B_i}}. (x_i:\cps{T_{i1}}) \rarr ((y_i:\cps{T_{i2}}) \rarr \cps{C_{i1}}) \rarr \cps{C_i}}\}$~.
                Here, we define $\Sigma$ to be
                $\{\repi{\op_i: \forall \rep{X_i: \rep{B_i}}. (x_i:T_{i1}) \rarr ((y_i:T_{i2}) \rarr C_{i1}) \rarr C_i}\}$,
                Then, it holds that $\tau_s = \cps{\Sigma}$.
                That is, we have
                \begin{enumrm}[resume]
                    \item\llabel{sub-t'-2} $\jdsub{\Gamma}{\tau'}
                    {\forall \alpha. \cps{\Sigma} \rarr ((x_r:\cps{T_r}) \rarr \cps{C_r}) \rarr \tau}$~.
                \end{enumrm}

                On the other hand, by the IH of \lref{ty-cpsc}, we have
                \begin{enumrm}[resume]
                    \item\llabel{ty-c} $\jdty{\cpsinv{\Gamma}}{c}{C_0}$ and
                    \item\llabel{sub-C0} $\jdsub{\Gamma}{\cps{C_0}}{\tau'}$
                \end{enumrm}
                for some $C_0$.
                By Lemma \ref{lem:assume-atm}, w.l.o.g., we can assume that
                $C_0 = \tycomp{\Sigma_0}{T_0}{\tyctl{x_0}{C_{01}}{C_{02}}}$~.
                Then, by Lemma \ref{lem:cps:trans} with \lref{sub-t'-2} and \lref{sub-C0}, we have
                \begin{itemize}
                    \item $\jdsub{\Gamma}{\forall \beta. \cps{\Sigma_0} \rarr ((x_0:\cps{T_0}) \rarr \cps{C_{01}}) \rarr C_{02}}
                    {\forall \alpha. \cps{\Sigma} \rarr ((x_r:\cps{T_r}) \rarr \cps{C_r}) \rarr \tau}$~.
                \end{itemize}
                Then, by inversion, we have
                \begin{itemize}
                    \item $x_0 = x_r$,
                    \item $\jdsub{\Gamma, \alpha}{\cps{\Sigma}}{\cps{\Sigma_0}}$,
                    \item $\jdsub{\Gamma, \alpha, h:\cps{\Sigma}}{\cps{T_0}}{\cps{T_r}}$,
                    \item $\jdsub{\Gamma, \alpha, h:\cps{\Sigma}, x_r:\cps{T_0}}{\cps{C_r}}{\cps{C_{01}}}$,
                \end{itemize}
                and
                \begin{enumrm}[resume]
                    \item\llabel{sub-t} $\jdsub{\Gamma, \alpha, h:\cps{\Sigma}, k:(x_r:\cps{T_0}) \rarr \cps{C_{01}}}{\cps{C_{02}}}{\tau}$~.
                \end{enumrm}
                By Lemma \ref{lem:cps:rm-nonrfn}, we have
                \begin{itemize}
                    \item $\jdsub{\Gamma, \alpha}{\cps{\Sigma}}{\cps{\Sigma_0}}$,
                    \item $\jdsub{\Gamma, \alpha}{\cps{T_0}}{\cps{T_r}}$, and
                    \item $\jdsub{\Gamma, \alpha, x_r:\cps{T_0}}{\cps{C_r}}{\cps{C_{01}}}$~.
                \end{itemize}
                Then, by \ref{lem:cps:cpsinv-sub}, we have
                \begin{itemize}
                    \item $\jdsub{\cpsinv{\Gamma}}{\Sigma}{\Sigma_0}$,
                    \item $\jdsub{\cpsinv{\Gamma}}{T_0}{T_r}$, and
                    \item $\jdsub{\cpsinv{\Gamma}, x_r:T_0}{C_r}{C_{01}}$~.
                \end{itemize}
                Therefore, by subsumption on \lref{ty-c}, we have
                \begin{enumrm}[resume]
                    \item\llabel{ty-c-2} $\jdty{\cpsinv{\Gamma}}{c}{\tycomp{\Sigma}{T_r}{\tyctl{x_r}{C_r}{C_{02}}}}$~.
                \end{enumrm}
                Thus, by \rulename{T-Hndl} with \lref{ty-cr}, \lref{ty-ci} and \lref{ty-c-2}, we have
                \[
                    \jdty{\cpsinv{\Gamma}}{\expwith{h}{c}}{C_{02}}~.
                \]
                Also, by Lemma \ref{lem:cps:rm-nonrfn} and \ref{lem:cps:rm-unused-tvar} with \lref{sub-t}, we have
                \[
                    \jdsub{\Gamma}{\cps{C_{02}}}{\tau}~.
                \]
                Now we have the conclusion with $C = C_{02}$.
        \end{description}
    \end{enumit}
\end{proof}

\begin{corollary}[Backward type preservation (for closed expressions)] \quad
    \begin{itemize}
     \item If\, $\jdty{\emptyset}{\cps{v}}{\tau}$, then
           there exists some $T$ such that
           $\jdty{\emptyset}{v}{T}$ and
           $\jdsub{\emptyset}{\cps{T}}{\tau}$.
     \item If\, $\jdty{\emptyset}{\cps{c}}{\tau}$, then
           there exists some $C$ such that
           $\jdty{\emptyset}{c}{C}$ and
           $\jdsub{\emptyset}{\cps{C}}{\tau}$.
    \end{itemize}
\end{corollary}
\begin{proof}
    Immediate from Theorem \ref{thm:cps-backward}
    since $\emptyset$ is obviously cps-wellformed.
\end{proof}

\bibliographystyle{unsrtnat}
\bibliography{main}